\RequirePackage{fix-cm}
\documentclass[twocolumn]{svjour3}          % twocolumn
\smartqed  % flush right qed marks, e.g. at end of proof
\usepackage{amsmath}
\usepackage{graphicx}
\usepackage[T1]{fontenc}  
\usepackage[utf8,latin1]{inputenc} % % Commented out by DS - should be in std latex distrib
\usepackage{epsfig}
\usepackage{txfonts}
\usepackage{amssymb}
\usepackage{siunitx}
\usepackage{natbib}        
\usepackage{lipsum}
\usepackage{hyperref}
\usepackage{threeparttable,tablefootnote}

\usepackage{url}
\usepackage{color}
\usepackage{bm}
\usepackage{sgl-commands}

\newcommand{\vbeta}{\mbox{\boldmath$\beta$}}
\newcommand{\vtheta}{\mbox{\boldmath$\theta$}}
\newcommand{\valpha}{\mbox{\boldmath$\alpha$}}

\newcommand{\half}{\mbox{$\frac12$}}

\newcommand*\diff{\mathop{}\!\mathrm{d}}
\newcommand{\rhalf}{\mbox{$r_{1/2}$}}
\newcommand{\rg}{\mbox{$r_{\rm g}$}}

\newcommand{\chapintro}{Chapter 1} %{Saha et al. (in preparation)}
\newcommand{\chapgal}{Chapter 2} %{Shajib et al. (in preparation)}
\newcommand{\chapdm}{Chapter 6} %{Vegetti et al. (in preparation)}
\newcommand{\chaptimedelays}{Chapter 9} %{Birrer et al. (in preparation)}
\newcommand{\chapfinding}{Chapter 2} %{Lemon et al. (in preparation)}
\newcommand{\chapsn}{Chapter 5} %{Suyu et al. (in preparation)}

\makeatletter
\newcommand{\ion}[2]{%
\relax\ifmmode
\ifx\testbx\f@series
{\mathbf{#1\,\mathsc{#2}}}\else
{\mathrm{#1\,\mathsc{#2}}}\fi
\else\textup{#1\,{\mdseries\textsc{#2}}}%
\fi}

\makeatother

\journalname{Space Science Reviews}
\begin{document}

\title{Microlensing of strongly lensed quasars}

\titlerunning{Strong and microlensing of quasars}        % if too long for running head

\author{G.~Vernardos$^{1,2,3}$ \and D.~Sluse$^{4}$ \and D.~Pooley$^{5}$ \and R.~W.~Schmidt$^{6}$ \and M.~Millon$^{1,7}$ \and L.~Weisenbach$^{8}$ \and V.~Motta$^{9}$ \and T.~Anguita$^{10,11}$ \and P.~Saha$^{12}$ \and M.~O'Dowd$^{2,3,13}$ \and A.~Peel$^{1}$ \and P.~L.~Schechter$^{14,15}$
}

\authorrunning{Vernardos, Sluse, Pooley, et al.} % if too long for running head

\institute{
$^1$Institute of Physics, Laboratory of Astrophysics, Ecole Polytechnique F\'ed\'erale de Lausanne (EPFL), Observatoire de Sauverny, 1290 Versoix, Switzerland
\\
$^{2}$Department of Astrophysics, American Museum of Natural History, Central Park West and 79th Street, NY 10024, USA
\\
$^{3}$Department of Physics and Astronomy, Lehman College of the City University of New York, Bronx, NY 10468, USA
\\
$^4$STAR Institute, Quartier Agora - All\'ee du six Aout 19c, B-4000 Li\'ege, Belgium
\\
$^5$Department of Physics and Astronomy, Trinity University, San Antonio, TX 78212, USA
\\
$^6$Astronomisches Rechen-Institut, Zentrum f\"ur Astronomie der Universit\"at Heidelberg, M\"onchhofstr. 12-14, D-69120 Heidelberg, Germany
\\
$^7$Kavli Institute for Particle Astrophysics and Cosmology and Department of Physics, Stanford University, CA 94305 Stanford, USA
\\
$^8$Institute of Cosmology and Gravitation, University of Portsmouth, Dennis Sciama Building, Burnaby Road, Portsmouth, PO1 3FX, UK
\\
$^9$Instituto de F\'{\i}sica y Astronom\'{\i}a, Facultad de Ciencias, Universidad de Valpara\'{\i}so, Avda. Gran Breta\~na 1111, Valpara\'{\i}so, Chile
\\
$^{10}$Instituto de Astrof\'{i}sica, Facultad de Ciencias Exactas, Universidad Andres Bello, Av. Fernandez Concha 700, Las Condes, Santiago, Chile
\\
$^{11}$Millennium Institute of Astrophysics, Nuncio Monse\~nor S\'otero Sanz 100, Providencia, Santiago, Chile
\\
$^{12}$Physik-Institut, University of Zurich, Winterthurerstrasse 190, 8057 Zurich, Switzerland
\\
$^{13}$The Graduate Center of the City University of New York, 365 Fifth Avenue, New York, NY 10016, USA
\\
$^{14}$Massachusetts Institute of Technology, Department of Physics, Cambridge, MA 02139 USA
\\
$^{15}$Massachusetts Institute of Technology, Kavli Institute for Astrophysics and Space Research, Cambridge, MA 02139 USA
}

%\email{veronica.motta@uv.cl}
%\email{millon@stanford.edu}
%\email{tanguita@gmail.com}
%\email{tanguita@gmail.com}
%\email{weisluke@alum.mit.edu}
%\email{psaha@physik.uzh.ch}    

\date{Received: date / Accepted: date}

%\input{xxx}

\maketitle

\begin{abstract}
Strong gravitational lensing of quasars has the potential to unlock the poorly understood physics of these fascinating objects, as well as serve as a probe of the lensing mass distribution and of cosmological parameters.
In particular, gravitational microlensing by compact bodies in the lensing galaxy can enable mapping of quasar structure to $<10^{-6}$ arcsec scales.
Some of this potential has been realized over the past few decades, however the upcoming era of large sky surveys promises to bring this promise to full fruition.
In this chapter, we review the theoretical framework of this field, describe the prominent current methods for parameter inference from quasar microlensing data across different observing modalities, and discuss the constraints so far derived on the geometry and physics of quasar inner structure.
We also review the application of strong lensing and microlensing to constraining the granularity of the lens potential, i.e. the contribution of the baryonic and dark matter components, and the local mass distribution in the lens, i.e. the stellar mass function.
Finally, we discuss the future of the field, including the new possibilities that will be opened by the next generation of large surveys and by new analysis methods now being developed.
\end{abstract}

\label{chapter_qso}

%%%%%%%%%%%%%%%%%%%%%%%%%%%%%%%%%%%%%%%%%%%%%%%%%%%%%%%%%%%%%%%%%%%%%%%%%%%%%%%%%%%%%%%%%%%%%%%%%%%%%%%%%%%%%%%%%%%%%%%%%%%%%%%%%%%%%%%%%%%%%%%%%%%%%%%%%%%%%%%%%%%%%%%%%%%%%%%%%%%%%%%%%%%%%%%%%%%%%%%%%%%%%%%%%%%%%%%%
\section{Introduction}
\label{sec5:intro}

\begin{figure*}
    \includegraphics[width=\textwidth]{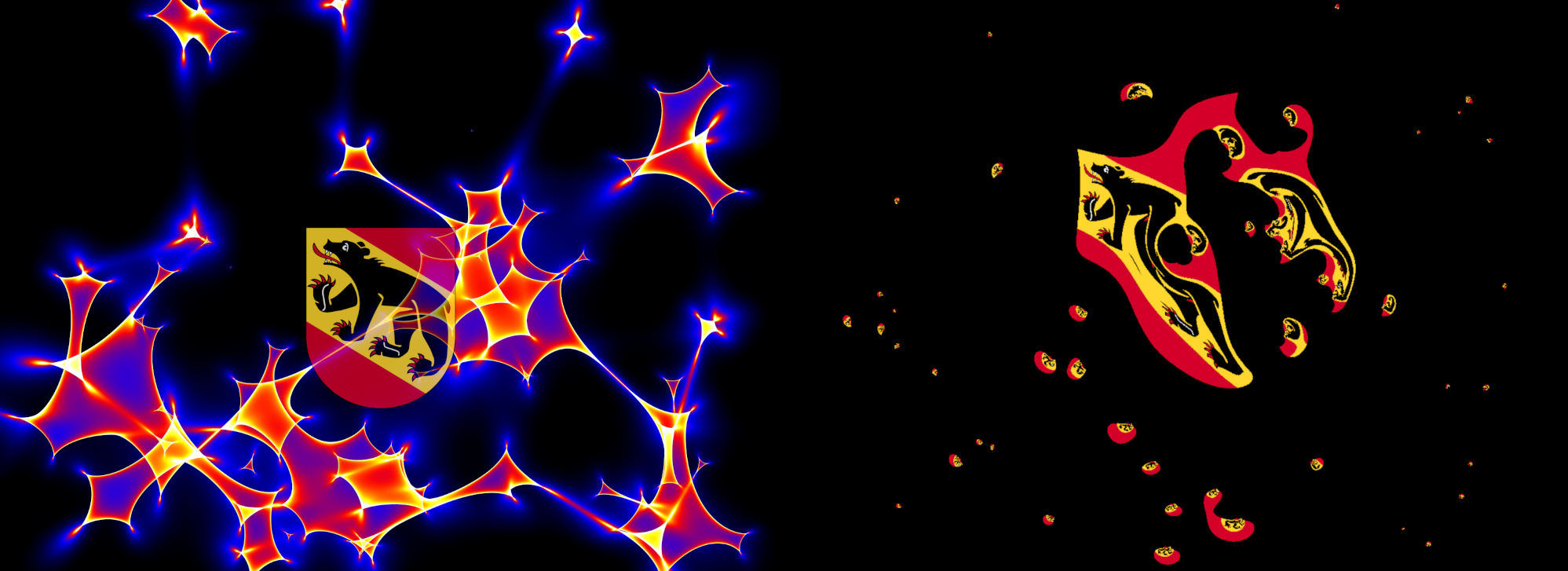}
    \caption{Illustration of the effect of microlensing on a source, in this case, the coat-of-arms of Bern, Switzerland, the seat of the International Space Science Institute, where the ``Strong Gravitational Lensing'' workshop was held between 18-22 of July 2022 that lead to the writing of this review. \textbf{Right:} A collection of microlenses on the image plane (the same as the one shown in \figref{fig5:microimages_loops}) produces magnified and distorted micro-images of the source. \textbf{Left:} The corresponding magnification pattern is shown on the source plane (the same as the one shown in \figref{fig5:example_mag_map}). For an interactive version of this figure visit: \href{https://austinpeel.github.io/gladius-microlensing/}{https://austinpeel.github.io/gladius-microlensing/}}
\label{fig5:micro_bern}
\end{figure*}

Microlensing is a phenomenon that allows us to probe very small spatial and mass scales in the Universe that are otherwise inaccessible by current and next-generation instruments.
The underlying physical principles of gravitational light deflection are the same as in the case of strong lensing by galaxies and clusters, however, the main difference is that these deflections happen in such small angular scales ($\sim10^{-6}$ arcsec) that the resulting multiple images of a source are unresolved.
The only remaining quantity with an observable effect is the total magnification, i.e. the total flux received from all the unresolved microimages (the centroid of the images can be affected too, but this requires higher sensitivity to detect).
A demonstration of the phenomenon is shown in \figref{fig5:micro_bern}.
Given our solid theoretical understanding of microlensing, we can use it as a tool that is, in many ways, more powerful than any telescope we could build in the next several decades.
With this tool, we can learn both about the background source and the lens itself (the massive galaxy along the line of sight).
In this case, the type of sources that we examine are quasars, although microlensing can also affect lensed supernovae (see \chapsn) and other more 'exotic' sources like Fast Radio Bursts \citep{bib5:Lewis2020}, Gamma-Ray Bursts \citep{bib5:Mao1993}, and gravitational waves \citep{bib5:Diego2019}.
In this review, we present the state-of-the-art of the field of quasar microlensing and expand on previous works by \citet{bib5:Schneider1992}, \citet{bib5:Schneider2006}, and \citet{bib5:Schmidt2010}.

\begin{figure*}
    \includegraphics[width=\textwidth]{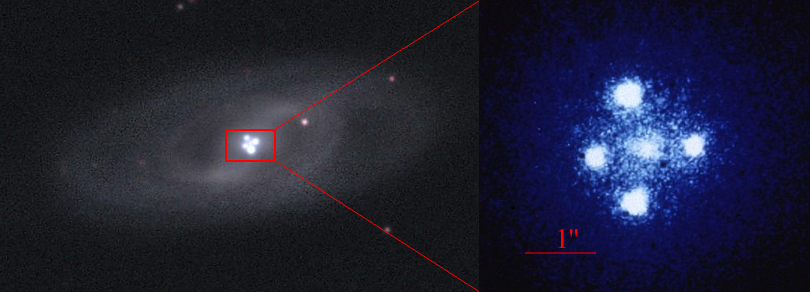}
    \caption{Multiply imaged quasar Q 2237+0305. \textbf{Left:} the lensed quasar seen through the central regions of a foreground spiral galaxy-lens (credits: J. Rhoads, S. Malhotra, I. Dell'Antonio, NOAO/WIYN/NSF). \textbf{Right:} zoom on the bulge of the galaxy-lens seen in the centre, surrounded by multiple images of the background quasar produced by gravitational lensing (credits: NASA, ESA, and STScI).}
\label{fig5:einstein_cross}
\end{figure*}

The first discovery of a gravitational lens, the doubly imaged quasar Q 0957+561 \citep{bib5:Walsh1979}, was followed by the pioneering works of \citet{bib5:Chang1979}, \citet{bib5:Gott1981}, \citet{bib5:Young1981}, and \citet{bib5:Chang1984} who suggested that light from the multiple images could be further affected by the presence of stellar mass objects near the line of sight.
Indeed, microlensing was confirmed a decade later in the quadruply lensed Q 2237+0305 \citep{bib5:Irwin1989}, also known as Huchra's lens \citep{bib5:Huchra1985} or ``The Einstein Cross'', shown in \figref{fig5:einstein_cross}.
These, and the studies of \citet{bib5:Paczynski1986} and \citet{bib5:Kayser1986}, set the foundations of the field of quasar microlensing.

\begin{figure*}
    \centering
    \includegraphics[width=0.95\textwidth]{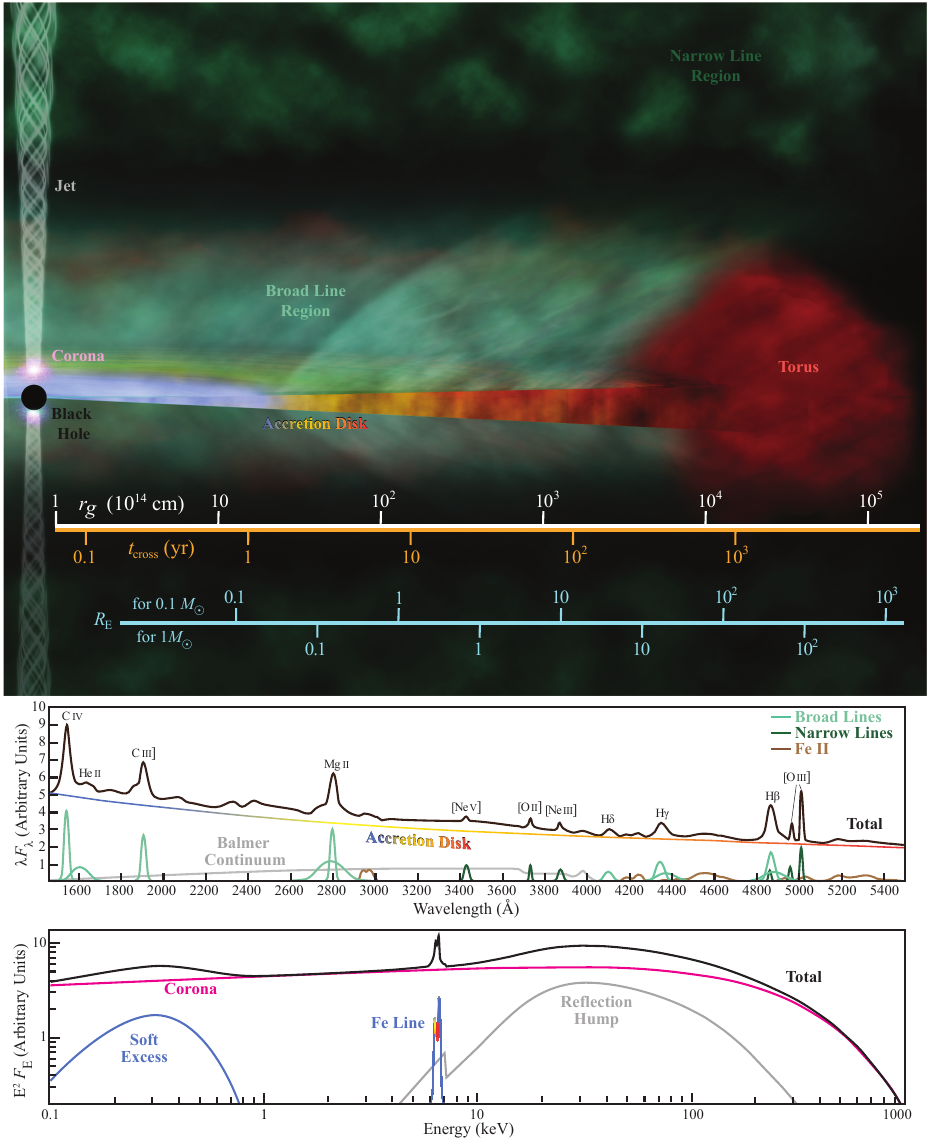}
    \caption{
    \textbf{Top:} Schematic view of the structure of a quasar with the main regions that can be probed by microlensing color-coded as follows, starting from the smallest/innermost (left to right): X-ray corona (purple), accretion disc (blue to red), Broad Line Region (lighter green), dust torus (red), and Narrow Line Emission (darker green). For completeness, the black hole and the jet are also shown. Different characteristic lengths are shown in logarithmic scale in the bottom of this panel: the gravitational radius, $\rg$, for a $7 \times 10^{8}$ M$_{\odot}$ black hole (see \eqref{eq5:gravitational_radius}), the corresponding time it takes to cross this length, $t_{\rm cross}$, for an effective velocity of 500 km/s (see \secref{sec5:velocity_model}), and the Einstein radius on the source plane, $\re$, for lens and quasar located at redshifts of 0.5 and 2 respectively (see \eqref{eq5:einstein_radius_source}). The latter is given for two different masses of the microlenses. Inspired from fig. 1 of \cite{bib5:Moustakas2019}.
    \textbf{Middle:} Composite UV-optical spectrum of an AGN with a few main emission lines identified. Its different components have been color-coded to match the corresponding regions shown in the top panel. \textbf{Bottom:} Same as the middle panel but for the X-rays.}
    \label{fig5:sketch}
\end{figure*}

Although throughout this review we refer to the ``quasar'' as if it was a single, well-defined object, a quasar, or Active Galactic Nucleus (AGN, we use both terms indistinctly), is a composite of several different structures around a Super-Massive Black Hole (SMBH).
Each of these regions produces its own signature radiation and together they sum up to give an astrophysical object that is bright across the whole electromagnetic spectrum.
Through decades of dedicated work \citep{bib5:Netzer2015,bib5:Padovani2017}, a general picture of the various quasar regions has been inferred.
However, given the small physical scales around the black hole ($\sim$ ld) and the large distances to these objects ($z \gtrsim 0.5$), the dimensions and exact geometry of these regions remain largely unresolved by any current or planned telescope.
With microlensing, we can {\it measure} and place direct constraints on the sizes and locations of the emitting regions of a quasar at a cosmological distance on scales from micro- to nano-arcsec.
In \figref{fig5:sketch} we summarize our current understanding of quasar structure and highlight the microlensing effect on its different components.

What we can learn about the composition of galaxies from microlensing is as important as it is inaccessible by any other means.
While every astronomer knows that a galaxy is ``a gravitationally bound system of stars, stellar remnants, interstellar gas, dust, and dark matter'', establishing the relative proportions of each component is one of the great challenges of extragalactic astronomy.
In particular, the contribution of either dark matter and stellar mass is subject to much disagreement and uncertainty (see section 4 in \chapgal).
For example, there are important works that measure stellar mass-to-light ratios for elliptical galaxies \citep[see][and references therein]{bib5:Cappellari2016}, but buried within these papers the authors invariably caution that uncertainties in the faint end of the stellar mass function, where the stars are practically invisible, render their results uncertain by a factor of two.
Because microlensing is sensitive only to mass, it can be used to determine the amount of mass in individual stars, including stellar remnants, brown dwarfs, and red dwarfs that are too faint to produce any photometric or spectroscopic signatures.
In other words, microlensing is an independent and direct method to determine the graininess of the gravitational potential.

In order to prepare the unfamiliar reader with the data and methods used in microlensing studies, in the remaining of this section we summarise some general theoretical and observational properties. 
In \secref{sec5:background_theory} we set the foundations of the theory of microlensing as well as our best current theoretical understanding of the structure of quasar emission regions.
Section \ref{sec5:data_analysis} presents the different approaches to analyze how microlensing manifests across time and wavelength.
The application of these methods to real data leads to measurements of quasar structure and lens galaxy mass, which are reviewed in Sects. \ref{sec5:quasar_results} and \ref{sec5:lensing_galaxy_results} respectively.
We conclude in \secref{sec5:future_prospects} with the promising future of quasar microlensing, which will be revolutionized by the upcoming all-sky surveys of this decade, and the challenges that need to be addressed in order to deliver groundbreaking scientific results.

\subsection{General theoretical properties and facts}
Microlensing is caused by compact astrophysical objects with dimensions much smaller than their Einstein radii\footnote{Here we assume that the reader is familiar with basic concepts of gravitational lensing, like the Einstein radius, critical lines, and caustics - see \chapintro~for an overview.}, $\tae$ (i.e. stars and stellar remnants), as opposed to smoothly varying mass distributions over large scales (e.g. galaxies and clusters).
In fact, this radius, which is typically of the order of $10^{-6}$ arcsec for lensed quasars (hence the prefix ``micro'', see also \citealt{bib5:Kayser92}), is critical because it determines the scale length of the phenomenon.
The rule to remember is that the smaller (larger) a quasar emitting region with respect to $\tae$, the stronger (weaker) the amplitude of microlensing.

The optical depth to microlensing is dramatically increased along the line of sight to a quasar that is already strongly lensed by a foreground galaxy (see \figref{fig5:einstein_cross}).
Hence, we almost always expect a population of compact objects acting as microlenses and in very few cases single, isolated point masses.
Such populations create caustic features of very high magnification, often superimposed in a non-linear fashion, but because flux has to be conserved (e.g. compared to a smooth mass sheet with the same total mass) there need to be large swathes of de-magnified regions on the source plane as well.
Depending on where the source lies within this caustic pattern it can be magnified or de-magnified.

Due to the relative velocity of observer, lens, microlenses, and quasar, which is of the order of 1 $\tae$ per decade \citep{bib5:Mosquera2011b}, we expect to see significant variations ($\sim$1 mag) of the microlensing magnification as the quasar emitting regions cross caustics and areas of demagnification.
Moreover, because the quasar region(s) emitting at any given wavelength can vary in size, there is a chromatic dependence of this (de)magnification.
It is important to note that quasars are intrinsically variable objects, therefore, in order to obtain the microlensing signal (in a single epoch or as a function of time) one needs to cleanly subtract this intrinsic variability after taking into account the time delays per multiple image due to the presence of the lensing galaxy (see \chaptimedelays).

\subsection{Observational considerations}
From an observer's point of view, one may wonder which kind of data are needed in order to use microlensing for one of the above mentioned science applications.
The size of the source with respect to $\tae$, together with the dynamic nature of the phenomenon may guide the answer to this question.
The wavelength range to consider covers almost the whole electromagnetic spectrum, from the innermost X-ray corona and accretion disc to the Broad Line Region (BLR, see \figref{fig5:sketch}), while the time scales when large variations are expected are in general shorter for the most compact regions.
When the quasar is crossing a caustic, the magnification increases rapidly with time and therefore nightly cadence is required to capture in detail how the different quasar regions respond.
Under a ``normally'' microlensed quasar, i.e. not undergoing a high magnification event, the daily cadence requirement can be relaxed and weekly observations should suffice for long-term microlensing signals. 
In practice, the existing observational resources are quite restricted both in wavelength coverage and cadence (e.g. season gaps).

Over the years, widely different observations have been performed to capture the various manifestations of microlensing variability: from decade-long monitoring to multiwavelength snapshots and spectra.
The limited resources have resulted in two main strategies, each with its own advantages and drawbacks: either long-term weekly monitoring but, with very few exceptions, in a single band, or multi-wavelength snapshots at a given moment in time.
For example, light curves provide more data points to constrain quasar structure and lens mass but require an additional model component (the effective velocity) and a good measurement of the time delay, while snapshots can constrain the accretion disc temperature profile but are prone to other systematic biases and limitations, like contamination from massive substructures (e.g. anomalous flux ratios, see \chapdm) and application to image pairs with negligible, or known, time delays.
Although very information-rich on quasar structure, data for high magnification events have so far been scarce due to their rarity and quick evolution.
Requiring almost daily cadence, these have been almost exclusively observed in a single system (the Einstein cross).

%%%%%%%%%%%%%%%%%%%%%%%%%%%%%%%%%%%%%%%%%%%%%%%%%%%%%%%%%%%%%%%%%%%%%%%%%%%%%%%%%%%%%%%%%%%%%%%%%%%%%%%%%%%%%%%%%%%%%%%%%%%%%%%%%%%%%%%%%%%%%%%%%%%%%%%%%%%%%%%%%%%%%%%%%%%%%%%%%%%%%%%%%%%%%%%%%%%%%%%%%%%%%%%%%%%%%%%%
\section{Background and theory}
\label{sec5:background_theory}

In this section, we present the theoretical background that is specific to microlensing, as opposed to strong lensing in general.
For an overview of the basic principles and theory of lensing we refer the reader to \chapintro.

\subsection{Light deflection}
\label{sec5:theoretical_description_microlensing}
To study the lensing effect of an ensemble of
compact objects in the lens plane on a background quasar, we
start again by setting up the lens equation
(\chapintro), but this
time for many objects:
\begin{equation}
\bang  =  \tang - 
\frac{\dds}{\dd \ds} \sum_{i=1}^{N} \hat{\boldsymbol{\alpha}}_i (\tang),
\label{eq5:lenseq}
\end{equation}
where we are summing over the deflection angles $\hat{\boldsymbol{\alpha}}_i$ due to $N$ compact objects with masses $M_i$
situated at $\tang_i$ in the lens plane.

In addition to the surface mass density at the quasar image position due to compact objects within a radius $\theta$:
\begin{equation}
    \kappastar=\Sigma_\ast (<\theta)/\Sigmacr,
\end{equation}
with
$\Sigmacr=\frac{c^2}{4 \pi G} \frac{\ds}{\dds \dd}$
(\chapintro),
we need to include the effect of a 
constant surface density of matter $\kappac$
in the lens plane to account for a smooth dark matter component (also expressed in units of $\Sigmacr$).
The total surface mass density is then:
\begin{equation}
\label{eq5:total_kappa}
    \kappa = \kappastar + \kappac .
\end{equation}
Finally, we need to consider the effect of shear due to
the tidal field of the lensing galaxy at the position of the quasar image
(\chapintro). Using Einstein's formula for the deflection
(\chapintro) the lens equation due to 
$N$ compact masses $M_i$ in the presence of a
mass sheet $\kappac$ and shear $\gamma=\sqrt{\gamma_1^2+\gamma_2^2}$ becomes:
\begin{equation} 
\bang = \left( \begin{array}{cc} 1 - \kappac -\gamma_1 & -\gamma_2 \\ -\gamma_2 & 1- \kappac + \gamma_1 \end{array} \right) \tang  + \frac{4 G}{c^2} \frac{\dds}{\dd \ds} \sum_{i=1}^{N} M_i \frac{\tang_i - \tang}{|\tang_i - \tang|^2}.
\label{eq5:microlensing_equation}
\end{equation}
For a homogeneous
mass distribution with surface density $\kappastar$, the second term becomes
$\propto -\kappastar \tang$.
This equation can also be written as:
\begin{equation} 
\bang = 
\left( \begin{array}{cc} 1 - \kappac -\gamma_1 & -\gamma_2\\
-\gamma_2 & 1- \kappac + \gamma_1
\end{array}
\right) \tang +
\sum_{i=1}^{N} {\tae}_i^2 \frac{ \tang_i - \tang}{|\tang_i - \tang|^2} ,
\label{eq5:microlensing_equation_Einstein_radius}
\end{equation}
where we used the Einstein radii ${\tae}_i$ of the compact objects in the lens plane
(\chapintro).
This is a fundamental quantity for microlensing as it defines the scale length of the phenomenon.
Thus, we define it here again in the image plane, as in the equation above:
\begin{equation}
    \tae= \sqrt{ \frac{4GM}{c^2} \frac{D_{ds}}{D_{d}D_{s}}}
    \label{eq5:einstein_radius_image}
\end{equation}
and its projection on the source plane:
\begin{equation}
    \re= \sqrt{ \frac{4GM}{c^2} \frac{D_{s}D_{ds}}{D_{d}}},
    \label{eq5:einstein_radius_source}
\end{equation}
in units of length.

How can we distribute $N$ masses such that a certain $\kappastar$ is achieved?
From the homogeneous mass case
$\bang\propto - \kappastar \tang$ (\eqref{eq5:microlensing_equation})
and $\bang \propto - ({\sum_{i=1}^N {\tae}_i^2}/{\theta^2})~\tang$ in \eqref{eq5:microlensing_equation_Einstein_radius}, it can be derived that
the normalized surface density in compact objects $\kappastar$ equals
the summed ``Einstein circles'' divided by the area in which the point masses are distributed:
\begin{equation}
\label{eq5:stellar_kappa}
    \kappastar = \frac{\pi {\tae}^2 }{\pi \theta^2} \sum_{i=1}^N m_i 
    = \frac{\pi \sum_{i=1}^{N} m_i}{A_{enc}}
\end{equation}
with $m_i = M_i/\msol$. In the last term, the area
$A_{enc}$, in which the $N$ point masses are distributed, has been normalized by the Einstein radius for one solar mass $\tae^2$ squared.
To create a mass ensemble of compact objects with surface density $\kappastar$, one needs to distribute
point masses with summed masses $m_i$:
\begin{equation}
    \sum_{i=1}^N m_i = \frac{\kappastar A_{enc}}{\pi}
\end{equation}
in the area $A_{enc}$.

It is customary to also rewrite the microlensing mapping from
\eqref{eq5:microlensing_equation_Einstein_radius}
in a normalized way by dividing by $(1-\kappac)/\sqrt{|1-\kappac}|$ (for $\kappac\neq1$).
Then, one can define the normalized lens-plane and source-plane coordinates:
\begin{equation}
\boldsymbol{z} = \frac{\tang}{\tae / \sqrt{|1-\kappac|} } \quad \mathrm{and} \quad \boldsymbol{\zeta} = \frac{1}{1-\kappac}~ \frac{\bang}{\tae/\sqrt{|1-\kappac|}}.
\label{eq5:unit_source_plane}
\end{equation}
The normalized lens equation follows as \citep{bib5:Paczynski1986,bib5:Kayser1986}:
\begin{equation} 
\boldsymbol{\zeta} = 
\left( \begin{array}{cc} 1 - g_1 & -g_2\\
-g_2 & 1 + g_1
\end{array}
\right) \boldsymbol{z} + sign\left(\frac{\kappastar}{1-\kappac}\right)
\sum_{i=1}^{N} m_i \frac{ {\boldsymbol{z}}_i - \boldsymbol{z}}{|\boldsymbol{z}_i - \boldsymbol{z}|^2} ,
\label{eq5:normalized_lens_equation}
\end{equation}
where $g_1$ and $g_2$ are called the components
of the reduced shear (see \chapintro).
Employing the signum function is due to \citet{bib5:Paczynski1986},
who used it because it shows that only the normalized surface mass density $\kappastar/(1-\kappac)$ and the reduced shear $g$
are needed to describe a quasar microlensing situation 
(note that $sign(\kappastar)>0$ always).
We note that these two quantities, normalized surface mass density and reduced shear, are also known as ``effective'' convergence and shear and are further discussed in \secref{sec5:caustic_structures_macro_parameters}.

Often in the literature, the compact-object surface density $\kappastar$
and the smooth surface density $\kappac$ are both given in addition to the shear $\gamma$.
It should be noted, however, that the above described degeneracy between the three quantities
holds (e.g. \citealt{bib5:Saha2000}).
In the case of $\kappac>1$ (sometimes called over-focussing) the deflection due
to the individual masses is
formally counted as repulsive. However, as Paczynski put it this is just a
``trick'' to make equation \eqref{eq5:normalized_lens_equation} simpler.
In the following we shall assume $\kappac<1$ for simplicity because this is the most
common scenario for observations (but see also \citealt{bib5:Dobler2007}).

\subsection{Magnification for ensembles of point masses}
\label{sec5:magnification}
The magnification of a microlensed image is denoted by the
symbol $\mu$. An image can be magnified, $|\mu|>1$, or
demagnified, $|\mu|<1$.
In addition the image can be mirror-inverted, which
corresponds to negative magnification $\mu <0$.
The sign of the magnification $\mu$ is also called parity (see also \chapintro).

For the lens mapping described by the lens equation introduced above, the magnification of a point source
is given by the inverse of the Jacobi determinant (see \chapintro):
\begin{equation}
    A = {\rm det}~J = {\rm det}~\left( \frac{\partial \zeta}{\partial \boldsymbol{z}} \right).
    \label{eq5:microlensing_determinant}
\end{equation}
The total magnification can be found by taking the sum:
\begin{equation}
    \mu_{\rm tot} = \sum_{j=1}^{M}
    \frac{1}{|A|_{\boldsymbol{z}=\boldsymbol{z}_j}},
    \label{eq5:magnification_sum}
\end{equation}
over all $M$ images at locations $\boldsymbol{z}_j$, or $\boldsymbol{\theta}_j$, on the lens plane.
Calculating the magnification for extended sources, such as a quasar
accretion disc, can proceed by splitting up the source into sub-sources
\cite[e.g.][]{bib5:Witt1994}.
However, in \secref{sec5:microlensing_maps} it is shown how to calculate the magnification
using the ray-shooting method. This latter technique is the most widely used approach.

As an example, the magnification $\mu$ for a constant sheet of matter $\kappac>0$ with shear $\gamma$ is given by (see also \chapintro):
\begin{equation}
\label{eq5:macro_magnification}
    \mu = \frac{1}{(1-\kappac)^2-\gamma^2}.
\end{equation}
We note that this is often referred to as the ``macro-magnification'' because $\kappac$ and $\gamma$ can be attributed to the lensing galaxy's macroscopic potential at the location of the multiple images.
It can be seen from this equation that for positive parity images the magnification is always $|\mu|>1$.
For negative parity the image can be magnified, $|\mu|>1$, or demagnified, $|\mu|<1$.

\subsection{Complex notation for microlensing} 
\label{sec5:complex_notation}
The normalized lens equation \eqref{eq5:normalized_lens_equation} describes
a mapping from the lens plane $\boldsymbol{z}$ to the source plane $\boldsymbol{\zeta}$.
Starting with \citet{bib5:Bourassa1973,bib5:Bourassa1975}, complex numbers
were used to describe this mapping. While they partly treated real and imaginary parts
separately, \citet{bib5:Witt1990} introduced a consequent complex notation.

Normalized positions in the lens plane are represented by $z=x + i y$ and in the source plane by $\zeta=\xi+ i \eta$.
The shear can be described as a complex quantity a well, $g=g_1+i g_2 = |g| \exp^{2 i \varphi}$.
Since for complex numbers $z/|z|^2=1/\bar{z}$ (the bar indicates the complex conjugate), the normalized
lens equation \eqref{eq5:normalized_lens_equation} can be re-written as:
\begin{equation}
    \zeta = z - g \bar{z} + 
    \sum_{i=1}^{N} \frac{m_i}{\bar{z}_i- \bar{z}}.
\label{eq5:complex_lens_equation}
\end{equation}
The complex formulation has the advantage that it can be well treated by
a computer language that can deal with complex numbers (e.g. Fortran, C, python).
For example, the magnification of an image at $z$ can be calculated using the inverse of the
complex version of \eqref{eq5:microlensing_determinant} \citep{bib5:Witt1990}:
\begin{eqnarray}
    {\rm det}~J &= &\left(\frac{\partial \zeta}{\partial z} \right)^2
             - \frac{\partial \zeta}{\partial \bar{z}}
                        \overline{\frac{\partial \zeta}{\partial \bar{z}}} \nonumber\\
    & &         = 1 - \left(-g + \sum_{i=1}^{N} \frac{m_i}{(\bar{z}_i - \bar{z})^2}\right)
                      \left(-\bar{g} + \sum_{i=1}^{N} \frac{m_i}{({z}_i - {z})^2}\right).
    \label{eq5:complex_jacobi_determinant}
\end{eqnarray}
In the last step, the corresponding derivatives of
\eqref{eq5:complex_lens_equation} were calculated with respect to $z$, $\bar{z}$ rather than the more usual pair
$x$, $y$ (i.e. the complex plane) are used. They have the same properties as ``normal'' derivatives (i.e. linearity, product rule, chain rule, conjugation), but can be used 
more efficiently with the complex lens equation.

\begin{figure*}
    \centering
    \includegraphics[width=0.99\textwidth]{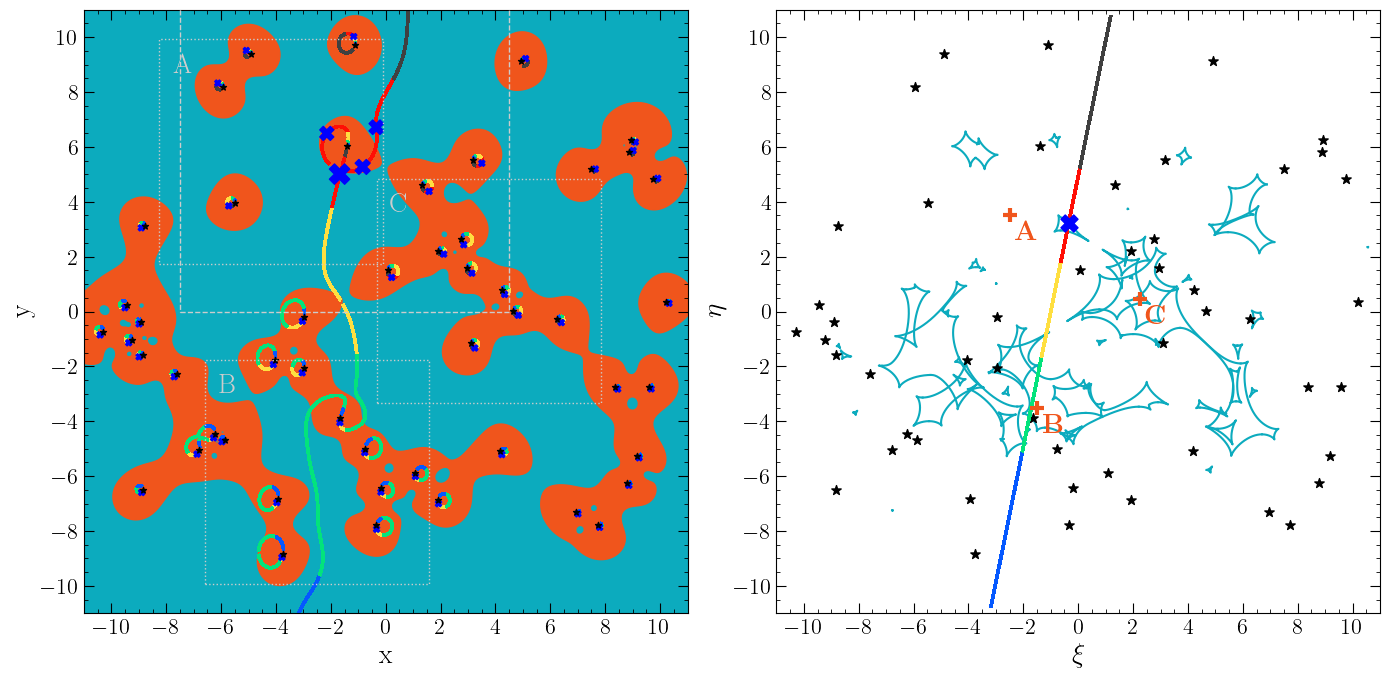}
    \caption{
    \textbf{Left:} Example images of a straight line lensed by an ensemble of 50 point masses (star symbols) distributed randomly within a circle of radius $11.2$ $\tae$ without shear. \textbf{Right:} corresponding source plane and caustic curves.
    Each image of the line can consist of multiply imaged parts, as indicated by the colored segments.
    The main image line passes through 2 stars.
    The image plane is covered by regions of negative (orange) and positive (blue) parity separated by the critical lines.
    There is at least one image close to every star, but those images with negative parity (micro-saddles) are usually highly demagnified and can get arbitrarily close to the point mass positions when the source is far away.
    The small gaps in the image lines are numerical artifacts near the critical lines (see also \secref{sec5:caustics_of_the_deflectors}).
    The micro-images for the particular position $\boldsymbol{\zeta}=(-0.35, 3.20)$ on the source line (right panel) are marked with blue crosses.
    Approx. $98$ per cent of the total magnification of the source, $\mu=6.7$, is due to the images on the main track (see also Fig. \ref{fig5:microimages_parity}).
    For these images, the symbols are scaled to the corresponding magnification (the other symbol sizes remain fixed for clarity).
    Coordinates in both panels are in units of the Einstein radius (the dashed rectangle corresponds to \figref{fig5:microimages_parity} and the labelled ones - A,B,C - as well as the corresponding source plane locations on the right panel, to \figref{fig5:microimages_on_map}).
    }
    \label{fig5:microimages_loops}     
\end{figure*}

\subsection{Finding the micro-images}
\label{sec5:micro-images}
The micro-images of a point source at a given position produced by a large number of point mass lenses
can be found using explicit search algorithms in the lens plane (e.g. \citealt{bib5:Paczynski1986,bib5:Saha2011}).
This can be done very efficiently using the complex notation.

The maximal number of micro-images that can be observed of a point source at $\zeta$
has been studied thoroughly \citep{bib5:Witt1990,bib5:Witt1991,bib5:Petters1992,bib5:Petters2001}.
\citet{bib5:Witt1990} has shown that all images of a point source at $\zeta$ can
be calculated by complex
conjugation of \eqref{eq5:complex_lens_equation}, by multiplying with
$\prod_{i=1}^N (z_i - z)$, and by re-inserting \eqref{eq5:complex_lens_equation}
for $z$. The result is a large polynomial of degree $(N+1)^2$ that only depends on
$\bar{z}$.
Taking the conjugate of the complex roots of this polynomial
yields all possible $(N+1)^2$ solutions - for $g=0$ the maximal number of images is $N^2+1$.
Not all of those solutions, however, also satisfy the real-valued lens-equation
\eqref{eq5:normalized_lens_equation} and a check needs to be performed.
This polynomial-technique to determine the solutions
of the normalized lens equation is very successfully used in the field of planetary microlensing,
where $N$ is only a few \citep[binary, triple system etc, e.g.][]{bib5:Bozza2010}.

Another elegant procedure to find all images corresponding to a point source is due to \citet{bib5:Witt1993} \citep[see also ][for a different implementation of the same idea]{bib5:Lewis1993}.
\citet{bib5:Witt1993} shows that all images of a point source can be found by following a straight source track from far away from the
ensemble of point masses to the position of interest $\zeta$:
\begin{itemize}
\item For a source position $\zeta_0$ very far away
from the ensemble of $N$ point masses there exists one image close to the source and $N$ images close
to the masses $m_i$.
\item To find these images one first needs to solve the lens equation iteratively to find the far image
(except for $|g|>1$, see \citealt{bib5:Witt1993}):
\begin{equation}
    z_{[k+1]} = \zeta_0 + g\,\bar{z}_{[k]}
               - \sum_{i=1}^{N} \frac{m_i}{\bar{z}_i - \bar{z}_{[k]}},
\end{equation}
where $k$ is the number of iteration.
\item The other images near the stars are found from:
\begin{equation}
    z \approx z_i - m_i/\left[ \bar{\zeta}_0 - \bar{z}_i + g\,z_i - \sum_{j=1,j\neq i}^N
    \frac{m_j}{z_j-z_i} \right].
\end{equation}
\item After all the initial image positions are found, the paper goes on to show how to find all
images of a straight line, and thus effectively all images for all source positions $\zeta$ on this line.
\end{itemize}
An example is given in the left panel of \figref{fig5:microimages_loops}, where a straight line on the source plane is lensed into a main image of a ``wiggly'' line and many small ``loops'' around the microlenses.
All images consist of single or multiply imaged parts of the straight line on the source plane.
Depending on the configuration, loops can sometimes combine to form bigger loops and the main line can also be connected to stars, as can be seen on the figure.

The total magnification is actually dominated by only a few images
\citep[e.g. see][]{bib5:Rauch1992,bib5:Wambsganss1992,bib5:Schechter2002,bib5:Granot2003,bib5:Saha2011}.
This is apparent in the left panel of \figref{fig5:microimages_loops} and illustrated further in \figref{fig5:microimages_parity}.
The overwhelming majority of the micro-images are faint negative-parity images (saddle-points of the arrival time) near each star.
There are formally also images coincident with the stars; these are maxima of the arrival time, and are so demagnified that they are negligible.
The important images are a small number of micro-minima and equal number of micro- saddle points, whose properties we describe in \secref{sec5:micro_image_swarm_properties}.

\subsection{Caustics of the deflectors}
\label{sec5:caustics_of_the_deflectors}
In the case of quasar microlensing, the combination of many point masses and a shear creates a complicated pattern of critical curves. 
The critical curves can be determined as the places in the lens plane where the determinant \eqref{eq5:microlensing_determinant} vanishes.
A small area in the lens plane would be mapped to a point, identifying a locus of formally infinite magnification in the source plane.

Using the complex formalism above, \citet{bib5:Witt1990} has shown that \eqref{eq5:complex_jacobi_determinant} can be used to work out
the location for all critical curves in the lens plane.
Because the determinant vanishes, the expression in brackets can be written as:
\begin{equation}
    -g + \sum_{i=1}^{N} \frac{m_i}{(\bar{z}_i-\bar{z})^2} = e^{i \varphi}.
\end{equation}
Similar to \secref{sec5:micro-images}, multiplying by $\prod_{i=1}^N (\bar{z}_i - \bar{z})^2$ yields a polynomial of degree $2N$.
All critical curves can be found by solving for all the roots $\varphi=0$ and then tracing all critical curves for $0<=\varphi<=2\pi$.
This can be easily achieved using a simple root-finder like the Newton-Raphson method.

In the left panel of \figref{fig5:microimages_loops} the critical lines for the $N=50$ stars were calculated in this way
by starting at the $2\,N = 100$ complex roots and following the critical curves for $\varphi$ between $0$ and $2\pi$.
The corresponding caustic lines in the source plane can be found by mapping the critical lines using the microlensing equation \eqref{eq5:complex_lens_equation} (or \eqref{eq5:microlensing_equation} and \eqref{eq5:microlensing_equation_Einstein_radius}) and are shown in the right panel of \figref{fig5:microimages_loops}.
The caustic lines separate regions of differing image multiplicity; whenever a source crosses a caustic, two images either appear or disappear.
The source position denoted by the cross symbol in \figref{fig5:microimages_loops} is inside the slightly elongated astroid caustic,
so that two additional images appear, as seen in the left panel of the same figure and in \figref{fig5:microimages_parity}.

The total magnification of a point source moving along the straight line in the right panel of \figref{fig5:microimages_loops} can be calculated as a function of position or time.
Characteristic spikes of extreme magnification are expected whenever the source is crossing a caustic.
Such a plot of magnification against time, also known as a ``light curve'', is shown in \figref{fig5:microimages_light_curves} and is examined in detail in \secref{sec5:time_varying_ML} as it is a very important microlensing observable.

\subsection{Properties of the ``swarm'' of microimages}
\label{sec5:micro_image_swarm_properties}
Finding all the microimages, albeit is feasible as described in \secref{sec5:micro-images}, is quite costly.
However, there is a number of key properties of the ``swarm'' of microimages, such as their number, distribution, and individual magnifications, that can be obtained more easily.
It is helpful to begin our examination of these properties with the time delay surface:
\begin{equation}
    \tau = \half(\vtheta - \vbeta)^2-\psi(\vtheta),
\end{equation}
where $\psi$ is some lensing potential (see \chapintro).
In the absence of any gravitational lensing, this surface is a circular paraboloid with a single extremum and one image is seen at the position of the source with magnification $\mu = 1$.
Under the influence of some background gravitational potential $\psi_B(\vtheta)$ with constant second derivatives (i.e. constant $\kappa$ and $\gamma$), the time delay surface transforms into an elliptical or hyperbolic paraboloid.
Whether the image, which corresponds to the observed macro-image, is a minimum or a saddle point then depends on the curvature of the time delay surface. 

\begin{figure}
    \centering
    \includegraphics[width=0.5\textwidth]{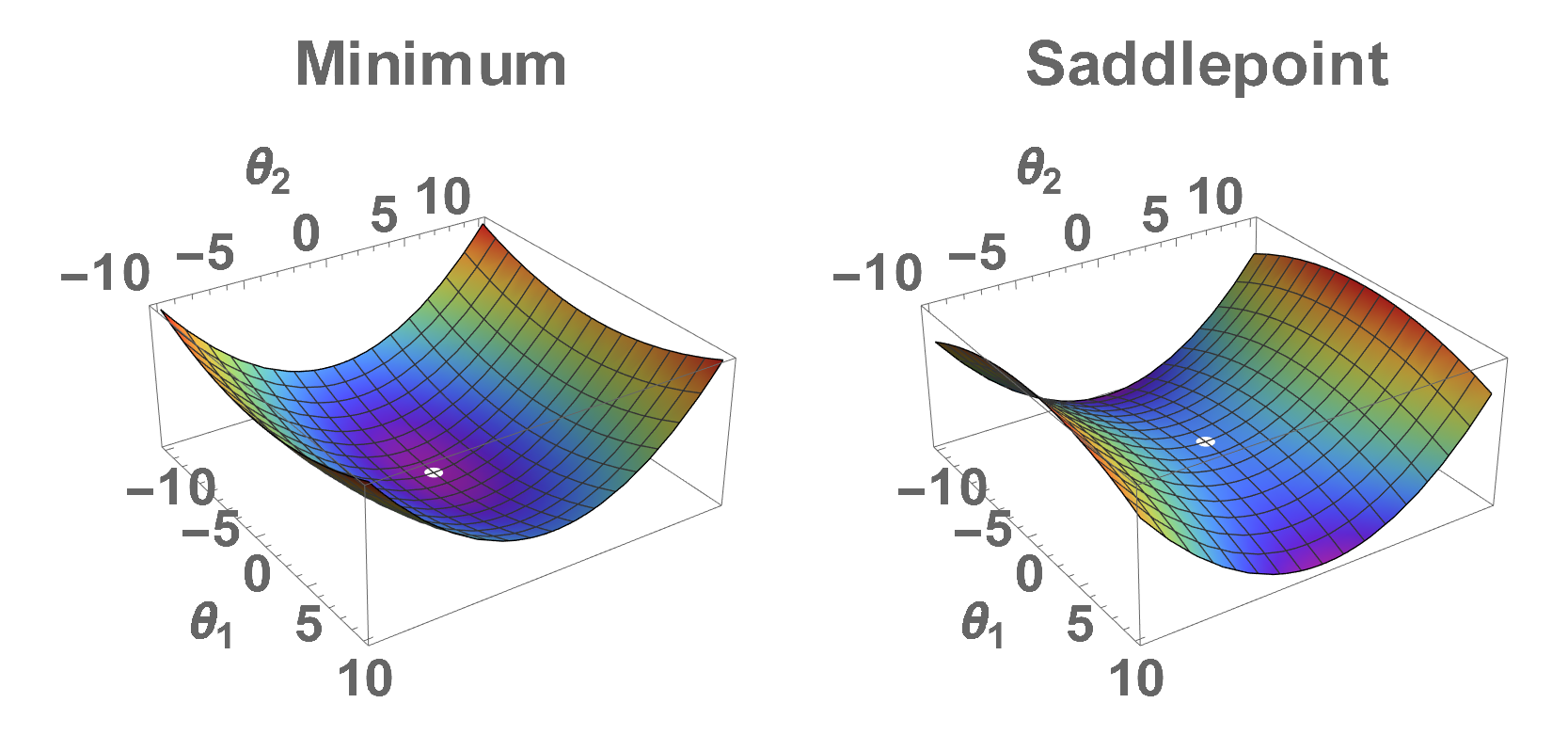}\\
    \includegraphics[width=0.5\textwidth]{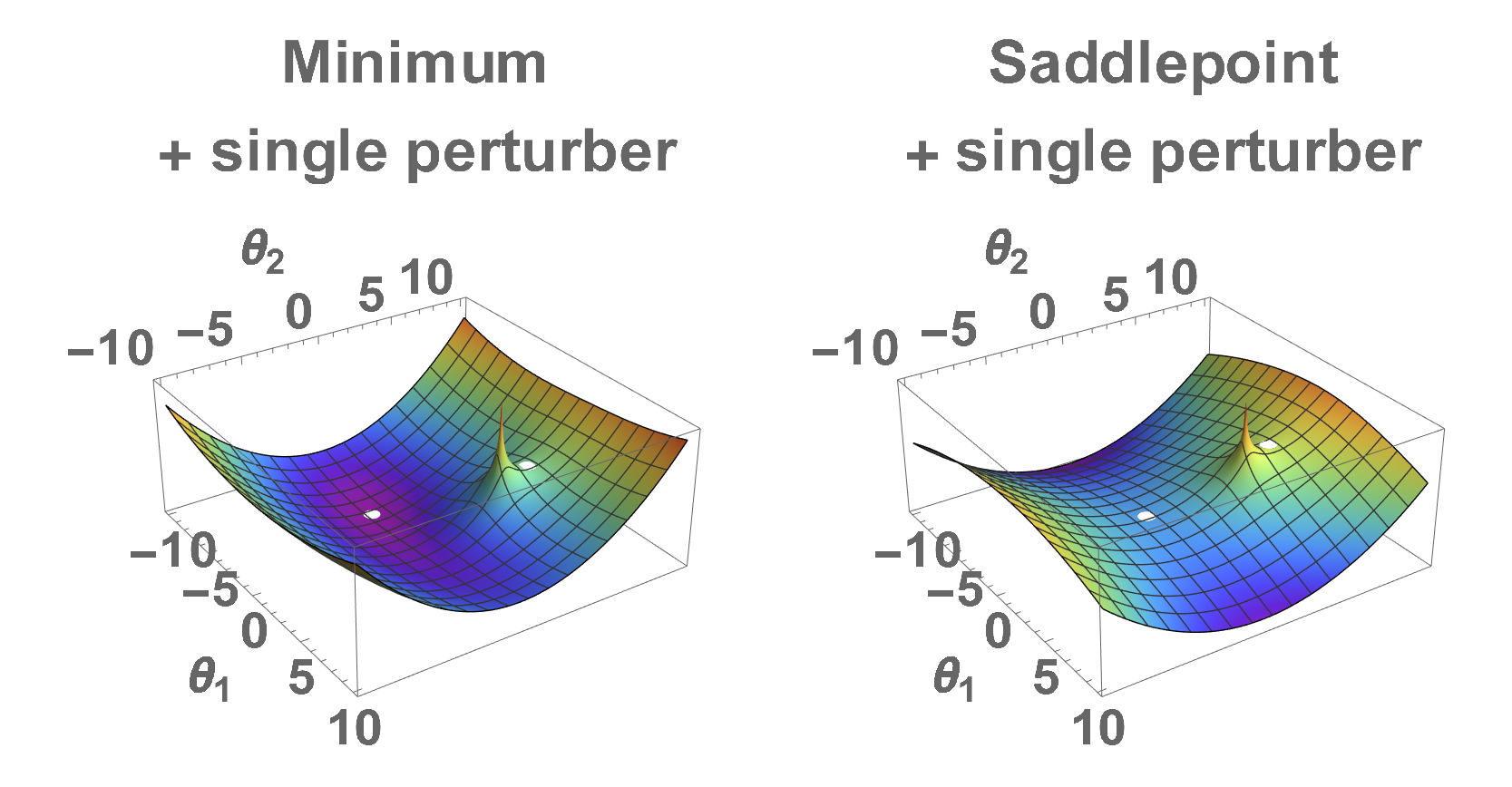}
    \caption{\textbf{Top:} The time delay surface for a minimum (left) and saddle point image (right). \textbf{Bottom:} Same but with the addition of a point-mass perturber. The white circles mark the locations of the micro-images.}
    \label{fig5:time_delay}
\end{figure}

Point masses introduce logarithmic terms into the gravitational potential (see \chapintro).
If the point masses have some convergence $\kappastar$ the time delay surface becomes:
\begin{equation}
    \tau = \half(\vtheta - \vbeta)^2-\psi_B(\vtheta) -\tae^2\sum_{i=1}^N m_i\ln|\vtheta-\vtheta_i| - \half(-\kappastar)\vtheta^2,
\end{equation}
where a term due to a negative constant surface mass density equal to the mass in stars is included to conserve the total convergence.
In the limit of few deflecting masses, the time delay is largely unperturbed, as shown in \figref{fig5:time_delay}.
There are scattered logarithmic spikes, but the likelihood of one lying close to the line-of-sight to the source is low.

The locations of the microimages satisfy the lens equation, which we re-write here as:
\begin{equation}
    \vbeta = \vtheta - \valpha_B(\vtheta) - \valpha_*(\vtheta),
\end{equation}
where $\valpha_{\rm B}(\vtheta)$ and $\valpha_*(\vtheta)$ are the deflection angles due to the potential $\psi_{\rm B}$ and the microlenses (with the negative constant convergence correction) respectively.
For constant second derivatives of the potential $\psi_{\rm B}$ (constant $\kappa$ and $\gamma$), the first two terms of the lens equation become proportional to $\vtheta$.
For microimages far away from the source position, $\vbeta$, these terms become very large.
Consequently, the sum of the deflections due to the microlenses (see \eqref{eq5:lenseq}) must also become very large.
Because point mass deflections are proportional to $1/|\vtheta-\vtheta_i|$, the location of the microimage is required to be very close to some microlens so that a single term in the sum dominates.
It can be shown that the magnifications of these far away micro-images behave as $\mu\propto|\vtheta|^{-4}$ \citep{bib5:Paczynski1986,bib5:Schneider1987a}, and they are therefore faint saddle points.

\begin{figure}
    \centering
    \includegraphics[width=.45\textwidth]{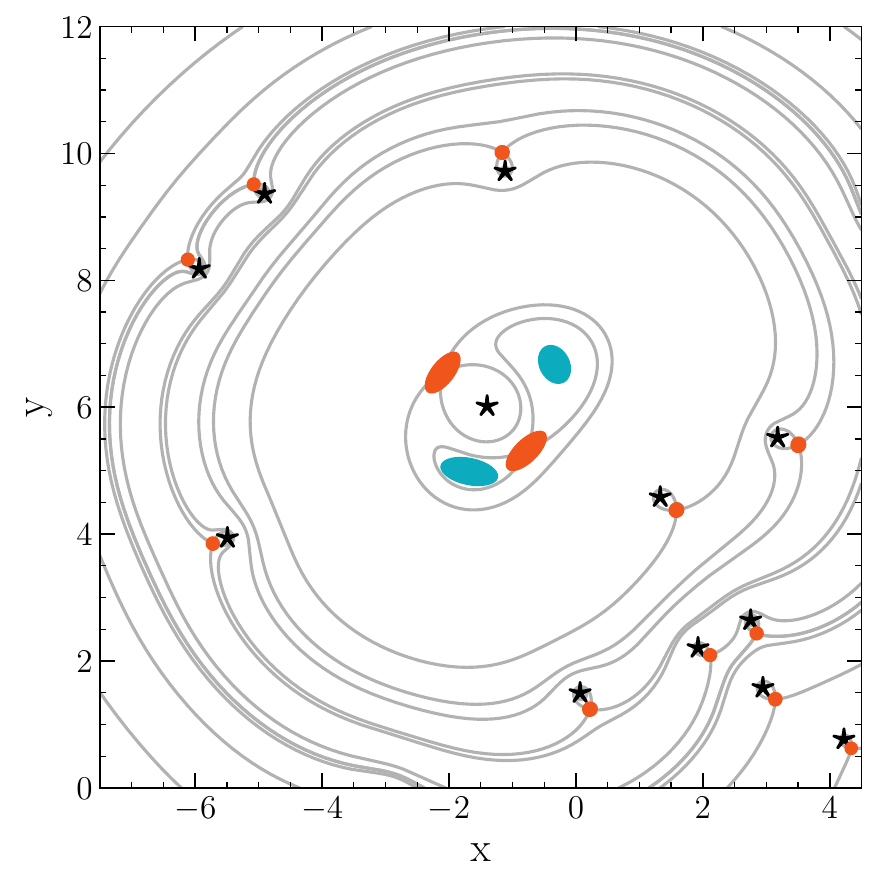}
    \caption{Zoomed-in region around the brightest images in the left panel of \figref{fig5:microimages_loops}.
    The gray curves are saddle-point contours of the arrival time, at whose self-intersection points saddle-point micro-images form (red ellipses).
    The micro-minima (blue ellipses) and positions of the microlenses (black stars) are also shown.
    The latter coincide with the locations of zero-brightness micro-maxima.
    The size and shapes of the ellipses are indicative of the magnification tensor.
    }
    \label{fig5:microimages_parity}
\end{figure}

For larger values of $\kappastar$, microlenses are more likely to lie closer to the macro-image and affect it.
So long as there is external shear, a single point mass close to the macro-image position can split it into four, instead of just two, extra images (counting the infinitely de-magnified micro-maximum) if the source lies within the star's caustic.
As $\kappastar$ increases, there is no longer a single micro-image that can be associated with the macro-image, which is split into an increasingly more complex ``swarm'' of microimages.
\figref{fig5:microimages_parity} shows the microimages along with contours of the light travel time that pass through micro-saddle-points.
Close to the macro-image position, the micro-images are not restricted to lie very close to an associated microlens anymore; the deflection term of the macro-potential is small enough to be counterbalanced by the sum of the deflections due to all the microlenses without any specific term dominating.
Instead, there is an effective region within which they tend to lie that determines the size of the micro-image swarm.

\subsubsection{The size of the micro-image swarm}
\label{sec5:size_of_swarm}
The magnification of a point source located at $\vbeta$ is \citep{bib5:Venumadhav2017}:
\begin{equation}
    \label{eq5:point_source_mag}
    \mu(\vbeta) = \int \delta(\vtheta - \valpha_B(\vtheta)-\valpha_*(\vtheta) - \vbeta) \diff^2 \vtheta.
\end{equation}
This can be seen by utilizing properties of the Dirac delta function; \eqref{eq5:point_source_mag} can be rewritten as a sum:
\begin{equation}
    \label{eq5:point_source_mag_sum}
    \mu(\vbeta) = \int \sum_{i} \mu(\vtheta_i)\delta(\vtheta - \vtheta_i)\diff^2 \vtheta,
\end{equation}
where the $\vtheta_i$ are the locations of the microimages. 
The deflection angle $\valpha_*(\vtheta)$ due to the point masses in \eqref{eq5:point_source_mag} can be viewed as a random variable that changes based on different realizations of the random point mass positions.
Averaging over all such realizations, we get:
\begin{equation}
\label{eq5:mean_magnification}
    \langle\mu(\vbeta)\rangle = \int \delta(\vtheta - \valpha_B(\vtheta)-\valpha_* - \vbeta)p(\valpha_*)\diff^2 \valpha_* \diff^2 \vtheta,
\end{equation}
where $p(\valpha_*)$ is the probability density function of the deflection angle.
By transforming coordinates from the image plane to the source plane with the change of variable $\vbeta' = \vtheta - \valpha_B(\vtheta) - \vbeta$, we arrive at:
\begin{equation}
    \langle\mu(\vbeta)\rangle = \int \delta(\vbeta' - \alpha_*)p(\valpha_*) \mu_B(\vbeta' + \vbeta)\diff^2 \valpha_* \diff^2 \vbeta'.
    \label{eq5:mean_mag_general}
\end{equation}
Performing the integral over $\valpha_*$, this simplifies to:
\begin{equation}
    \langle\mu(\vbeta)\rangle = \int p(\vbeta') \mu_B(\vbeta' + \vbeta) \diff^2 \vbeta'.
    \label{eq5:mean_mag_general2}
\end{equation} 
This resulting expression is a convolution of the microlens deflection probability density with the background magnification model.
What was once a point source has, on average, been ``smeared out'' into a source with a profile that looks like $p(\valpha_*)$.

The form of the probability density function of $\valpha_*$ was first worked out by \citet{bib5:Katz1986}, while \citet{bib5:Schneider1992} provide it in a slightly different form and \citet{bib5:Petters2009a} present a more thorough mathematical treatment.
This function is isotropic and depends only on the magnitude of the deflection angle, not the direction.
It is a combination of a bivariate normal distribution, whose width depends on the number of point masses $N$ as:
\begin{equation}
    \sigma_* = \tae \kappa_*^{1/2} \big[ \ln( 2e^{1-\gamma_E} N^{1/2} ) \big]^{1/2},
    \label{eq5:sigma_alpha}
\end{equation}
where $\gamma_E\approx0.577$ is the Euler-Mascheroni constant, with a tail that behaves as:
\begin{equation}
    p(\valpha_*)=\frac{\tae^2\kappastar}{\pi|\valpha_*|^4},
\end{equation} 
for large values of $\valpha_*$.

\begin{figure*}
    \centering
    \includegraphics[width=\textwidth]{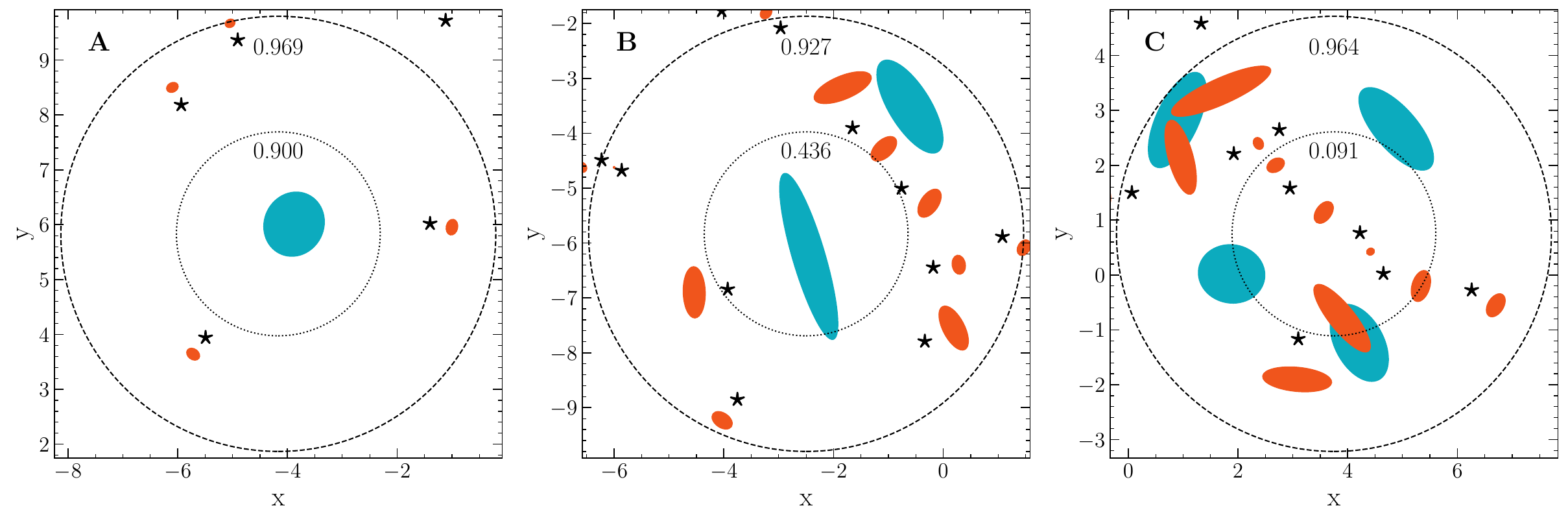}
    \caption{
    Microimages corresponding to a point source located at the three source positions indicated in the right panel of \figref{fig5:microimages_loops}.
    The blue (orange) ellipses represent the micro-minima (micro-saddles).
    The sizes of the ellipses scale logarithmically with magnification in order to show a greater dynamic range, while their orientation shows the micro-shear from the stars.
    The dashed and dotted ellipses (circles in this case because there is no macro-shear) denote the isophotes of 90 per cent and $1\sigma_*$ respectively, while the numbers indicate the actual fraction of the total magnification that lies within them.}
    \label{fig5:microimages_on_map}
\end{figure*} 

To determine the size of the image swarm for a point source, one can integrate $p(\valpha_*)$ over a portion of the source plane that contains a large fraction of the flux.
If we only integrate over a region with width $\sigma_*$, the probability density function behaves like a normal distribution and the resulting isophote in the image plane contains microimages that on average constitute only $39\%$ of the flux; in order to include $90\%$ of the flux (on average), integrating over a region with width $2.15\sigma_*$ is required.
We can transform \eqref{eq5:sigma_alpha} for $\sigma_*$, which depends on the number of point masses $N$, into a function of the macro-parameters only.
By considering the tail of the probability density function, one finds that $99\%$ of the average flux is contained within a circle of radius $10 \kappastar^{1/2} \tae$.
The resulting size of the image swarm is then given by transforming this circle to the image plane, resulting in an ellipse with axes given by:
\begin{equation}
    \label{eq:image_swarm_ellipse}
    r_\pm = \frac{10\kappastar^{1/2}\tae}{|1-\kappa\pm\gamma|}.
\end{equation}

Figure \ref{fig5:microimages_on_map} shows various micro-image configurations for the three source positions indicated in \figref{fig5:microimages_loops}.
The magnification lying within the marked contours is slightly different than the expected average due to sample variance of the source and microlens positions.

The size of the image swarm determines the allowable area where bright microimages can appear. This in turn affects the possible astrometric shifts from microlensing. Large astrometric shifts can come about only due to caustic crossings and the creation or annihilation of pairs of microimages, as otherwise the microimages only change positions smoothly. \cite{bib5:Treyer2004} found that strong astrometric microlensing shifts are seen in systems with $\kappa=\gamma=0.4$ and $\kappa=\gamma=0.6$, while weak shifts are seen in systems with $\kappa=\gamma=0.2$ and $\kappa=\gamma=0.8$. This is easily explained by the axis ratio of the image swarm ellipse for each set of parameters, which additionally provides the reason for a dependence on the direction of the external shear noted in \cite{bib5:Treyer2004}. The maximum allowable astrometric shift is roughly given by \eqref{eq:image_swarm_ellipse} and is therefore on the order of 10s of micro-arcseconds. In more extreme cases of magnification from, e.g., cluster critical curves, the astrometric shifts from microlensing may approach more observable milli-arcsecond levels.

For an extended source with some profile $I(\vbeta)$, the microimages in general simply result in a broadening of the source profile.
The width of $p(\valpha_*)$, which acoording to the previous discussion is of order $\kappastar^{1/2}\tae$, is added in quadrature to the width of the source $\sigma_{\rm I}$ \citep{bib5:Dai2021}, resulting in an effective width:
\begin{equation}
    \sigma_{\rm eff} \approx \sqrt{ \sigma_{\rm I}^2 + \kappastar\tae^2 }.
\end{equation}

\subsubsection{The number of micro-images}
\label{sec5:number_of_microimages}
While every point mass has an associated saddle point image, the total number of images for a field of $N$ point masses with external shear cannot exceed $5N$ \citep{bib5:Khavinson2004}.
This implies there are at most $2N-1$ micro-minima and $3N-1$ micro-saddles, plus one additional micro-image of either parity corresponding to the macro-minimum or saddle \citep{bib5:Petters2009b}.
In practice, for macro-images that are not near a macro-critical curve (i.e. outside the extreme high magnification regime), the number of expected micro-minima (extra image pairs) is fairly low, with some simulations suggesting an approximately Poisson distribution \citep{bib5:Granot2003}. 

\begin{figure}
    \centering
    \includegraphics[width=0.467\textwidth]{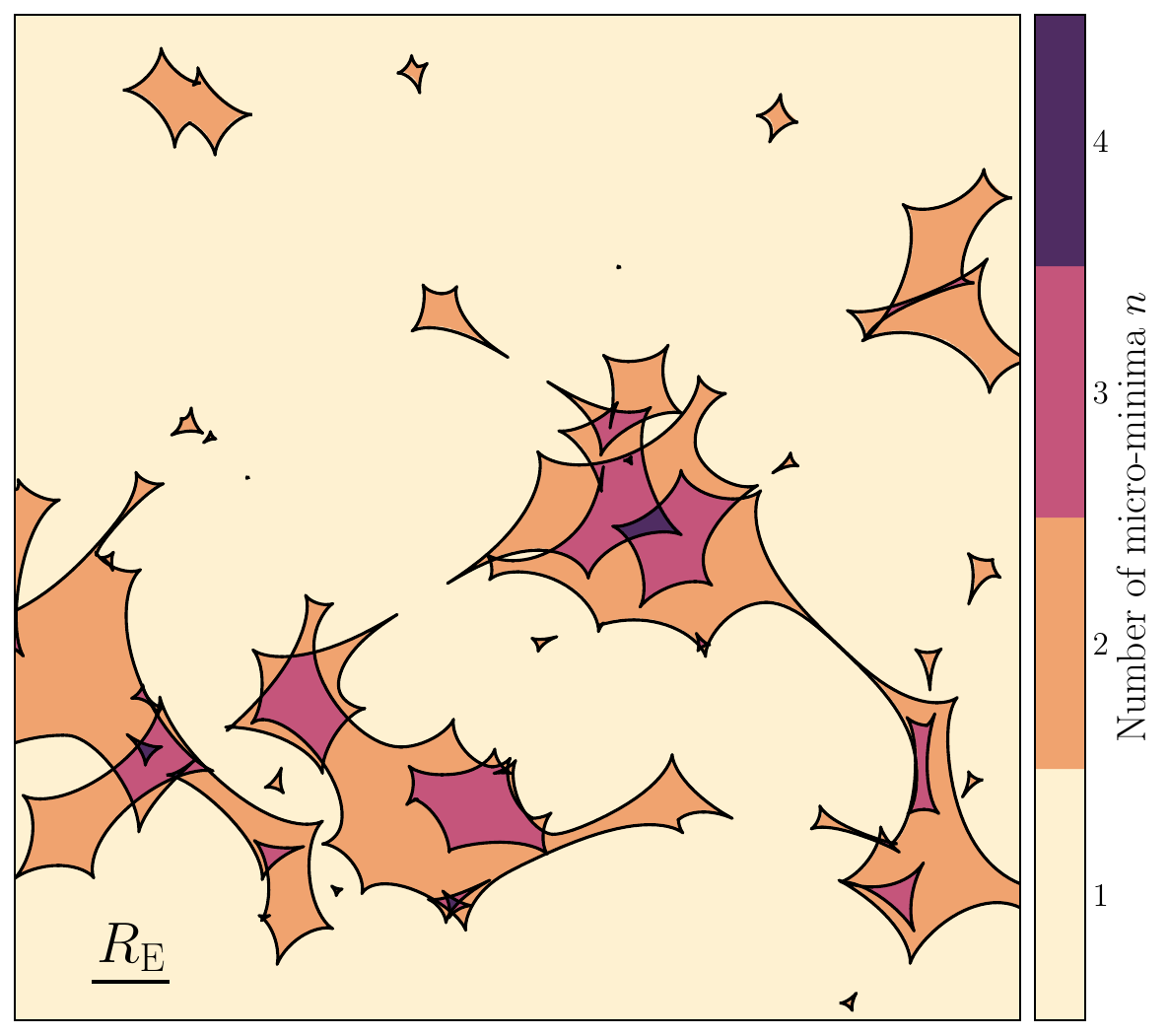}
    \caption{Caustics (black lines, same as in the right panel of \figref{fig5:microimages_loops}) and corresponding number of micro-minima from each region they enclose. The region of the source plane shown here is the same as in \figref{fig5:example_mag_map}.}
    \label{fig5:num_minima}
\end{figure}

\begin{figure}
    \centering
    \includegraphics[width=0.45\textwidth]{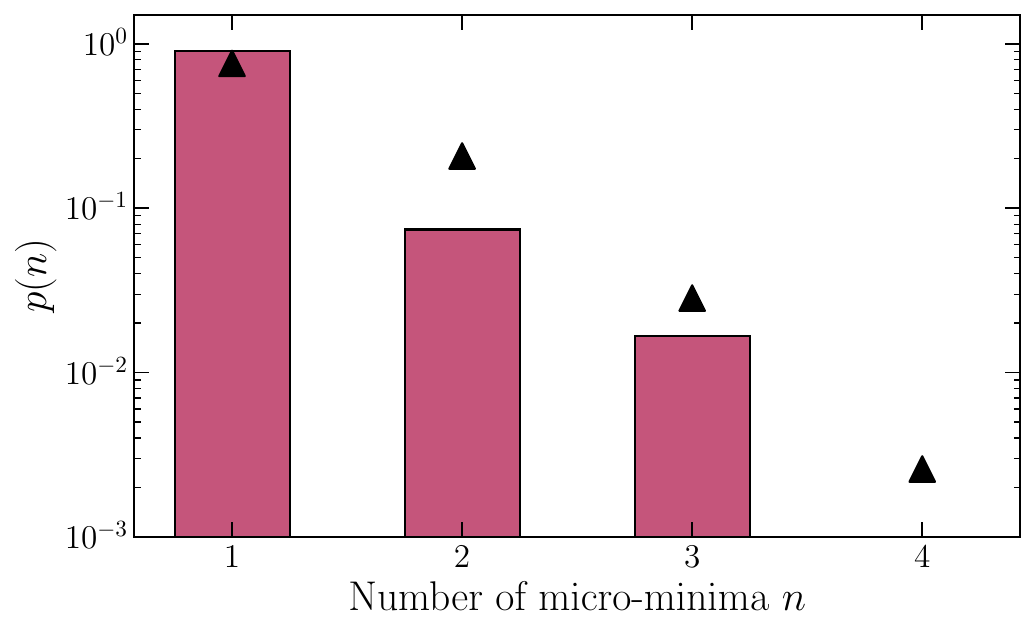}
    \caption{Histogram showing the distribution of the number of micro-minima. The black triangles denote the expected probabilities for a Poisson distribution with a mean $\langle n\rangle - 1$ (as the image is a macro-minimum).} 
    \label{fig5:num_minima_hist}
\end{figure}

The micro-critical curves split the image plane into regions of positive and negative parity, as shown in \figref{fig5:microimages_loops}.
The multiplicity with which regions of positive parity overlap when mapped back to the source plane gives the mean number of positive parity images $\langle n\rangle$, for which analytic formulas are given in \citet{bib5:Wambsganss1992,bib5:Granot2003}. 
Figure \ref{fig5:num_minima} shows the caustic pattern of \figref{fig5:microimages_on_map}, color-coded to show the number of micro-minima.
The histogram of the resulting distribution of the number of microminima is shown in \figref{fig5:num_minima_hist}.

\begin{figure}  
    \centering  
    \includegraphics[width=0.49\textwidth]{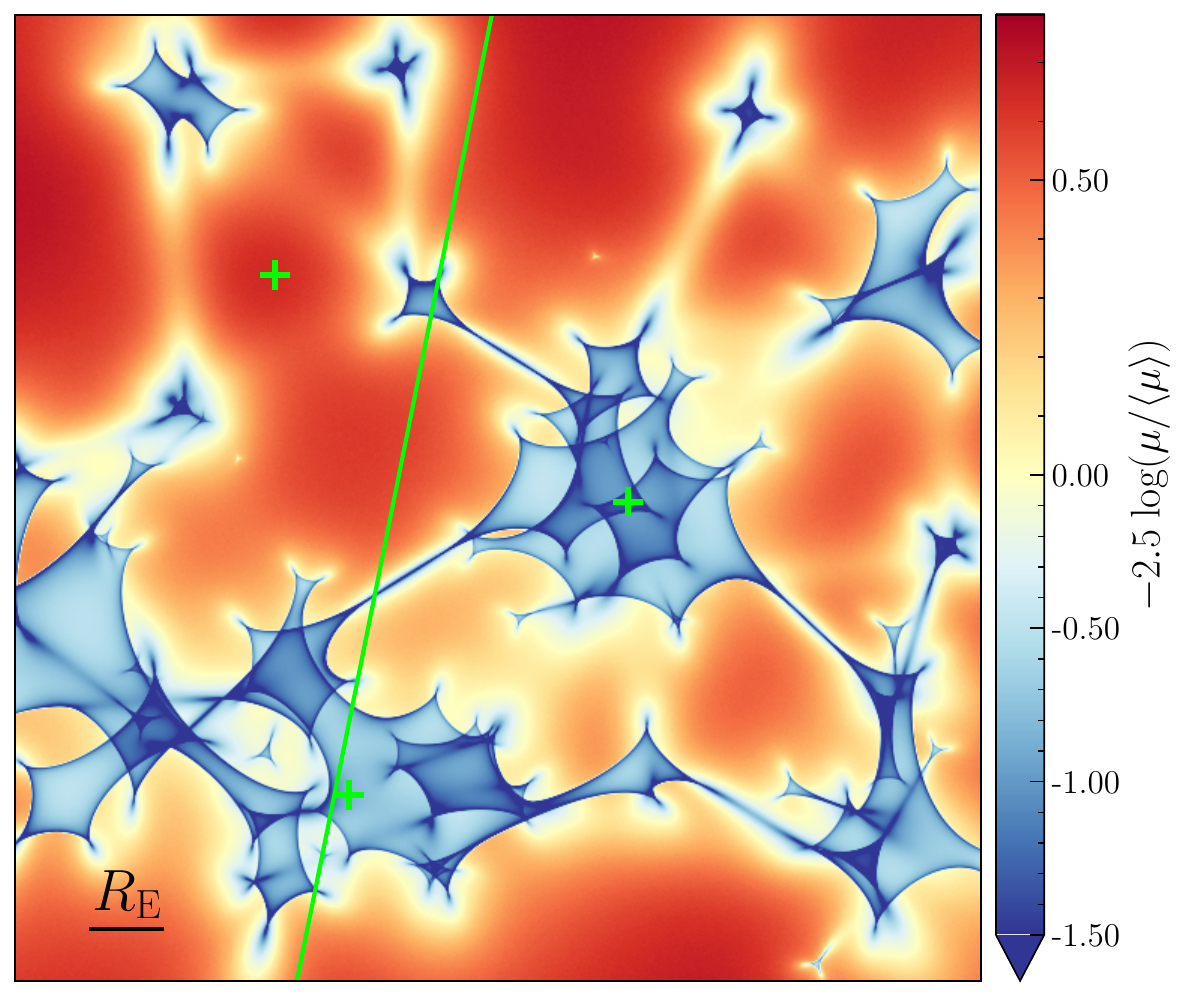}
    \caption{Magnification map corresponding to the microlens configuration shown in Fig. \ref{fig5:microimages_loops}, generated by the inverse ray-shooting code \textsc{GPUD} \citet{bib5:Thompson2010,bib5:Thompson2014}.
    The scale is in magnitudes over the mean magnification, $\langle \mu \rangle$, which approaches the macroscopic one given by Eq. \ref{eq5:macro_magnification} (a function of the local convergence and shear) as the number of rays increases.
    The field of view is $13\times13 \, \tae$ on the source plane and includes the track from the right panel of Fig. \ref{fig5:microimages_loops} (green line) and the locations used to produce Fig. \ref{fig5:microimages_on_map} (green crosses).
    Corresponding light curves for different accretion discs are shown in Fig. \ref{fig5:microimages_light_curves}.}
    \label{fig5:example_mag_map}
\end{figure}

Finally, by extending arguments found in \citet{bib5:Wambsganss1992,bib5:Granot2003}, it can be shown that the average magnification of a single micro-minimum is:
\begin{equation}
    \langle \mu_{\rm mm}\rangle = \frac{\mu_{\rm macro}}{\langle n\rangle} P_{\mu>0}(x_{img}),
\end{equation}
where $\mu_{\rm macro}/\langle n \rangle$ is the macro-magnification from \eqref{eq5:macro_magnification} divided by the expected number of micro-minima, and $P$ is the probability that location $x_{\rm mm}$ is a minimum and not a saddle-point.

\subsection{Microlensing maps}
\label{sec5:microlensing_maps}
A very useful tool for microlensing studies is the magnification field on the source plane induced by an ensemble of microlenses.
This can be used together with a model for the source to match simulated light curves and flux ratios with observations, as will be discussed in \secref{sec5:data_analysis}.
To calculate this magnification, we have to transform \eqref{eq5:magnification_sum}, the sum of magnifications due to all the microlenses, from the lens to the source plane. 
This is hardly practical, and becomes quickly intractable as the number of microlenses increases.
However, there has been a number of techniques developed specifically to optimize this task, which we review here.

\begin{figure}
\centering
\includegraphics[width=0.45\textwidth]{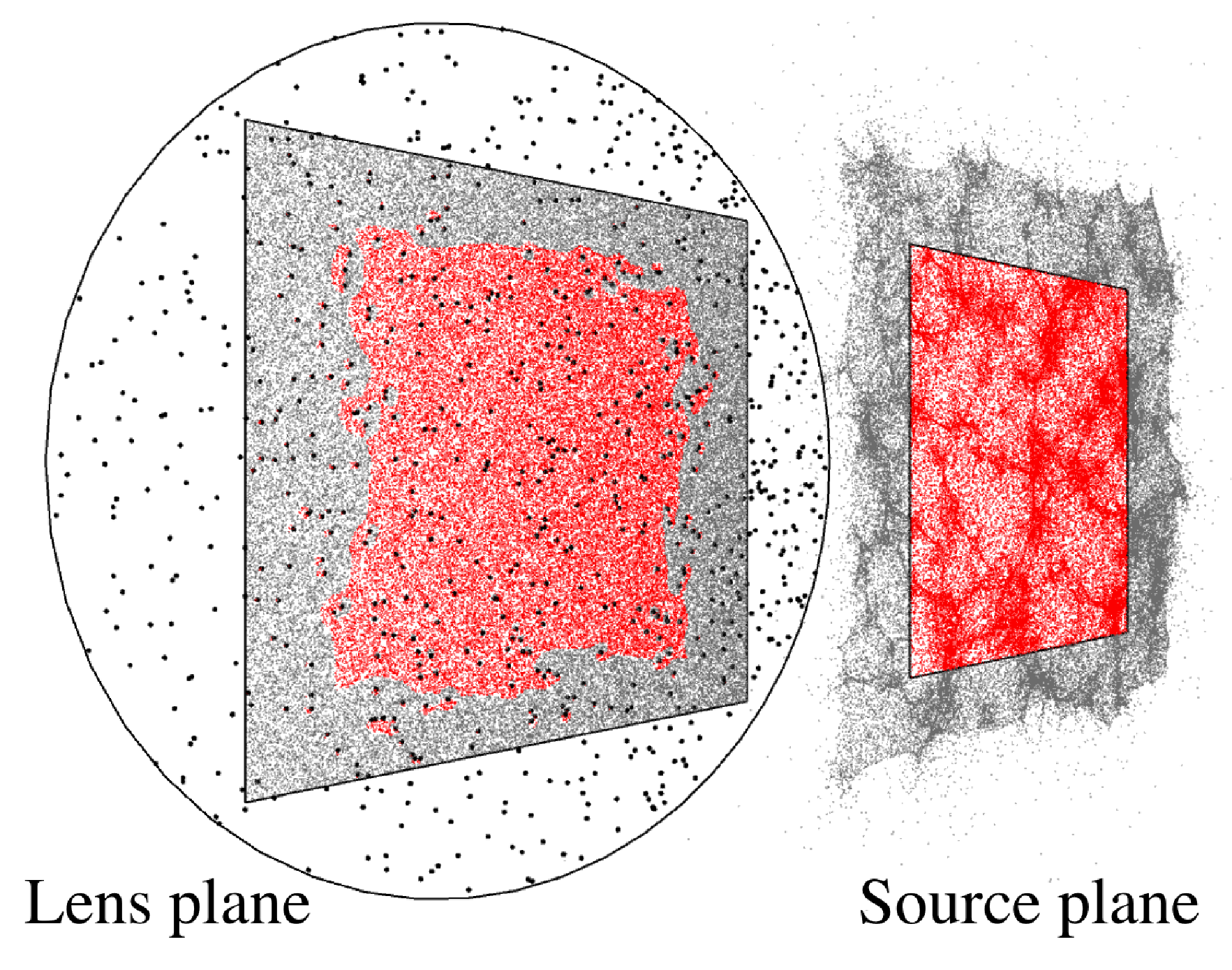}
\caption{Schematic representation of ray-shooting. Microlenses (black dots) are distributed in a circle of area A on the lens plane. A grid of light-rays (grey and red dots) is projected backwards from the observer through the lens plane, where the deflection by each lens is calculated for each ray using \eqref{eq5:microlensing_equation_Einstein_radius}, and mapped onto the source plane. A rectangular area on the source plane is selected (red points) away from edge effects, divided into pixels, and used to calculate the magnification from \eqref{eq5:pixellated_magnification}. Taken from \citet{bib5:Vernardos2014b}.}
\label{fig5:inverse_ray_shooting}     
\end{figure}

Our starting point is \eqref{eq5:microlensing_equation_Einstein_radius}, where the image positions $\boldsymbol{\theta}$ map uniquely to a position on the source plane, $\boldsymbol{\beta}$ - a one-to-one mapping.
As we have seen, the inverse is not true and a source can have multiple images.
Finding these solutions to the lens equation is a complex problem, if not impossible \citep[see][for analytical solutions in a few simple cases]{bib5:Schneider1992}.
Therefore, an alternative approach is to proceed in the inverse manner and calculate the total magnification as the sum of the intensity of all the microimages in a finite region of the source plane, as shown in \figref{fig5:example_mag_map}.
This technique, called inverse ray-shooting, was introduced by \citet{bib5:Kayser1986} and comprises propagating a grid of rays  backwards from the observer through the lens plane, where each ray is deflected using \eqref{eq5:microlensing_equation_Einstein_radius}, to the source plane, where its final position is mapped.
By dividing the source plane into pixels, counting the number of rays reaching each pixel, $N_{ij}$, and comparing it to the number of rays that would have reached that pixel if there was no lensing taking place, $N_{\mathrm{rays}}$, we obtain an estimate for the magnification, i.e. a pixelated magnification map:
\begin{equation}
\label{eq5:pixellated_magnification}
\mu_{ij} = N_{ij}/N_{\mathrm{rays}}.
\end{equation}
This relation is an approximation because it does not take into account light rays from outside the defined grid that could be deflected inside the region of the source plane under examination.
In practice, by choosing a grid of rays large enough with respect to the size of the source plane, the effect of such extreme deflections can be neglected.
The above outlined procedure is schematically shown in \figref{fig5:inverse_ray_shooting}.

Performing inverse ray-shooting directly is a computationally expensive procedure, with a forbidding cost for large scale applications \citep[see][for some benchmarks]{bib5:Bate2012,bib5:Vernardos2014b}.
Hence, efficient approximate implementations have been developed, as well as direct approaches taking advantage of hardware acceleration, which we outline below.

\subsubsection{The tree code}
One way of reducing the calculations required to produce magnification maps is by approximating the sum of all the deflections due to the microlenses.
The key idea here is that the deflection angle of a given light ray due to a given microlens is inversely proportional to the distance between the two, in an analogous way to gravitational N-body problems.
In the hierarchical tree code approach developed by \citet{bib5:Wambsganss1990a,bib5:Wambsganss1999}, microlenses that are further away from a light ray are grouped together in cells, or pseudo-particles, whose size and total mass depends on the distance to the light ray.
This approximation greatly reduces the final number of deflections that need to be calculated per light ray and, subsequently, increases the speed of the calculations, at the cost of a slight computational overhead and a loss in accuracy.
This method has been used extensively in the past to constrain properties of the background quasar and the lens \citep{bib5:Rauch1991,bib5:Keeton2006,bib5:Bate2008,bib5:Pooley2012,bib5:Dai2018,bib5:Hutsemekers2019}.
Due to its efficiency and the fact that it was the only widely available code for producing magnification maps for a long period of time, this approach can be considered as the ``industry standard'' in quasar microlensing.

Although the tree code has led to a tremendous speed-up of the calculations compared to direct ray-shooting, there are a few disadvantages.
The required accuracy of the code - the level at which a number of microlenses at a given distance from a light
ray are grouped into a pseudo-particle or treated individually - is something to be determined empirically by running the code multiple times until the results are acceptable.
Hence, a certain amount of expert knowledge is required to implement and use it.
The memory overhead of the tree data structure used to group the microlenses into cells may grow too large with the number of microlenses for a single processor to handle.
For this reason, a parallel version has been developed \citep{bib5:Garsden2010} that overcomes this by distributing the computations across many nodes on a computer cluster.
Using this approach, magnification patterns of up to a billion microlenses can be computed in realistic timescales.

\subsubsection{Polygon Mapping}
Another approach to speed up the solution of \eqref{eq5:pixellated_magnification} is to reduce the number of rays shot, $N_{\mathrm{rays}}$, as does the inverse polygon mapping technique developed by \citet{bib5:Mediavilla2006,bib5:Mediavilla2011b,bib5:Shalyapin2021}.
Here, the lens plane is divided into cells that are mapped onto the source plane using the lens equation.
By counting the portions of the area of the cells that are within a given source plane pixel, as opposed to the standard
inverse ray-shooting that counts rays, an estimate of the magnification is obtained.
The main advantage of this method is the increased computational efficiency, i.e. there are no unused rays that got deflected outside the source plane region of interest, while there is an improved resolution around caustics as well.
However, it has its own computational overhead due to finding the correct tiling of the lens plane and the large number of cells required near critical lines.
Another advantage of this method is that the error on its estimate of the number of rays per pixel follows $N_{ij}^{-3/4}$, which is smaller than the Poisson error $N_{ij}^{-1/2}$ of randomly shot rays.
In fact, the former is a direct result of using a regular grid of rays \citep[instead of random positions][]{bib5:Kayser1986}.
Selected examples of studies employing the inverse polygon mapping technique are \citet{bib5:Guerras2013a,bib5:Jimenez2014,bib5:Rojas2014,bib5:Esteban2020}.

\begin{figure*}
    \centering
    \includegraphics[width=0.95\textwidth]{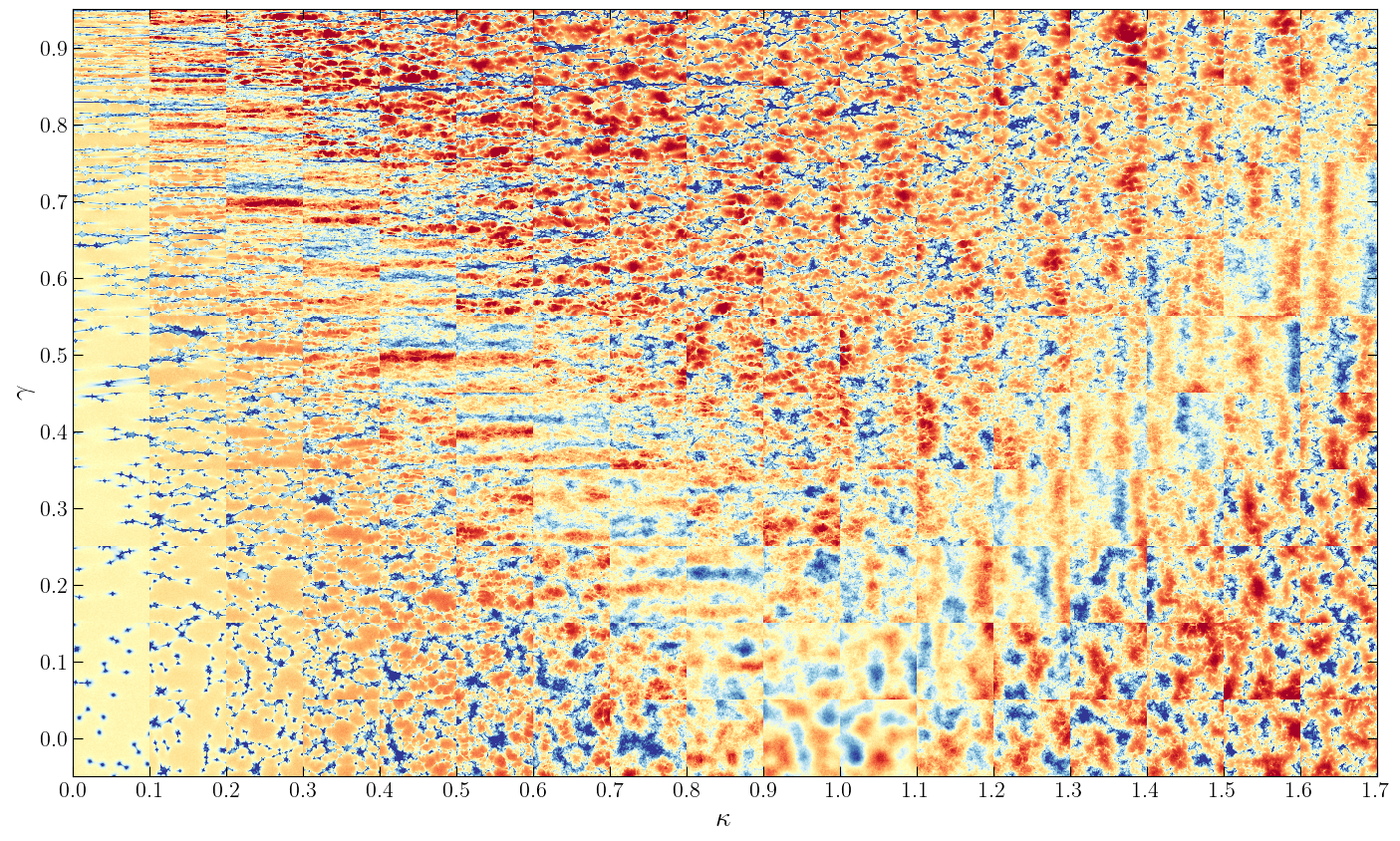}
    \caption{Tiling of the $\kappa-\gamma$ parameter space by $17 \times 10$ magnification maps with $\kappac=0$ from the GERLUMPH database. The central point of each tile corresponds to the $\kappa-\gamma$ values that were used to generate it. The caustics become denser as the number density of microlenses increases along the x-axis, while they appear increasingly stretched due to higher shear along the y-axis (the stretching always occurs along the x-axis of the maps themselves, see \figref{fig5:configuration} for the proper alignment of the maps in a real lensed system). The color scale is the same as in \figref{fig5:example_mag_map}.}
    \label{fig5:map_tiles}
\end{figure*}

\subsubsection{Direct approaches}
Numerous scientific problems, whose solution has so far been possible only through complex approximations, if available, can now be revisited due to the advent of Graphics Processing Units (GPU) and their design focused on massively parallel processing \citep[see the general brute-force and algorithm analysis techniques described in][]{bib5:Barsdell2010,bib5:Fluke2011}.
Although originally developed as specialized hardware to accelerate the generation of graphics on the computer screen, particularly for the game industry, the GPU architecture together with the emergence of the notion of general purpose programming libraries, has allowed for speed-ups of $\mathcal{O}(10)$ to $\mathcal{O}(100)$ for various algorithms.
This is the case for inverse ray-shooting, whose direct implementation is ``embarassingly parallel'' : each deflection by a single microlens is independent of any other microlenses and the deflections for each light ray are independent of all the other light rays.
Such an algorithm - GPUD - has been implemented by \citet{bib5:Thompson2010,bib5:Thompson2014}, with an obvious advantage being its simplicity and ease to modify and maintain.
Another advantage is the benefit from Moore's law \citep{bib5:Moore1965} for GPUs - a doubling of the speed every 1-2 years due to hardware improvements - without any software modification \citep[see fig. 2 in][]{bib5:Vernardos2014b}, which has ceased since 2005 for CPUs.
GPUD was used to carry out GERLUMPH, the largest parameter space exploration of microlensing maps discussed in Sect. \ref{sec5:caustic_structures_macro_parameters}, which enabled a series of previously unfeasible studies \citep{bib5:Vernardos2018,bib5:FoxleyMarrable2018,bib5:Vernardos2019a,bib5:Vernardos2019b,bib5:Neira2020}.
Recent developments by \citet{bib5:Zheng2022} can increase the speed of GPUD by 100 in the regime where it is the slowest, i.e. large numbers of lenses and high resolution maps.

The Fourier-based approach by \citet{bib5:Kochanek2004} separates the long- and short-range effects of the microlenses on ray deflections using a particle-particle/particle-mesh (P3M) algorithm, assumes the lens and source planes to be spatially periodic, and approximates the long-range deflections via a Fourier transform.
As a consequence, the edge effects in the source plane are removed and more of the magnification map is usable (higher efficiency).
However, there is an upper bound on the size of the generated magnification maps (approx. 8192 pixels on a side) due to the increasing amount of memory required by the Fourier transforms. 
Selected examples of studies employing the Fourier-based method are \citet{bib5:Poindexter2008,bib5:Chartas2009,bib5:Dai2010,bib5:MacLeod2015,bib5:Morgan2018,bib5:Cornachione2020b}.

\subsubsection{Combined approaches}
Apart from GPUD's direct approach to inverse ray-shooting, the tree and polygon based algorithms can also benefit from the massive GPU parallelization, at the cost of more complex algorithm development. 
A spectacular example is the Teralens\footnote{\url{https://github.com/illuhad/teralens}} algorithm, which follows the same general principles as the tree code of \citet{bib5:Wambsganss1999} but has a dramatically different algorithmic design tuned for maximum parallel efficiency.
One can also envisage merging the tree-based method that approximates the lens deflections with shooting polygons instead of rays, since the two approximate different parts of the lens equation.
Recently, \citet{bib5:JimenezVicente2022} achieved this by combining the inverse polygon method with the fast multiple method \citep{bib5:Greengard1987}, an algorithm similar in concept to a tree or particle-mesh algorithm.
Implementing this approach on the GPU would result in the fastest conceivable magnification map generating code.
Finally, alternative GPU/CPU parallel architectures could provide another path forward, requiring minimal modifications for speeding up existing CPU codes \citep{bib5:Chen2017}.

\subsection{How caustic structures respond to macro parameters}
\label{sec5:caustic_structures_macro_parameters}
The stellar-mass compact object populations that cause microlensing are unobservable - one would require to measure masses and positions of hundreds to thousands of such objects within galaxies at cosmological distances, orders of magnitude below the resolution of any conceivable telescope.
Hence, we have to resort to a statistical description of such populations at the location of the observed quasar multiple images that undergo microlensing.
This is based on local properties of the macroscopic mass and light distribution of the lensing galaxy.
The three major macro-parameters statistically defining a microlensing population are:
\begin{itemize}
    \item the total convergence, $\kappa$, defined in \eqref{eq5:total_kappa},
    \item its compact matter component, $\kappastar$, related to the number of microlenses via \eqref{eq5:stellar_kappa}, and
    \item the shear, $\gamma$.
\end{itemize}
The $\kappa,\gamma$ are introduced and linked to the lens potential within the general formalism presented in Sect. 3 of \chapintro, while specific choices of lens models and partitioning the mass between a smooth/dark and a compact/stellar component are discussed in Sects. 2 and 3 of \chapgal. 
Although the shear is a vector, its direction becomes important only when comparing to other multiple images, for example when generating light curves (e.g. see \figref{fig5:configuration}).
Therefore, without loss of generality, we can align the shear with the x-axis, leading to $\gamma_2 = 0$ in \eqref{eq5:microlensing_equation_Einstein_radius} - a useful trick to maximize the efficiency in using magnification maps.
This leads to all three parameters being scalar fields, functions of the position on the lens plane (an example of how these three parameters define microlensing populations which in turn produce corresponding magnification maps and microlensing observables is shown in \figref{fig5:configuration}).

\begin{figure}
    \centering
    \includegraphics[width=0.44\textwidth]{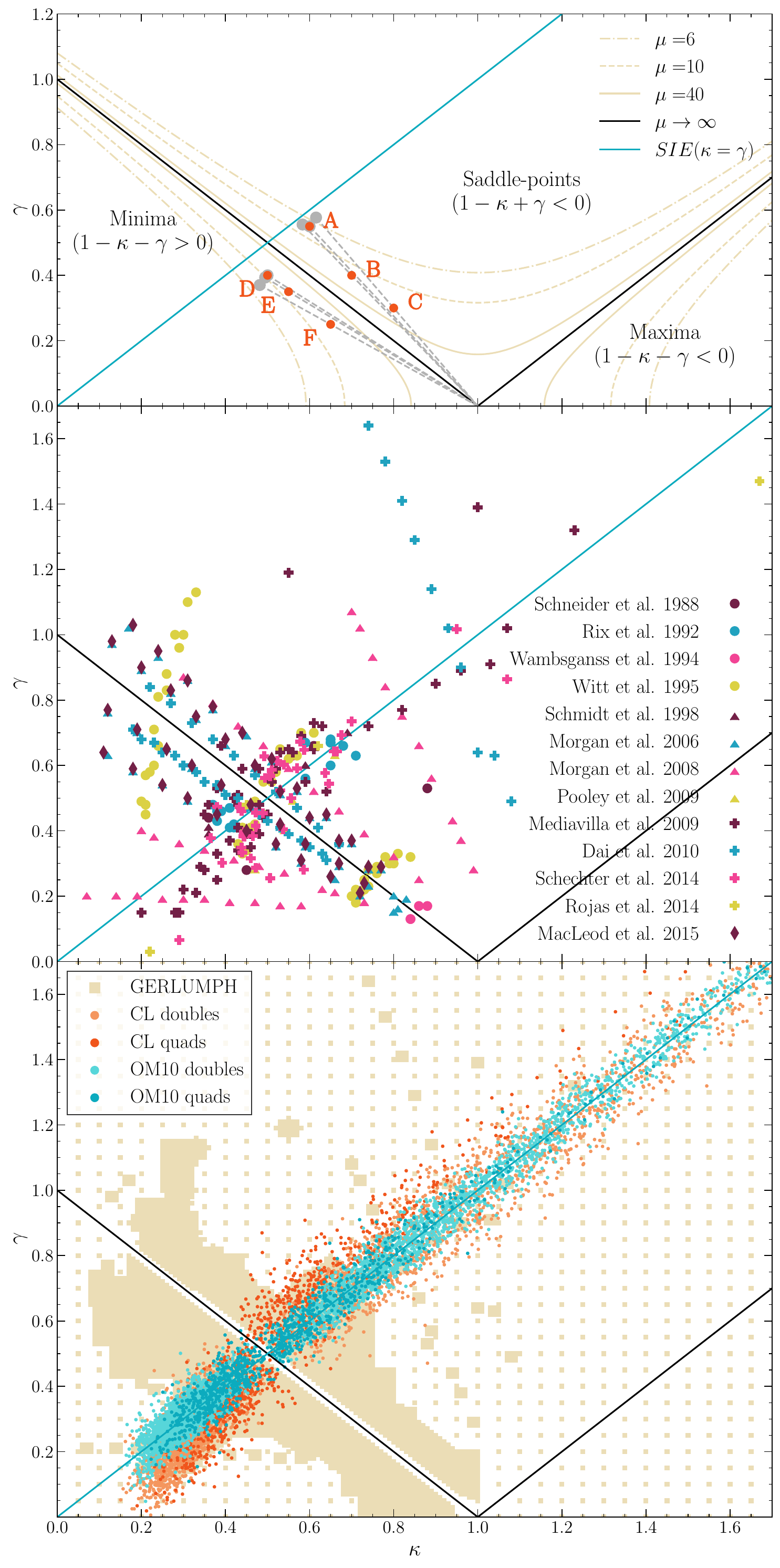}
    \caption{Theory (top), macromodels (middle), and simulations (bottom) shown in the $\kappa - \gamma$ parameter space. The critical line, i.e. where the magnification given by \eqref{eq5:macro_magnification} diverges dividing the parameter space to the minimum, saddle-point, and maximum image regions in direct correspondence to the macromodel multiple images, and the $\kappa=\gamma$ line, i.e. the locus of a Singular Isothermal Ellipsoid (SIE) macromodel, are shown in all panels for reference. \textbf{Top:} Selected locations with $\kappac=0$ (orange points) have the same $\kappaeff,\gammaeff$ with any other location along the lines to (1,0) (grey dashed lines) with $\kappac$ given by \eqref{eq5:effective_trans}. Because locations B,C and E,F lie very close to the lines emanating from A and D respectively, they have very similar $\kappaeff,\gammaeff$ values (grey points) for suitable $\kappac$ (this is illustrated in \figref{fig5:equivalent_maps} in more detail). Magnification contours from \eqref{eq5:macro_magnification} are also indicated. \textbf{Middle:} Selected macromodels from the literature - see \url{https://gerlumph.swin.edu.au/macromodels/} for an interactive version of this plot. \textbf{Bottom:} Two simulated populations of lensed quasars are shown, one from \citet[][]{bib5:Oguri2010} and an identical one generated with lens potential ellipticity and external shear distributions with lower and higher mean values respectively \citep[0.15 and 0.1 as opposed to 0.3 and 0.05, based on][]{bib5:Luhtaru2021}. The footprint of the GERLUMPH database is also shown.
    }
    \label{fig5:pspace}
\end{figure}

Figure \ref{fig5:map_tiles} demonstrates the effect of different macro-parameters on the magnification maps in the case where the whole mass is in the form of microlenses, i.e. $\kappac=0$ in \eqref{eq5:microlensing_equation_Einstein_radius}.
The density of the caustics is proportional to the total magnification (and consequently the number of microlenses from \eqref{eq5:magnification_sum}), whose contours are shown in the top panel of \figref{fig5:pspace}.
The effect of the shear is also striking, stretching the caustics and producing stratified maps with elongated ``filaments'' of caustics interspaced with demagnification troughs.
Interestingly, for super-critical regions where $\kappa>1$ (these correspond to highly demagnified saddle-point or maximum images), the anisotropic magnification (stretching) due to the shear changes direction by $\pi/2$ due to the $1-\kappa$ term being negative (see Eq. 76 of \chapintro).
This is clearly seen as the horizontal caustic structures for $\kappa<1$ become vertical for $\kappa>1$.

These diverse caustic structures have an impact on the microlensing effect present in observations, i.e. through light curves and magnification probability distributions of flux ratios (see the next section).
\citet{bib5:Vernardos2013} have explored the latter case, producing a similar tiling of the parameter space as \figref{fig5:map_tiles} but with magnification distributions (see their fig. 4). 
In summary, they found that along the critical curve (where magnification diverges), the distributions are centered around the macro-magnification, extend in a narrow area around it, $0.2 < \mathrm{log} (\mu/\mu_{\rm macro}) < 5$, where $\mu_{\rm macro}$ is given by \eqref{eq5:macro_magnification}, and appear symmetric (in log space).
Away from the critical curve, the distributions generally become wider and asymmetric, mostly due to the appearance of a high magnification tail and/or a secondary peak.

\begin{figure*}
    \centering
    \includegraphics[width=\textwidth]{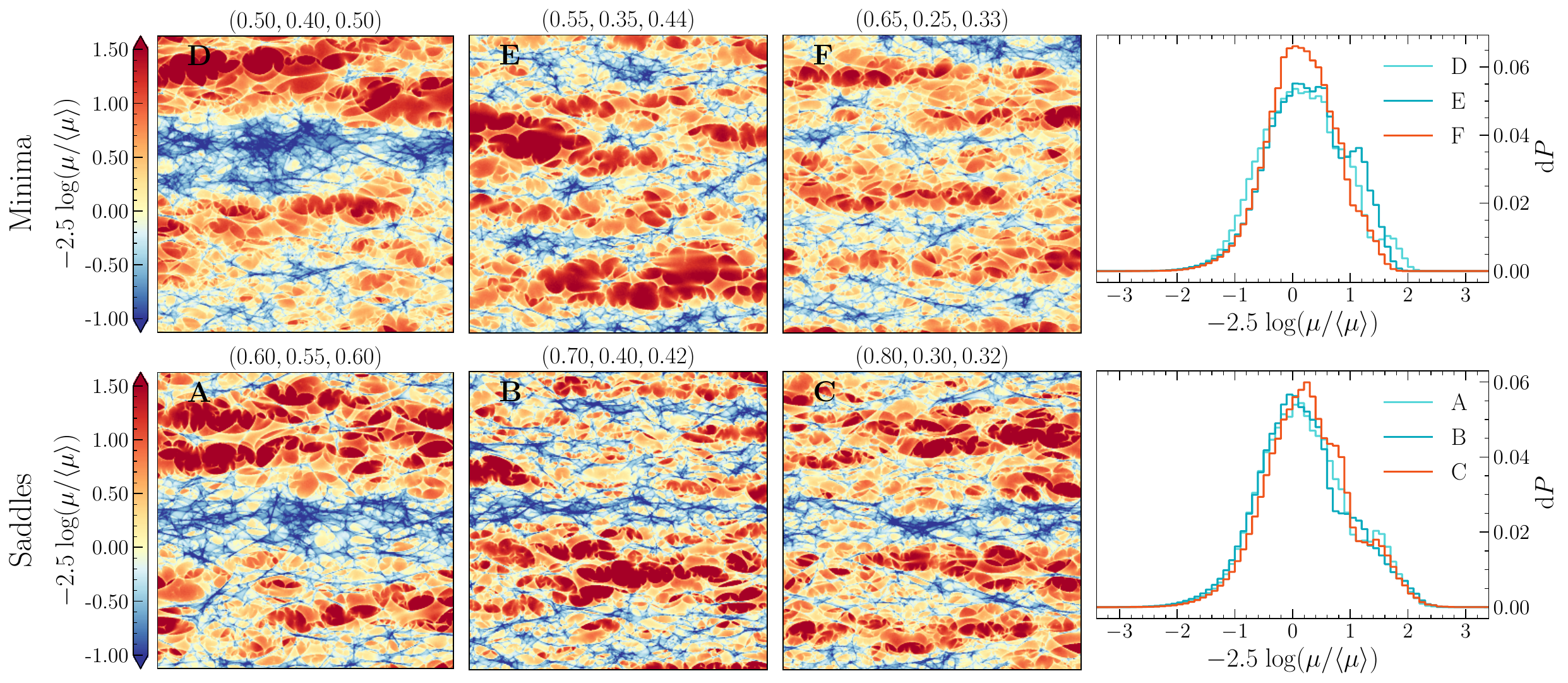}
    \caption{Magnification maps and histograms corresponding to the locations marked on the top panel of \figref{fig5:pspace}. The $\kappa,\gamma, \kappac$ values are given above each map and correspond to $\kappaeff,\gammaeff = (0.6\pm0.02,0.55\pm0.02)$ and $(0.5\pm0.02,0.4\pm0.02)$ for the saddle-points and minima respectively (grey points on the top panel of \figref{fig5:pspace}). Although the $\kappa,\gamma, \kappac$ of these maps are very different, their magnification histograms are very similar, making them `equivalent' maps \citep[per image type, see][]{bib5:Vernardos2014a}. The minor differences between the histograms can be attributed to the sample variance of examining a finite source plane region. The same cannot be readily said for light curves resulting from these maps.}
    \label{fig5:equivalent_maps}     
\end{figure*}

Including an additional convergence component due to smooth matter has fundamental implications that trace back to the mass-sheet degeneracy (\chapintro).
The effect is two-fold: on one hand, the number of microlenses is reduced since now there is less mass in the form of compact objects (see \eqref{eq5:stellar_kappa}), while on the other hand, the resulting caustic networks, which are now less dense, are isotropically magnified by the uniform mass-sheet described by $\kappac$ (see \eqref{eq5:microlensing_equation_Einstein_radius}).
An example of such maps is shown in \figref{fig5:equivalent_maps}.

Figure \ref{fig5:map_tiles} is actually showing one slice of a three dimensional parameter space - the third dimension being $\kappastar$.
As \citet{bib5:Paczynski1986} has shown, the $\kappa,\gamma,\kappastar$ parameter space can be transformed to an equivalent, two-dimensional parameter space of effective convergence and shear:
\begin{equation}
    \label{eq5:effective_trans}
    \kappaeff = \frac{\kappastar}{1-\kappa+\kappastar}, \quad \gammaeff = \frac{\gamma}{1-\kappa+\kappastar}.
\end{equation}
In the above equation, $\kappaeff$ is entirely due to compact matter.
The effect of the above transformation can be understood in the following way, as also illustrated in the top panel of \figref{fig5:equivalent_maps}.
Starting from a given location $\kappa_0,\gamma_0$ with $\kappac=0$ (locations A and D), maps with $\kappac>0$ lie in the third dimension in the traditional $\kappa,\gamma,\kappastar$ parameter space.
But in the effective space they lie along straight lines defined by $\kappa_0,\gamma_0$ and (1,0).
In terms of the magnification distribution, the above mass-sheet transformation leads to statistically equivalent maps \citep{bib5:Vernardos2014a}, i.e. we expect the same microlensing deviations from the macro-magnification, as shown in \figref{fig5:equivalent_maps}. Minor differences arise due to sample variance from using a finite field of stars to generate maps for a finite source plane region. Such differences can be minimized by, e.g., averaging over multiple maps or using a larger region in the source plane.
In terms of light curves, however, there are other slight differences: we expect less peaks because there are fewer caustics to cross, but also longer intervals for the variability because caustics cover now a bigger area on the source plane.

\subsection{Quasar structure overview}
\label{sec5:quasar_structure_overview}
Because the microlensing signal depends on the size and structure of the lensed source, we aim in this section at providing an overview of the characteristics of the AGN emission regions over the electromagnetic spectrum, e.g. shape and size of the X-ray corona, size and temperature profile of the accretion disc, geometry of the broad line region, etc.
The generally accepted structure of a quasar has been pieced together by painstaking work over many decades to account for all of the various emission and absorption features that have been seen across the electromagnetic spectrum.  
Figure~\ref{fig5:sketch} shows a schematic representation of the different components. 

A scale commonly used as ``unit length'' when it comes to describing inner region of AGN is the gravitational radius $\rg$, also called Schwarzschild radius: 
\begin{equation}
\rg = \frac{2GM_{\rm BH}}{c^2},
\label{eq5:gravitational_radius}
\end{equation}
where $M_{\rm BH}$ is the mass of the black hole, and $G$ and $c$ are the gravitational constant and speed of light.
This radius defines the event horizon of a (non-spinning) black hole.
For a typical mass of $M_{\rm BH}=10^8 \, M_{\odot}$, we have $\rg \sim 3 \times 10^8$ km that corresponds to $\sim 2$ AU, $10^{-2}$ lt-days, or $\sim 10^{-5}$ pc.
In the perspective of microlensing applications, this should be compared to the typical Einstein radius of a microlens in the source plane \citep[taken as the average from ][see also  \eqref{eq5:einstein_radius_source}]{bib5:Mosquera2011b}, {$\re = 2.5 \times 10^{16} \sqrt{ \langle M \rangle/0.3 \mathrm{M}_\odot}$\,cm $\sim $ 9.7 $\sqrt{\langle M\rangle/0.3 \mathrm{M}_\odot} $\,l-d $\sim 840 \, \sqrt{\langle M\rangle/0.3 \mathrm{M}_\odot} \; \rg$ (for $M_{\rm BH}=10^8 \, \mathrm{M}_{\odot}$).
Finally, another quantity of interest from general relativity typically associated with the inner edge of the disc is the radius of the innermost stable circular orbit of a photon, $R_{\rm ISCO} = \alpha \rg$, where $\alpha=3$ for a non-rotating black hole and $9/2$ ($1/2$) for a maximally pro- (retro-) grade spinning one.

Our discussion will ``follow the photons'' from their production in the accretion disc through their encounters with other material in the quasar on their way to the observer.  
The aim is to help the reader associate features in the electromagnetic radiation with physical structures in the quasar.
We do not aim to give a complete description of every aspect of quasar emission; instead we focus on those most relevant to microlensing.

The hot material in the accretion disc produces an abundance of optical/UV photons. 
Some of those encounter a population of high-energy electrons in the innermost regions near the black hole and are inverse-Compton upscattered to X-ray energies; this is the ``primary'' X-ray emission. 
Some of this energy will reach the observer directly and result in an X-ray continuum spectrum, while the rest of the emission from this X-ray corona will interact with other parts of the quasar and give rise to reflection features, the most prominent of which being the Fe K$\alpha$ emission line at 6.4 keV. 
This line can have two components: a broad component that comes from the inner part of the accretion disc, and a narrow component that comes from material farther out (i.e., the broad-line region or torus).

The matter accretion onto the central supermassive black hole can reach a rate of a few solar masses per year and radiate with an efficiency ranging between 5.7\% in the non spinning case and 43\% for high spin \citep{bib5:Thorne1974}, much larger than nuclear fusion ($\sim 0.8$\%). 
This energy release causes the accretion disc to heat up and emit a power-law continuum radiation from UV to optical ranges. 
This emission arises from the inner part of the disc that corresponds to distances from the central black hole ranging from several to tens of astronomical units. 
The high energy continuum from the inner disc ionizes the gas in the surrounding area, producing broad emission lines characteristic of quasar optical spectra (Fig.~\ref{fig5:sketch}). 

Based on the variety of observed line widths and ratios, AGN were originally classified into various types and subtypes.
In the nineties, a unification scheme based on orientation has been proposed to explain this phenomenology in a coherent, geometric way \citep{bib5:Antonucci1993,bib5:Urry1995}. 
The model postulates that obscuring material in the equatorial plane (assumed to be a dust torus\footnote{The validity of this paradigm, with dust being in a torus in the equatorial plane, is currently under scrutiny by the community as interferometric observations of dust in AGN are also consistent with a polar emission.}) and orientation with respect to the line of sight are causing this diversity. 

There is growing evidence that the physical properties of AGN also play a role in their spectral appearance. 
In particular, variation in the accretion rate, which is suspected to be related to the launch of radio jets\footnote{Only approximately 10\% of AGN are radio-loud and show synchrotron emission as evidence for a jet.
Other AGN, however, are not radio-silent, but the origin of their radio emission is not yet fully established.} \citep{bib5:Laor2008, bib5:White2015}, also impacts the spectral appearance of AGN \citep{bib5:Marziani2001,bib5:Shen2014,bib5:Elitzur2014}. 

The following subsections describe more quantitatively our knowledge of the AGN emission regions that are the most relevant for microlensing studies, as shown in \figref{fig5:sketch}.
As direct imaging is not possible, we briefly explain the methods used to build up our understanding of the unified model of AGN structure.
Our journey into the heart of AGNs starts with the most compact regions, which are also the most susceptible to microlensing, and ends at the interface between the AGN and its host galaxy, namely the torus and the narrow line region. 

\subsubsection{The most compact emission}
\label{sec5:Xray_overview}
After the first X-ray satellites were launched in the 1970s (Uhuru, Ariel 5, SAS-3, OSO-8), the X-ray spectral properties of quasars and (suspected) black hole X-ray binaries became known, and much work was done to understand their features.
In particular, the X-ray continuum power-law spectrum seen in quasars and the high energy emission of black hole X-ray binaries was explained as the Compton upscattering (also called ``inverse Compton scattering'' or just ``Compton scattering'' in the literature) of lower energy photons by a population of hot, energetic electrons \citep[e.g.][]{bib5:Shapiro1976, bib5:Katz1976, bib5:Pozdnyakov1976, bib5:Galeev1979, bib5:Sunyaev1979, bib5:Sunyaev1980}.
As more and higher quality observations of quasars and Seyfert galaxies were obtained by the HEAO I and then EXOSAT observatories, the remarkable similarity of the X-ray spectra \citep[a single power-law continuum with spectral index of $\sim$0.7][]{bib5:Rothschild1983, bib5:Mushotzky1984, bib5:Turner1989} suggested a common origin.
After the Ginga satellite detected the Fe emission line and a broad hump reflection feature \citep{bib5:Rothschild1983} that were predicted by \citet{bib5:Guilbert1988} and \citet{bib5:Lightman1988, bib5:Mushotzky1984, bib5:Turner1989}, a ``two-phase'' model AGN was proposed by \citet{bib5:Haardt1991, bib5:Haardt1993} in which a hot, tenuous corona exists above the accretion disc and Compton upscatters UV and optical photons from the disc to X-ray energies with a power-law continuum spectral shape.
The corona is thought to be supplied with energetic particles accelerated by the magnetic fields anchored in the accretion disc \citep[e.g.][]{bib5:Haardt1994, bib5:Dimatteo1998,bib5:Merloni2001}.
For decades, the geometry and extent of the corona was largely unknown, with arguments that  it may be patchy, extend over parts of the inner disc, and vary with time \citep[e.g.][]{bib5:Gallo2015,bib5:Wilkins2015}.
However, spectral and timing studies suggest a compact, centrally located corona \citep[e.g.][]{bib5:Brenneman2006, bib5:Fabian2009, bib5:Fabian2013,bib5:Parker2015}.  Microlensing has strongly constrained the X-ray emitting corona to be compact.

\subsubsection{The accretion disc}
\label{sec5:qso_accretion_disc}
The powerhouse of AGN emission originates from within 10-1000 $\rg$  (i.e. 0.1-10 lt-days for a black hole mass $M_{BH} \sim 10^8 M_{\odot}$) of the central supermassive black hole.
This structure is thought to be a geometrically thin (in the vertical direction) but optically thick disc that is heated locally by the dissipation of gravitational binding energy through accretion of matter \citep[the ``thin-disc'' model, ][]{bib5:LyndenBell1969,bib5:Shakura1973}. 
Ignoring relativistic effects, the temperature profile for such a thin-disc can be expressed as \citep[e.g.][]{bib5:Zdziarski2022}: 
\begin{equation}
T(R) \propto R^{-\beta} \left( 1 - \sqrt{\frac{R_{\rm in}}{R}} \right)^{1/4}, 
\label{eq5:temperature_AD}
\end{equation}
where $R_{\rm in}$ is the inner edge of the disc, and $\beta = 3/4$.
The inner edge effects are commonly ignored \citep[see][regarding the impact of disc truncation and wind]{bib5:Zdziarski2022}, and the dependence $T\propto R^{-\beta}$ is often kept as a first order generalisation of the thin-disc.

For this model, the radius at which the temperature coincides with the rest wavelength of the observations ($k T = h_p c/ \lambda_{\rm rest}$) is:
\begin{equation}
    R_{\lambda}^{\rm{flux}} \simeq \frac{3.4\times10^{15}}{\sqrt{\cos i}} \frac{D_s}{r_H} \left( \frac{\lambda}{\mu \rm m} \right)^{3/2} \left( \frac{\rm{zpt}}{3631 \, \rm Jy} \right)^{1/2} 10^{-0.2 (m-19)} h^{-1} \, \rm cm,
    \label{eq5:disc_radius_from_flux}
\end{equation}
where $r_H$ is the Hubble radius, $i$ is the disc inclination angle, $\rm{zpt}$ is the zero point of the AB magnitude system\footnote{This can be modified to use another wavelength, for instance, for HST F814W filter $\lambda=0.814 \, \mu \rm m$ and $\rm{zpt}=2409 \, \rm Jy$.}, and $m$ is the intrinsic magnitude of the source \citep{bib5:Morgan2010}.
A handy quantity for comparison with microlensing studies is the half-light radius of the disc.
\cite{bib5:Cornachione2020a} refers to the latter as ``luminosity size'', and its expression, based on the observed specific flux per wavelength (in a particular band), $F_{\lambda, \rm{obs}}$, is given by: 
\begin{align}
R_{\lambda, 1/2}^{\rm{flux}} & = C(\beta) \, R_{\lambda}^{\rm{flux}}  \nonumber \\
  & = C(\beta) \left( \frac{\lambda^5_{\rm{obs}} D^2_L  F_{\lambda, \rm{obs}}}{4\pi h c^2 \cos(i)(1+z)^4} \right)^{1/2}  \times \left({\int_{u_{\rm in}}^{\infty} \frac{u du}{\exp{(u^\beta)} -1}} \right)^{-1/2},      
\label{eq5:luminosity_size}
\end{align}
where $D_L$ is the luminosity distance to the quasar, $u = R / r_\lambda$, and the factor $\cos(i)$ accounts for disc inclination.
The factor $C(\beta)$ is required to transform the radius $R_\lambda^{\rm{flux}}$ into a half-light radius and is given in \cite{bib5:Li2019} as a numerical solution to  
\begin{equation}
   \int_{u_{\rm in}}^{C(\beta)} \frac{u du}{\exp{(u^\beta)} -1}  = \frac{1}{2} \int_{u_{\rm in}}^{\infty} \frac{u du}{\exp{(u^\beta)} -1}.      
   \label{eq5:Cbeta}
\end{equation}
For the standard disc, we have $C(\beta=3/4) = 2.44$.
On the other hand, knowing the mass of the central black hole, $M_{\rm{BH}}$, and the luminosity, $L$, the disc half-light radius\footnote{We have converted the commonly quoted disc size $R_{\lambda,1/2}^{\rm{BH}}$ into a half-light radius by multiplying it by 2.44 (\eqref{eq5:Cbeta}).} can be calculated as:
\begin{equation}
    R_{\lambda, 1/2}^{\rm{BH}}=2.37\times10^{16} \left( \frac{\lambda_{\rm rest}}{\mu \mathrm{m}} \right)^{4/3} \left( \frac{M_{\rm BH}}{10^9 \mathrm{M}_{\odot}} \right)^{2/3} \left( \frac{L}{\eta L_{\rm E}} \right)^{1/3} \, \mathrm{cm},
    \label{eq5:radius_MBH}
\end{equation}
where $L_E$ is the Eddington luminosity and $\eta$ is the accretion efficiency \citep[][]{bib5:Marconi2003,bib5:Graham2007,bib5:Graham2016}.
Typical values for these parameters are $L/L_{\rm E} \sim 1/3$ and $\eta=0.1$ \citep{bib5:Kollmeier2006,bib5:Shen2008,bib5:Edelson2015}.
Although both methods assume the same model, $R_{\lambda, 1/2}^{\rm{flux}}$ estimations are a factor of $\sim2-3$ smaller than $R_{\lambda, 1/2}^{\rm{BH}}$ \citep{bib5:Collin2002,bib5:Pooley2007,bib5:Morgan2010} and cannot be reconciled neither through uncertainties in the measured black hole mass, nor, in the case of strongly lensed quasars, through lensing magnification \citep[e.g.][]{bib5:Mosquera2011a}. 

Extended versions of the standard thin-disc model have been formulated \cite[see e.g.][]{bib5:Abramowicz2013, bib5:Middleton2015, bib5:Lasota2015} that include general relativistic corrections, radiative transfer in the disc atmosphere, black hole spin, or disc winds \citep{bib5:Novikov1973, bib5:Thorne1974, bib5:Hubeny2000, bib5:Sadowski2011, bib5:Davis2011}. 
While the standard accretion disc remains the dominant paradigm, a growing number of observations challenge at least its universality.
A non exhaustive list of alternative models that have been proposed includes: advection dominated accretion discs \citep{bib5:Ichimaru1977, bib5:Narayan1994}, slim discs \citep{bib5:Abramowicz1988}, inhomogeneous accretion \citep{bib5:Dexter2011}, magnetically arrested discs \citep{bib5:Zamaninasab2014}, discs with modified viscosity laws to account for magnetization \citep{bib5:Grzedzielski2017}, torn discs \citep{bib5:Nixon2012, bib5:Hall2014}, and puffy discs \cite{bib5:Lancova2019, bib5:Wielgus2022}. 

Constraints on the accretion disc temperature profile have been achieved over the last decade by continuum reverberation mapping. 
This technique consists in measuring how the UV-optical continuum responds to variations of the X-ray emission. 
Measuring the time lag of this response, $\tau$, in multiple bands can be translated into an increase in size of the emission region with wavelength.
The data accumulated for a growing number of AGN indicate that the disc size varies as expected for the thin-disc model of \cite{bib5:Shakura1973}.
Uncertainties on the slope of the temperature profile are, however, still too large to robustly rule out alternatives. 
In addition, the disc size is found to be larger than predicted by the model~ \citep[e.g.][]{bib5:Edelson2015, bib5:Cackett2020, bib5:Guo2022}. 
Diffuse continuum emission, originating from the inner part of the BLR \citep[such as Balmer and Paschen emission; ][]{bib5:Korista1995, bib5:Korista2001, bib5:Gardner2017} may explain the excess UV-optical size in several reverberation mapped systems, but other sources of diffuse emission (e.g. scattering, thick gas clouds, etc) may also be present in AGN \citep[e.g.][]{bib5:Lawrence2012, bib5:Lawther2023}.
All in all, our theoretical understanding of accretion discs is under tension.
On the one hand, theoretical developments indicate that the accretion disc should not be universally described by a thin-disc model.
On the other hand, the thin-disc cannot explain all the observations: the size is found to be larger than expected but its dependence on wavelength follows the $R \propto \lambda^{4/3}$ relation from the standard model.
Unfortunately, firm conclusions are hard to draw due to the blending of the disc emission with diffuse pseudo-continuum emission of debated nature.

\subsubsection{Intermediate sizes: the broad line region} 
\label{sec5:BLR}
One of the most salient features detected in UV-optical quasar spectra are broad emission lines. 
They arise mostly from hydrogen and helium recombination lines, but also permitted and semi-forbidden lines, such as \ion{C}{iv}, and \ion{C}{iii]}, and also complex multiplets from \ion{Fe}{ii}.
These lines, observed over 7 orders of magnitude in AGN luminosities, arise from the Broad Line Region, BLR, a region of radius ranging from $\sim$ 4 up to 100 times larger than the accretion disc, and possibly originating in the form of a clumpy wind launched from the latter \citep[see \figref{fig5:sketch} and ][]{bib5:Elitzur2014, bib5:Czerny2017}. 
The BLR is an important probe of the physical conditions (e.g. gas density, hydrogen column density, metallicity, ...) prevailing in the direct vicinity of the black hole. 
It is composed of a large number of clouds or of an inhomogeneous/clumpy wind of gas, photoionized by the continuum emission.
As is the case for the accretion disc, the BLR is also too compact to be spatially resolved with current telescopes. 
Therefore, our knowledge of the geometry (e.g. disc-like, bi-conic, bowl-like) and kinematics (Keplerian rotation, inflow, outflow) of BLR gas is yet limited despite of more than fifty years of investigation. 
The physical conditions in the BLR are explored through advanced photoionization codes, such as the state-of-the-art code CLOUDY \citep{bib5:Ferland2013}.
This code enables one to reproduce most of the line ratios observed in AGN and has improved our understanding of the chemical evolution of AGN host galaxies \citep{bib5:Nagao2006}. 
Some challenges remain, however, such as reproducing the iron multiplet or the ratio between optical and UV \ion{Fe}{ii} emission, which shows a large scatter at the population level \citep[][]{bib5:Ferland2009, bib5:Sarkar2021}. 
Our inability to accurately reproduce this major component of AGN spectra is likely linked to uncertainties on the distribution and kinematics of the iron emission in the BLR \citep{bib5:Ferland2009}.

Most of our knowledge on the structure of the BLR comes from the reverberation mapping technique \citep{bib5:Blandford1982}. 
This technique has enabled the measurement of the luminosity-weighted distance, $R_{\rm BLR}$, of the BLR from the continuum for the H$\beta$ line in about 70 local AGN \citep{bib5:Peterson1985, bib5:Horne1991, bib5:Kaspi2000, bib5:Bentz2013, bib5:Bentz2015}. 
This distance scales with the square root of the quasar luminosity, confirming that the BLR gas is mostly photoionized. 
Time-lags for UV emission lines have been harder to obtain. 
The \ion{Mg}{ii} lags have been measured for about 50 AGNs and the luminosity-size relation agrees with the one obtained for H$\beta$ \citep{bib5:Homayouni2020, bib5:Yu2022}. 
Time lags for higher ionization lines, such as \ion{C}{iv} or \ion{He}{ii} have been measured for about 50 systems \citep{bib5:Peterson2005, bib5:Kaspi2007, bib5:Grier2019} and found to be systematically shorter than for H$\beta$, indicating a stratification of the BLR (i.e. high ionization gas is found closer to the continuum).
The line broadening of the (optical) \ion{Fe}{ii}, as well as reverberation mapping data of a small sample of local Seyfert, indicate that this blend arises from a region at least as large as the Balmer BLR \citep[e.g.][]{bib5:Barth2013, bib5:Hu2015}.

Information on the geometry and kinematics of the BLR is more difficult to achieve. 
Direct modeling of the lines using radiative transfer codes can successfully reproduce their shapes \citep{bib5:Murray1997, bib5:Borguet2010, bib5:Higginbottom2014}.
However, the same line shape can be reproduced by a variety of geometries and kinematics of the emitting region, limiting the usefulness of this approach in constraining BLR structure.
Spectro-polarimetry of emission lines has been a useful complementary probe, revealing, for instance, the rotating disc-structure of H$\alpha$ emission in some Seyfert galaxies \citep{bib5:Smith2005}.
But such measurements are limited to significantly polarized AGN, and results arguably depend on the (poorly known) location of the scattering region at the origin of the polarization.
Velocity-resolved delay maps, i.e. measurements of the reverberation time-lag for various velocity slices in the line, are another indirect probe of BLR structure \citep{bib5:Horne1991, bib5:Peterson1999}. 
Results obtained for a few dozens of systems are often difficult to interpret due to the unknown transfer function that encodes the time-delay distribution across the broad line as a function of the line-of-sight velocity \citep[][and references therein]{bib5:Villafana2022}.
The more direct method developed by \cite{bib5:Pancoast2011} does not require knowledge of the transfer function but has other limitations, especially in presence of complex variability features \citep[see e.g.][]{bib5:Li2013, bib5:Pancoast2014, bib5:Bentz2021}. 
Overall, the data indicate a disc-like geometry for the region emitting the H$\beta$ line (which is the best studied line) but suggest a surprising diversity of BLR kinematics, including Keplerian discs, inflows and outflows.

\subsubsection{Further out: The torus, NLR, and the radio domain}
\label{sec5:torus_NLR_radio}
Above $\sim1.5 \mu\rm{m}$ (rest-frame), the contribution of the accretion disc to the continuum emission becomes subdominant compared to the emission from the dust.
%(Fig.~\ref{fig5:sketch}). 
Observations of dozens of local AGN indicate that the near/mid-infrared dust emission arises from two regions: a ``compact'' region characterised by hot/warm dust temperature ($T \sim 1400$\,K), and a colder component with $T \sim 300$\,K that may be 100 to 1000 times more extended than the hotter emission \citep{bib5:Kishimoto2011a}. 
Interferometric studies of local AGN suggest that the ratio of effective areas between these two components is of the order of 400, such that cold emission dominates above typically $7\,\mu$m, while hot emission peaks around $2\,\mu$m \citep{bib5:Kishimoto2011b}. 
The latter component is the only one compact enough to be susceptible to microlensing. 
Dust reverberation mapping studies show that the $K-$band reverberation radius (dominated by dust emission) scales with the square root of the luminosity, establishing a solid dependence of the dust emission on AGN luminosity \citep[e.g.][]{bib5:Suganuma2006, bib5:Koshida2014, bib5:Yang2020}. 
This reverberation radius can be used as the foundation of a ring-like model of the torus by fitting it as a function of luminosity.
\citet{bib5:Kishimoto2007} derived the scaling of the inner radius of the torus as a function of the luminosity in the V-band, i.e. $\nu L_\nu(5500\,\text{\AA})$, as:
\begin{equation}
    R_{\rm in} = 0.47 \left( \frac{ 6 \nu L_\nu(5500\,\text{\AA})}{10^{46} \rm {erg /s}} \right) \rm{pc}
\end{equation}
The remaining model components are the surface brightness and outer radius.
The surface brightness is observationally unknown but theoretical arguments favour a power-law decrease \citep{bib5:Barvainis1987}.
Interferometric data suggest that $R_{\rm out}/R_{\rm in} \leq 2$ at 2.2\,$\mu$m and reach a factor of several at longer wavelengths \citep{bib5:Kishimoto2007}. 

Finally, we may stress that the geometry and clumpiness of the dust emitting region is yet debated.
A torus-like structure has been favoured for decades as it provides a strong support to the orientation-based AGN manifestations within the framework of the unified model. 
However, a disc-like ring with dust winds launched from it is suggested in some systems instead of a torus \citep[e.g.][]{bib5:gravity2020}.
Overall, a lot of open questions remain regarding the dust emitting region.
Depending of the AGN luminosity and torus properties, the near-/mid-infrared AGN emission region can be small enough to be slightly affected by microlensing.   

The narrow emission lines (with FWHM $\leq 800\kms$) commonly detected in quasar spectra either arise from the AGN host galaxy or from the Narrow Line Region (NLR). 
The gas in the NLR is photo-ionised, while narrow emission in the host can be associated, for example, with star formation and may therefore appear to be spatially offset with respect to the NLR (and possibly spatially resolved in high resolution images). 
The gas in the NLR is exterior to the torus, reaching tens to thousands of parsecs (depending on the luminosity), and can also be spatially resolved \citep{bib5:Bennert2002, bib5:Dempsey2018}. 
The frequent asymmetry in the shape of narrow lines, in particular of [\ion{O}{iii}]$\sim\lambda\lambda \, 4959,5007\,\text{\AA}$, indicates the common presence of outflowing material that can regulate stellar activity in the host \citep[e.g.][]{bib5:Speranza2021}. 
Due to these characteristics, the NLR is expected to be too large to undergo any microlensing. 

AGN emission at radio wavelengths reveals a dichotomy whose origin (and even existence) is still debated. 
About 15\% of AGN are radio-loud, while the remaining are qualified as radio-quiet. 
The emission from radio-loud quasars is mostly due to synchrotron emission. 
Interferometric radio data often reveal an unresolved core (at sub-parsec scales) and multiple components of a jet that sometimes extends over kiloparsecs. 
The strongest radio emission is often observed in blazars and flat spectrum systems, whose (often relativistic) jet is aligned to our line-of-sight.
Quasars classified as radio-quiet can still have some energy emitted in the radio.
Thermal free-free emission (i.e. bremsstrahlung) from either a stellar or AGN component is possible.
Additionally, magnetic heating of the corona, thin free-free emission from winds, or synchrotron emission from the base of a small jet or particles accelerated in shocks are also plausible mechanisms of this radio emission \citep[see e.g.][and reference therein]{bib5:Silpa2020}. 
The radio emission from AGN is commonly assumed to be insensitive to microlensing, but this complex picture of radio AGN does not fully imply this. 
The existence of radio variability on time scales of a few days \citep[e.g.][]{bib5:Biggs2018} indicates that regions compact enough to be microlensed should contribute a substantial fraction of the radio emission. 
Observation of microlensed radio-emission of AGN remains, however, elusive \citep{bib5:Koopmans2000, bib5:Biggs2023}.

\subsection{Microlensing and the quasar structure}
\label{sec5:ML_and_quasar_structure}
The main reason a detailed physical picture of AGN structure is not yet in place is our inability to spatially resolve their innermost regions. 
``Direct'' imaging of the vicinity of a supermassive black-hole has been possible only for a handful of nearby systems thanks to the Event Horizon Telescope, which required a tremendous technical and observational effort to turn our planet into a giant radio interferometer. 
This enabled the reconstruction of an image of the shadow of the black hole in M87$^\star$ and of the one lying in the center of our own Galaxy \citep{bib5:EHT2019, bib5:EHT2022}. 
While EHT has been used to observe some distant quasars ($z > 0.5$) at 20 $\mu$arcsec resolution at 230 GHz \citep{bib5:Jorstad2023}, we are still far from reaching such a spatial resolution for a sizeable sample of systems over the whole electromagnetic spectrum. 
Our most powerful near-infrared interferometers resolve AGN on scales of $\sim$ 1\,pc, corresponding to emission from the ``dust torus'' \citep{bib5:gravity2021}, but we are yet far from getting a comprehensive view of the innermost parsec. 

\begin{figure*}
	\centering
	\includegraphics[width=0.9\textwidth]{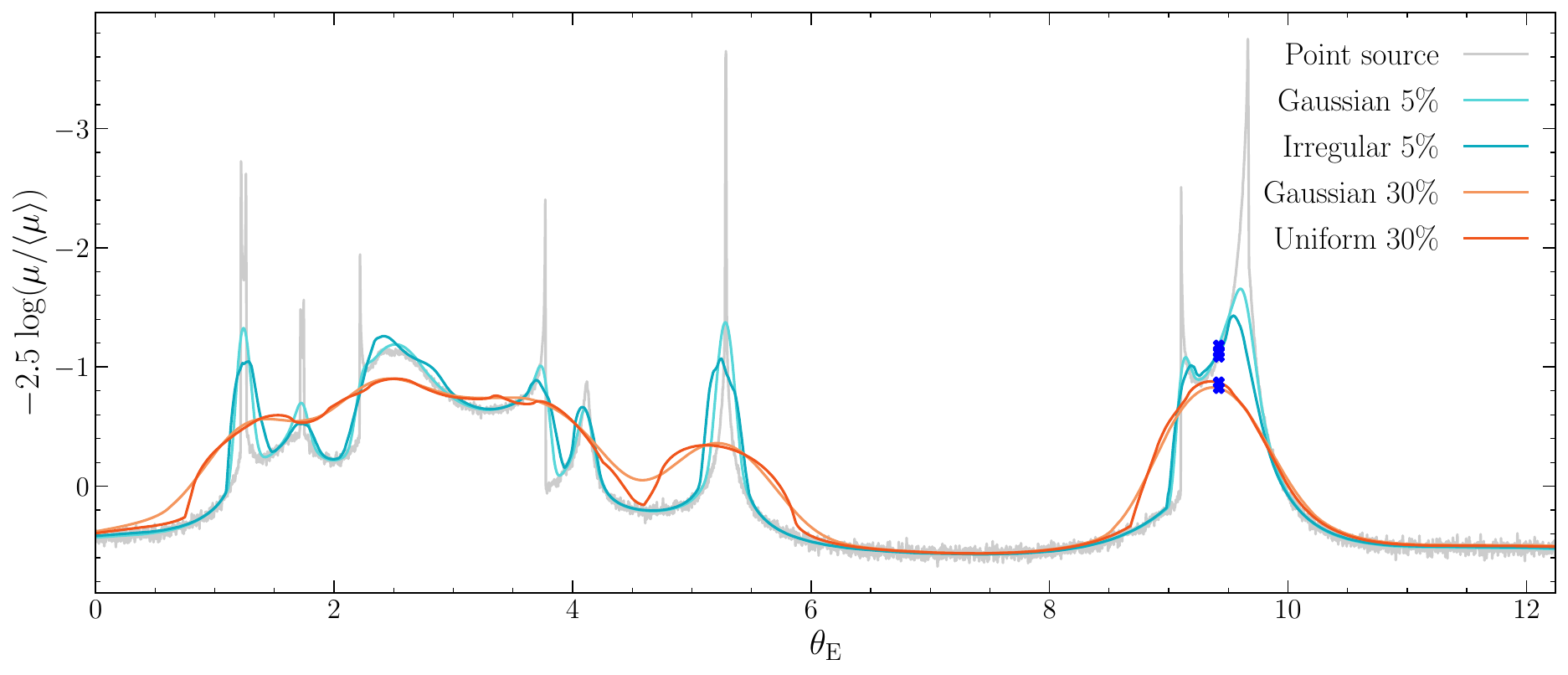}
	\caption{
		Light curves for sources of different size and shape moving along the trajectory shown in the right panel of \figref{fig5:microimages_loops} and in \figref{fig5:example_mag_map} (from bottom to top).
		When the point source (gray curve) enters and exits the areas bordered by caustic lines, characteristic double-horned profiles can be seen.
		Larger sources lead to smoother curves without the extremely magnified peaks, as it can be seen by the Gaussian-shaped sources with half-light radii at 5 (blue) and 30 (red) per cent of the Einstein radius.
		The shape of the source plays a secondary role and only modulates small scale features in the light curves, as demonstrated by a uniform and irregular disc (the latter being simply a greyscale, black background photo of the face of one of the authors) with the same half-light radii as the Gaussian profiles.
		Statistical errors (noise) introduced by the inverse ray-shooting technique in the magnification map pixels can be seen for the point source, which are averaged out as soon as the source has a finite size of just a few pixels.
		The source position marked with the blue cross in the right panel of Fig. \ref{fig5:microimages_loops} is also indicated here.
	}
	\label{fig5:microimages_light_curves}     
\end{figure*}

The fortunate match between the microlensing Einstein radius and the size of the otherwise unresolved AGN regions represents an opportunity to shed light on quasar structure. 
Microlensing (de)magnification directly scales with source size (see next section), turning variable signals in time and wavelength into an astrophysical ruler that enables a sensitive measurement of the heart of quasars at multiple scales. 
For instance, it is the only technique that enables a measurement of the size of the X-ray continuum emission. 
Contrary to reverberation mapping that gets hampered by relativistic time dilation at high redshift,, microlensing does not rely on observing the intrinsic quasar variability.
It is therefore particularly suitable to probe the size of distant quasar emitting regions, nicely complementing reverberation mapping measurements in the local Universe.
Differential microlensing between regions of different sizes also provides a tool to study finer properties, like the temperature profile of the disc or the geometry and kinematics of the BLR.
Here as well, microlensing nicely complements standard techniques (e.g. reverberation mapping, photoionization, line shape modeling) that rely on different working assumptions.  
At the scale of $\gtrsim 0.1 pc$, the radial structure of the hot torus may also potentially be constrained by microlensing (or its absence of). 
The possibility of using microlensing to zoom in the radio and sub-milimeter emission regions is less clear, but observations hint that microlensed effects in these wavelengths may not be immediately excluded. 
A presentation of the main results obtained with microlensing techniques is given in \secref{sec5:quasar_results}.

%%%%%%%%%%%%%%%%%%%%%%%%%%%%%%%%%%%%%%%%%%%%%%%%%%%%%%%%%%%%%%%%%%%%%%%%%%%%%%%%%%%%%%%%%%%%%%%%%%%%%%%%%%%%%%%%%%%%%%%%%%%%%%%%%%%%%%%%%%%%%%%%%%%%%%%%%%%%%%%%%%%%%%%%%%%%%%%%%%%%%%%%%%%%%%%%%%%%%%%%%%%%%%%%%%%%%%%%
\section{Methods}
\label{sec5:data_analysis}
Microlensing offers two main methods to probe quasar structure and the partition of matter in the lens.
The first one, known as the ``single-epoch method'' (described in \secref{sec5:single_epoch}), takes advantage of the differential microlensing occurring between regions of different sizes.
For instance, our basic understanding that inner (outer) parts of accretion discs are hotter (cooler) and hence emit in bluer (redder) wavelengths (see \figref{fig5:sketch} and Eqs. \ref{eq5:disc_radius_from_flux} and \ref{eq5:radius_MBH}) postulates that microlensing magnification measured at the same epoch will depend on wavelength.
As we can anticipate from the structure of AGNs (\secref{sec5:quasar_structure_overview}), the wavelength dependence of microlensing is not restricted to the disc.  
The second method, known as the ``light curve method'' (described in \secref{sec5:time_varying_ML}), analyses the amplitude and rate of the time variability of the microlensing effect.
The timescale of variability is shorter for the smaller regions, explaining why it is mostly applied to study the X-ray (corona) and optical continuum (disc) regions.
Other methods, generally sharing concepts with the single-epoch and light curve methods, exist, but they are usually tailored to a particular type of data or scenario.
A non-exhaustive overview of some selected such methods is given in \secref{sec5:exotic_methods}.

All these techniques require identifying the presence of microlensing in the data and measuring its amplitude.
In order to infer properties of the source or the lens, a forward modeling method is generally followed that simulates microlensing data and compares them to the observations. 
Section \ref{sec5:general_considerations} presents important aspects that need to be considered upon designing the simulations, while \secref{sec5:observational_considerations} explains how the microlensing signal can be extracted from the data in the most common cases.

\subsection{General modeling considerations}
\label{sec5:general_considerations}
Microlensing variability depends most crucially on the size of different emitting regions of the quasar with respect to the Einstein radius of the microlenses (see \eqref{eq5:einstein_radius_source}).
In order to simulate the magnification for any size and shape of the source one simply needs to convolve a magnification map with the source's brightness profile.
The magnification range produced by the stars in the lens galaxy increases as the source becomes more compact, which is imprinted on all kinds of microlensing data, i.e. flux ratios, light curves, spectra, and high magnification events (see \tabref{tab5:microlensing_variability}).
An example for the case of light curves is shown in \figref{fig5:microimages_light_curves}, where we see a clear dependence primarily on the size of the source - short dramatic changes and smooth extended variations for small and large sources respectively - and to second order on its shape.
Another example is illustrated in \figref{fig5:mpds_configuration}, where the magnification probability distribution is shown for a "point source" ($< 0.0025 \, \tae$, see below), and three Gaussian brightness profiles of increasing size.
In fact, it has been shown \citep{bib5:Mortonson2005,bib5:Vernardos2019a} that it is mainly the half-light radius of the source, $\rhalf$, that determines the extent of microlensing effects, while its detailed projected two-dimensional shape plays a secondary role (as long as $\rhalf$ is kept the same).
For this reason, a common choice is a Gaussian luminosity profile for the source ($\rhalf = 1.18 \sigma$).
In a similar way, a common choice of the dependence of size on wavelength, particularly for the accretion disc, is the parametric model:
\begin{equation}
    \rhalf = r_0 \left( \frac{\lambda_{\rm{rest}}}{\lambda_0} \right)^{1/\beta},
    \label{eq5:parametric-sizes}
\end{equation}
where $\lambda_0$ and $r_0$ are a reference wavelength and size, and $\beta$ is the slope of the temperature profile.
This allows for more flexibility, as opposed to, for example, the thin-disc model that has a fixed temperature slope of $\beta = 3/4$ (and depends on other physical parameters like the black hole mass, etc, see \eqref{eq5:radius_MBH}).

\begin{figure}
	\centering
	\includegraphics[width=0.5\textwidth]{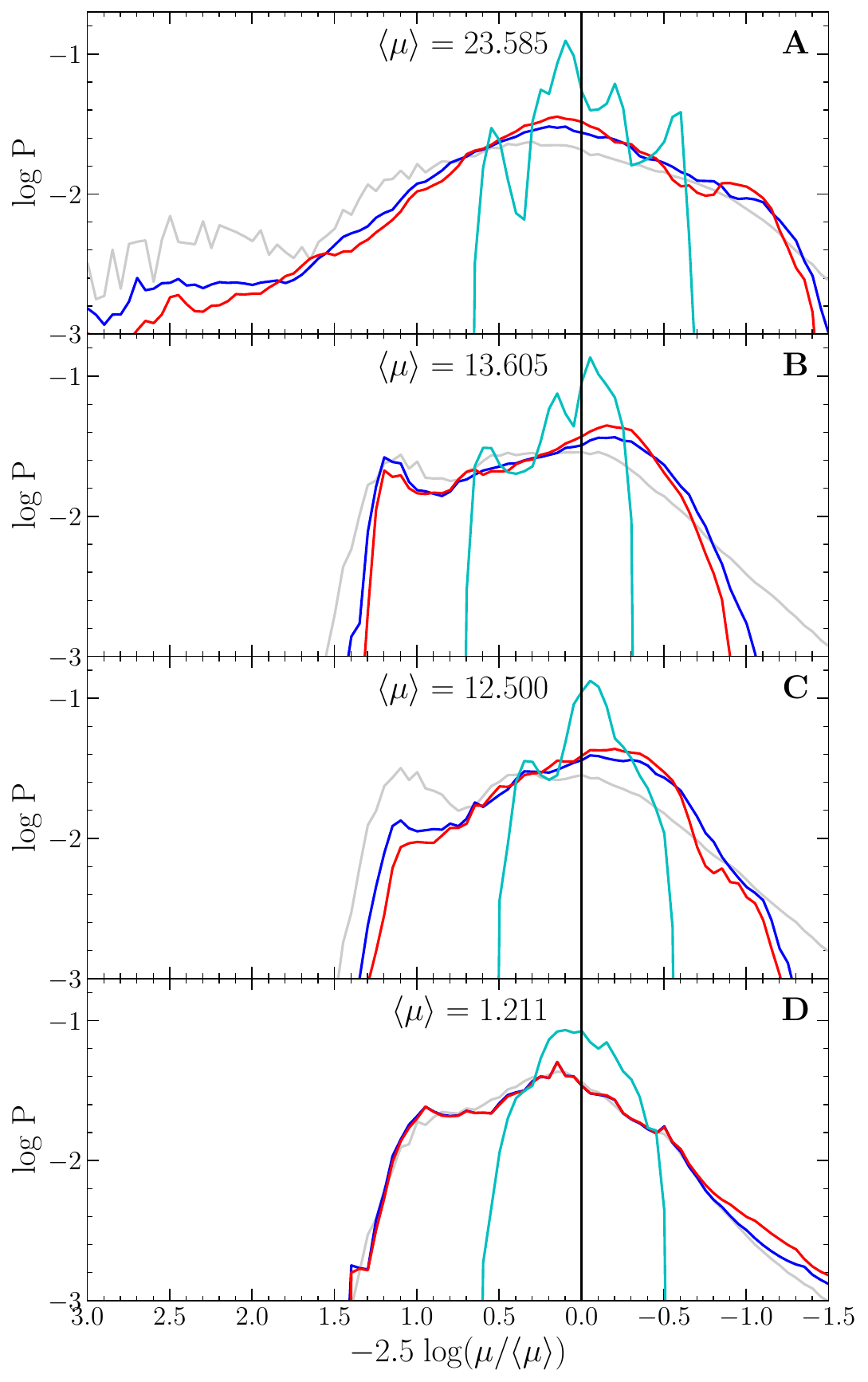}
	\caption{
		Magnification probability distributions from the maps corresponding to the multiple images of RXJ 1131$-$1231 (shown in \figref{fig5:configuration}) for a point source ($< 0.0025 \, \re$, grey lines) and three Gaussian brightness profiles with sizes of 0.117 (blue), 0.2 (red) and 1.6 (cyan) $\re$, respectively (matching the source sizes shown in \figref{fig5:spec_micro}). The vertical lines indicate the macromagnification $\langle \mu \rangle$. We can see that the larger the source the smaller the extent of the deviations from the macromagnification due to microlensing. However, in this example we used the same $25^2$-$\tae$, $10,000^2$-pixel maps for all the sources, which have a sufficiently large resolution compared to the size of the smallest sources (blue and red circles on \figref{fig5:spec_micro}) but are probably not wide enough to result in unbiased magnifications for the largest source. The high resolution is not needed in this case because of the smoothness of the chosen Gaussian profiles.}
	\label{fig5:mpds_configuration}
\end{figure}

\begin{figure*}
	\centering
	\includegraphics[width=\textwidth]{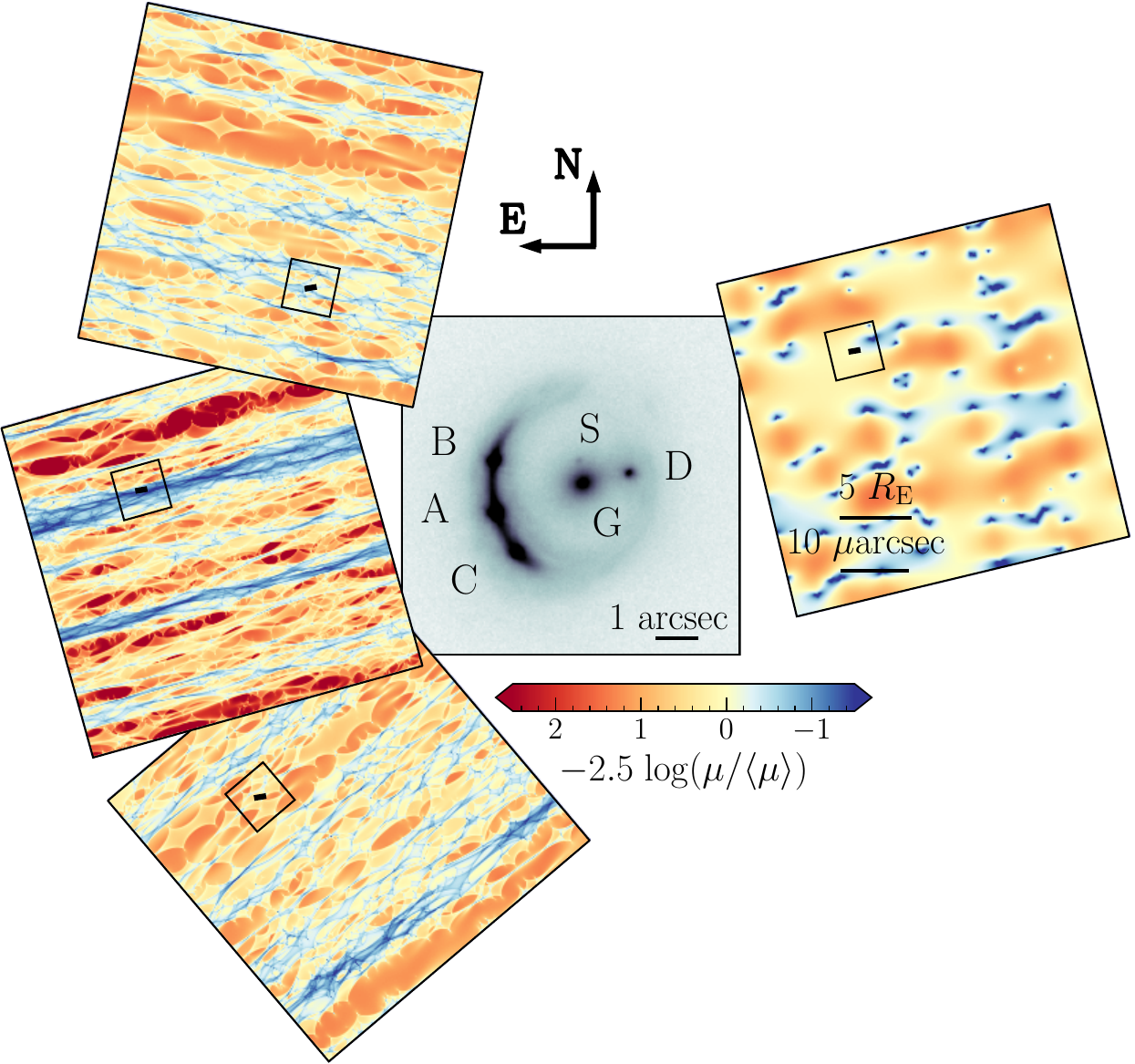}
	\caption{Lens model of quadruply lensed quasar RXJ 1131$-$1231 taken from \citet{bib5:Chen2016} and corresponding magnification maps at the locations of the quasar multiple images. The maps are consistent with the mass model, i.e. have the corresponding $\kappa$, $\kappastar$, $\gamma$ \citep[see table 1 in][]{bib5:Vernardos2022molet} and are aligned with the local total shear vector. A trial trajectory for a quasar moving (from right to left) across the source plane is shown on each map (black line with length of $\approx 0.5 \, \tae$; the corresponding microlensing light curves are shown in \figref{fig5:lcurves_configuration}). Adapted from \citet{bib5:Vernardos2022molet}.}
	\label{fig5:configuration}
\end{figure*}

A macroscopic model of the lensing galaxy defines the general properties of microlensing variability.
It is the starting point of almost all of the methods used to infer physical quantities from a microlensing signal. 
The macromodel values of $\kappa,\gamma,\kappastar$ at the locations of the observed multiple images (see also \chapgal) are used to obtain corresponding microlensing magnification maps, as illustrated in \figref{fig5:configuration} for a quad.
Such maps are the indispensable tool to model microlensing variability and are computed with one of the techniques presented in \secref{sec5:microlensing_maps}.
Because the $\kappa,\gamma,\kappastar$ parameters only define the microlenses at the population level, e.g. they do not define the positions and masses of the microlenses uniquely, each map is one of many\footnote{As many as the different configurations of the microlens positions and the ways $\kappastar$, the mass density in microlenses, can be distributed to them - by all measures, infinite.} random and equivalent realizations that can be used interchangeably.
This is important because maps have a finite size and resolution that can be limiting depending on the application, hence using equivalent maps can help. In addition, averaging over multiple realizations of maps can help avoid the sample variance in the histogram from using a single map.

\begin{figure}
	\centering
	\includegraphics[width=0.45\textwidth]{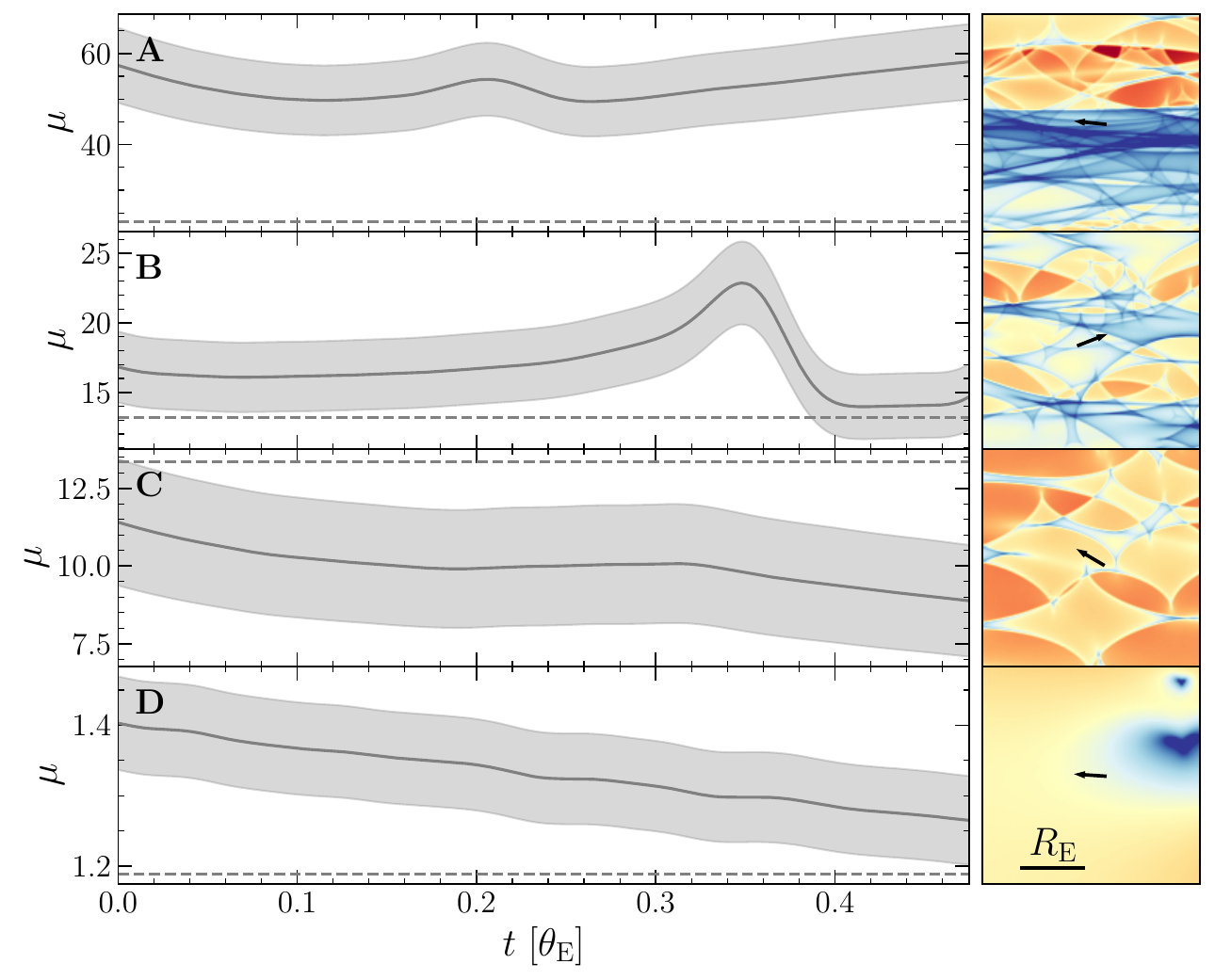}
	\caption{Realization of magnification as a function of time for each multiple image of RXJ 1131$-$1231 and corresponding zoomed-in region of the magnification maps shown in \figref{fig5:configuration}. The source trajectory is indicated by the arrow (same as the trajectories in \figref{fig5:configuration}). The horizontal dashed lines correspond to the macromagnification of each image. The shaded areas indicate the Poisson error \citep[after convolution,][]{bib5:Vernardos2015} of the magnification resulting from the numerical inverse ray-shooting technique \citep{bib5:Kayser1986} that was used to generate the maps.}
	\label{fig5:lcurves_configuration}
\end{figure}

A map constitutes a finite region of the source plane and has to be large enough to be representative, i.e. cover enough ``troughs'' and clusters of superimposed caustics (see \figref{fig5:map_tiles}) in order not to be biased towards lower or higher magnification\footnote{This is required for statistical studies, e.g. flux ratio or light curve analysis, but does not need to be the case for high magnification events, which are usually dominated by a single caustic crossing that has well-defined, universal properties \citep{bib5:Fluke1999}.}.
The higher the resolution of the map compared to $\tae$, the better the pixels approximate the size of a point source.
For pixel sizes of $< 0.006 \, \tae$ (practically point sources), \citet{bib5:Vernardos2013} found that the minimum size for unbiased maps is $24\times24$ $\tae$.
For larger sources, there are two reasons for having an adequately sized map: avoiding convolution edge effects and covering enough area for unbiased magnifications, i.e. mitigating the sample variance (e.g. studies of the BLR usually use maps whose size can be up to 50-200 $\tae$).
Statistical studies need to sample the stochastically varying magnification field in an unbiased way.
It should be noted that locations that are within a source-size from each other become correlated via convolution\footnote{That is, if the source is placed in these locations it will overlap with itself. This applies equally to tracks across maps. For practical purposes, the full extent of the source could be replaced by some fraction of its half-light radius.}.
This correlation may be desirable when studying time series, whose values are not independent from each other.
The number of sources that can fit in a map of given size without overlap decreases with the source size and the more overlap between source positions the more correlated their magnification values become.
The issue merely reflects the already mentioned problem of sample variance: if a few, long-range features dominate a magnification map, even after smoothing via convolution with a large source, then larger or statistically equivalent maps should be used in order to obtain unbiased magnifications.

Once the map size is set, then the minimum pixel size (resolution) should be small/high enough so that the smallest source would still be covered by a few map pixels. 
This rule-of-thumb is to avoid extreme (de)magnifications that can appear if a source is equal to or smaller than a map pixel\footnote{A map pixel is simply a finite area in the source plane that indiscriminately collects rays that are shot from the image plane. If, for example, the pixel happens to contain a caustic, which is a line of infinitesimal width, the very high resulting magnification is attributed to the entire pixel and any source that may lie within it, regardless of its exact position.}.
Similarly, because each pixel is a numerical approximation of the magnification by a finite number of rays, statistical (Poisson) fluctuations average out over a few pixels.
Both of these effects can be seen in the light curves shown in Figs. \ref{fig5:microimages_light_curves} and \ref{fig5:lcurves_configuration}.
In fact, the pixel size should be small enough to resolve any part of the source that contains a large fraction of the total flux (e.g. a hotspot within an accretion disc).
Conversely, if there are mostly small brightness gradients over the parts of the source that emit most of the flux, then there is no need for high resolution.
This is the case for the magnification of the BLR, which is smoothed out due to its extent over several $\tae$.
In the examples shown in \figref{fig5:mpds_configuration}, the probability distributions for the largest source could be biased and using multi-scale, multi-resolution maps cannot be avoided, at the cost of making sure they are consistent with each other.

\subsection{General observational considerations}
\label{sec5:observational_considerations}
At any given time, there is a high probability that microlensing takes place in at least one image of a lensed quasar due to the high optical depth for microlensing.
This probability is the largest for quads, where there is almost always one image that is subject to substantial microlensing \citep{bib5:Witt1995}.
The presence of microlensing can be assessed by measuring fluxes, in one or more bands, or spectral ratios, at a single or at multiple epochs, as summarized in \tabref{tab5:microlensing_variability}, and then comparing these measurements to the microlensing-free case.
The latter serves as a ``micro-lensing zero-point'' or ``no-microlensing baseline'', and has to be determined from ancillary data or models.
However, it is difficult to obtain this for single images, as well as disentangling the intrinsic variability of the quasar, without additional information (e.g. the time delay, see below).
For this reason, multiple images are considered in pairs.

For an image $i$, the magnification from the macro-model $M_i$ (\eqref{eq5:macro_magnification}), can be used to derive the baseline magnification ratio between a pair of images ($i,j$), i.e. $M = M_i / M_j$. 
There is, however, no a priori guarantee that the macromodel captures the full complexity of the system.
For instance, the presence of a nearby satellite galaxy may produce a flux ratio anomaly compared to a standard macro-model (e.g. \chapdm).
If the latter is not accounted for, the estimated amplitude of microlensing from the data will be biased.
For that reason, one can alternatively derive the amplitude of microlensing directly from the data, as explained in \secref{sec5:microlensing_from_images} and \secref{sec5:microlensing_from_spectra}.

The requirement to compare pairs of lensed images implies that the interpretation of the microlensing signal is in general ambiguous: either one image is magnified or the other one is demagnified.
It can also be that microlensing is taking place simultaneously in both images in a pair.
The comparison of multiple pairs of images in quads helps resolving this ambiguity.
The combination of different kinds of data, such as time-series and multi-wavelength data, may also be helpful for determining a plausible microlensing scenario in a lensed system. 

\begin{table*}
	\centering
	\caption{Summary of the most common types of data sets used depending on the microlensing application. Temperature and time delay are denoted as $T$ and $\Delta t$.}
	\label{tab5:microlensing_variability}
    \begin{threeparttable}
	\begin{tabular}{llll}
		% Header
		& Long-term variability   & Snapshots  & High Magnification Events      \\
		\hline
		% First row
		\begin{tabular}[c]{@{}l@{}}Observational\\ characteristics\end{tabular} & 
		\begin{tabular}[c]{@{}l@{}}Decade-long monitoring\\ Weekly cadence\\ Single band\end{tabular} &
		\begin{tabular}[c]{@{}l@{}}Fixed moment in time\\ Multiple bands / spectrum\end{tabular} &
		\begin{tabular}[c]{@{}l@{}}Month-long monitoring\\ Daily cadence\\ Multiple bands / spectrum\end{tabular} \\
        \hline
		% Second row
        \begin{tabular}[c]{@{}l@{}}Science\\ applications\end{tabular} &
		\begin{tabular}[c]{@{}l@{}}Accretion disc size \\ Baryonic/dark matter ratio\\ Mean microlens mass\end{tabular} &
		\begin{tabular}[c]{@{}l@{}}Accretion disc $T$ profile\\ Accretion disc size \\ BLR structure$+$kinematics \\ Baryonic/dark matter ratio\\ Mean microlens mass\end{tabular} &
		\begin{tabular}[c]{@{}l@{}}Detailed disc geometry\\and $T$ profile\\ BLR structure$+$kinematics\end{tabular} \\
		% Third row
        \hline
        \begin{tabular}[c]{@{}l@{}}Advantages\\ /drawbacks\end{tabular} &
		\begin{tabular}[c]{@{}l@{}}Additional velocity model\\ Needs a measured $\Delta t$\end{tabular} &
		\begin{tabular}[c]{@{}l@{}}Needs an independently\\ measured $\Delta t$\end{tabular} &
		\begin{tabular}[c]{@{}l@{}}Rare (1 per decade/system)$\dagger$\\ Hard to predict\end{tabular} \\
		\hline
	\end{tabular}
	\begin{tablenotes}\footnotesize
		\item [$\dagger$] Except in the singular case of Q2237$+$0305 that has a very high effective velocity.
	\end{tablenotes}
    \end{threeparttable}
\end{table*}

One can distinguish three main strategies for carrying out observations suitable for microlensing studies (Table~\ref{tab5:microlensing_variability}): snapshot data, long-term monitoring, and targeted monitoring triggered based on specific criteria. 
The single-epoch method (see \secref{sec5:single_epoch}) is observationally the simplest one, as it consists of one-time observations of a system (snapshots) in many wavelengths.
Flux ratios are assumed to have been corrected for time delays, otherwise, it is necessary to simulate the impact of intrinsic variability, either by theoretical models (e.g. Vernardos et al. 2023, submitted) or by means of Monte-Carlo simulations. 
Long-term monitoring data have commonly been a by-product of campaigns aiming at measuring the time delay between lensed images \citep[e.g.][and \chaptimedelays]{bib5:Tewes2013}.
Once corrected for the time delay, these light curves can be used to extract microlensing signals.
Surveys observing regularly the same region of the sky, like the Rubin Observatory Legacy Survey of Space and Time (LSST), will yield an increase in the availability of such data in the next decade. 
Finally, there is a category of suitably timed observations targeting specific systems that are undergoing High Magnification Events (HME).
These are generally defined as sharp increases of brightness (e.g. $>1$ mag in the optical) in one of the images in a pair over a period of a few weeks to months or even years, owing to a very compact region of the quasar, most likely very close to the SMBH, going through a very high magnification region.
These events are usually attributed to single caustic crossings, although more complex situations can produce similar events (see Neira et al. 2023, submitted, and \figref{fig5:microimages_loops}).
Triggering such follow-up observations is not easy due to the large number of systems that have to be monitored in order to capture the onset of an event.
Although a handful of candidate HMEs have been captured by dedicated campaigns, we expect a few hundred events per year to be observed by surveys like LSST \citep{bib5:Neira2020}.

\subsubsection{Measuring microlensing from imaging data}
\label{sec5:microlensing_from_images}
Good proxies of the microlensing-free flux ratios can be measured in regions that are considered large enough to be free of microlensing, i.e. tens of $\tae$ \citep[see ][for an estimate of the predicted amplitude of microlensing fluctuations for large sources]{bib5:Refsdal1997}.
As explained in \secref{sec5:torus_NLR_radio}, a sweet-spot may be the radio and mid-infrared ranges.
While those regions may be the ones less affected, it is not guaranteed that they are totally free of microlensing.
For instance, microlensing is suspected to occasionally occur at radio wavelengths \citep{bib5:Koopmans2000, bib5:Biggs2023}, but also radio emission may be spatially offset from the optical disc or blend multiple sources of emission only discernible with very high resolution interferometry \citep[e.g.][]{bib5:Hartley2019, bib5:Paschenko2020}.
In the mid-infrared range, one needs to disentangle between different emission regions to derive flux ratios totally free of microlensing \citep{bib5:Stalevski2012b, bib5:Sluse2013}.
Rest-frame emission with $\lambda > 11 \mu$m is probably the least susceptible to microlensing (\secref{sec5:torus_NLR_radio}), but is observationally out of reach for quasars at $z \gtrsim 2$. When the source size gets too large, one should also account for finite size effects arising because the macro-magnification is not constant over the source profile. In some rare situations, the source may even cross a macro-caustic introducing more dramatic changes of flux ratios but generally accompanied by spatially resolved emission.

When spectra are available, NLR flux ratios may also be used (see \secref{sec5:microlensing_from_spectra}), however, they may sometimes deviate from the baseline and become spatially resolved \citep[e.g.][]{bib5:Sluse2007}. 
Because of small asymmetries in the emission region, there can be a luminosity centroid that is slightly offset compared to the emission from the accretion disc. 
The differential magnification between the disc and the NLR due to strong lensing may yield a small bias on flux ratio measurements. 
Even in the case where a NLR brightness peak is in projection coincident with the location of the disc, biases can occur due to the way the flux ratio is obtained.
The measurement  needs to be performed over the whole resolved emission from the NLR, and a correction for source size effects associated to strong lensing has to be applied. 

It is important to keep in mind that other effects can yield a deviation of monochromatic flux ratios from the baseline.
The intrinsic quasar variability, as well as the differential extinction caused by different amounts of dust along the path of individual lensed images, inevitably introduce additional uncertainties.
The former can be accounted for by observing a system at multiple epochs separated by the time delay, while the latter requires multi-wavelength data to be corrected for (see \secref{sec5:extinction}).

\subsubsection{Measuring microlensing in spectra}
\label{sec5:microlensing_from_spectra}

\begin{figure*}
% DS: Fig. from Figure_ML_spectra_ISSI_book.ipynb 
    \centering
\includegraphics[width=0.86\textwidth]{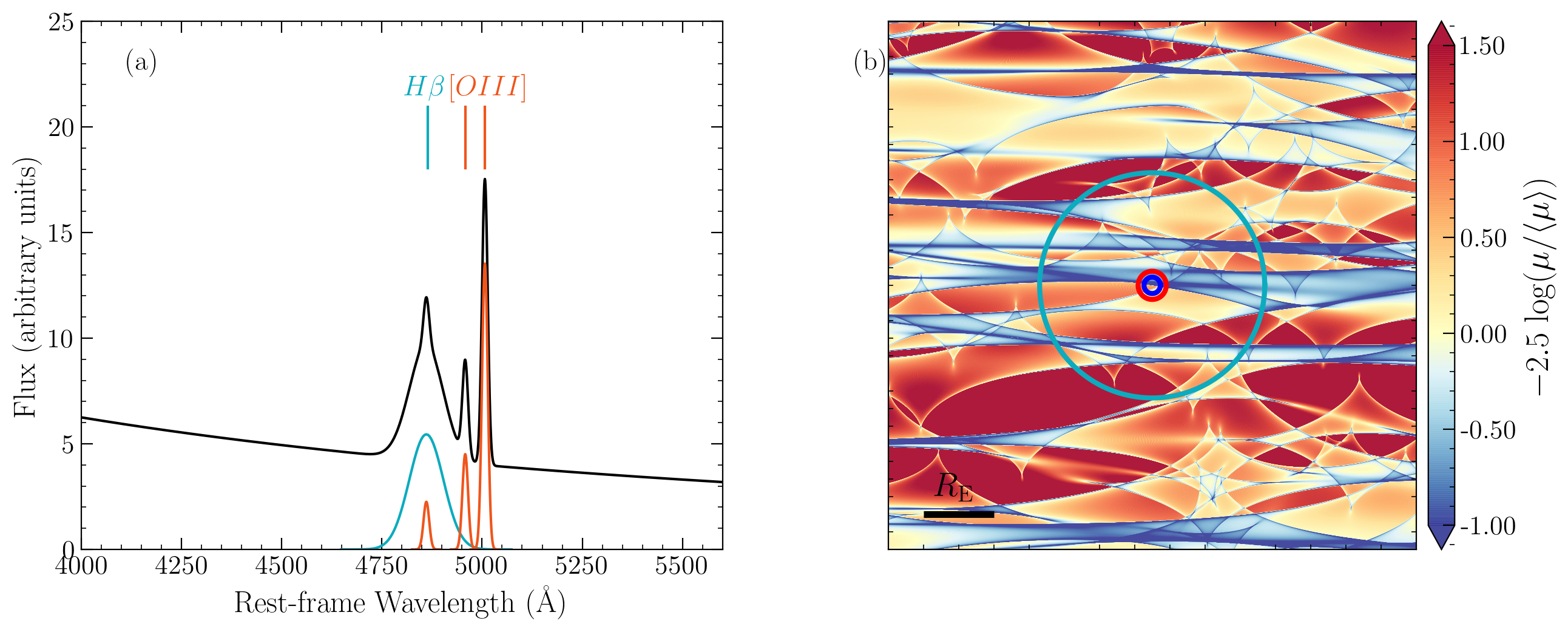} \\
\includegraphics[width=0.86\textwidth]{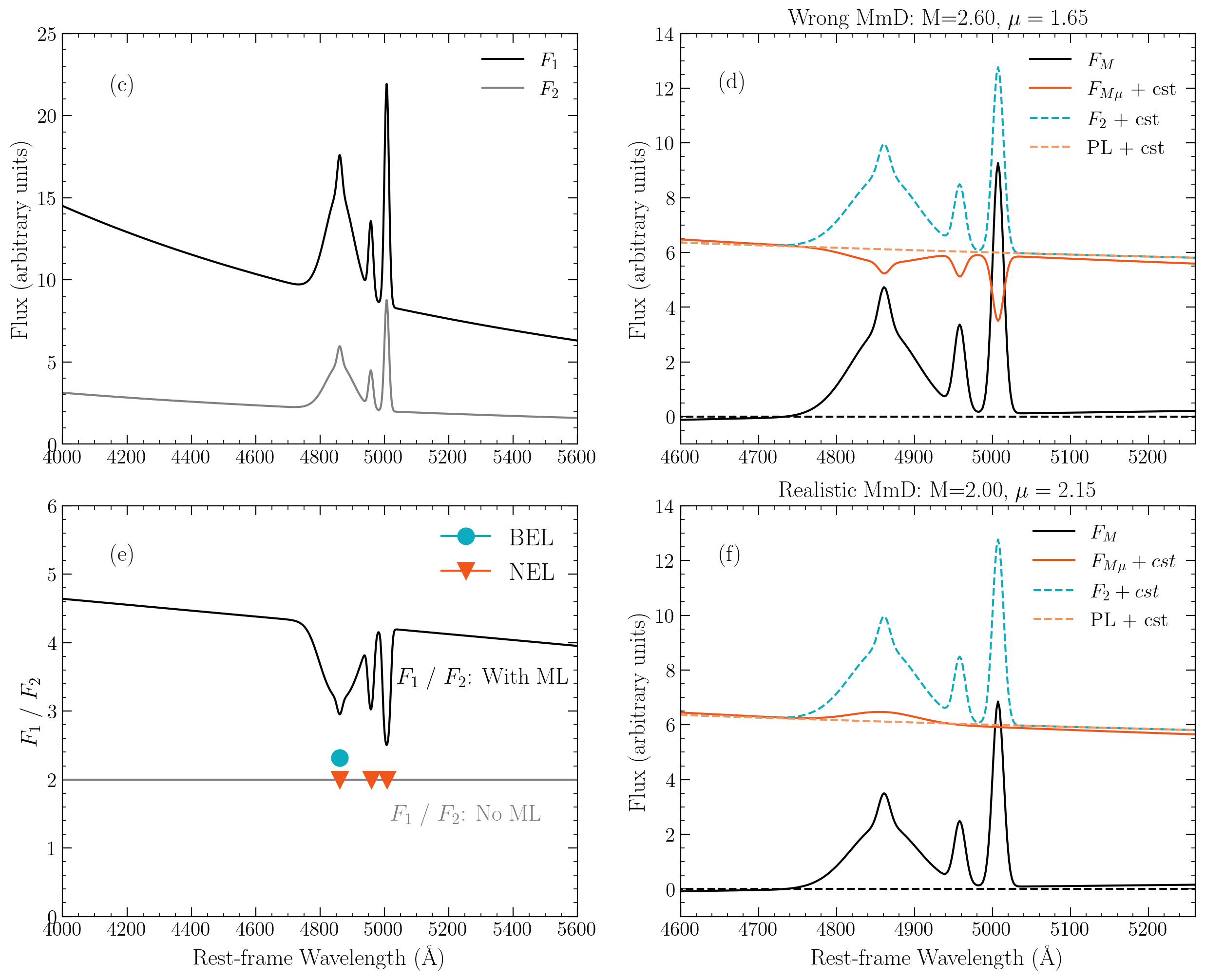} \\

\caption{Illustration of the spectral deformation introduced by microlensing on an AGN spectrum. {\bf{Panel (a)}}: The rest-frame range 4000-6000~\AA~is shown. The emission components considered are a power-law continuum, a broad $H\beta~\lambda$ 4864~\AA~emission arising from the BLR, and 3 narrow emission lines arising from the NLR and associated to $H\beta$ and [\ion{O}{iii}$]~\lambda\lambda$4959, 5007. The broad line components (i.e. after continuum subtraction) are shown with colored solid lines. {\bf{Panel (b):}} The colored circles correspond to \rhalf\, of the continuum emission at 4000~\AA~(blue; $\rhalf = 0.117\, \re$), 6000~\AA~(red; $\rhalf = 0.2\, \re$), and the BLR (turquoise; $\rhalf = 1.6\, \re$), overlaid on the magnification map of image A of RXJ 1131$-$1231 (\figref{fig5:configuration}). The NLR is much larger than the map ($\rhalf = 48\,\re$) and hence not shown. The micro-magnification of the different components are $\mu_{\rm {blue}} = 2.46$, $\mu_{\rm {red}} = 1.86$, $\mu_{\rm{BLR}}=1.16$. The solid black horizontal bar corresponds to $1 \re$. {\bf{Panel (c):}} The spectra of two lensed images $F_1$ and $F_2$. The macro-magnification ratio between the two images is fixed to $M = M_1/M_2 = 2$. {\bf{Panel (e):}} Spectral ratio for the microlensing situation displayed (black), and in absence of microlensing (grey). The ratios derived from the narrow lines (triangles) and from the broad lines (circles) are also shown. {\bf{Panels (d \& f)}}: Result of the MmD decomposition method, explained in \secref{sec5:microlensing_from_spectra}. This method derives $F_{M\mu}$ ($F_M$), which corresponds to the fraction of the flux that is (not) microlensed. {\bf{Panel (d)}} shows the decomposition when an incorrect pair ($M$, $\mu$) is used. {\bf{Panel (f)}} shows a realistic decomposition achieved by minimizing the appearance of the NEL emission in $F_{M\mu}$. Due to the chromatic microlensing of the continuum, we have $F_{M} < 0$ for $\lambda < 4700$~\AA~and $F_{M} > 0$ for $\lambda < 4700$~\AA. The value $\mu = 2.15$ retrieved is $\mu_{\rm {blue}} > \mu > \mu_{\rm {red}}$. It corresponds to $\mu$ at $\lambda \sim 4900$~\AA. Detrending the chromaticity of the continuum microlensing prior to the MmD can be performed to achieve a more precise decomposition.}
    \label{fig5:spec_micro}
\end{figure*} 

The spectra of quasars combine emissions arising from regions of different sizes, which are therefore subject to a different amount of microlensing.
Consequently, the shape of a microlensed spectrum will be deformed compared to the intrinsic one.
In general, at a given epoch, the smallest region will be more affected than more extended regions\footnote{However, this is not strictly always true. While less frequent, one can find specific configurations of the source with respect to the microlensing network, where, for example, a more compact emitting region can be less magnified because of being located in a demagnification trough compared to a more extended emission region that can cover several caustics \citep[e.g.][]{bib5:Hutsemekers2023}.} 
This basic rule helps to anticipate the shape of a spectral deformation.

The simplest diagnostic for the presence of microlensing in spectra consists of calculating spectral ratios between pairs of gravitationally lensed images (see also \figref{fig5:spec_micro}).
Let us consider the spectra of two lensed images of an AGN, $F_{1}(\lambda)$ and $F_2(\lambda)$, as observed in the optical rest-frame range at two epochs separated by the time delay.
In absence of microlensing, those two images are simply magnified by the macro-magnification $M_1$ and $M_2$.
The spectra may contain emission from the power-law continuum\footnote{For the sake of simplicity, we ignore here pseudo-continuum emission from, for example, the AGN host galaxy or from \ion{Fe}{ii}, which are known to be present in AGN spectra.}, which arises from the compact accretion disc with size of typically $\sim 0.1 \,\tae$, emission from broad lines originating from the BLR with size $\sim 2 \,\tae$, and emission from narrow emission lines that arise from regions with size $\sim 50 \,\tae$ (\figref{fig5:sketch}). 
In a no-microlensing scenario, we should have $M(\lambda) = F_1 / F_2 = M_1 / M_2$ being a constant across wavelengths.
If extinction due to the lensing galaxy is present, $M(\lambda)$ is not constant anymore and differs from the macro-magnification M (\secref{sec5:extinction}).
 
Figure~\ref{fig5:spec_micro}c shows an example of spectral deformation expected for a pair of lensed images due to microlensing.
For simplicity, this figure displays the case where $F_2$ is virtually free of microlensing (i.e. $\mu \sim 1$), and the continuum of $F_1$ is microlensed by $\mu \sim 2.15$ at the wavelength of $H\beta$. 
The spectral ratio $F_1 / F_2$ (\figref{fig5:spec_micro}e) reveals two features.
First, a dip is visible at the wavelengths corresponding to the emission lines.
This can be easily understood in terms of the expected relative sizes of the emitting regions: since the continuum comes from the smallest region, it is on average more microlensed than the broad and narrow lines, resulting in more continuum than line flux.
Second, a chromatic slope is seen because of differential magnification of the continuum itself: bluer emission arises from the inner, smaller part of the disc and is more magnified than redder.
The strength of this chromatic effect depends mostly on the amplitude of microlensing and on the relative change of size of the disc as a function of wavelength.
The flux ratio in the narrow lines will be the closest to $M$ due to the large size of the NLR.
It is important to realise that an accurate measurement of $M$ based on the narrow lines (when unresolved by strong lensing) cannot be achieved by simply looking at the spectral ratios due to the presence of both continuum and line flux at the same wavelength.
Instead, it should be based on the integrated flux in the lines, i.e. after continuum subtraction [see panel (e) of \figref{fig5:spec_micro}].
In order to better characterise the potential role of variability, one can also compare flux ratios of the same lensed images at multiple epochs. 
This method, in combination with quantitative measurements on the spectral ratio of pairs of images, has been used by \citet{bib5:Popovic2005} to disentangle microlensing from intrinsic variability and millilensing in several systems. 

The spectral ratios mostly provide a simple and quick diagnostic of the presence and amplitude of microlensing.
A more quantitative criterion of differential microlensing between the continuum and the lines can be obtained by measuring the line equivalent width:
\begin{equation} 
	W = \int_{\lambda_1}^{\lambda_2} \left(\frac{F_\lambda}{\mathcal{F}} - 1 \right) \,d\lambda,
	\label{equ:EW}
\end{equation}
where $\mathcal{F}$ is the continuum flux at the level of the line and $F_\lambda$ the total observed spectral emission at wavelength $\lambda$.
Because it involves a normalisation by the continuum, it is easy to verify that $W$ measured in two lensed images will differ only if the amplitude of microlensing is different in the line and in the continuum \citep[e.g.][and Vernardos et al. 2023, submitted]{bib5:Lewis1998b,bib5:Sluse2015}. 

Another method to derive microlensing signals from spectral deformations consists of modeling the pairs of spectra with the same intrinsic components, such as the sum of a power law continuum and of a series of Gaussian/Lorentzian line profiles that represent the emission (and/or absorption) components, as depicted in panels (d) and (f) of \figref{fig5:spec_micro}. 
The comparison of the flux ratios in each of these components may reveal the presence of microlensing \citep{bib5:Sluse2007, bib5:Sluse2011}.
The drawback of this method is that there is no guarantee that each of these components effectively represents a spatially different emission region (e.g. a single-peaked Gaussian line profile could result from the superposition of a double-peaked emission from a Keplerian emission and of a Gaussian profile arising in a wind). 
It is also possible that the intrinsic emission profile will get deformed due to microlensing, such that a model valid for a non-microlensed line will fail for a microlensed one. 

To cope with the limitation of the previous method, \citet{bib5:Sluse2007} have proposed isolating the flux arising from regions affected by microlensing from the flux arising from larger, microlensing-free regions by linearly combining pairs of spectra.
This method, called Macro-micro decomposition (MmD), is similar to the one proposed by \cite{bib5:Angonin1990} and \cite{bib5:ODowd2015}. It assumes that one spectrum is minimally microlensed (hereafter $F_2$) and is a good template of the intrinsic spectrum. 
If $\mu$ is the average microlensing magnification of the continuum in the wavelength range of the line, $M = M_1 / M_2$ is the macro-magnification ratio, $F_{M\mu}$ the microlensed part of the flux, and $F_M$ the non-microlensed part, then one can write: 
\begin{eqnarray}
	F_1 & = & M \times (\mu \, F_{M\mu} + F_M) \\
	F_2 & = & F_M + F_{M\mu} 
\end{eqnarray}
These two equations can easily be combined to isolate $F_M$ and $F_{M\mu}$: 
\begin{eqnarray}
	F_M \ & = & \frac{-A \;}{A - M} \; \; \left( \frac{F_1}{A} - F_2 \right), \label{eq5:FM}
	\\ 
	F_{M\mu}
	& = & \frac{M}{A - M } \; \; \left( \frac{F_1}{M} - F_2 \right),  \label{eq5:FMmu}
\end{eqnarray}
where we have defined $A = M \times \mu$.
This method is illustrated in panels (d) and (f) of \figref{fig5:spec_micro}. 
In addition, any decomposition should verify $F_M > 0$ and $F_{M\mu} > 0$.
In general, these two equations can be verified for a range of (positive) values of ($M , \mu$).
The procedure described below enables the derivation of the values of those parameters that minimize the amount of microlensing in the broad line.
First, one determines the product, $A = M \times \mu$, by measuring the flux ratio $F_1 / F_2$ in a region that contains only continuum emission.
This implies that $F_M = 0$ in the continuum. 
Once $A$ is derived, $M$ is fined-tuned until the flux $F_{M\mu}$ mimics as closely as possible a continuum emission.
$F_{M\mu}$ can deviate from a pure continuum emission if other regions are microlensed (e.g. the broad lines).
Note that $M$ can also be set a priori from the macro-model or from ancillary imaging data as explained in \secref{sec5:microlensing_from_images}.
To first order, this method also generally works when the second lensed image is also microlensed \citep[see appendix C of][]{bib5:Sluse2012}.
In that case, only the relative microlensing between the pair of images is retrieved.
When more than two lensed images are observed, the comparison between multiple pairs of images generally enables one to identify which image is minimally microlensed. 
There is a number of situations where the above decomposition does not work.
For instance, when the line flux is magnified only for some velocity bins while others are demagnified, e.g. the blue wing of the line is magnified while the red wing is demagnified.
The method also fails when some velocity bins are more (de-)magnified than the underlying local continuum. 
A description of the method when both absorption and emission lines are present can be found in \cite{bib5:Hutsemekers2010}. 
If applied to pairs of spectra that have not been corrected by the time delay, intrinsic variability could mimic microlensing-induced line deformations, and/or chromatic microlensing. 
These effects remain in principle small for time delays $\lesssim$ 40 days \cite{bib5:Yonehara2008, bib5:Sluse2012}.   

\cite{bib5:Guerras2013a} proposed another method to measure microlensing from spectra.
It consists in deriving the amount of microlensing present in a line by calculating the difference of magnitude between the line wings and core. 
First, one needs to locally fit the continuum flux based on two regions on each side of the line and free of pseudo-continuum emission.
From the continuum subtracted spectrum, one can measure the magnitude $m_{1,2} = -2.5 \log(F^{\rm{line}}_{1, 2})$ of the line, either in its core or in the line wings. 
The differential microlensing between the wing and core of the line is given by: 
\begin{equation}
	\Delta m \  =  (m_1 - m_2)_{\rm wings} - (m_1 - m_2)_{\rm core}
\end{equation}
As shown by \cite{bib5:Guerras2013a}, $\Delta m$ is naturally corrected for differential extinction, which can be assumed to be constant over the wavelength range covered by the line. 
$\Delta m$ exactly corresponds to the microlensing of the line only if the line core is effectively free of microlensing. 

\citet{bib5:Braibant2017} have introduced several indicators to quantify the effect of differential magnification between the wings and core of a line and/or line asymmetry. 
They can be calculated using the ratio between the line profiles in pairs of lensed images, are independent of the shape of the line profile, and can be calculated on both simulated or observed emission lines. 
These indicators are useful in establishing quantitative diagnostics of line deformations \citep[see][]{bib5:Braibant2017}, and can be calculated even for moderate signal-to-noise ratios.
Calculating such indicators instead of using the full profile may however be seen as a loss / degradation of the signal.
Instead, one may rather use continuum subtracted line profiles in pairs of spectra to derive microlensing as a function of velocity: $\mu(v) = F^{\rm{line}}_1 / (M * F^{\rm line}_2$). 
In the situation where the line is also contaminated by pseudo-continuum emission (such as \ion{Fe}{ii}), a model of the latter needs to be subtracted as well and uncertainties propagated to the estimate of $\mu(v)$.
The use of $\mu(v)$ is only possible for spectra with a sufficient signal-to-noise ratio.
Even in that case, $\mu(v)$ may need to be truncated at high velocity because the line flux becomes too low to enable a reliable estimate of $\mu$.

\subsubsection{The role of extinction}
\label{sec5:extinction}
 Extinction caused by dust along the line of sight to different lensed images can lead to a different no-microlensing baseline and disguise itself as chromatic microlensing (wavelength dependence).
To mitigate this, let us consider that the intrinsic flux of an AGN lensed image at a given wavelength, $\lambda$ (observer's frame), and time, $t$, can be expressed as $m_0(\lambda,t)$ (in mag).
An observed lensed image $i$ will be magnified by $M_i(\lambda,t)$ (the total magnification includes the macrolens and microlensing that could vary with wavelength and time), and have a time delay $\Delta t_i$, resulting in the observed spectrum \citep{bib5:Falco1999}:
\begin{equation}
\begin{split}
    m_i(\lambda,t)  =  m_0\left(\frac{\lambda}{1+z_S},t-\Delta t_i \right) -2.5 \log \left[ M_i \left(\frac{\lambda}{1+z_L},t \right) \right] \\
    + E_i R_i \left(\frac{\lambda}{1+z_L} \right) + E_{\rm Gal} R_{\rm Gal} (\lambda) + E_S R_S \left(\frac{\lambda}{1+z_S} \right),
    \end{split}
\end{equation}
where $E_i$, $E_{\rm Gal}$, $E_S$ are the image $i$ extinctions, $E(B-V)$, produced by the lens galaxy, our Galaxy, and the source host respectively, and $R_i(\lambda)$, $R_{\rm Gal}(\lambda)$, and $R_S(\lambda)$ indicate the corresponding extinction laws.
This equation can be simplified considering several assumptions.
First, the terms associated with our Galaxy and the host galaxy extinction can be considered as negligible because the separation between rays at the source and at the observer is considered small and thus the amount of extinction should be the same for all images.
On the contrary, the contribution of the lens galaxy should be more important because the images pass through regions separated by kiloparsecs, which are likely to contain different amounts of dust \citep{bib5:Cardelli1988}.
Finally, assuming that the magnification is independent of $\lambda$ and $t$ (i.e. no microlensing), the source spectrum is independent of $t$ and time delays, and the lens galaxy extinction curve is the same for all images, we can estimate the magnitude difference between two images $i$ and $j$ as \citep{bib5:Falco1999}:
\begin{equation}
   m_i(\lambda) - m_j(\lambda)= -2.5 \log \left( \frac{M_i}{M_j} \right) + (E_i-E_j) R \left(\frac{\lambda}{1+z_L} \right),
\end{equation}
which depends only on the constant macro-magnification ratios, $M_i/M_j$, the extinction difference, $E_i-E_j$, and the extinction curve in the lens rest frame, $R(\lambda/(1+z_L))$.
The previous assumptions might not hold for the smallest regions of the AGN (e.g. the continuum could be affected by chromatic microlensing magnification, (see \figref{fig5:spec_micro}), thus it is better to use narrow emission lines (which are only affected by the macrolensing magnification and extinction).
For example, the doubly imaged SBS0909+523 system was used to estimate the extinction curve of the lens galaxy and confirm the lens redshift by detecting the 2175 \AA \, blue bump \citep{bib5:Motta2002, bib5:Mediavilla2005}, while also estimating chromatic microlensing in the system.

\subsection{The single-epoch method}
\label{sec5:single_epoch}
Here we consider the application of the single-epoch method to photometric data. 
These consist of single or combined exposures in a narrow or broad band, from which flux measurements for point sources, i.e. the quasar multiple images, are extracted.
In this case, the flux between a pair of macroscopically observed multiple images $A$ and $B$, $\Delta m^{\rm{obs},k}_{\rm{AB}}=m_{\rm{B}}^{\rm{obs},k}-m_A^{\rm{obs},k}$ where $k$ denotes the photometric band used, is measured with respect to a baseline that is believed unaffected by microlensing, $\Delta m^{\rm{ref},k}_{\rm{AB}}=m_{\rm{B}}^{\rm{ref},k}-m_{\rm{A}}^{\rm{ref},k}$.
The microlensing signal can then be defined as:
\begin{equation}
\Delta m^k_{\rm{AB}}=\Delta m^{\rm{obs},k}_{\rm{AB}} - \Delta m^{\rm{ref},k}_{\rm{AB}}.
\end{equation}
We note that the no-microlensing baseline may or may not depend on the band/wavelength, for example if there is extinction or if it is taken to be the macromodel flux ratio respectively (see \secref{sec5:observational_considerations}).

Simulated microlensing flux ratios (or more precisely magnification ratios) can be obtained from the magnification maps corresponding to images $A$ and $B$ given a model of the source light profile.
The former are defined mainly by the $\kappa,\gamma,\kappastar$ parameters, but can also depend on other assumptions like the mass function of the microlenses, which we can collectively denote as $\boldsymbol{\eta}_{m}$.
The latter essentially boils down to a way of determining $\rhalf$ of the source as a function of wavelength.
This radius may be calculated based on a physical model of the source, for instance, in the case of an accretion disc one can use \eqref{eq5:radius_MBH} that depends on the black hole mass, luminosity and accretion efficiency, or be purely phenomenological (for example \eqref{eq5:parametric-sizes}).
We can collectively denote all the free parameters of the source as $\boldsymbol{\eta}_s$.
Simulated microlensing flux ratios as a function of wavelength are in fact a function of these free model parameters:
\begin{equation}
    \Delta m^{\rm{mod}}_{\rm{AB}}(\lambda_{\rm{rest}}) \equiv \Delta m^{\rm{mod}}_{\rm{AB}}(\lambda_{\rm{rest}} | \boldsymbol{\eta}_m, \boldsymbol{\eta}_s).
    \label{eq5:flux_ratio_sim_definition}
\end{equation}
To obtain the microlensing flux ratio in any photometric band one should in principle integrate over a range of wavelengths and take into account the band's response function.
In practice, only the central wavelength of the band can be considered and the dependence on $\lambda_{rest}$ can be replaced by the index $k$.
Finally, to obtain the ratios between images $A$ and $B$ we randomly select locations on the convolved maps and divide their magnification values pairwise.
We note that these locations have to be the same across the different bands considered.

The observed and simulated microlensing flux ratios can be compared through a chi-squared statistic:
\begin{equation}
    \chi^2_n = \sum_i \sum_{j \; > i} \sum_k^K \left[ \frac{\Delta m_{i j}^{\rm{obs},k} - \Delta m_{i j}^{\rm{ref},k} - \Delta m^{\rm{mod},k,n}_{ij}(\boldsymbol{\eta}_m,\boldsymbol{\eta}_s)}{\sigma_{i j}^k}  \right]^2 \, ,
\label{sec5:eq_chi2k}
\end{equation}
where $i,j$ is $1-4$ for a quadruply lensed quasar, $\sigma_{i j}^k$ is the error associated with the observed flux ratios, $K$ is the number of bands, and $n$ denotes a given pair of locations between two magnification maps.
We can thus obtain the final likelihood for a given combination of parameters $\boldsymbol{\eta}_m,\boldsymbol{\eta}_s$ as the sum:
\begin{equation}
    \mathcal{L}(\boldsymbol{d}| \boldsymbol{\eta}_m,\boldsymbol{\eta}_s) = \sum_n^N e^{-\chi^2_n/2},
    \label{eq5:likelihood_k}
\end{equation}
where $\boldsymbol{d}$ represents the flux ratio observations between all image pairs across all bands and we use $N$ trials of simulated flux ratios (i.e. pairs of locations between maps).

The likelihood above can be used in a Bayesian setup with the probability density of the posterior given by:
\begin{equation}
    \frac{\diff P}{\diff \boldsymbol{\eta}_m \diff \boldsymbol{\eta}_s} \propto \mathcal{L}(\boldsymbol{d} | \boldsymbol{\eta}_m,\boldsymbol{\eta}_s) \frac{\diff p}{\diff \boldsymbol{\eta}_m} \frac{\diff p}{\diff \boldsymbol{\eta}_s},
    \label{eq5:single_epoch_bayes}
\end{equation}
where $p$ are the priors on the free parameters of the problem - common choices being uniform or logarithmic.
In this way, the posterior probabilities for any parameter of the model, e.g. the dependence of size on wavelength ($\nu$, for an accretion disc) or $\kappastar$, can be obtained by marginalizing over all the remaining free parameters of the model \citep[e.g.][]{bib5:Bate2018}.

Finally, one can extend this method to constrain a common underlying source model with data from multiple systems \citep[e.g.][who rescaled the size of the accretion disc by the black hole mass]{bib5:Jimenez2015a}.
The final joint likelihood is obtained by the product:
\begin{equation}
    \frac{\diff P}{\diff \boldsymbol{\eta}_s} \propto \prod^L_l \frac{\diff p_l}{\diff \boldsymbol{\eta}_s} \int \mathcal{L}(\boldsymbol{d}_l | \boldsymbol{\eta}_{m,l},\boldsymbol{\eta}_s) \frac{\diff p_l}{\diff \boldsymbol{\eta}_{m,l}} \diff \boldsymbol{\eta}_{m,l} \;,
    \label{eq5:joint_likelihood_k}
\end{equation}
where each of the $L$ observed lensed systems has its own map parameters $\boldsymbol{\eta}_m$ (e.g. the $\kappa,\gamma$ for each image) and may have its individual priors $p_l$ for the source parameters $\boldsymbol{\eta}_s$.

\subsubsection{Advantages and drawbacks}
The main advantage of the single-epoch method is that both data and simulations are easy to handle.
One needs snapshots of a lensed system in one or more bands, which are easy to schedule and obtain, and the method can be applied to several systems simultaneously.
Creating pairs of simulated microlensing magnification ratios for different sources is straightforward.
The parameter space of the model\footnote{In almost all cases, the free parameters consist of the source parameters and $\kappastar$ (see Sections \ref{sec5:quasar_results} and \ref{sec5:lensing_galaxy_results}), while $\kappa,\gamma$ are assumed to be known from the macromodel \citep[although they can be included in the analysis, as in][]{bib5:Vernardos2018}.} is usually smooth and sampling the bulk of the probability does not rely heavily on priors.
However, the resulting probability distribution can be quite extended and in some cases can lead only to upper limits for some parameters (usually the accretion disc size, see \secref{sec5:accretion_disc}).
\citet{bib5:Bate2018} have shown that this method can have poor constraining power depending on the chromaticity amplitude, i.e. by how much the flux ratio between different bands varies, and the offset with respect to the no-microlensing baseline.
Other difficulties include controlling systematic uncertainties like extinction, the no-microlensing baseline (see \secref{sec5:observational_considerations}), leakage of broad line emission in the wavelength range covered by a broad band filter (see also \secref{sec5:thermal_slope}), and quasar intrinsic variability.
To mitigate the latter, configurations with small time delays like crosses or close image pairs are preferred.
Although one could still apply this method to any image pair with a known time delay, two snapshots would be required in this case correctly spaced in time.

\subsection{The light curve method}
\label{sec5:time_varying_ML}
In a lensed quasar, the observer, lensing galaxy, micro-lenses, and background emission regions are not static.
The combined velocity of all the different components makes quasar microlensing a dynamic phenomenon in timescales that vary from weeks to decades.
Several lensed quasars have been monitored for more than a decade, mainly for time-delay cosmography applications (see \chaptimedelays) and mostly in a single photometric band.
Once time delays are measured, it is possible to shift the observed light curves of each image by the corresponding delay and subtract them pair-wise to cancel out the intrinsic variations of the quasar.
The resulting difference curves are the observed microlensing signal, i.e. the ratio of microlensing magnification between the pair of multiple images, which contains valuable information on quasar structure and lensing galaxy mass partition and IMF.
An example of such data for the doubly lensed quasar Q J0158$-$4325 is shown in \figref{fig5:lcs}.

\begin{figure*}
    \centering
    \includegraphics[width=0.95\textwidth]%{figures/J0158_split_Euler4_GP_noresiduals_fit.pdf}
    %\caption{\textbf{Top:} R-band light curves of the doubly lensed quasar Q J0158$-$4325 observed by the Leonhard Euler 1.2m Swiss Telescope. The blue and orange lines show a Gaussian Process regression used to interpolate between the data points along with its uncertainties (shaded envelopes). \textbf{Bottom:} Difference light curve between images B and A, shifted by the measured time delay \citep[$\Delta t_{AB} = 22.7$ d, image A leading][]{bib5:Millon2020a}. The horizontal dashed line indicates the no-microlensing baseline (see \secref{sec5:observational_considerations}) as predicted by the macromodel of \cite{bib5:Morgan2008a}. This system displays strong microlensing effects as indicated by the magnitude difference between the two images that steadily increases by $>1$ mag over the 13 years of the monitoring campaign. Figure adapted from \cite{bib5:Millon2022}.
    {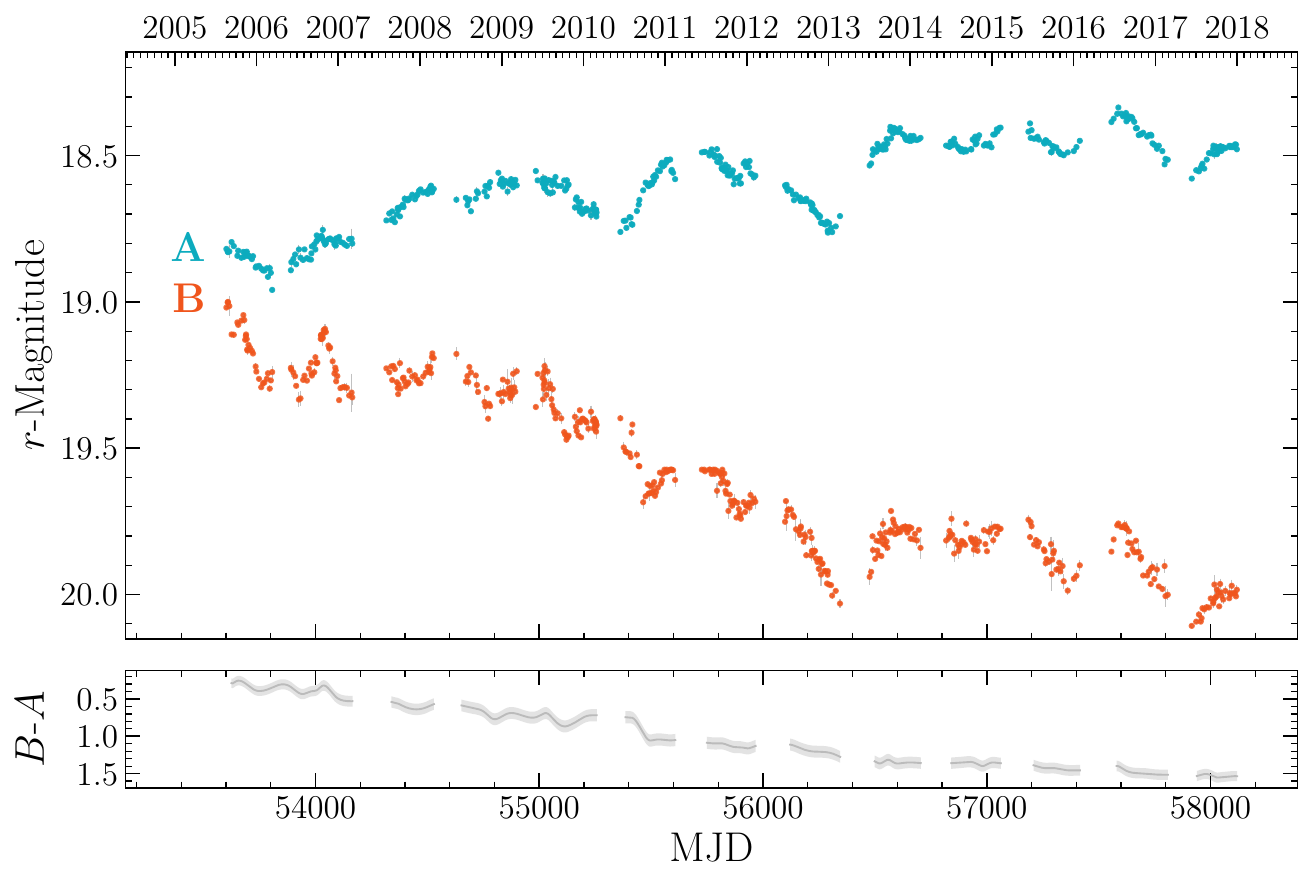}
    \caption{Luminosity variation of the doubly lensed quasar J$\,$0158-4325 in the r-band, from 13 years of monitoring at the Euler Swiss Telescope in La Silla Observatory. \textbf{Top:} Light curves of the two lensed images, $A$ (minimum) and $B$ (saddle). Measurement errors are indicated by the grey vertical lines (mostly smaller than the size of the points). \textbf{Bottom:} A spline model of the $A$ image light curve subtracted from another spline model of $B$ \citep[see also fig. 3 in][]{bib5:Paic2022}. Importantly, the $B$ spline model is shifted by the best time-delay estimate \citep[$\Delta t_{AB} = 22.7$ d, image A leading][]{bib5:Millon2020a}. This system displays strong microlensing effects as indicated by the magnitude difference between the two images that steadily increases by $>1$ mag over the 13 years of the monitoring campaign.}
    \label{fig5:lcs}
\end{figure*}

\subsubsection{The effective velocity model}
\label{sec5:velocity_model}
To describe time-varying microlensing it is necessary to define a model for the effective projected velocity, which results in the background source moving along trajectories or ``tracks'' on top of a magnification pattern.
These tracks can then be used to extract magnification values as a function of time that translate into a light curve.
In fact, this is the biggest difference and complication with respect to the single-epoch method from the simulations' point of view.
The two additional degrees of freedom, the length and direction of the simulated trajectories, are critically increasing the number of simulations required (see next section).

Let us define the effective velocity of the source, $\boldsymbol{\upsilon}_e$, as the vector sum of the transverse velocities of the observer, $\boldsymbol{\upsilon}_o$, microlenses, $\boldsymbol{\upsilon}_\star$, lensing galaxy, $\boldsymbol{\upsilon}_l$, and source, $\boldsymbol{\upsilon}_s$ \citep[see e.g.][]{bib5:Kayser1986,bib5:Kochanek2004,bib5:Neira2020}:
\begin{equation}
    \label{eq5:veff}
    \boldsymbol{\upsilon}_e = \frac{\boldsymbol{\upsilon}_o}{1+z_l}\frac{D_{ls}}{D_{ol}} + \frac{\boldsymbol{\upsilon}_\star+\boldsymbol{\upsilon}_l}{1+z_l}\frac{D_{os}}{D_{ol}} + \frac{\boldsymbol{\upsilon}_s}{1+z_s},
\end{equation}
where time is measured in the observer's rest frame and length on the source plane.
These velocity components are also schematically shown in \figref{fig5:transvel}.
The only fully measurable of these vector quantities is $\boldsymbol{\upsilon}_o$, which is determined relative to the Cosmic Microwave Background velocity dipole ($\boldsymbol{\upsilon}_\mathrm{CMB}$):
\begin{equation}
    \label{eq5:vcmb}
    \boldsymbol{\upsilon}_0 = \boldsymbol{\upsilon}_\mathrm{CMB} - (\boldsymbol{\upsilon}_\mathrm{CMB} \cdot \hat{z}) \, \hat{z},
\end{equation}
where $\hat{z}$ is the direction of the line-of-sight.
The unknown peculiar velocities of both the lensing galaxy and the background quasar can be considered as uniformly random in direction with magnitude in the source plane drawn from a normal distribution $\mathcal{N}(0,\sigma_g)$ with:
\begin{equation}
    \label{eq5:vgauss}
    \sigma_g = \left[ \left( \frac{\sigma_l^{\mathrm{pec}}}{1+z_l}\frac{D_{os}}{D_{ol}} \right)^2 + \left( \frac{\sigma_s^{\mathrm{pec}}}{1+z_s}\right)^2 \right]^{1/2},
\end{equation}
where $\sigma_l^{\mathrm{pec}}$ and $\sigma_s^{\mathrm{pec}}$ are the standard deviations of the peculiar velocity distributions of the lens and the source respectively.
A proper treatment of the final velocity component, the individual velocities of the microlenses, would require the use of computationally expensive ``moving patterns'',  where the individual movement of each star is taken into account (see \figref{fig5:micro_bern}).
However, \cite{bib5:Kundic1993} showed that this relative movement was equivalent to an increase in the magnitude of the effective velocity.
\cite{bib5:Wyithe2000a} showed that the magnitude of this effect can be approximated by a ``bulk velocity'' of the micro-lenses such that:
\begin{equation}
    \label{eq5:vstar}
    \upsilon_\star = \sqrt{2} \epsilon \sigma_\star,
\end{equation}
where $\sigma_\star$ is the velocity dispersion at the lens center and $\epsilon$ is an efficiency factor that depends on local $\kappa$ and $\gamma$ values.
\begin{figure*}
    \centering
    \includegraphics[width=1.0\textwidth]{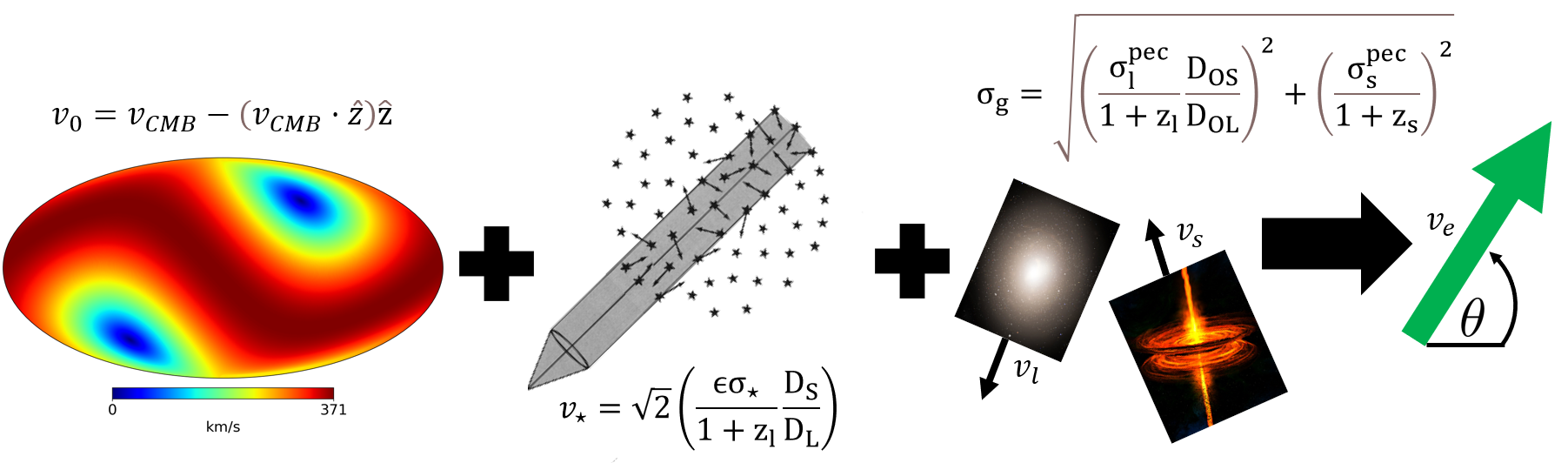}
    \caption{Schematic representation of the components of the relative transverse velocity defined in equation \ref{eq5:veff}. \textbf{Left:} the magnitude and direction of the transverse CMB dipole velocity, i.e. the velocity of the observer. \textbf{Middle:} the individual velocities (arrows) of the microlenses (star symbols) that lie along the line of sight (grey cone). \textbf{Right:} the random peculiar velocities of lensing galaxy and quasar.}
    \label{fig5:transvel}
\end{figure*} 
\begin{figure*}
    \centering
    \includegraphics[width=1.0\textwidth]{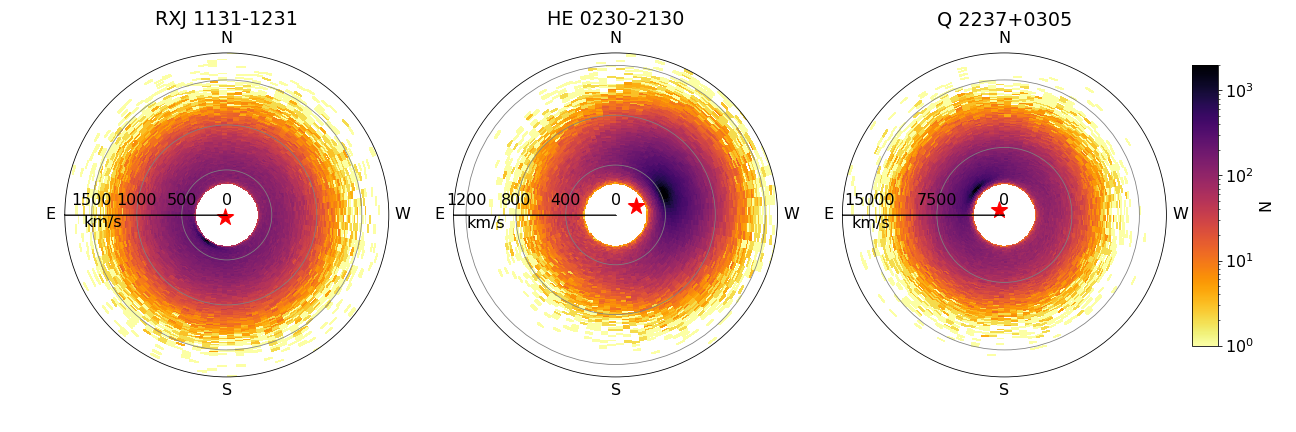}
    \caption{Histograms of simulated effective velocities for three known lensed quasar systems. The minimum possible value for the velocity (inner radius of the ring-like structure) is defined by $\boldsymbol{\upsilon}_\star$, which depends on the lensing galaxy's velocity dispersion. The peaks of the distributions are skewed towards the direction defined by the fixed velocity component $\boldsymbol{\upsilon}_l$. The star symbol denotes the direction and magnitude of this component.}
    \label{fig5:veldist}
\end{figure*} 
Since in the general case only $\boldsymbol{\upsilon}_o$ is known, a direction and magnitude for the final effective velocity vector need to be drawn from a distribution, as defined by equation \ref{eq5:veff}.
By construction, this distribution is skewed towards the fixed direction of $\boldsymbol{\upsilon}_o$ and has a minimum magnitude defined by $\boldsymbol{\upsilon}_\star$.
As an example, \figref{fig5:veldist} shows histograms of effective velocity realizations for three well-known lensed quasar systems.

Once an effective velocity distribution and the duration of an observing period have been established, one can finally define tracks of the source traveling through the magnification pattern.
These tracks have a length given by the duration of the observations times the velocity modulus and a direction defined by the vector sum of the velocity components (see \figref{fig5:configuration}).
Extraction of the pixel values by superposition of this track on top of a magnification pattern convolved with the chosen quasar model results in a simulated light curve (see \figref{fig5:lcurves_configuration}).
Unlike static microlensing, in this case, in addition to the strength, also the steepness and the timescales of the variations shown in the light curve will depend on the specific model of the source (see \figref{fig5:microimages_light_curves}).

\subsubsection{Fitting light curves}
\label{sec5:lcve_analysis}
The standard approach to analyse light curve data consists of fitting simulated light curves to difference curves and was introduced by \citet{bib5:Kochanek2004}.
To simplify a bit the equations, we focus on a single photometric band light curve and a single system.
In this case, we define the time series that constitute a difference light curve between a pair of macroscopically observed multiple images A and B as $\Delta m^{\rm{obs},t}_{\rm{AB}} = m^{\rm{obs},t}_{\rm{B}} - m^{\rm{obs},t}_{\rm{A}}$, where $t$ denotes one of $T$ measurements from the difference light curve.
The no-microlensing baseline is $\Delta m^{\rm{ref}}_{\rm{AB}}=m_{\rm{B}}^{\rm{ref}}-m_{\rm{A}}^{\rm{ref}}$ and does not depend on time.
The microlensing signal can then be defined as:
\begin{equation}
\Delta m^t_{\rm{AB}}=\Delta m^{\rm{obs},t}_{\rm{AB}} - \Delta m^{\rm{ref}}_{\rm{AB}}.
\end{equation}

Before generating simulated microlensing light curves, a number of hyper-parameters need to be defined for the magnification maps, $\boldsymbol{\eta}_m$, and the source, $\boldsymbol{\eta}_s$, in the same way as for the single-epoch method.
Here, we also need to include the hyper-parameters that are related to the effective velocity, i.e. its magnitude and direction, which we denote as $\boldsymbol{\eta}_v$.
Finally, the simulated microlensing light curves that are sampled from trial trajectories on convolved maps are a function of these free model parameters:
\begin{equation}
    \Delta m^{\rm{mod}}_{\rm{AB}}(t) \equiv \Delta m^{\rm{mod}}_{\rm{AB}}(t | \boldsymbol{\eta}_m, \boldsymbol{\eta}_s, \boldsymbol{\eta}_v).
    \label{eq5:lc_sim_definition}
\end{equation}
The same notes as for the single-epoch method apply here, i.e. we can replace a broad photometric band by its central wavelength, we can randomly select starting locations for the trajectories on the convolved maps and divide their magnification values pairwise, and these locations have to remain the same if we are considering data in many bands.
However, the pair of trajectories must have the same effective velocity and the orientation of each map, or more precisely the orientation of the local shear at the position of each multiple image, has to be taken into account, as illustrated in \figref{fig5:configuration}.

The observed and simulated microlensing light curves can be compared through a chi-squared statistic:
\begin{equation}
    \chi^2_n = \sum_i \sum_{j \; > i} \sum_t^T \left[ \frac{\Delta m_{i j}^{\rm{obs},t} - \Delta m_{i j}^{\rm{ref}} - \Delta m^{\rm{mod},t,n}_{ij}(\boldsymbol{\eta}_m,\boldsymbol{\eta}_s,,\boldsymbol{\eta}_v)}{\sigma_{i j}^t}  \right]^2 \, ,
\label{sec5:eq_chi2t}
\end{equation}
where $i,j$ is $1-4$ for a quadruply lensed quasar, $\sigma_{i j}^t$ is the error associated with each point in the difference light curve, and $n$ denotes a given pair of trajectories between two magnification maps.
The observant reader may have noticed the similarity between this and \eqref{sec5:eq_chi2k}, however, we note that because of $T \gg K$ and the larger parameter space due to $\boldsymbol{\eta}_v$, inference of the target physical parameters becomes much harder in this case.
We can define a likelihood function in the same way as in \eqref{eq5:likelihood_k} using $N$ pairs of simulated light curves, and cast it in a Bayesian framework to obtain:
\begin{equation}
    \frac{\diff P}{\diff \boldsymbol{\eta}_m \diff \boldsymbol{\eta}_s \diff \boldsymbol{\eta}_v} \propto \mathcal{L}(\boldsymbol{d} | \boldsymbol{\eta}_m,\boldsymbol{\eta}_s,\boldsymbol{\eta}_v) \frac{\diff p}{\diff \boldsymbol{\eta}_m} \frac{\diff p}{\diff \boldsymbol{\eta}_s} \frac{\diff p}{\diff \boldsymbol{\eta}_v},
    \label{eq5:lc_bayes}
\end{equation}
where $\boldsymbol{d}$ is now difference light curve data, i.e. a set of measurements $\Delta m^{\rm{obs},t}$ and their associated uncertainties, $\sigma^t$ for one or more pairs of multiple images.
The velocity model presented in \secref{sec5:velocity_model} can serve as the prior for $\boldsymbol{\eta}_v$ here.
The posterior probabilities for any parameter of the model can be obtained by marginalizing over all the other remaining free parameters.
Extending \eqref{sec5:eq_chi2t} to include multiple wavelengths and systems (as in \eqref{eq5:joint_likelihood_k}) is left as an exercise for the reader.

\subsubsection{Advantages, drawbacks, and other approaches}
% First paragraph on advantages
The light curve method is quite appealing because of the larger amount of data - hundreds as opposed to a single epoch - that, in principle, should result in more constraining power over the source and lens galaxy models. 
That alone is a sufficiently strong argument to consider this a major advantage over other methods.
Of course, the first step for the applicability of this method is the availability of the data, meaning that the required monitoring observations for periods of years should have been scheduled and taken place, a demanding and expensive task.
LSST will dramatically improve on the availability of microlensing light curves by monitoring thousands of lensed quasars for at least a decade.
However, the potential advantages of the light curve method come at the cost of complicated calculations, additional model degeneracies, and some ``hidden'' complexity in the data.

% Complicated microlensing signal
The signal contained in the difference light curves can in fact be much more complex an amplitude modulation as the source is crossing the magnification field produced by the microlenses, as shown in \ref{fig5:microimages_light_curves}.
A rapid flickering on a timescale of weeks to months is often observed \citep{bib5:Schild1996,bib5:Hjorth2002,bib5:Schechter2003,bib5:Millon2020b}, which is too fast to be explained by the transverse velocity models described in \secref{sec5:velocity_model}, as it can be seen in \figref{fig5:lcs}.
Several explanations have been put forth to explain these fast microlensing variations: \cite{bib5:Schild1996} proposes that a population of planetary mass objects act as microlenses, whereas \cite{bib5:Blackburne2010} attribute this flickering to a variation of the accretion disc size over time, \cite{bib5:Gould1997,bib5:Wyithe2002a,bib5:Schechter2003,bib5:Dexter2011} invoke inhomogeneities or broad absorption clouds in relativistic motion in the accretion disc being magnified by microlensing, and \citet{bib5:Millon2022} propose a secondary black hole in orbit within the disc.
A different explanation simply invokes the combination of reverberation of the BLR and microlensing, as introduced by \cite{bib5:Sluse2014} \citep[see also eq. 12 in][]{bib5:Paic2022}.
Because the accretion disc and the BLR have different physical sizes, they are affected differently by microlensing.
In addition to this, the BLR is expected to reverberate the intrinsic variability of the quasar with some delay due to its location further away from the SMBH \citep[and some smoothing due to its size, see][]{bib5:Cackett2021}.
This leads to a difference in contrast between the reverberated and the direct signal that can introduce variations in the difference curves on the same timescale as the quasar intrinsic variations.
These variations are much faster than the usual microlensing timescale and could explain the complex signals present on such shorter timescales.

% Degeneracies and systematics
When a source moves through a caustic network, whose characteristic size is described by $\tae$, there is a three-way degeneracy that is introduced between the source size, effective velocity, and the mass of the microlenses.
The relation between these parameters is not exact, but can be understood by the following examples.
If we disregard any velocity component, then the source size and microlens mass are obviously degenerate through the ratio $\rhalf/\re$ that determines the extent of microlensing effects.
For a given effective velocity, a small source moving in a sparse caustic field (few high-mass microlens) and a large source moving through a dense field of caustics (many small-mass microlenses) will result a similar small-amplitude microlensing signal.
A similar signal would result from a slow and small or a fast and large source moving through a fixed-mass caustic network.
Finally, for a fixed source size, small $\tae$ and low velocity lead to magnification similar to having a large $\tae$ and high velocity.
Eventually, if the observations are long enough then these degeneracies are easier to break, but this is not always the case.
Therefore, priors on these three parameters are often imposed, with the velocity model presented in \secref{sec5:velocity_model} being a commonly used one \citep{bib5:Kochanek2004}.
Other sources of systematic bias in resulting measurements have not been adequately or at all explored \citep[see][for a study of such biases for the single-epoch technique]{bib5:Bate2018}.
For example, these could be the peak-to-peak amplitude in the light curves as a function of wavelength and the deviation of the mean magnification from the no-microlensing baseline.

A more practical, but possibly the main disadvantage of this method, is its very high computational cost, as the probability of finding a trajectory that is a good-fit to the data is decreasing with the length of the light curve.
To mitigate this, authors have divided full light curves into separately-fitted segments \citep{bib5:Poindexter2010a} or averaged data within a season \citep{bib5:Morgan2012}.
Another common simplifying assumption is to assume that microlensing is taking place in only one of the two images and thus simulate trajectories on a single map \citep{bib5:Kochanek2004}.
With the advent of machine learning methods, \cite{bib5:Vernardos2019b} have investigated the use of Convolutional Neural Networks to infer the source size from the microlensing features seen in the difference curves.
Although promising, careful thinking needs to go into the design of the training set for such methods in order to reflect the complexity of real world scenarios as closely as possible.

\subsection{Other microlensing data analysis methods}
\label{sec5:exotic_methods}
Apart from the two general methods to analyze microlensing light curve and flux ratio data described until now, there is a number of other more specialized approaches tailored to specific situations and data.
Here we give a short summary of these methods and refer the reader to the corresponding papers for more details.

\subsubsection{Measuring SMBH properties from HMEs}
Observing a lensed quasar at any random moment, either by obtaining flux ratios or through monitoring, can be used to constrain its overall size (the half-light radius) as a function of wavelength.
However, when a microlensing HME takes place there is much more information to be extracted on the immediate vicinity of the SMBH, e.g. its event horizon, mass, and spin.
The magnification during such events can often be described analytically, which can make modelling simpler (see also \secref{sec5:source_reconstruction}).
\citet{bib5:Mediavilla2015} use three such events in Q 2237$+$0305 whose fine structure at the peak they attribute to the Innermost Stable Circular Orbit (ISCO) at $\approx 3$ gravitational radii.
This result is obtained by fitting both classical and relativistic accretion disc models convolved with an analytic function for the caustic magnification (a straight fold combined with a linear term to factor in long-range magnification gradients due to neighbouring microlenses, i.e. not responsible for this particular caustic).
Their relativistic model includes the effect of beaming, Doppler shift, and gravitational redshift, which effectively improves their fit to observed data.
\citep{bib5:Best2022} on the other hand adopt a machine learning approach to measuring the ISCO, which they train with thousands of simulated microlensing light curves and accretion disc models that include relativistic beaming, Doppler shifts, and lensing from the central SMBH.

\subsubsection{Modelling X-ray line deformation}
Several authors have studied with increasing sophistication how the shape of the Fe K$\alpha$ reflection line changes as a result of microlensing.
The emitted 6.4 keV photons in the inner regions of the disc are affected by the gravitational redshift, bending, Doppler broadening, and relativistic beaming, resulting in a well-known two peaked profile \citep{bib5:Fabian1989,bib5:Laor1991}.
During an HME or caustic crossing, some regions of the disc producing these photons will be magnified more than others, resulting in additional peaks \citep{bib5:Heyrovsky1997, bib5:Popovic2006, bib5:Jovanovic2009} and edges \citep{bib5:Neronov2016} to the profile.
As pointed out by \citep{bib5:Krawczynski2017}, this may explain the Fe K$\alpha$ line variations observed in RX J1131$-$1231 \citep{bib5:Chartas2012}.  

The above mentioned studies used methods similar to those described below in \S\ref{sec5:BLR_modeling}.
The work by \citet{bib5:Ledvina2018} first makes maps at representative inclinations of the Fe line intensity and energy shift ($g$-factor) around a quasar (in a $60\times 60 \rg$ region) and then sweeps a high-magnification, narrow, linear fold caustic across those maps.
At specific moments in time, the total observed (microlensed) Fe K$\alpha$ line profile is constructed.
One of the most prominent microlensing-induced spectral features is seen as the caustic sweeps across the inner disc: a sharp and highly magnified peak dominates the spectral profile.
The location of the peak changes in time as the caustic moves as different parts of the disc get highly magnified (e.g., their figures 2 and 3).

\subsubsection{Modeling broad emission line deformations} 
\label{sec5:BLR_modeling}

\begin{figure*}
    \centering
    \includegraphics[width=\textwidth]{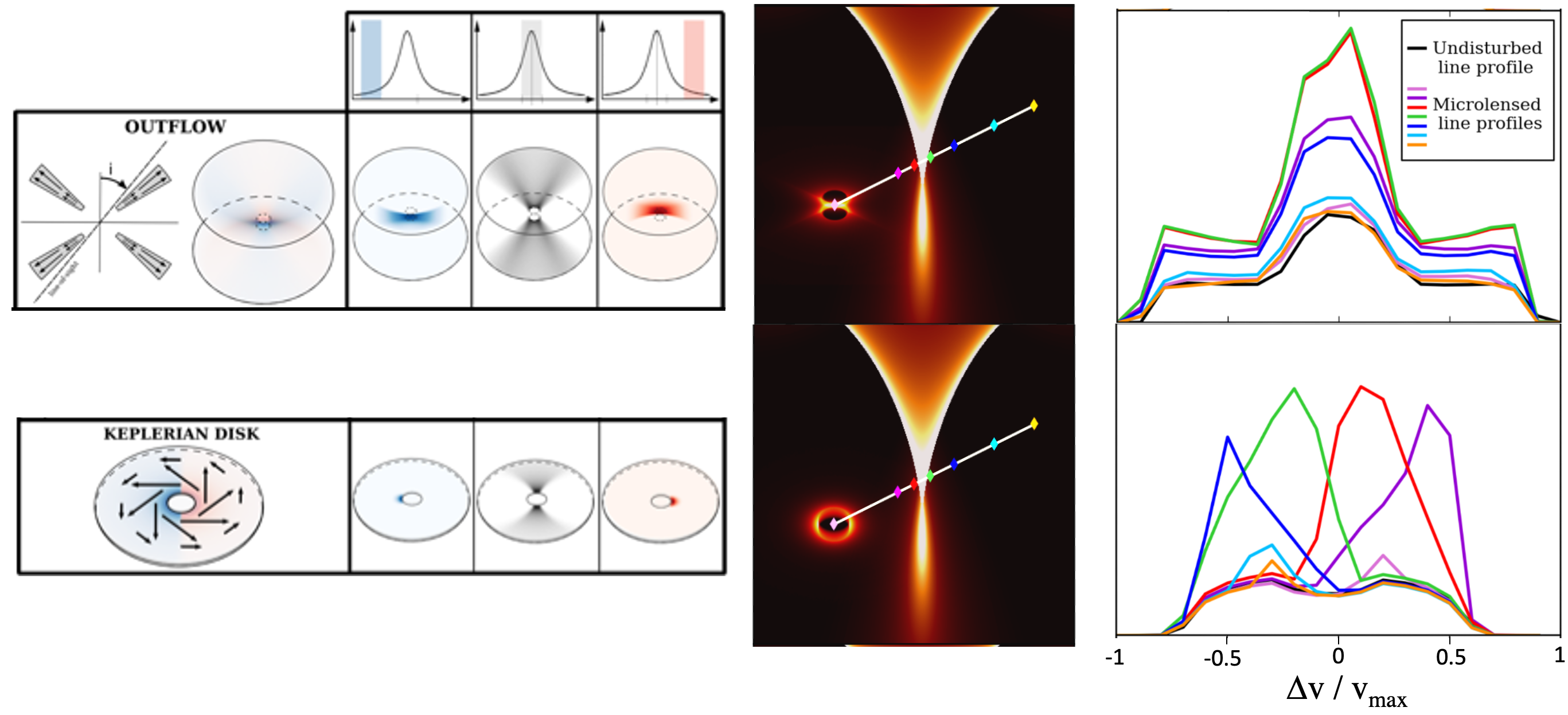}
    \caption{Illustration of broad line deformation produced by microlensing for two different models of the broad emission line region, with $\rhalf \sim 0.3 R_E$ (the size chosen to be smaller than expected BLR size to highlight line deformations). Two models are sketched the polar wind (top) and the keplerian disc (bottom) The left most panel shows the two models at intermediate inclination. The blue and red colors indicate the location of the approaching and receding gas, corresponding to blue and red velocity components of the line. The right panel depicts the line deformation corresponding to different location of the BLR with respect to the caustic, as shown in the middle pannel (colored losanges corresponding to the colored line profiles on the right). For the depicted event, the microlensing of the Keplerian disc yields asymmetric red/blue line profile distortions while microlensing of the biconical outflow is characterized by symmetric wings/core distortions. Adapted from \cite{bib5:Hutsemekers2017}.}
    \label{fig5:BLR_deformation}
\end{figure*}

Because the broadening of an emission line is due to different velocity components of the emitting material that arise from spatially distinct regions, it is necessary to calculate the microlensing signal for an ensemble of luminosity profiles, each associated to a velocity slice in the BLR, as shown in \figref{fig5:BLR_deformation}. 
These luminosity profiles are subsequently convolved with microlensing maps to calculate the line deformation.
This is similar to the procedure followed to model the temperature profile of the accretion disc, but instead of calculating the disc luminosity profile at different wavelengths (corresponding to different temperatures), one calculates the luminosity profile of the BLR for different velocities. 
A simple power-law decrease of the BLR brightness with distance to the central black hole is assumed, as well as simple geometries, such as a keplerian disc, a polar or equatorial wind, and various velocity fields able to reproduce emission line profiles \citep[e.g.][]{bib5:Murray1995, bib5:Murray1997}.
To simulate such profiles, one could use analytical models \citep[e.g.][]{bib5:Abajas2002} or radiative transfer simulations that can produce the light profile of the BLR for arbitrary velocity slice(s) \citep[e.g.][]{bib5:Braibant2017}. 
For radiative transfer models, the photo-ionizing source of light is supposed to be the quasar accretion disc, assumed to be smaller and have the same inclination as the BLR. 
The microlensing of the continuum is calculated together with the line deformation to ensure that the model can self-consistently reproduce the data. 
The exact luminosity profile of the disc is not critical (see \secref{sec5:general_considerations}) and a uniform profile has commonly been considered. 
The need for the simulations to reproduce both the continuum and the BLR emission implies that the resolution of the microlensing maps should be high enough to sample the disc.

Once deformed line profiles have been calculated for a variety of BLR models, they can be compared to the data in different ways.  
For example, \citet{bib5:ODowd2011} used the ratio between the microlensed and un-lensed profiles, which is proportional to the amplitude of microlensing as a function of the gas velocity (i.e. $\mu(v)$), \citet{bib5:Abajas2007} calculated relative changes of the line FWHM, while \citet{bib5:Braibant2017} integrated the flux over a fraction of the line profile to quantify the red-blue asymmetry and its wing-to-core magnification/deformation.
In order to minimize the loss of information inevitable by the use of such ``integrated'' quantities, which may even lead to overfitting, \citet{bib5:Hutsemekers2021, bib5:Hutsemekers2023} used the velocity dependent microlensed amplitude, $\mu(v)$ (\secref{sec5:microlensing_from_spectra}).
This choice minimizes the loss of information but requires high signal-to-noise data and in the case of too simple BLR models the method may be unable to reproduce the observed signal at all. 

The comparison of models to data can proceed through a standard $\chi^2$ statistic.
For instance, one can compare the amplitude of the simulated microlensing signal in the continuum, $\mu_{\rm C}^{\rm mod}$, and the profile of the line deformation, $\mu_{{\rm B}}^{\rm mod} (v)$, for an ensemble of velocity bins, $\Delta v_k$ (see e.g. \secref{sec5:microlensing_from_spectra} for methods enabling to measure those quantities). 
If $\boldsymbol{\eta}_B$ and $\boldsymbol{\eta}_C$ are the ensemble of model parameters of the BLR and the disc (i.e. continuum emission), then the total $\chi_n^2$ associated to a model realisation $n$ can be calculated as the sum of the corresponding BLR and continuum terms:
\begin{equation}
\chi^2_n = \sum_k \left[ \frac{\mu^{\rm obs}_{{\rm B}}( \Delta v_k) - \mu^{\rm mod}_{{\rm B}} (\Delta v_k | \boldsymbol{\eta}_B)} { \sigma^{\rm obs}_{\rm B}(\Delta v_k)} \right]^2 + \left[ \frac{ \mu^{\rm obs}_{\rm C} - \mu^{\rm mod}_{\rm C}(\boldsymbol{\eta}_C)} {\sigma^{\rm obs}_{\rm C}} \right]^2 .
\end{equation}
We note that the first term in the above equation is not unique.
Instead of $\mu_{{\rm B}} (\Delta v_k)$, one can use other measurements of line deformations like the ``integrated'' indicators described above.
The inference on the model parameters can be cast in a Bayesian framework as described in Sections \ref{sec5:single_epoch} and \ref{sec5:lcve_analysis} for single- and multi- epoch data respectively.

Existing works show that the largest constraining power arises when the amplitude of microlensing is large. 
For low amplitudes, the whole range of possible line deformations are very similar for any BLR model \citep{bib5:Braibant2017}. 
On the other hand, there are indications that more complex BLR models than the three listed above (i.e. keplerian disc, polar and equatorial wind) may be needed to accurately reproduce the line deformations observed in some systems \citep{bib5:Hutsemekers2019,bib5:Hutsemekers2021}.

\subsubsection{Extensions to the single-epoch technique}
A single flux ratio measurement obtained from a snapshot of a lensing system at a given band has little constraining power; compared to the degrees of freedom of the microlensing models (source structure and mass in the lens) the problem is under-constrained.
To mitigate this, flux ratios obtained at different epochs could be used, however, successful incorporation of such additional data requires attention to both the time separation between observations and the analysis method used.
If enough time has passed between observations to result in the source having moved to an entirely different region of the magnification map, then the measurements can be considered independent, otherwise their covariance needs to be estimated and considered in, for example, $\chi^2$ fits.
These time intervals are shorter for smaller sources because magnification is less correlated on short distances through convolution.

In the X-rays, where the source is expected to be the smallest, \citet{bib5:Pooley2012} consider such flux ratios observed in multiple epochs as independent measurements.
\citet{bib5:Guerras2020} used $\approx 10$ flux ratio measurements in the X-rays for each of four systems over periods of 5-15 years.
This timescale is most likely longer than the source-size crossing time in the X-rays.
They did not consider these measurements as independent, but used the mean and a second-order moment of their distribution instead, effectively compressing the X-ray light curve to these two summary statistics.
Their simulated flux ratio distributions where obtained along tracks on magnification maps whose length - an important parameter that is proportional to the effective velocity - was arbitrarily chosen and not fitted for.

\citet{bib5:Fian2016,bib5:Fian2018,bib5:Fian2021a} have followed a similar approach to measure the accretion disc size by collapsing observed difference light curves to flux ratio probability distributions.
However, instead of measuring the mean and width of the distribution, they calculate its distance from the one obtained from magnification maps for different source sizes.
While the method has been applied to very long light curve data (between 10 and 20 years), the microlensing variability observed over that period is not guaranteed to be representative of the full magnification map.
Indeed, the corresponding tracks on the maps may not be longer than a few $\tae$ - the exact value depending on the unknown effective velocity.
More work may be needed to demonstrate that this caveat does not bias the final results of this method.

\subsubsection{The microlensing time-delay as a probe for accretion disc size}
In addition to the well known macroscopic time delay in lensed quasars, which can be used as an independent way to perform cosmography and measure the Hubble constant (see \chaptimedelays), \cite{bib5:Tie2018a} introduced the microlensing time delay that can extend or shorten the arrival times of the signal in the different multiple images by a few days in an uncorrelated way.
The origin of this delay is due to the combination of the driving variability mechanism of the quasar, assumed to be a lamppost, and how microlensing can magnify different parts of the accretion disc.
Based on the fact that this effect depends on the source size, which in turn depends on wavelength, \citet{bib5:Chan2021} proposed using measured differences in the time delays between different bands to constrain the accretion disc size.
Their method is complementary to traditional curve shifting techniques, which consider the amplitude of microlensing only, but, unlike these methods, they take into account the distortion and delay caused by the convolution of the lamppost variability with the light profile of the accretion disc and the caustic network.

The method requires a set of measured time delays in different bands with respect to some baseline with an uncertainty of 0.1 days, which is about 10 times better than the highest precision delays achieved nowadays.
Additional assumptions are that the quasar varies as a lamppost-like model and its accretion disc brightness profile shape \citep[more precisely, its spatial derivative][]{bib5:Tie2018a} is known.
Forward modelling and fitting of the time delays as a function of wavelength, and consequently half-light radius of the disc, follows.
The value of the micro-magnification of a multiple image can be used to drastically speed up the calculations by pre-selecting the matching locations on magnification maps.
Using multi-band light curves of a fiducial quadruply imaged quasar, the authors were able to forecast that the measured accretion disc size from such data would be within a factor of 0.2 and 2 of the truth.
This method provides measurements that are independent of the choice of Initial Mass Function for the microlenses in the lens galaxy (see also \secref{sec5:mean_mass_imf}).

\subsubsection{Reconstructing accretion disc profiles}
\label{sec5:source_reconstruction}
The methods to measure accretion discs discussed until now correspond to the so-called forward problem: microlensing observables are calculated and fitted to data based on disc models prescribed a priori.
However, the problem can be approached in an inverse way as well, i.e. reconstructing an unknown disc profile from the measured data.
An observed microlensing light curve results from the convolution of a source profile with the magnification field of the microlenses.
Hence, the situation is similar to deconvolution which requires solving an ill-posed inverse problem with the use of regularization to avoid overfitting and stabilize against noise.

The first to adopt this treatment were \citet{bib5:Grieger1988,bib5:Grieger1991}, followed by \citet{bib5:Mineshige1999}, \citet{bib5:Agol1999}, and \citep{bib5:Bogdanov2002}.
These studies made two main assumptions: they described the source with a one-dimensional ``strip'' profile and focused on HMEs.
The first assumption is justified by the fact that the observed magnification depends mostly on the brightness distribution of the source along the direction perpendicular to the caustic \citep[see fig. 1 in][]{bib5:Mineshige1999}.
On long time scales, of the order of the $\tae$ crossing time, the source can be affected by several microlenses that lead to superimposed caustics and a complex and non-linear magnification field (see \secref{sec5:time_varying_ML}).
But in the case of short HMEs, the magnification field corresponds to that of a caustic crossing, which is simpler and can be described analytically to a good approximation \citep{bib5:Gaudi2002a,bib5:Gaudi2002b}.
This simplifies the problem of deconvolution considerably but can lead to some problems, e.g. the caustic can be discontinuous in the case of a cusp, it can have significant curvature, and magnification can vary along its direction \citep{bib5:Agol1999}.
It turns out that the latter two can be resolved if the source has a size of 10\% of $\tae$ \citep{bib5:Grieger1988}.

This method has been demonstrated to be stable on simulations using basic regularization schemes \citep[see][ for a review]{bib5:Benning2018}, showing potential of resolving the accretion disc structure within a few gravitational radii from the central SMBH.
When it comes to real observations, the best and only system to provide the necessary HMEs has been the quadruply lensed quasar Q 2237$+$0305.
Its accretion disc was found to have a brightness profile consistent with the thin-disc model \citep{bib5:Agol1999,bib5:Bogdanov2002,bib5:Koptelova2007}.

\subsubsection{Time varying equivalent widths of non-lensed quasars}
Looking for signatures of lensing in the variability of non-lensed quasars is an idea that has been explored by several authors over decades \citep{bib5:PressGunn1973,bib5:Canizares1982,bib5:Schneider1993,bib5:Zackrisson2003,bib5:Bruce2017,bib5:Hawkins2022}.
Vernardos et al. (2023, submitted) revisit this in the context of variable emission line equivalent widths.
The data, in this case, are two measured equivalent widths of the same line obtained by spectroscopic observations of the same system taken $\approx 10$ years apart.
The authors develop a probabilistic model for these measurements using the magnification of a single compact microlens (similar to Galactic microlensing models).
Once the most promising candidates are selected from spectroscopic surveys and re-observed to confirm that lensing is indeed responsible for their varying equivalent widths, the microlens mass and effective velocity can be measured.

\subsubsection{Astrometric Microlensing}

The variations in flux due to microlensing are correlated with astrometric shifts due to the individual microimages \citep{bib5:Lewis1998c}. In particular, the creation or annihilation of pairs of microimages during a caustic crossing event can give rise to astrometric shifts on the order of 10s of micro-arcseconds \citep{bib5:Treyer2004}. While the detection of such shifts have been long predicted, they have yet to be confirmed observationally on extra-galactic scales. They may, however, contaminate studies of proper motions of lensed AGN from Gaia \citep{bib5:Makarov2022}. Astrometric shifts from stellar microlensing have been observed however; see e.g. \cite{bib5:McGill2023}.

%%%%%%%%%%%%%%%%%%%%%%%%%%%%%%%%%%%%%%%%%%%%%%%%%%%%%%%%%%%%%%%%%%%%%%%%%%%%%%%%%%%%%%%%%%%%%%%%%%%%%%%%%%%%%%%%%%%%%%%%%%%%%%%%%%%%%%%%%%%%%%%%%%%%%%%%%%%%%%%%%%%%%%%%%%%%%%%%%%%%%%%%%%%%%%%%%%%%%%%%%%%%%%%%%%%%%%%%
\section{Quasar Results}
\label{sec5:quasar_results}

\subsection{Black hole tomography with HMEs}
\label{sec5:black_hole}
Due to the rarity of observations of HMEs in the currently known lensed quasars, there are few results regarding their very promising use for constraining the inner accretion disc and the black hole.
HMEs have almost exclusively been observed in one system: the Einstein cross (\figref{fig5:einstein_cross}). %, a Type 1 quasar located at $z = 1.695$ with $\lambda L_{\lambda} (1450\AA) \sim 10^{45.5}$ erg/s.
This system, due to its exceptionally low lens redshift ($z = 0.039$) that results in short variability time scales, has been the target of several intensive monitoring campaigns \citep{bib5:Ostensen1996,bib5:Udalski2006,bib5:Goicoechea2020}.
A few HMEs in this object have been captured in almost 3-decades worth of data and two indicative examples are shown in \figref{fig5:events_in_Q2237} \citep[see fig. 1 of][for another example]{bib5:Eigenbrod2008b}.
Study of such HMEs has allowed to place the inner edge of the quasar accretion disc at ~3 $R_\mathrm{g}$ \citep{bib5:Mediavilla2015} and detect a possible warped disc geometry \citep{bib5:Abolmasov2012a}.
Several monitoring programs have provided a rich dataset for this system, but such rapid variability is a statistical outlier.

For the vast majority of lensed quasars, a typical HME is expected in roughly every 20 years\footnote{This estimate is provided by \citet{bib5:Mosquera2011b}.
However, as introduced in \secref{sec5:observational_considerations}, HME is an overarching term for any variation with a high-enough magnification - see Neira et al. 2023 submitted, for more examples.}.
Therefore, monitoring of tens of systems for several years would be required in order to observe a statistically meaningful sample of HMEs.
Such monitoring has been performed by the COSMOGRAIL program for ~30 systems over more than a decade.
Although its main goal has been measuring the time delays between the macroscopic quasar images, microlensing signals can be extracted from the observed light curves as a by-product.
A few potential HMEs can be spotted in the final data \citep{bib5:Millon2020a}, albeit none unambiguously due to lack of observations during its peak.
The GLENDAMA archive also provides a compilation of monitoring data from various telescopes and filters for 9 systems \citep{bib5:GilMerino2018}, where a few events can be identified.

The majority of observations and analyses have employed single-band photometry, which is a cheaper and more viable approach for long-term monitoring.
If daily observations are taken during a HME, then such single-band data should be enough to provide valuable insights to quasar structure.
However, the real wealth of information on accretion disc structure, which is accessible only during the event's peak, can be extracted through multi-band, or even better, spectroscopic observations.
With existing facilities, such observations are expensive and impractical for long monitoring periods, therefore, predictions and early detection of the onset of HMEs and their peaks are mandatory to trigger multi-wavelength follow-up.

\begin{figure}
    \centering
    \includegraphics[width=.5\textwidth]{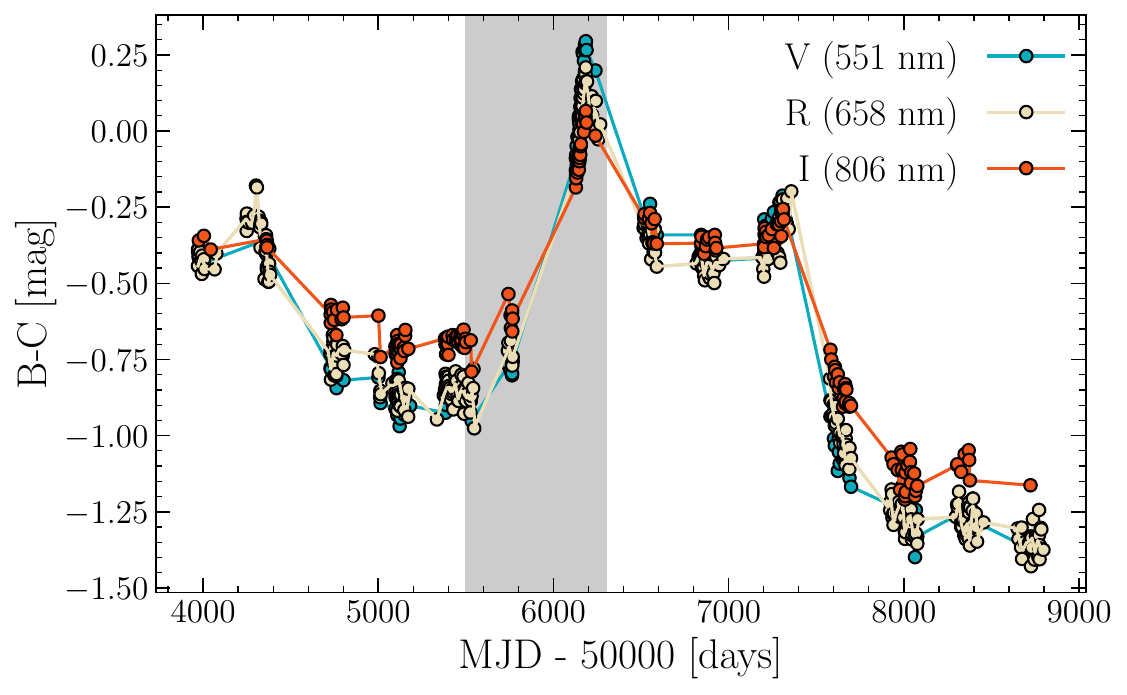}
    \caption{Difference light curve between images B and C of Q 2237$+$0305 (i.e. the Einstein Cross) observed in the VRI bands. The double-peaked signal is typical of the source entering and exiting a single diamond caustic (e.g. as in Galactic microlensing events). Notice how the brightest band changes from I (red) to V (blue) within the grey-shaded area - a crucial indicator of an imminent HME. Adapted from \citet{bib5:Goicoechea2020}.}
\label{fig5:events_in_Q2237}
\end{figure}

\subsubsection{HME prediction and triggers}
An alert system has been used by OGLE \citep{bib5:Wozniak2000} and a triggering mechanism for multi-band follow-up has been proposed by \citet{bib5:Wyithe2000b}.
However, such prediction algorithms have not been used extensively and as of yet there has not been any clear prediction and detailed follow-up of a HME as it unfolds, i.e. predicting the peak from light curves like those shown in \figref{fig5:events_in_Q2237} and scheduling detailed observations accordingly (spectroscopy, X-ray spectra and imaging, etc).
This could also be attributed to using only just a single band as the base signal from which to predict HMEs.
Every HME is preceded by a rise in magnification in the light curve, but such a rise could be due to other reasons too, e.g. improperly subtracted intrinsic variability, reverberated variability from the BLR \citep{bib5:Sluse2013, bib5:Paic2022}, or just long-term microlensing (see \secref{sec5:observational_considerations}).  
Microlensing, however, has a distinct signature as a function of wavelength: larger (redder) sources are expected to get magnified sooner than smaller (bluer) ones \citep{bib5:Young1981,bib5:Schneider1987a}, which can be beautifully seen in the VRI band monitoring data of the Einstein cross shown in \figref{fig5:events_in_Q2237}.
Therefore, having such multi-wavelength information - even in two bands - is paramount for obtaining a clear and coherent signal that will lead to early and robust predictions of the onset of HMEs.

%%%%
\subsection{Corona}
\label{sec5:corona}
The first indication from microlensing that the X-ray emitting corona is fairly compact came from the study of {\it Chandra} observations of 10 quadruply lensed quasars by \citet{bib5:Pooley2007}.
They showed that, in almost every case where a flux ratio anomaly is present, it is more extreme in X-rays than in the optical.
This indicates that the X-rays must arise from a region much smaller than the optical half-light radius.
\citet{bib5:Pooley2007} qualitatively conclude that the X-ray emission region size is consistent with the inner edge of the accretion disc or smaller.  

Quantitative results first came when the light curve method \citep{bib5:Kochanek2004} was applied to multiple {\it Chandra} and {\it HST} observations of PG 1115$+$080 by \citet{bib5:Morgan2008b}.
They found that the X-ray half-light radius (rest frame 1.4--21.8 keV) is $\log{(r_\mathrm{1/2,X}/\mathrm{cm})} = 15.6^{+0.6}_{-0.9}$.
For a black hole mass of $1.2\times 10^{9} M_\odot$, this is consistent with the inner edge of the disc or smaller.
As {\it Chandra} has observed and followed-up on more lensed quasars and the light curve method has been applied to many of them, it is clear that a compact size for the X-ray emitting region is common for all systems observed to date, typically within 10 $\rg$ or less.
A summary of the results so far is given in \tabref{tab5:corona_results}.

\begin{table}
	\centering
	\caption{X-ray Microlensing Results on the size of the X-ray corona.}
	\label{tab5:corona_results}
	\begin{tabular}{llll}
		% Header
		System & $\log{(r_\mathrm{1/2,X}/\mathrm{cm})}$   & Reference \\
		\hline
		% data
		PG 1115+080 & $15.6^{+0.6}_{-0.9}$ & \citet{bib5:Morgan2008b}\\
		HE 1104$-$1805& $<15.3$ & \citet{bib5:Chartas2009}\\
		RX J1131$-$1231 & $14.4^{+0.2}_{-0.3}$ & \citet{bib5:Dai2010}\\
		Q J0158$-$4325 & $<14.6$ & \citet{bib5:Morgan2012}\\
		Q 2237+0305 & $15.5^{+0.3}_{-0.3}$ & \citet{bib5:Mosquera2013}\\
		HE 0435$-$1223 & $<14.8$ & \citet{bib5:Blackburne2014}\\
		SDSS J0924+0219 & $<15$ & \citet{bib5:MacLeod2015}\\
	\end{tabular}
\end{table}

%%%%
\subsection{Accretion disc}
\label{sec5:accretion_disc}

\subsubsection{Inner edge of the disc}
\label{sec5:FeKalpha}
The Fe K$\alpha$ reflection feature is one of the best probes we have of the inner edge of the accretion disc.
Detection and possible microlensing of this emission feature has been reported for H1413+117 \citep{bib5:Oshima2001, bib5:Chartas2004, bib5:Chartas2007}, MG J0414+0534 \citep{bib5:Chartas2002}, QSO 2237+0305 \citep{bib5:Dai2003, bib5:Chen2012}, SDSS J1004+4112 \citep{bib5:Ota2006, bib5:Chen2012}, RXJ 1131$-$1231 \citep{bib5:Chartas2012}, QJ 0158$-$4325, HE 0435$-$1223, SDSS J0924$+$0219, and HE 1104$-$1805 \citep{bib5:Chen2012}.  
In almost all cases, the equivalent width of the Fe K$\alpha$ line is much larger for these lensed quasars than it is for comparable, non-lensed AGN.
This indicates substantial microlensing and suggests a compact size for that region, comparable to or even smaller than the X-ray continuum region \citep[i.e. the corona, e.g.][]{bib5:Chen2012}.  

\citet{bib5:Chartas2017} apply their $g$-distribution method to the Fe lines seen in RX J1131$-$1231, QJ 0158$-$4325, and SDSS J1004+4112.
In this method, the observed Fe line shifts are due to microlensing caustics near the inner edge of the disc, and the distribution of the shifts can provide estimates of the size of the innermost stable circular orbit (ISCO), spin, and inclination ($i$, with zero being perpendicular to the line-of-sight) of the disc.
To date, results have only been obtained on RX J1131$-$1231, for which \citet{bib5:Chartas2017} report values of $r_{\mathrm{ISCO}} \lesssim 8.5 \rg$ and $i \gtrsim 55^\circ$.

Strong amplitude of microlensing has been found in the \ion{Fe}{iii} multiplet visible in rest-frame UV (i.e. range [2039-2113]\,\AA) suggesting that emission arises in the direct vicinity of the disc and in a region almost as compact as the latter \citep{bib5:Guerras2013b, bib5:Shalyapin2014, bib5:Fian2021b}.
By combining the SMBH mass inferred from the redshift and broadening of this line, with microlensing-based estimates of BLR sizes, \cite{bib5:Mediavilla2020} constrained the virial factor, $f$, of several low ionization lines in an ensemble of 10 quasars.

\subsubsection{Size}
\label{sec5:AD_size}

Just a year after the first detection of a microlensing-induced variability in Q 2237$+$0305 \citep{bib5:Irwin1989}, \citet{bib5:Wambsganss1990b} compared the observed variations to simulated light curves and predicted the Optical-UV size of its accretion disc radius to be $< 2\,10^{15}$\,cm. 
This early result showed consistency with the expectations of the thin-disc theory.
However, further analysis of this and another $\sim$20 systems showed that the sizes measured from microlensing are systematically larger by a factor of 3-5 than the expectations from the thin-disc model \citep[see][for a recent compilation]{bib5:Cornachione2020b}. 
These results were obtained from the analysis of the microlensing signal in optical light curves (see \secref{sec5:lcve_analysis}), focusing either on a single HME \citep{bib5:Anguita2008, bib5:Eigenbrod2008b, bib5:Mediavilla2015} or using all variations in the difference light curves over a long period of time \citep{bib5:Hainline2012, bib5:Hainline2013, bib5:MacLeod2015, bib5:Morgan2008b, bib5:Morgan2012, bib5:Morgan2018, bib5:Cornachione2020a}.
Measurements based on single-epoch observations also generally suggest larger disc sizes \citep[e.g.][]{bib5:Blackburne2010, bib5:Jimenez2012}. 
Although it seems that this method consistently results in larger sizes than predicted by the thin-disc theory, no consensus on the physical implications for quasar accretion discs has been reached. 

Before discussing possible explanations for this discrepancy, it is important to remember that single-band multi-epoch analysis can constrain only a combination of the slope of the temperature profile of the disc, $\beta$, and its size, $r_s$ (\eqref{eq5:temperature_AD}). 
This means that this tension should be investigated in the $\beta$ - $r_s$ plane, shown in \figref{fig5:disc_sizes}. 
In this perspective, a possible solution could be a shallower slope than predicted by the thin-disc theory \citep{bib5:Cornachione2020b}. 
However, slopes are more prone to systematic uncertainties than absolute source sizes, as discussed in the next subsection (\secref{sec5:thermal_slope}). 

\begin{figure}
    \centering
    \includegraphics[width=.5\textwidth]{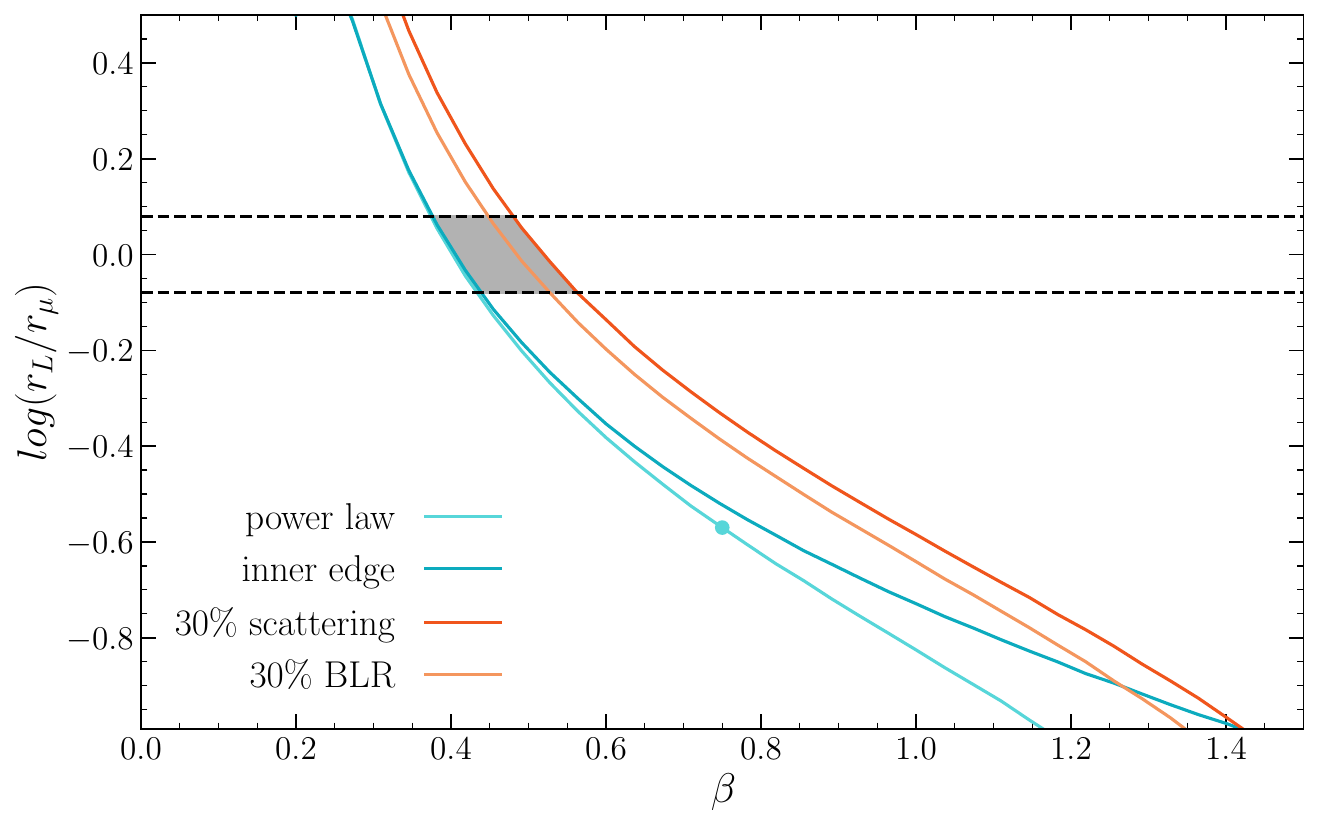}
    \caption{Ratio between the luminosity half-light radius $r_L \equiv R_{\lambda, 1/2}^{\rm flux}$ (\eqref{eq5:luminosity_size}) and the half-light radius measured from microlensing $\rhalf$ ($\equiv r_\mu$ on the figure). Discs described by a power-law temperature profile are represented by the light blue line whereas the dot indicates the particular case of $\beta = 3/4$, corresponding to the thin-disc model. Alternative disc models including an inner edge or contamination from the BLR or scattered light are also plotted. These alternative models, however, are insufficient to reconcile luminosity and microlensing sizes. For this to happen, measurements should fall in the shaded grey region, which means that the slope of the temperature profile should be reduced drastically compared to the thin-disc model. Figure reproduced from \citet{bib5:Morgan2010} and \citet{bib5:Cornachione2020a}.}
\label{fig5:disc_sizes}
\end{figure}

Several alternative explanations have been proposed to reconcile the disc size discrepancy.
One suggestion to explain the larger sizes inferred by microlensing is the inhomogeneous accretion model proposed by \cite{bib5:Dexter2011}.
However, the level of inhomogeneities required by this model to match the observations would need to be unrealistically high. 
Another plausible explanation could be that part of the UV-continuum emission is actually coming from a much more extended region, hence biasing the microlensing measurement. 
For example, \citet{bib5:Hutsemekers2015} presented spectropolarimetric observations of the BAL lensed quasar H~1413+117 that showed two distinct sources of continuum emission. 
Coupling the observations with microlensing simulations, they noted that these regions differ significantly in their size with a compact unpolarized emission coming directly from the disc and a much larger non-microlensed continuum coming from an extended region located along the polar axis. 
The fraction of total flux arising from that extended emission may however be less than $\sim$40 per cent.
This would reduce the microlensing source size by less than a factor two \citep{bib5:Dai2010, bib5:Sluse2015}.

Alternatively, \cite{bib5:Abolmasov2012b} proposed a super-Eddington accretion rate for these specific systems, leading to the formation of an optically thick envelope scattering the radiation emitted from the disc.
This effect would make the apparent disc size larger and independent of wavelength. 
Similarly, low-density scattering atmospheres could produce non-black-body emission spectra \citep[see the model by][]{bib5:Hall2018}, which would result in a flatter spectral energy distribution than a thin-disc. 
This could result in a broken power-law temperature profile, increasing the apparent half-light radius of the disc. 
A similar broken power-law profile could also appear from iron opacity \citep{bib5:Jiang2016}. 

The disc wind model \citep[e.g.][]{bib5:Li2019} with mass outflows could also explain the discrepancy. 
The principle is again the same: the disc wind is flattening the radial temperature profile that increases the source size measured by microlensing.
It might also be that the solution to this problem is related to our detailed understanding of the origin of X-ray emission and its interaction with the disc.
\citep{bib5:Papadakis2022} calculated that for a disc illuminated by the X-ray corona, energy is injected into the disc increasing the temperature mainly at larger radii.
Therefore, X-ray illuminated discs will have an increased apparent size.   
Finally, we note that both changes in the disc structure and the presence of unaccounted extended emission could be at play at the same time, as proposed by \cite{bib5:Zdziarski2022}. 
These authors showed that a combination of disc truncation at an inner radius larger than the innermost stable circular orbit, disc winds, and local color corrections to the disc black-body emission could explain larger discs. 

\subsubsection{Thermal Slope}
\label{sec5:thermal_slope} 
Any measurement of accretion disc sizes as a function of wavelength using multi-band or spectroscopic observations (single-epoch or monitoring) allows one to infer the thermal slope of the accretion disc (because $r_\lambda \propto \lambda^{1/\beta}$ or $T\propto R^{-\beta}$; \eqref{eq5:temperature_AD} and \secref{sec5:qso_accretion_disc}).
In 2008, several studies using the techniques described in \secref{sec5:data_analysis} obtained the first results for Q 2237$+$0305 \citep{bib5:Anguita2008, bib5:Eigenbrod2008a, bib5:Mosquera2009} but also for a few other systems \citep{bib5:Poindexter2008, bib5:Bate2008, bib5:Floyd2009}.
These results showed consistency with the expectation for a thin-disc model, i.e. $\beta=3/4$ \secref{sec5:qso_accretion_disc}. 
The error bars of these measurements, were too large to rule out alternative disc models. 
Thanks to multi-epoch data gathered for a sample of 11 systems, \cite{bib5:Morgan2010} alerted that, even if the size of the disc increased as $M_{BH}^{2/3}$, in agreement with the thindisc model, their too large absolute sizes implied thermal slopes in conflict with that model.
The results from \citet{bib5:Blackburne2011}, who used single-epoch multi-band data from X-ray to NIR for 12 lensed quasar systems, are in agreement with \citet{bib5:Morgan2010}. 
They infer a variety of thermal slope measurements, but most of them significantly steeper that expected ($\beta > 3/4$); even consistent with no size variation with wavelength at all (i.e. $1/\beta \rightarrow 0$). 
Several additional studies thereafter also found steeper than expected slopes \citep[$\beta \gtrsim 1$, ][]{bib5:Munoz2011, bib5:Motta2012, bib5:Jimenez2014, bib5:Munoz2016}. 
On the other hand, a few studies using either the single-epoch \citep{bib5:Rojas2014,bib5:Bate2018} or the light curve method \citep{bib5:Cornachione2020a} found slopes shallower than expected ($\beta \lesssim 0.5$).

At the time of writing, the tension between different microlensing measurements still exists. 
There are, however, several possible avenues being explored to explain the contradicting results.
A first attempt at explaning this investigates the impact of flux not arising from within the disc. 
As one may notice by looking at an AGN spectrum (\figref{fig5:sketch}), broad band data can include some fraction of emission from the broad emission lines.
The contribution to the total flux from the two regions will vary in each lensed quasar image due to differential microlensing (see \figref{fig5:microimages_light_curves}). 
To minimise contamination, \cite{bib5:Jimenez2014} have specifically used data originating from narrow band imaging, spectroscopy, and broad-band imaging free of salient emission lines. 
This sound observational strategy does not however correct for the contribution arising from less salient lines, like the \ion{Fe}{ii} pseudo-continuum, Balmer continuum emission or other pseudo-continua, which are present in different amounts in AGN spectra\footnote{The impact of this ``blended'' emission on slope measurements is not specific to microlensing as it also affects reverberation mapping \citep[e.g.][]{bib5:Cackett2020}.}.
\citet{bib5:Sluse2015} measured the disc temperature profile of H1413$+$117 from single-epoch data and shown that not correcting for the observed extended continuum pushes the measured slope towards steeper temperature profiles. 
Due to the different amount of non-disc emission among quasars, but also of its possible variability as a function of wavelength (and possibly time), it is difficult to predict a generic impact on the measured slope. 
The latter effectively corresponds to a ratio of sizes, and may be thus prone to any systematic uncertainty affecting the reference source size. 
To mitigate this effect, \cite{bib5:Cornachione2020b}, have chosen to explore the inconsistency between the source size and luminosity size $r_L$ (\eqref{eq5:luminosity_size}, \figref{fig5:disc_sizes}) as a function of $\beta$.

A second explanation is systematic biases and uncertainties in the methods.
For the single-epoch technique, \citep{bib5:Bate2018} showed that low amplitudes of chromaticity, i.e. $< 0.4$ mag, defined  as the difference of microlensing magnification between the reddest and bluest wavelengths, may yield slopes that are steeper than expected. 
On the other hand, the selection of observations displaying large amplitude of microlensing may systematically favour smaller sizes.
Overall, simulations suggest that selection effects need to be carefully assessed when interpreting single-epoch measurements \citep{bib5:Bate2018, bib5:Guerras2020}. 
Monitoring data may be less prone to biases but the impact of sampling, number of bands, and length of monitoring on the results still needs to be quantified.
A related source of systematic uncertainty is the one associated to the macro-model and specifically to the exact value of the macro-magnification.
This can be uncertain by a factor of a few in each multiple image (see \secref{sec5:observational_considerations}).
This may impact the theory size of the source (see \secref{sec5:qso_accretion_disc}), which depends on the magnification-corrected size of the source, i.e. $R_\lambda^{\rm BH} \propto L^{1/3}$ \citep[see e.g.][]{bib5:Cornachione2020c}.
A biased estimate of the black hole mass may have an effect when comparing microlensing sizes to theory as $R_\lambda^{\rm BH} \propto M_{\rm BH}^{2/3}$ \citep{bib5:Morgan2018}. 

Finally, a third explanation is that intrinsic scatter of disc properties between AGN naturally takes place.
For instance, the accretion efficiency, $\eta$, directly impacts the source size (\eqref{eq5:radius_MBH}), but depends on the black hole spin \citep[e.g.][]{bib5:Thorne1974}. 
However, as discussed by \citet{bib5:Morgan2010}, the efficiency alone may be insufficient explaining all the microlensing data. 
On the other hand, a non universal disc model is also a possibility, with a slim disc or an advection dominated flow occuring for high or low Eddington accretion rates respectively \citep[e.g.][]{bib5:Lasota2023}.
Therefore, it could be that the various disc models invoked to explain over-sized discs (see discussion in \secref{sec5:AD_size}) are in fact the case for different subsamples of systems. 

In any case, none of the above solutions may be sufficient on its own to explain all microlensing measurements. 
Large multi-epoch surveys of the sky, like LSST, but also targeted works through e.g. spectroscopic monitoring, in combination with a careful account for potential sources of systematic uncertainties, may enable to design the optimal experiment for revealing the properties of the accretion disc.

%%%%
\subsection{Broad Line Region}
\label{sec5:broad_line_region}

Microlensing of the BLR has long been thought to be a marginal effect \citep{bib5:Schneider1990, bib5:Lewis1998a}. After the early measurements of the size of BLR with reverberation mapping at the end of the 1990s, several authors have suggested that BLR microlensing may be larger than previously though \citep{bib5:Abajas2002, Lewis2004}. This has been soon after confirmed with the first striking line deformation observed by \cite{bib5:Richards2004} in the large separation lensed quasar J1004$+$4112. It is now securely detected in more than 30 AGN and suspected to be visible in at least 80\% of the microlensed AGN \citep{bib5:Sluse2012, bib5:Guerras2013a}. Often, only a small fraction ($\lesssim$ 10\%) of the line flux is found to be microlensed, but extreme cases have been observed where almost the whole line is magnified by ML \citep{bib5:Richards2004, bib5:Keeton2006, bib5:Sluse2007, bib5:Sluse2011}. Except in the situation where some velocities are magnified and other are demagnified \citep{bib5:Hutsemekers2023}, the amplitude of microlensing of the broad lines provides a robust estimate of the projected and luminosity weighted BLR size. Information on the structure and kinematics of the BLR is encoded in the microlensing induced deformation of the emission line profile (\figref{fig5:BLR_deformation}). A variety of line deformations are observed, affecting either differently each wing of the line \citep{bib5:Richards2004, bib5:Braibant2014}, or symmetrically the whole line, with however the higher velocities being more magnified than the lower ones. That finding supports a BLR kinematics where the highest velocity component takes place closer to the central engine. Overall, the diversity of broad emission line deformations is qualitatively supported by simulations. The following subsections review important insights on the BLR obtained for various emission lines.  

\subsubsection{High ionization lines} 
\label{sec5:high_ionization_lines}

\cite{bib5:Guerras2013a} have shown, from the analysis of microlensing in image pairs in 16 strongly lensed systems that the size of high ionization lines (\ion{O}{vi}]~$\lambda\lambda$1035, \ion{Ly}{$\alpha$}+\ion{N}{v}~$\lambda\lambda$ 1216, \ion{Si}{iv}+\ion{O}{iv}~$\lambda\lambda$ 1400 and \ion{C}{iv}$~\lambda\lambda$ 1549) is substantially smaller than that of the lower ionization lines, providing a confirmation of the stratification of the BLR expected from photo-ionization models, and observed in reverberation mapping data. 
This result has recently been confirmed from an expanded sample of systems by \cite{bib5:Fian2021b}. Large sample studies \citep{bib5:Sluse2012, bib5:Guerras2013a}, also reveal frequent occurrence of microlensing in one of the line wings, indicating that the high ionization BLR may not have a spherically symmetric geometry.

Detailed studies of individual systems have provided further insights on the size and structure of the BLR for specific quasars. 
To-date, the best studied system has been the Einstein cross. \cite{bib5:ODowd2011} have shown that the microlensing-induced line deformation observed in that system favour a gravitationally dominated dynamics over an accelerating outflow. 
Furthermore, constraints on the size and geometry of the region emitting \ion{C}{iv} have obtained owing to the spectroscopic monitoring carried out by \citet{bib5:Eigenbrod2008a}. 
On the one hand, the microlensing-induced size has been found to be compatible with the $R_{BLR}-L$ relation measured in local AGN \citep{bib5:Sluse2011, bib5:Hutsemekers2021}. 
On the other hand, forward models of line deformations observed for 3 geometries and kinematics models (and a range of orientations with respect to the observer) ruled out an equatorial wind model, and favoured a keplerian disc over a biconical polar wind \citep{bib5:Hutsemekers2021}. 
In addition, inclination of $\sim 40\deg$ of BLR had been favoured by these authors, in agreement with the Type 1 properties of this system, and previous constraints on the inclination derived from microlensing of the accretion disc \citep{bib5:Poindexter2010a}. 
Another system displaying salient line deformation is the large separation quad  J1004$+$4112. 
The recurrent asymmetric blue-wing enhancement of \ion{C}{iv} observed in that system has been qualitatively reproduced by \cite{bib5:Abajas2007} using a biconical geometry.
The microlensing interpretation of the atypical microlensing signal observed in this system \citep{bib5:Lamer2006} has been challenged by \cite{bib5:Green2006} who has proposed a non-microlensing explanation of the data, where each strongly-lensed image is crossing different regions of warm outflowing material. 
As shown by several authors \citep[e.g.][]{bib5:Lamer2006, bib5:Hutsemekers2023}, the choice of normalisation of the line plays an important role in the apparent recurrence of the blue wing enhancement and in the quantification of the velocity dependant amplitude of microlensing. 
Using as baseline the flux ratios from radio data \citep{bib5:Hartley2021}, \cite{bib5:Hutsemekers2023} have shown that a microlensing-induced deformation of the line stable over 15 years exists, with the red wing being demagnified while the blue wing is strongly magnified. 
This scenario is supported by simulations that enable them to show that a keplerian disc or an equatorial wind provide a good description of the BLR. 

\subsubsection{Low ionization lines}
\label{sec5:low_ionization_lines}

Emission lines requiring lower ionisation energy, such as Balmer lines (in e.g. H$\alpha\,\lambda 6562.8 $ and H$\beta\,\lambda 4861.3$) and \ion{Mg}{ii}$\,\lambda\lambda$2798, are observed at larger wavelengths than higher ionization lines. 
Because most lensed AGNs lie at  $z> 1.5$, those lines have been less regularly observed and scrutinized for microlensing than higher ionization ones. \ion{Mg}{ii} is found to be on average less microlensed than higher ionization lines, which means that their sizes are a few times larger than high ionization lines \cite{bib5:Guerras2013a, bib5:Fian2021b}. 
This is in agreement with the expected stratification of the BLR with increasing ionization degree. 

Symmetric line deformations are the most common, but asymmetric deformation are occasionally observed. 
There are no statistical constraints on line deformations of Balmer lines, but both symmetric and asymmetric deformation of these lines  have been reported \citep{bib5:Sluse2007, bib5:Braibant2016, bib5:Braibant2014}. 
For two systems (Q 2237$+$0305 and HE 0435$-$1223), a forward modeling of the observed deformation has been performed (cf. \secref{sec5:BLR_modeling}), strongly favouring a keplerian disc geometry \citep{bib5:Braibant2016, bib5:Hutsemekers2019, bib5:Hutsemekers2021}. 
A half-light radius $\rhalf(\rm{H}\beta) = 47 \pm 19$ lt-days has been derived for Q 2237$+$0305, about an order of magnitude below the expectation from $R_{BLR} - L$ relation derived from reverberation mapping \citep{bib5:Grier2017}. 
\cite{bib5:Hutsemekers2021} have proposed that this could be related to the high accretion rate estimated for that system. 

The comparison of line deformations based on data obtained at the same epoch (or pairs of data accounting for the time delay) provide further insights on the existence of a common BLR structure for various atomic species. 
To our knowledge, there are only a handful of systems where both \ion{Mg}{ii} and a Balmer line have been scrutinized. 
In HE 0435$-$1223, \ion{Mg}{ii} and H$\alpha$ are found to be similarly microlensed \citep{bib5:Braibant2014}. 
For WGD2038$-$4008, both lines are found to be free of microlensing deformation \citep{bib5:Melo2021}. 
This contrasts with the results obtained for RXJ 1131$-$1231 where a very broad and spatially compact emission has been identified only in \ion{Mg}{ii} \citep{bib5:Sluse2007}. 
While differences are also observed between these two lines in SBS0909$+$532, this is at two very different epochs \cite{bib5:Mediavilla2005, bib5:Guerras2013a}). Therefore, the microlensing configuration could have varied substantially between the two observations, precluding to draw robust conclusions. 

The \ion{C}{iii]}$~\lambda 1908.73$ line, sometimes classified as intermediate ionization line, is commonly present is lensed AGNs spectra but suffers from a strong blending with \ion{Al}{iii} and \ion{Si}{iii} lines, which complicates the identification of microlensing induced line deformation. 
Nevertheless, substantial microlensing is often identified in this line, with appearance of blue/red asymmetries similar to what is found for \ion{C}{iv} \citep{bib5:Sluse2012, bib5:Fian2021b}. 
This line may therefore arise from a region smaller than the one emitting low-ionization lines. 
In their analysis of the spectro-photometric monitoring data of Q 2237$+$0305, \cite{bib5:Sluse2011} found an emitting size of \ion{C}{iii]} emission region compatible with the one of \ion{C}{iv}, but with hints of a different structure in the two regions. 

\subsubsection{UV-Optical Iron lines} 
\label{sec5:UV_optical_ionization_lines}

The Iron emission blend covers almost the whole UV to optical range (\secref{sec5:BLR}). 
\cite{bib5:Guerras2013b} performed the only systematic study of microlensing of UV \ion{Fe}{ii} in a sizeable sample of 13 quasars.
They focused on the range [2050-2650]\,\AA\, and identified strong microlensing in 4 systems. 
They derived a size of the corresponding emitting region of several light-days, comparable to the accretion disc size. 
Evidence for compact UV \ion{Fe}{ii} emission has also been reported in Q 2237+0305 \citep{bib5:Sluse2011}, H1413+117 \citep{bib5:ODowd2015} (in [1590-1680]\,\AA) and in J1131-1231 (in [3080-3540]\,\AA) by \cite{bib5:Sluse2007}. 
The low redshift of the latter system enabled those authors to compare microlensing of the UV and optical \ion{Fe}{ii}. 
They found a microlensing of optical \ion{Fe}{ii} comparable to that of the Balmer lines, at the exception of the region [4630-4800]\,\AA~potentially arising from a more compact region. 

% text moved to 4.3.1 because of compactness - Suggestion from Ref. 1
%Microlensing of the \ion{Fe}{iii} multiplet in the range [2039-2113]\,\AA \ has also been scrutinized with microlensing.
%Strong amplitude of microlensing has been found in this this multiplet, suggesting that \ion{Fe}{iii} may arise in the direct vicinity of the disc and in a region almost as compact as the latter \citep{bib5:Guerras2013b, bib5:Shalyapin2014, bib5:Fian2021b}. 
%By combining the SMBH mass inferred from the redshift and broadening of this line, with microlensing-based estimates of BLR sizes, \cite{bib5:Mediavilla2020} constrained the virial factor, $f$, of several low ionization lines in an ensemble of 10 quasars. 

\subsubsection{Broad absorption lines}
\label{sec5:broad_absorption_lines}

Broad absorption lines ($FWHM > 2000\,\kms$) are present in 10 to 20\% of quasars at $2 < z < 4$ , and are predominantly observed in resonance lines of ionized species such as \ion{C}{iv}, \ion{Si}{iv} or \ion{N}{v} \citep{bib5:Allen2011}. 
The blueshifts from $1000\,\kms$ to up to 20\% of the speed of light displayed by the absorber may be the signature of outflowing gas. 
At X-ray wavelengths, BALs quasars are generally X-ray weak, probably due to absorption\footnote{We may note that mini-BAL, characterised by a small level of UV absorption, and narrow absorption line quasars, characterised by absorption width with $FWHM < 500 \,\kms$, generally show different X-ray properties \citep{bib5:Hamann2013, bib5:Chartas2021}.} \citep{bib5:Gallagher2006}. 
The idea of using  microlensing for probing the structure of the outflowing gas in BAL quasars has been proposed very early, but yet applied to a limited number of systems \citep[e.g.][]{bib5:Hutsemekers1993, bib5:Hutsemekers1994, bib5:Lewis1998a, bib5:Chelouche2005}. 

The best-studied system is  H1413$+$117, a.k.a. the cloverleaf. 
The main insights on the geometry of the outflow come from the decomposition of the line profiles using the MmD decomposition (see \secref{sec5:microlensing_from_spectra}). 
The analysis of this system  \citep{bib5:Hutsemekers2010, bib5:ODowd2015} reveals (a) a nearly black absorption of \ion{C}{iv} in the microlensed flux, and unveils the underlying line emission in the non microlensed component; (b) an onset velocity of the absorption of about 1500\,$\kms$ for \ion{C}{iv} and a larger velocity onset of \ion{Si}{iv} and \ion{Al}{iii}; (c) part of the \ion{C}{iv} emission is reabsorbed over a small wavelength range.
These characteristics support radial changes in the absorbing material, which can be interpreted as a two components outflowing wind: one component which may be co-spatial with the emission, and another one, more distant, that partially re-absorb the BEL emission \citep{bib5:Borguet2010}. 
The ionization dependence of the onset velocity of the BAL can be interpreted in the context of the disc$+$wind model.  

%%%%
\subsection{Torus}
\label{sec5:torus}

Due to limited access to the mid-infrared range from the ground and to the low resolution of mid-infrared instrumentation in the last decades, observations of lensed quasars in the mid-infrared have been rather limited. 
\cite{bib5:Agol2000} reports the first measurements of the flux ratios in that range for the Einstein Cross. 
The flux ratios measured at 8.9 and 11.7 $\mu$m were found to be in agreement with those measured at radio wavelengths, but relative uncertainties reach $\sim$20\%. 
Nevertheless, this observation has been used to rule out the presence of (compact) synchrotron emission in the mid-infrared, and support the idea that the emission in that range is dominated by dust arising from a region too large to be microlensed \citep{bib5:Agol2000, bib5:Wyithe2002b}.
Several additional observations of the Einstein Cross have been performed over the years \citep[see][for a census]{bib5:VivesArias2016}, in broad agreement with the early measurements of \cite{bib5:Agol2000}. 
The absence of microlensing yields $R_{1/2} \gtrsim 200 \sqrt{< M> /0.3 M_\odot }$ lt-days at about 11\,$\mu$m \citep{bib5:VivesArias2016}. 
Due to this large size, the few other observations of lensed quasars in the mid-infrared have instead been used to search for flux ratio anomalies, interpreted as microlensing by substructures of mass $M\sim 10^6-10^8 M_\odot$ \citep[e.g.][]{bib5:Chiba2005, bib5:Minezaki2009}. 

Although microlensing in the mid-infrared range is small, it may not always be negligible. 
Observations of 6 lensed quasars at $2.2\,\mu$m and $3.8\,\mu$m by \cite{bib5:Fadely2011} show that microlensing can occur. 
Since this typically corresponds to $\sim 1\,\mu$m rest-frame, this may be explained by the yet substantial contribution of accretion disc emission. 
Observations of unlensed quasars as well as simulations however suggest that, for lower luminosity systems, the hot component of the torus (peaking around 2.2 $\mu$m rest-frame) may be microlensed in some systems \citep{bib5:Stalevski2012b, bib5:Sluse2013}. 
Upcoming observations of lensed quasars with the James-Webb Space Telescope is expected to shed new lights on the presence of microlensing in wavelengths ranges dominated by dust emission. 

%%%%
\subsection{Scattered and polarized emission}
\label{sec5:scattered_and_polarized_emission}

Polarization of UV-optical light in AGNs is observed at levels ranging between zero and 4-5\%, rare systems sometimes reaching polarization levels of 10-20\%. 
Broad absorption line AGNs, radio-loud systems and blazar generally display the larger linear polarization degrees. 
In the UV-optical, the polarization arises predominantly from scattering of the light by electrons, and maybe dust. The exact location of that region (equatorial/polar, inside/outside the BLR, etc), and even its ubiquity in all AGN are yet totally opened questions. Because microlensing re-weight the spatial distribution of the emission in the inner regions of AGNs, it is expected to produce (de-)polarization of AGNs, hence modifying both the polarization degree and angle of microlensed images \citep{bib5:Belle2000, bib5:Hales2007, bib5:Kedziora2011}. 

The first observational evidence for polarization differences induced by microlensing required the high resolution capabilities of the HST targeting the BAL quasar H1413$+$117 \citep{bib5:Chae2001}.
It took more than a decade before obtaining resolved polarimetric and spectropolarimetric observations of individual images of lensed quasars from the ground, again for H1413$+$117 \citep{bib5:Hutsemekers2010, bib5:Hutsemekers2015, bib5:Sluse2015}. 
Those studies have shown a twist of the polarisation angle of image D, which displays slowly varying microlensing since about 25 years. 
This observation can be explained with the presence in that BAL of two spatially separated regions producing orthogonal polarization: an equatorial region which is microlensed, and a more extended polar region. 
Resolved polarised observations of the two large separation lensed quasars J1004$+$4112 and Q0957$+$561 have been presented by \cite{bib5:Popovic2020, bib5:Popovic2021}. 
Similarly, changes in the polarisation degree and angles are observed in those objects due to microlensing. 
This supports the existence of a compact a equatorial scattering region in systems which are not BALs. 

In addition to an imprint of microlensing in the polarisation signal, spectroscopic data of microlensed BAL have revealed the presence of extended diffuse emission separated from the accretion disc \citep{bib5:Sluse2015, bib5:Hutsemekers2020}. 
This discovery has been facilitated by the presence of almost black absorption covering the emission from the accretion disc but not the most extended continuum. 
Differential microlensing has revealed non-microlensed continuum emission contributing to up to 40\% of the observed continuum. 
Despite of contributing to a large fraction of the observed continuum, this emission is insufficient to explain the larger size of the accretion disc unveiled by microlensing studies \citep[Sect.~\ref{sec5:accretion_disc}; ][]{bib5:Sluse2015}. 
The presence of non-microlensed (pseudo-)continuum emission in non-BAL systems is yet to be demonstrated, as well as its exact origin. 
If this emission is generically present in AGNs, a small flickering of microlensed lightcurves could be expected at wavelengths free of broad emission lines. 
While such flickering is observed in numerous systems (see Sect.~\ref{sec5:lcve_analysis}), it is not yet demonstrated that it is caused by extended continuum emission.

%%%%%%%%%%%%%%%%%%%%%%%%%%%%%%%%%%%%%%%%%%%%%%%%%%%%%%%%%%%%%%%%%%%%%%%%%%%%%%%%%%%%%%%%%%%%%%%%%%%%%%%%%%%%%%%%%%%%%%%%%%%%%%%%%%%%%%%%%%%%%%%%%%%%%%%%%%%%%%%%%%%%%%%%%%%%%%%%%%%%%%%%%%%%%%%%%%%%%%%%%%%%%%%%%%%%%%%%
\section{Lensing Galaxy Results}
\label{sec5:lensing_galaxy_results}
What can be constrained uniquely by microlensing about the lens galaxy is the partition of its mass into baryons (compact) and dark (smooth) matter, and how the baryonic component is further partitioned between the microlenses, i.e. its mass function.
As we have already seen in \secref{sec5:caustic_structures_macro_parameters} (see also \figref{fig5:map_tiles}), the values of $\kappa,\gamma,\kappastar$\footnote{Separating the shear into a smooth and compact component does not make any difference in the lens equation (see \eqref{eq5:microlensing_equation}), as opposed to the mass density. Note that if there is an external shear component then it would have to be subtracted in order to link the local values of the shear at the positions of the multiple images to the second derivative of the lens mass distribution.} at the positions of the multiple images, which are directly linked to the local values of the second derivatives of the lens potential (see \chapintro), define microlensing variability.
The total mass in baryons, defined by $\kappastar$, can be further distributed in compact objects in different ways.
Microlensing magnification is generally more sensitive to the mean microlens mass, but information on a mixture of masses and the IMF can be extracted under specific conditions.

\subsection{Measuring the baryonic/dark matter content under different constraints from macromodels}
\label{sec5:measuring_kappastar}
In principle, the $\kappa,\gamma,\kappastar$ could all be constrained as free parameters of a microlensing model fitted to observables, but in practice, only a single free parameter is used.
This can be either $\kappastar$ itself, with $\kappa,\gamma$ derived separately from a macromodel, or a parameter partitioning the total mass between a stellar and dark mass profile (usually the mass-to-light ratio), leading to a sequence of covariant $\kappa,\gamma,\kappastar$.
A lens macromodel is a mass distribution/lens potential fitted to the imaging data of a lensed quasar.
Lens modelling requires some form of the mass profile to be chosen first, e.g. SIE, power-law, etc, and proceeds by making specific assumptions for the (unknown) extended source brightness profile, including the quasar point source, and solving the lens equation to fit the lensed light components to the data (see \chapgal).
It is straightforward to calculate the values of $\kappa,\gamma$ at the multiple image positions from the resulting lens potential.
One could additionally make use of the fact that light traces mass, fit a two-dimensional brightness profile to the observed lens light (e.g. a Sersic profile, spline polynomials, etc), assume some mass-to-light ratio, and thus compute a stellar-mass component to incorporate in the mass model.
In combination with a dark matter component, e.g. a Navarro-Frenk-White profile, $\kappastar$ can be provided straightforwardly, albeit not in a completely unbiased way (see below).

Both flux ratio and light curve microlensing data can be used to constrain the local graininess of matter, i.e. the value of $\kappastar$, and both require simulated magnification maps.
These maps are generated for given $\kappa,\gamma,\kappastar$ combinations derived from some macromodel and used to extract observable quantities.
In this case, all the other parameters of the model - accretion disc profile, effective velocity, microlens mass - are treated as nuisance parameters and marginalised over.
By fitting the model to the data, the most likely $\kappastar$, or more precisely $\kappa,\gamma,\kappastar$ combination, can be found.
We present the results on $\kappastar$ from flux ratios and light curves below.

\subsubsection{Flux ratios}
The idea of constraining $\kappastar$ from microlensing flux ratios was originally proposed by \citet{bib5:Schechter2002}, who found that a smooth mass component in addition to microlenses can increase variability especially for saddle-points\footnote{\citet{bib5:Schechter2002} showed this for a single $\kappa,\gamma$ combination.
This result was later extended to the entire parameter space by \citet{bib5:Vernardos2014a}.}.
\citet{bib5:Bate2007} and \citet{bib5:Congdon2007} found that a finite source can in fact cause the same broadening of microlensing magnification distributions, requiring the two parameters - source size and $\kappastar$ - to be studied simultaneously.
\citet{bib5:Mediavilla2009} studied flux ratios in 29 image pairs obtaining their $\kappa,\gamma$ from a SIS plus external shear macromodel and found an overall $5_{-3}^{+9}$ per cent of matter in stars (at 90 per cent confidence) for a given source size \citep[their result is positively correlated with the source size and probably contains various other systematic effects and biases, see the discussion in][]{bib5:Mediavilla2009}.
In a later study, \citet{bib5:Jimenez2015a} and \citet{bib5:Jimenez2015b} took into account the source size simultaneously with $\kappastar$ and found consistently 20 per cent of matter in stars at $1-2$ effective radii for a sample of 18 lenses from flux ratios in the optical and X-rays respectively.
They used the more realistic SIE plus external shear macromodels from \citet{bib5:Schechter2014}, which are known to produce smaller magnifications.
\citet{bib5:Bate2011} found $20-50$ per cent of matter in stars within $5-10$ kpc of the lens center for two systems after marginalizing over the source properties (size and temperature profile) and $\kappastar$ for Q 2237\allowbreak$+$0305.
This was an expected outcome for this system because the multiple images are located so close to the lens center that the stellar mass component is comparable to, or even dominates over, smoothly distributed dark matter \citep[see references in ][for the macromodels they used]{bib5:Bate2011}.
\citet{bib5:Pooley2012} found 7 per cent of matter in stars at a mean distance of $6.6$\,kpc from the lens center by analyzing an ensemble of 14 quadruply lensed quasars and a decreasing trend of $\kappastar$ with distance.
They derived their $\kappa,\gamma$ from a SIS plus external shear macromodel \citep[see also][]{bib5:Pooley2007,bib5:Blackburne2011}.

The results summarized above are prone to systematic errors in measurements and degeneracies in modelling that are specifically associated with flux ratio snapshots.
As discussed previously (see \secref{sec5:accretion_disc}), analyzing flux ratios with microlensing is more robust for close image pairs that have almost the same macro-magnification and small time delays.
But properly deblending the image fluxes can be tricky precisely because of the proximity of the images, while unaccounted for intrinsic source variability can lead to $\kappastar$ overestimates \citep{bib5:Mediavilla2009}.
Any overlap of the observed wavelengths (e.g. for a broad band) with emission lines, especially broad lines that are known to undergo microlensing as well, could contaminate the measured flux ratios.
The extent of the chromatic variation and the level of the no-microlensing baseline, which are known to affect source measurements \citep{bib5:Bate2018}, could also play a role for measuring $\kappastar$, but this has not been investigated yet.
Finally, there are cases where the flux ratio data are simply not sufficient to lead to any conclusive $\kappastar$ results\footnote{For the accretion disc, \citet{bib5:Guerras2020} find that at least 4 measurements (or 4 pairs of images) should be used to obtain reliable estimations. Although not explicitly studied, something similar probably holds for measuring $\kappastar$ too.} \citep[e.g.][]{bib5:Bate2008,bib5:Floyd2009}.

\subsubsection{Light curves}
Constraining $\kappastar$ from light curve data has proceeded largely by applying the fitting method of \citet{bib5:Kochanek2004}.
Studies using this approach have adopted a different strategy to calculating $\kappa,\gamma$ from a macromodel and having $\kappastar$ as a free parameter.
Their mass model consists of a combination of a NFW halo and a concentric de Vaucouleurs component of the mass in stars, obtained by fitting the light of the lensing galaxy and a constant mass-to-light ratio \citep[see also][]{bib5:Lehar2000}.
The total mass being conserved, a single free parameter, $f_{\rm{ML}}$, describes the combination of the two components, from purely stars ($f_{\rm{ML}}=1$) to just a NFW halo ($f_{\rm{ML}}=0$).
Varying this parameter leads to different $\kappa,\gamma,\kappastar$ combinations, not just $\kappastar$ (and also a varying mass-to-light ratio across the lensing galaxy).
Examples of such families of models are shown in \figref{fig5:pspace} - the points from \citet{bib5:Morgan2008b}, \citet{bib5:Dai2010}, and \citet{bib5:MacLeod2015}.

Light curve data analysis favours low values of $\kappastar$, albeit not decisively.
\citet{bib5:Chartas2009} and \citet{bib5:Dai2010} find $f_{\rm{ML}}=0.2$ and 0.3 for lensed quasars HE1104$-$1805 and RXJ 1131$-$1231 respectively.
Modest trends supporting dark matter dominated models are found for PG1115\allowbreak$+$080 \citep{bib5:Morgan2008b}, QJ0158\allowbreak$-$4325 \citep{bib5:Morgan2012}, and WFI2026\allowbreak$-$4536 \citep{bib5:Cornachione2020b}, while measurements for Q0957\allowbreak$+$561 \citep{bib5:Hainline2012}, SBS0909\allowbreak$+$532 \citep{bib5:Hainline2013}, and J0924\allowbreak$+$0219 \citep{bib5:Morgan2006,bib5:MacLeod2015} have been inconclusive.

One of the biggest caveats of analyzing light curves is the very demanding computations \citep{bib5:Kochanek2004,bib5:Poindexter2010a} that have resulted in the analysis of only $\approx 10$ systems \citep{bib5:Cornachione2020a}.
However, the generally weak trends in $\kappastar$ are probably due to insufficient uncorrelated variability in the light curves to constrain the dark matter fraction \citep{bib5:Hainline2012}.
Including the time dimension, as opposed to snapshots, comes at the cost of additional parameters, mainly in relation to the effective velocity, but this alone does not seem to affect the measurements of $\kappastar$.

\subsubsection{Impact of macromodel uncertainty on measuring \texorpdfstring{$\kappastar$}{kappastar}}
Apart from the possible systematic biases and statistical uncertainties inherent to the analysis methods described above, any uncertainty of the macromodel itself can potentially have an impact on $\kappastar$ measurements.
\citet{bib5:Vernardos2014c} have shown that $\Delta\kappa,\Delta\gamma$ as small as 0.02 can lead to significant differences in the properties of microlensing magnification.
Macromodel biases and uncertainties of this magnitude are not uncommon and can be due to three reasons that we briefly list here.
First, they can be due to missing ingredients of the mass model, like perturbers, disc- or bar-like structures, or higher order moments in the potential \citep[e.g. see][for an example of how the latter affect time delays]{bib5:VanDeVyvere2022a}.
For example, not accounting for the nearby galaxy G2 and a few close-by perturbers in the model of WFI2033$-$4723 yields a relative change of $\kappa$ by up to 60 per cent, and of the absolute macro-magnification by up to a factor 2 for some of the lensed images \citep{bib5:Rusu2020}.
The presence of any other massive substructure ($>10^6 M_{\odot}$), whose impact on flux ratios can be dramatic \citep{bib5:Mao1998,bib5:Dalal2002,bib5:Metcalf2002}, can bias the values of $\kappa,\gamma$ locally, albeit with different characteristics compared to microlensing \citep{bib5:Inoue2016}.
Second, there is a fundamental degeneracy between the initial mass function (IMF) and the dark matter fraction \citep[e.g. see][]{bib5:Oguri2014,bib5:FoxleyMarrable2018}, which, in short, states that the same lens potential and light distribution can be produced by a low dark matter fraction and many low-mass (and less luminous) stars, or a high dark matter fraction and fewer but brighter stars.
Finally, the mass-sheet degeneracy \citep{bib5:Falco1985,bib5:Gorenstein1988}, i.e. scaling the lens galaxy mass distribution and adding a constant surface mass density (mass-sheet), leaves observables such as the image positions, shapes, and flux ratios, unchanged but affects the $\kappa,\gamma$.
Although there are ways to mitigate these effects, e.g. by breaking the mass-sheet degeneracy through modelling of kinematics data (see \chapgal) or explicitly modelling substructures (see \chapdm), the complexity of the true lens potential and the corresponding quality of a macromodel will eventually affect microlensing measurements.

\subsection{Microlens mass and the IMF}
\label{sec5:mean_mass_imf}
Microlensing variability depends on the relative size of micro-caustics on the source plane with respect to the emitting source.
A typical value for this size is $\re$ that is $\propto \mathrm{M/M_{\odot}}^{1/2}$ (\eqref{eq5:einstein_radius_source}), where M is the microlens mass. 
The masses for a population of microlenses are, in principle, free to vary in the broad range roughly from small star clusters (or even intermediate mass black holes) to sub-stellar objects like planets and moons.
In the latter case, one can speak of `nano-lensing'\footnote{The other extreme is usually referred to as `milli-lensing'. This is out of the scope of this review because the multiple images produced can be either separated enough to be detected or because any variability takes place on timescales of centuries or more. The reader is referred to \chapdm\,for more details.}, which is still relevant for quasars as long as there is a wavelength to observe in which their emitting region is accordingly small.
It turns out that microlensing variability (at a fixed source size) depends primarily on the microlens mean mass and less on the shape of the mass spectrum.
However, the IMF of the lensing galaxy can still be probed by measuring the microlens mean mass across different wavelengths (or by indirectly measuring an overall mass-to-light ratio for the galaxy lens, see below).
This is suggested by observations as well, for example, by the lack of correlation between the microlensing variability in optical and X-ray observations found by \citet{bib5:Mosquera2013} \citep[see also][]{bib5:Pooley2007}, where the former could be due to stellar microlenses and the later due to planetary mass nanolenses \citep[e.g.][]{bib5:Guerras2020}.

% First results from simulations
It has been known from early on \citep{bib5:Wambsganss1992a,bib5:Lewis1995,bib5:Wyithe2001,bib5:Congdon2007} that the expected microlensing magnification shows very little dependence on the shape of the mass spectrum of the microlenses.
Consequently, a widely adopted approach  has been using a fixed-mass microlens population to produce magnification maps.
\citet{bib5:Schechter2004} found that magnification histograms can be substantially different in the case of a mix of two such populations whose masses differ by more than a factor of ten but have a comparable total mass.
\citet{bib5:Esteban2020} further examine such bi-modal mass distributions and find that they can be replaced by a single population with mass equal to their geometric mean.
Both studies admit that they intentionally focus on extreme cases and that for extended sources the low mass component can effectively behave like a mass-sheet - a case equivalent to a single fixed-mass population with a different $\kappastar$ and the low-mass objects absorbed into a smooth component.

% Sub-stellar masses
The source appears the smallest in the X-rays and allows one to probe the sub-stellar mass range for the microlenses.
The innermost accretion disc is known to emit the relativistically blurred iron Fe K$\alpha$ line in the X-rays \citep{bib5:Reynolds2003}.
Lensed quasars can display shifts in the peak energy of this line over time, attributed to microlensing \citep{bib5:Chartas2017}.
\citet{bib5:Dai2018} and \citet{bib5:Bhatiani2019} modelled such shifts in three systems and interpreted them as due to the presence of a population of sub-stellar mass objects ($10^{-8} - 10^{-3}$ M$_{\odot}$).
They base their argument on the fact that the cross section of an X-ray source is too small for microlensing by stellar-mass objects.
This can be understood by considering the caustic density resulting from the same total mass density divided into a few stellar-mass microlenses compared to many more smaller mass ones.
As the caustic density increases with smaller masses so does the corss section for microlensing of an X-ray (very small) source \citep[see fig. 2 in][]{bib5:Bhatiani2019}.
The study of \citet{bib5:Guerras2020} similarly supports planetary masses for the microlenses using the standard deviation of distributions of uncorrelated X-ray flux ratios over time (i.e. a collapsed light curve).

% Stellar masses:
Observations in the optical are mostly affected by stellar-mass microlenses.
\citet{bib5:Jimenez2019} calculate the microlens mean mass of a sample of 24 lensed quasars to be $0.08 < \mathrm{M/M}_{\odot} < 0.21$ (at 68 per cent confidence) using optical and X-ray flux ratios.
They used fixed-mass simulations and marginalized over all the other free parameters of their model.
Nevertheless, there was still a degeneracy with the source size that was broken by using a size prior from reverberation mapping \citep{bib5:Mediavilla2017}.
Using light curve data for Q 2237$+$0305, \citet{bib5:Kochanek2004} and \citet{bib5:Poindexter2010a} measured $0.006 < \mathrm{M/M}_{\odot}< 0.2$ and $0.12 < \mathrm{M/M}_{\odot} < 1.94$ (at 68 per cent confidence) respectively.
Both studies highlight the importance of the velocity priors used to get these marginally consistent measurements.
The latter uses longer light curves and a velocity model that includes the random motions of the microlenses, which could explain the higher mass values found.
Under similar dependence on velocity priors, \citet{bib5:MacLeod2015} and \citet{bib5:Morgan2018} found $0.016 < \mathrm{M/M}_{\odot} < 0.8$ and $ 0.03 < \mathrm{M/M}_{\odot} < 0.44$ (at 68 per cent confidence) for J0924$+$0219 and WFI2033$-$4723 respectively, while \citet{bib5:Cornachione2020c} found $0.1 < \mathrm{M/M}_{\odot} < 1$ for Q0957$+$561 and a somewhat unexpected order of magnitude smaller mass range for SBS0909$+$532. 

% Higher mass
The interest in probing higher masses (up to 100 M$_{\odot}$) has been kept high due to observations towards the Galactic bulge by the MACHO experiment \citep{bib5:Alcock2000} and the recent detection of gravitational waves by LIGO/Virgo, attributed to high-mass ($\approx 40$ M$_{\odot}$) black hole mergers.
Interpreting the former observations is arguably model-dependent and can allow for more mass in the form of dark, compact objects \cite{bib5:Hawkins2015}.
\citet{bib5:Hawkins2020} reached this conclusion in order to explain microlensing variability in lensed quasar light curves as well.
However, only theoretical arguments were set forward to attribute this to stellar-mass black holes (and not e.g. stars) and microlensing observables were not modelled explicitly using such mass distributions.
A re-analyis of the same data by \citet{bib5:Awad2023} suggested that a standard explanation, where galaxies are composed of a population of stars with a standard IMF and smoothly distributed dark matter, cannot be ruled out.
Similarly, \citet{bib5:Esteban2022} found a negligible fraction of mass in $\approx 30\,$M$_{\odot}$ objects by explicitly using a mass spectrum in their magnification map simulations.

% The IMF
In addition to directly embedding a mass spectrum in microlensing models, there is also an indirect way the IMF can be probed through the total mass in microlenses.
This is based on breaking the IMF-dark matter fraction degeneracy \citep[e.g. see][]{bib5:Oguri2014}, i.e. the lens total mass ($\kappa$, coming from the macromodel) and the lens light distribution can be attributed to either a low dark matter fraction and many low-mass (and less luminous) stars, or a high dark matter fraction and fewer but brighter stars.
An example is the stellar mass fundamental plane \citep{bib5:Hyde2009}, which is obtained by converting surface brightness to stellar mass through an assumption on the IMF \citep[e.g. see][]{bib5:Kauffmann2003}.
The spectral data used to achieve this light-to-mass conversion suffer from negligible contributions from stellar remnants, brown dwarfs, and red dwarfs that are too faint, hence placing loose constraints on the lower mass part of the IMF.
\citet{bib5:Schechter2014} recalibrate the mean of the IMF (a Salpeter one in this case) to a higher mass based on microlensing measurements of X-ray flux ratios in 10 lensed quasars.
We note that this was an indirect measurement; although the magnification maps used had a microlens mass spectrum \citep[assumed to be a Kroupa IMF, e.g.][]{bib5:Blackburne2011}, its form was fixed and the free parameter of the model was in fact $\kappastar$.
\citet{bib5:Oguri2014} followed a similar approach to constrain the shape of the IMF, finding a preference for Salpeter over Chabrier \citep[see][for a detailed comparison of the methodology of the two studies]{bib5:Schechter2014}.
Finally, \citet{bib5:Vernardos2019a} presented a theoretical study of how optical flux ratios could be used to constrain the IMF.
They found that the IMF could be measured better in a sample of doubly-lensed quasars (due to their higher numbers) that additionally have higher ratios of Einstein to effective radius (i.e. their multiple images appear further away from the lens-galaxy light).

\subsection{Convergence and shear, \texorpdfstring{$\kappa,\gamma$}{kappa,gamma}}
As mentioned in the beginning of \ref{sec5:measuring_kappastar}, the values of $\kappa,\gamma$ can in principle be constrained alongside $\kappastar$ and the mean microlens mass from microlensing data.
However, in practice this is almost never the case due to the very high computational cost of creating and sampling from many magnification maps.
An exception is the study of MG0414$+$0534 by \citet{bib5:Vernardos2018}, where $\kappa,\gamma$ were treated as free parameters and constrained (over an area) from flux ratio data independently from any macromodel.
\citet{bib5:Vernardos2019b} tested a proof-of-concept that the values of $\kappa,\gamma$ could be measured from microlensing light curves as well.
Although promising, these studies have their own limitations and their applicability to a wide range of systems/data remains to be explored.
Achieving such measurements will constitute an entirely new probe of matter in galaxies and provide constraints to macromodels - $\kappa,\gamma$ are directly linked to the second derivatives of the lens potential - with potential application to galaxy evolution, time delay cosmography, etc.

%%%%%%%%%%%%%%%%%%%%%%%%%%%%%%%%%%%%%%%%%%%%%%%%%%%%%%%%%%%%%%%%%%%%%%%%%%%%%%%%%%%%%%%%%%%%%%%%%%%%%%%%%%%%%%%%%%%%%%%%%%%%%%%%%%%%%%%%%%%%%%%%%%%%%%%%%%%%%%%%%%%%%%%%%%%%%%%%%%%%%%%%%%%%%%%%%%%%%%%%%%%%%%%%%%%%%%%%
\section{Future Prospects and Open Questions}
\label{sec5:future_prospects}
From a curiosity applicable to only a handful of systems, microlensing has steadily matured over the years, both in terms of observations and methodology.
It can now complement well-established fields like reverberation mapping \citep[for AGN structure e.g. see][]{bib5:Cackett2021} and galaxy-galaxy lensing (for the mass partition between baryons/dark matter and the IMF, e.g. see \chapgal~and \chapdm). 
Microlensing constitutes an additional observational probe based on an ``orthogonal'' set of assumptions, e.g. it is not sensitive to the brightness profile of AGN emission in the same way as reverberation techniques, nor does it critically depend on the shape of the mass profile of the lens and the macroscopically observed brightness distribution of the lensed source (shape of the arcs, rings, etc).
As such, it can be combined with more traditional methods, for example, to study the kinematics and morphology of the BLR \citep{bib5:Garsden2011} and to break the degeneracy between the stellar IMF and the dark matter fraction \citep{bib5:Oguri2014}.
Moreover, microlensing provides the additional advantage of probing objects that may be hard to study with standard methods.
For instance, microlensing measurements are independent of the level of intrinsic variability of the source (a needed prior for the reverberation method to work), naturally probes objects at cosmic-noon (i.e. $1.5 < z < 2.5$), and thanks to macro-magnification, enables to study even faint systems at high redshift. 

Currently, several hundreds of lensed quasars are known (see \chapfinding) but suitable microlensing data have been obtained for only a few tens of them.
In the next decade, the field will be revolutionised by the advent of all-sky surveys, like LSST and Euclid, which are not only expected to discover thousands of new systems but will also provide high-quality data to perform microlensing analyses.
Microlensing holds a great potential in addressing major science questions by facing the challenge of the upcoming avalanche of data and subsequent adaptation and refinement of its analysis techniques.
The following subsections outline these three points in detail.

\subsection{Science potential}
There is a treasure trove of information to be mined for the central quasar engine from monitoring data and in particular from HMEs.
Once thousands of lensed quasars are discovered and regularly observed, we will be poised to start observing tens or even hundreds of such events.
During their peaks, the inner accretion disc and the flow of material through the ISCO are magnified and encoded in the microlensing signal.
With high cadence ($<$nightly) and signal-to-noise ratio data, especially in the X-rays and optical-UV wavelengths (could also include radio, see \secref{sec5:scattered_and_polarized_emission}), we will be able to perform a variety of tests on accretion disc theory and the theory of relativity.
%We may even be able to produce two-dimensional images of the SMBH, comparable to those from EHT, but for hundreds of quasars at cosmological distances.

Two main microlensing results for the accretion disc are its size and temperature profile, which are respectively found larger and shallower than expected.
More insights into these somewhat puzzling results would be gained by simply increasing the number of studied systems from the current 15 or so.
Even without new and better-adapted to ``big-data'' methods, simply turning the crank with existing ones would suffice for this goal.

Microlensing has enabled the measurement of the BLR half-light radius, in general agreement with expectations from the $R-L$ relations from reverberation mapping.
It has additionally confirmed its increase in size with ionization degree for tens of systems.
For a small subsample of objects, constraints on the BLR kinematics have been set by modeling the broad line deformations. 
There is a great potential in expanding such works to more systems.
The best observational setup for such work consists in spatially resolved spectroscopy, ideally from space or from the ground in combination with adaptive optics. 
The modeling of the kinematics is computationally more demanding, and may benefit from development of innovative modeling strategies to be massively applied to order(s) of magnitude more systems. 

Directly measuring the graininess of matter and by extension the mass density in stars can shed light into the interplay of dark matter with baryons throughout galaxy evolution.
This critical information is very hard to obtain with other methods and prone to degeneracies and assumptions.
The goal to achieve is measuring $\kappastar$ as a function of galactic radius within the lens or per galaxy type \citep[e.g.][]{bib5:Pooley2012}.
Such microlensing results could be combined with other modelling methods as priors, e.g. galaxy-galaxy lens modelling (see \chapgal), used to break degeneracies \citep[e.g.][]{bib5:Oguri2014}, or recalibrate scaling relations like the fundamental plane \citep[e.g.][]{bib5:Schechter2014}.
Again, just the sheer number of new measurements will be enough for gaining important new knowledge.

The expected large number of new observations will enable slicing into subcategories sharing similar physical properties (e.g. quasar black hole mass and Eddington ratio, lens galaxy type and/or age, etc).
This will be crucial for drawing a re-fined picture of AGN structure and to understand how their properties depend on key physical characteristics.
This will be particularly powerful in combination with improved methods (see below), which would make the most out of every single measurement, i.e. not just counting on large number statistics.

Finally, we would like to make a brief comment on the potential of performing cosmological studies with microlensing. 
Firstly, understanding their central engine can clarify whether quasars can be used as standard candles, e.g. with relations such as the one between the luminosity in the X-rays and UV light \citep{bib5:Risaliti2019}.
These are exactly the wavelengths where the quasar is the most prone and sensitive to microlensing.
And secondly, microlensing is the only probe for peculiar velocities of galaxies at high redshifts \citep[$z>0.1$,][]{bib5:Mediavilla2016}.
Such velocities can in turn be used to constrain the growth of structure \citep[e.g.][]{bib5:Koda2014} and shed more light into the $\sigma_{8}$ tension between early and late Universe \citep{bib5:diValentino2021}.

\subsection{Future observations}
The two main and invaluable sources for microlensing data in the next decade will be the all-sky surveys by the Vera Rubin observatory and the Euclid satellite, both starting to provide data within 2023.
The former, will provide $\sim800$ epochs across 6 bands over a decade for each patch of sky (on average), while the latter will carry out a survey at a much higher spatial resolution (space-based).
Both will provide data suitable for microlensing studies for thousands of systems without explicit need for follow-up: LSST will provide 10-year long light curves and Euclid flux ratios in the optical and infrared.
The LSST light curves in particular will not only be suitable for long-term microlensing variability studies (see \secref{sec5:time_varying_ML}) but they could also be used to predict the onset of HMEs.

Imaging and high cadence photometry will play a key role in probing the accretion disc, but a lot may be gained by obtaining spatially resolved spectroscopy.
This will be possible for much fainter than current systems, owing to the upcoming 30-m class telescopes.
Such data may be instrumental in deblending the disc emission from any other superimposed source of extended (pseudo)-continuum emission.
At the same time, BLR deformations may be studied in detail. 

In X-rays, almost all of the progress made in constraining the size of the corona, the inner disc radius, and the dark matter content of the lensing galaxies has been a result of the ability to determine the X-ray fluxes of each image of a lensed quasar.
This task requires sub-arcsecond spatial resolution in X-rays, and {\it Chandra} has been the only observatory with this capability.
Launched in 1999, the mission has made incredible contributions to this field, but the satellite is aging and the main detector has lost most of its effective area below 1 keV.
Currently, there is no mission with similar or better spatial resolution in X-rays that has been approved by any of the major space agencies.
The Astro2020 Decadal recommended that NASA plan for the launch of the next flagship X-ray mission no earlier than the late 2040s.
Such a mission would likely have {\it Chandra}-like or better spatial resolution.

In the interim, NASA Astrophysics has introduced a new line of satellites (Probes) the first of which is to be launched in the early 2030s and will be either a far-infrared mission or an X-ray mission.
One of the X-ray mission concepts (AXIS) aims to have spatial resolution similar to {\it Chandra} but with an order of magnitude more collecting area.
If such a mission was to meet these specifications, it would be able to make substantial progress in this field both by observing lensed quasars at higher redshifts / lower fluxes and by extending the X-ray light curves of {\it Chandra}-observed lensed quasars to over 30- and likely over 40-year baselines.

The presence of microlensing at radio wavelengths is still debated with only two claimed signatures in existing data.
There is a possible confirmation bias in not finding more cases.
A dedicated observing program is required to investigate the question in a systematic way.
In the near future, the Square Kilometer Array may become an ideal facility to confirm or rule out the existence of microlensing in the radio domain, and probe radio emission at even higher scales than possible with interferometry. 

The impact of microlensing on the polarisation of the quasar light has not yet been explored much, but important results on the presence of scattered light have been obtained.
A polarimetric imaging survey with a 2-m class telescope may enable to systematically study the presence of scattered light in a larger number of microlensed quasars, possibly contributing to understanding oversized accretion discs. Spectropolarimatry may naturally complement photometric data, in particular for the study of the structure of BAL quasars.
Tools for simulating polarized signal are basically in place, but data are required to develop the field. 

\subsection{Developing new methods}
For all microlensing applications, the increase in sample size will require novel analysis approaches.
New technologies like GPUs have already helped in understanding the magnification map parameter space that would have otherwise been inaccessible due to the high computational cost.
This remains true for the study of systematic biases in the analysis methods; a few works have explored this, albeit in a limited fashion, for the single-epoch method, while a similar study for the light curve method is hindered by its very resource-demanding computations.
This is especially true when long light curves most likely containing more complex signals are used; one such example could be the role of the dynamic nature of the microlensing map (individual $\boldsymbol{\upsilon}_{\star}$ terms per microlens in \eqref{eq5:veff}). 
Identifying and mitigating such biases will be essential for delivering groundbreaking science with microlensing.
One way to achieve this is by understanding how the length of observations, the amount of microlensing amplitude variations within it, and errors in the macromodel $\kappa,\gamma$ and magnification can affect resulting quasar structure measurements and correct for it, and similarly for determining the stellar matter fraction, $\kappastar$.
The alternative possibility of simply allowing for large number statistics to improve the precision of the results (without a guaranteed improvement on the accuracy due to possible systematic biases), although more straightforward to achieve, would prevent reaching the full potential of microlensing by, for example, studying subcategories of objects with key properties of interest (e.g. as a function of redshift).

Because the BLR covers a larger fraction of the microlensing map than the disc, its modeling might be affected by different systematic uncertainties than the latter.
The modeling of the BLR together with the accretion disc limits the range of events reproducing the data.
It may be worth investigating if this is associated with a gain in precision and accuracy on the retrieved sizes.
A drawback of the BLR modeling strategy is the need for microlensing maps that are sufficiently large for modeling the BLR, and sufficiently high resolution to constrain the disc.
Another challenge that arises with the increase in number of observational constraints (e.g. the microlensing in many velocity bins in the case of the BLR) is the difficulty for the models to reproduce the full signal down to the noise.
It often happens that only a few tens of realisations out of millions of trials successfully reproduce the data.
Simulations may help finding out if this is evidence for the need of more complex models, or if the observed events are simply rare.
Whatever the answer, the computational cost is large, and alternatives to the existing ``Monte-Carlo'' sampling are desired.
We note that this problem is not specific to BLR modeling, for instance, the models of high photometric precision light curves are generally unable to reproduce the data down to the noise.
In the latter case, however, there are good reasons to think that it is not the statistics to blame, but the model that ignores non-disc contributions to the broad band flux (e.g. BLR emission blended with disc emission). 

Machine Learning is another novel technology that could in fact help with achieving control over the systematics.
However, apart from a handful of exploratory works, machine learning methods applied to microlensing have remained largely untapped.
Key applications where such methods could make a difference would be fast generation of magnification maps and/or light curves for any $\kappa,\gamma,\kappastar$, and source size, or direct inference of the model parameters (source size and geometry, $\kappastar$, etc) from the data, avoiding expensive Bayesian likelihood estimations.

Last but not least, the unprecedented amount of data could be mined for the target information in novel ways that go beyond standard approaches, e.g. population studies, inclusion of more complex signals like reverberated flux and/or binary black holes, etc.
Finally, HMEs were mentioned before in the context of machine learning, however, regardless of implementation, a robust predictor and follow-up trigger for such events should be developed and put to the task - there is a unique chance to capitalize LSST data for this.

\subsection{Closing remarks}
We believe that with this review we have drawn a coherent and detailed picture of the current state of the field of quasar microlensing.
We would argue that both data and methods have evolved from circumstantial and basic to targeted and advanced, but not yet fully mature - the field is in its ``adolescence''.
This means that although first results are coherent and exciting, there is more work needed to control systematic biases and to be able to address the avalanche of data from all-sky surveys in the next decade.
We are confident that this will happen and microlensing will achieve its full potential to deliver groundbreaking scientific results.

\section*{Declarations}

\textbf{Competing Interests} The authors have no conflicts of interest to report.

\begin{acknowledgements}
We thank the International Space Science Institute in Bern (ISSI) for their hospitality and the conveners for organizing the stimulating workshop on ``Strong Gravitational Lensing''.
GV has received funding from the European Union's Horizon 2020 research and innovation programme under the Marie Sklodovska-Curie grant agreement No 897124.
GV's and MO's research was made possible by the generosity of Eric and Wendy Schmidt by recommendation of the Schmidt Futures program.
DS would like to thank Damien Hutsem\'ekers and Lorraine Braibant for discussions and material that helped shaping \figref{fig5:BLR_deformation}.
MM acknowledges the support of the Swiss National Science Foundation under grant P500PT\_203114.
VM acknowledges partial support from Centro de Astrof\'{\i}sica de Valpara\'{\i}so and ANID Fondecyt Regular 1231418.
TA acknowledges support from ANID-FONDECYT Regular Project 1190335, the Millennium Science Initiative ICN12\_009 and the ANID-BASAL project FB210003.
Thanks to F. Dux for producing \figref{fig5:lcs}.
\end{acknowledgements}

\bibliographystyle{aps-nameyear}
\bibliography{biblio}

\begin{thebibliography}{403}
% BibTex style file: aps.bst  (nameyear), 2013-04-23
\ifx \bisbn   \undefined \def \bisbn  #1{ISBN #1}\fi
\ifx \binits  \undefined \def \binits#1{#1} \fi
\ifx \bauthor  \undefined \def \bauthor#1{#1} \fi
\ifx \bjtitle  \undefined \def \bjtitle#1{\textrm{#1}}\fi
\ifx \batitle  \undefined \def \batitle#1{#1} \fi
\ifx \bctitle  \undefined \def \bctitle#1{#1} \fi
\ifx \bvolume  \undefined \def \bvolume#1{\textbf{#1}}\fi
\ifx \byear  \undefined \def \byear#1{#1} \fi
\ifx \bissue  \undefined \def \bissue#1{#1} \fi
\ifx \bfpage  \undefined \def \bfpage#1{#1} \fi
\ifx \blpage  \undefined \def \blpage #1{#1} \fi
\ifx \burl  \undefined \def \burl#1{#1} \fi
\ifx \doiurl  \undefined \def \doiurl#1{#1} \fi
\ifx \betal  \undefined \def \betal{et al.} \fi
\ifx \binstitute  \undefined \def \binstitute#1{#1} \fi
\ifx \beditor  \undefined \def \beditor#1{#1} \fi
\ifx \bpublisher  \undefined \def \bpublisher#1{#1} \fi
\ifx \bbtitle  \undefined \def \bbtitle#1{\textit{#1}} \fi
\ifx \bedition  \undefined \def \bedition#1{#1} \fi
\ifx \bseriesno  \undefined \def \bseriesno#1{#1} \fi
\ifx \blocation  \undefined \def \blocation#1{#1} \fi
\ifx \bsertitle  \undefined \def \bsertitle#1{#1} \fi
\ifx \bsnm \undefined \def \bsnm#1{#1} \fi
\ifx \bsuffix \undefined \def \bsuffix#1{#1} \fi
\ifx \bparticle \undefined \def \bparticle#1{#1} \fi
\ifx \barticle \undefined \def \barticle#1{#1} \fi
\ifx \botherref \undefined \def \botherref #1{#1} \fi
\ifx \url \undefined \def \url#1{#1} \fi
\ifx \bchapter \undefined \def \bchapter#1{#1} \fi
\ifx \bbook \undefined \def \bbook#1{#1} \fi
\ifx \bcomment \undefined \def \bcomment#1{#1} \fi
\ifx \oauthor \undefined \def \oauthor#1{#1} \fi
\ifx \citeauthoryear \undefined \def \citeauthoryear#1{#1} \fi
\ifx \texttildelow  \undefined \def \texttildelow{\symbol{126}} \fi
\def \endbibitem {}
\ifx \bconflocation  \undefined \def \bconflocation#1{#1} \fi

\bibitem[\protect\citeauthoryear{{Abajas} et~al.}{2002}]{bib5:Abajas2002}
\begin{barticle}
\bauthor{\binits{C.} \bsnm{{Abajas}}},
\bauthor{\binits{E.} \bsnm{{Mediavilla}}},
\bauthor{\binits{J.A.} \bsnm{{Mu{\~n}oz}}},
\bauthor{\binits{L.{\v{C}}.} \bsnm{{Popovi{\'c}}}},
\bauthor{\binits{A.} \bsnm{{Oscoz}}},
\batitle{The influence of gravitational microlensing on the broad emission
  lines of quasars}.
\bjtitle{The Astrophysical Journal}
\bvolume{576}(\bissue{2}),
\bfpage{640}--\blpage{652}
(\byear{2002}).
doi:\doiurl{10.1086/341793}
\end{barticle}
\endbibitem

\bibitem[\protect\citeauthoryear{{Abajas} et~al.}{2007}]{bib5:Abajas2007}
\begin{barticle}
\bauthor{\binits{C.} \bsnm{{Abajas}}},
\bauthor{\binits{E.} \bsnm{{Mediavilla}}},
\bauthor{\binits{J.A.} \bsnm{{Mu{\~n}oz}}},
\bauthor{\binits{P.} \bsnm{{G{\'o}mez-{\'A}lvarez}}},
\bauthor{\binits{R.} \bsnm{Gil{-}Merino}},
\batitle{Microlensing of a biconical broad-line region}.
\bjtitle{\apj}
\bvolume{658}(\bissue{2}),
\bfpage{748}--\blpage{762}
(\byear{2007}).
doi:\doiurl{10.1086/511023}
\end{barticle}
\endbibitem

\bibitem[\protect\citeauthoryear{Abolmasov and
  Shakura}{2012a}]{bib5:Abolmasov2012b}
\begin{barticle}
\bauthor{\binits{P.} \bsnm{Abolmasov}},
\bauthor{\binits{N.I.} \bsnm{Shakura}},
\batitle{Microlensing evidence for super-eddington disc accretion in quasars}.
\bjtitle{Monthly Notices of the Royal Astronomical Society}
\bvolume{427},
\bfpage{1867}--\blpage{1876}
(\byear{2012}a).
doi:\doiurl{10.1111/j.1365-2966.2012.21881.x}.
\burl{http://mnras.oxfordjournals.org/cgi/doi/10.1111/j.1365-2966.2012.21881.x}
\end{barticle}
\endbibitem

\bibitem[\protect\citeauthoryear{Abolmasov and
  Shakura}{2012b}]{bib5:Abolmasov2012a}
\begin{barticle}
\bauthor{\binits{P.} \bsnm{Abolmasov}},
\bauthor{\binits{N.I.} \bsnm{Shakura}},
\batitle{Resolving the inner structure of qso discs through
  fold-caustic-crossing events}.
\bjtitle{Monthly Notices of the Royal Astronomical Society}
\bvolume{423},
\bfpage{676}--\blpage{693}
(\byear{2012}b).
doi:\doiurl{10.1111/j.1365-2966.2012.20904.x}.
\burl{http://mnras.oxfordjournals.org/cgi/doi/10.1111/j.1365-2966.2012.20904.x}
\end{barticle}
\endbibitem

\bibitem[\protect\citeauthoryear{Abramowicz and
  Fragile}{2013}]{bib5:Abramowicz2013}
\begin{barticle}
\bauthor{\binits{M.A.} \bsnm{Abramowicz}},
\bauthor{\binits{C.P.} \bsnm{Fragile}},
\batitle{Foundations of black hole accretion disk theory}.
\bjtitle{Living Rev. Relativity}
\bvolume{16},
\bfpage{1}
(\byear{2013})
\end{barticle}
\endbibitem

\bibitem[\protect\citeauthoryear{Abramowicz et~al.}{1988}]{bib5:Abramowicz1988}
\begin{barticle}
\bauthor{\binits{M.A.} \bsnm{Abramowicz}},
\bauthor{\binits{B.} \bsnm{Czerny}},
\bauthor{\binits{J.P.} \bsnm{Lasota}},
\bauthor{\binits{E.} \bsnm{Szuszkiewicz}},
\batitle{Slim accretion discs}.
\bjtitle{The Astronomical Journal}
\bvolume{332},
\bfpage{646}
(\byear{1988})
\end{barticle}
\endbibitem

\bibitem[\protect\citeauthoryear{Agol and Krolik}{1999}]{bib5:Agol1999}
\begin{barticle}
\bauthor{\binits{E.} \bsnm{Agol}},
\bauthor{\binits{J.} \bsnm{Krolik}},
\batitle{Imaging a quasar accretion disk with microlensing}.
\bjtitle{The Astrophysical Journal}
\bvolume{524},
\bfpage{49}--\blpage{64}
(\byear{1999}).
doi:\doiurl{10.1086/307800}.
\burl{http://stacks.iop.org/0004-637X/524/i=1/a=49}
\end{barticle}
\endbibitem

\bibitem[\protect\citeauthoryear{Agol et~al.}{2000}]{bib5:Agol2000}
\begin{barticle}
\bauthor{\binits{E.} \bsnm{Agol}},
\bauthor{\binits{B.} \bsnm{Jones}},
\bauthor{\binits{O.} \bsnm{Blaes}},
\batitle{Keck mid-infrared imaging of qso 2237]0305}.
\bjtitle{The Astrophysical Journal}
\bvolume{545},
\bfpage{657}--\blpage{663}
(\byear{2000})
\end{barticle}
\endbibitem

\bibitem[\protect\citeauthoryear{Alcock et~al.}{2000}]{bib5:Alcock2000}
\begin{botherref}
\oauthor{\binits{C.} \bsnm{Alcock}},
\oauthor{\binits{R.A.} \bsnm{Allsman}},
\oauthor{\binits{D.R.} \bsnm{Alves}},
\oauthor{\binits{T.S.} \bsnm{Axelrod}},
\oauthor{\binits{A.C.} \bsnm{Becker}},
\oauthor{\binits{D.P.} \bsnm{Bennett}},
\oauthor{\binits{K.H.} \bsnm{Cook}},
\oauthor{\binits{N.} \bsnm{Dalal}},
\oauthor{\binits{A.J.} \bsnm{Drake}},
\oauthor{\binits{K.C.} \bsnm{Freeman}},
\oauthor{\binits{M.} \bsnm{Geha}},
\oauthor{\binits{K.} \bsnm{Griest}},
\oauthor{\binits{M.J.} \bsnm{Lehner}},
\oauthor{\binits{S.L.} \bsnm{Marshall}},
\oauthor{\binits{D.} \bsnm{Minniti}},
\oauthor{\binits{C.A.} \bsnm{Nelson}},
\oauthor{\binits{B.A.} \bsnm{Peterson}},
\oauthor{\binits{P.} \bsnm{Popowski}},
\oauthor{\binits{M.R.} \bsnm{Pratt}},
\oauthor{\binits{P.J.} \bsnm{Quinn}},
\oauthor{\binits{C.W.} \bsnm{Stubbs}},
\oauthor{\binits{W.} \bsnm{Sutherland}},
\oauthor{\binits{A.B.} \bsnm{Tomaney}},
\oauthor{\binits{T.} \bsnm{Vandehei}},
\oauthor{\binits{D.} \bsnm{Welch}},
The macho project: Microlensing results from 5.7 years of large magellanic
  cloud observations
\textbf{542},
281--307
(2000).
doi:\doiurl{10.1086/309512}
\end{botherref}
\endbibitem

\bibitem[\protect\citeauthoryear{Allen et~al.}{2011}]{bib5:Allen2011}
\begin{barticle}
\bauthor{\binits{J.T.} \bsnm{Allen}},
\bauthor{\binits{P.C.} \bsnm{Hewett}},
\bauthor{\binits{N.} \bsnm{Maddox}},
\bauthor{\binits{G.T.} \bsnm{Richards}},
\bauthor{\binits{V.} \bsnm{Belokurov}},
\batitle{A strong redshift dependence of the broad absorption line quasar
  fraction}.
\bjtitle{Monthly Notices of the Royal Astronomical Society}
\bvolume{410},
\bfpage{860}--\blpage{884}
(\byear{2011}).
doi:\doiurl{10.1111/j.1365-2966.2010.17489.x}
\end{barticle}
\endbibitem

\bibitem[\protect\citeauthoryear{Angonin et~al.}{1990}]{bib5:Angonin1990}
\begin{barticle}
\bauthor{\binits{M.-C.} \bsnm{Angonin}},
\bauthor{\binits{M.} \bsnm{Remy}},
\bauthor{\binits{J.} \bsnm{Surdej}},
\bauthor{\binits{C.} \bsnm{Vanderriest}},
\batitle{First spectroscopic evidence of microlensing on a bal quasar? the case
  of h1413+117}.
\bjtitle{Astronomy and Astrophysics}
\bvolume{233},
\bfpage{5}
(\byear{1990})
\end{barticle}
\endbibitem

\bibitem[\protect\citeauthoryear{Anguita et~al.}{2008}]{bib5:Anguita2008}
\begin{barticle}
\bauthor{\binits{T.} \bsnm{Anguita}},
\bauthor{\binits{R.W.} \bsnm{Schmidt}},
\bauthor{\binits{E.L.} \bsnm{Turner}},
\bauthor{\binits{J.} \bsnm{Wambsganss}},
\bauthor{\binits{R.L.} \bsnm{Webster}},
\bauthor{\binits{K.A.} \bsnm{Loomis}},
\bauthor{\binits{D.} \bsnm{Long}},
\bauthor{\binits{R.} \bsnm{Mcmillan}},
\batitle{The multiple quasar q2237 + 0305 under a microlensing caustic}.
\bjtitle{Astronomy \& Astrophysics}
\bvolume{480},
\bfpage{327}--\blpage{334}
(\byear{2008}).
doi:\doiurl{10.1051/0004-6361}
\end{barticle}
\endbibitem

\bibitem[\protect\citeauthoryear{Antonucci}{1993}]{bib5:Antonucci1993}
\begin{barticle}
\bauthor{\binits{R.} \bsnm{Antonucci}},
\batitle{Unified models for active galactic nuclei and quasars}.
\bjtitle{Annual Review of Astronomy and Astrophysics}
\bvolume{31},
\bfpage{473}
(\byear{1993})
\end{barticle}
\endbibitem

\bibitem[\protect\citeauthoryear{{Awad} et~al.}{2023}]{bib5:Awad2023}
\begin{botherref}
\oauthor{\binits{P.} \bsnm{{Awad}}},
\oauthor{\binits{J.H.H.} \bsnm{{Chan}}},
\oauthor{\binits{M.} \bsnm{{Millon}}},
\oauthor{\binits{F.} \bsnm{{Courbin}}},
\oauthor{\binits{E.} \bsnm{{Paic}}},
Probing compact dark matter objects with microlensing in gravitationally lensed
  quasars.
arXiv e-prints,
2304--01320
(2023).
doi:\doiurl{10.48550/arXiv.2304.01320}
\end{botherref}
\endbibitem

\bibitem[\protect\citeauthoryear{Barsdell et~al.}{2010}]{bib5:Barsdell2010}
\begin{barticle}
\bauthor{\binits{B.R.} \bsnm{Barsdell}},
\bauthor{\binits{D.G.} \bsnm{Barnes}},
\bauthor{\binits{C.J.} \bsnm{Fluke}},
\batitle{Analysing astronomy algorithms for graphics processing units and
  beyond}.
\bjtitle{Monthly Notices of the Royal Astronomical Society}
\bvolume{408},
\bfpage{1936}--\blpage{1944}
(\byear{2010}).
doi:\doiurl{10.1111/j.1365-2966.2010.17257.x}.
\burl{http://doi.wiley.com/10.1111/j.1365-2966.2010.17257.x}
\end{barticle}
\endbibitem

\bibitem[\protect\citeauthoryear{Barth et~al.}{2013}]{bib5:Barth2013}
\begin{botherref}
\oauthor{\binits{A.J.} \bsnm{Barth}},
\oauthor{\binits{A.} \bsnm{Pancoast}},
\oauthor{\binits{V.N.} \bsnm{Bennert}},
\oauthor{\binits{B.J.} \bsnm{Brewer}},
\oauthor{\binits{G.} \bsnm{Canalizo}},
\oauthor{\binits{A.V.} \bsnm{Filippenko}},
\oauthor{\binits{E.L.} \bsnm{Gates}},
\oauthor{\binits{J.E.} \bsnm{Greene}},
\oauthor{\binits{W.} \bsnm{Li}},
\oauthor{\binits{M.A.} \bsnm{Malkan}},
\oauthor{\binits{D.J.} \bsnm{Sand}},
\oauthor{\binits{D.} \bsnm{Stern}},
\oauthor{\binits{T.} \bsnm{Treu}},
\oauthor{\binits{J.H.} \bsnm{Woo}},
\oauthor{\binits{R.J.} \bsnm{Assef}},
\oauthor{\binits{H.J.} \bsnm{Bae}},
\oauthor{\binits{T.} \bsnm{Buehler}},
\oauthor{\binits{S.B.} \bsnm{Cenko}},
\oauthor{\binits{K.I.} \bsnm{Clubb}},
\oauthor{\binits{M.C.} \bsnm{Cooper}},
\oauthor{\binits{A.M.} \bsnm{Diamond-Stanic}},
\oauthor{\binits{S.F.} \bsnm{H{\"o}nig}},
\oauthor{\binits{M.D.} \bsnm{Joner}},
\oauthor{\binits{C.D.} \bsnm{Laney}},
\oauthor{\binits{M.S.} \bsnm{Lazarova}},
\oauthor{\binits{A.M.} \bsnm{Nierenberg}},
\oauthor{\binits{J.M.} \bsnm{Silverman}},
\oauthor{\binits{E.J.} \bsnm{Tollerud}},
\oauthor{\binits{J.L.} \bsnm{Walsh}},
The lick agn monitoring project 2011: Fe ii reverberation from the outer
  broad-line region.
Astrophysical Journal
\textbf{769}
(2013).
doi:\doiurl{10.1088/0004-637X/769/2/128}
\end{botherref}
\endbibitem

\bibitem[\protect\citeauthoryear{Barvainis}{1987}]{bib5:Barvainis1987}
\begin{barticle}
\bauthor{\binits{R.} \bsnm{Barvainis}},
\batitle{Hot dust and the near-infrared bump in the continuum spectra of
  quasars and active galactic nuclei}.
\bjtitle{Astrophysical Journal}
\bvolume{320},
\bfpage{537}
(\byear{1987}).
doi:\doiurl{10.1086/165571}
\end{barticle}
\endbibitem

\bibitem[\protect\citeauthoryear{Bate and Fluke}{2012}]{bib5:Bate2012}
\begin{barticle}
\bauthor{\binits{N.F.} \bsnm{Bate}},
\bauthor{\binits{C.J.} \bsnm{Fluke}},
\batitle{A graphics processing unit-enabled, high-resolution cosmological
  microlensing parameter survey}.
\bjtitle{The Astrophysical Journal}
\bvolume{744},
\bfpage{90}
(\byear{2012}).
doi:\doiurl{10.1088/0004-637X/744/2/90}.
\burl{http://stacks.iop.org/0004-637X/744/i=2/a=90?key=crossref.467615264b93049ab9ccddbf0bb38fec}
\end{barticle}
\endbibitem

\bibitem[\protect\citeauthoryear{Bate et~al.}{2007}]{bib5:Bate2007}
\begin{barticle}
\bauthor{\binits{N.F.} \bsnm{Bate}},
\bauthor{\binits{R.L.} \bsnm{Webster}},
\bauthor{\binits{J.S.B.} \bsnm{Wyithe}},
\batitle{Smooth matter and source size in microlensing simulations of
  gravitationally lensed quasars}.
\bjtitle{Monthly Notices of the Royal Astronomical Society}
\bvolume{381},
\bfpage{1591}--\blpage{1596}
(\byear{2007}).
doi:\doiurl{10.1111/j.1365-2966.2007.12330.x}.
\burl{http://mnras.oxfordjournals.org/cgi/doi/10.1111/j.1365-2966.2007.12330.x}
\end{barticle}
\endbibitem

\bibitem[\protect\citeauthoryear{Bate et~al.}{2008}]{bib5:Bate2008}
\begin{barticle}
\bauthor{\binits{N.F.} \bsnm{Bate}},
\bauthor{\binits{D.J.E.} \bsnm{Floyd}},
\bauthor{\binits{R.L.} \bsnm{Webster}},
\bauthor{\binits{J.S.B.} \bsnm{Wyithe}},
\batitle{A microlensing study of the accretion disc in the quasar mg
  0414+0534}.
\bjtitle{Monthly Notices of the Royal Astronomical Society}
\bvolume{391},
\bfpage{1955}--\blpage{1960}
(\byear{2008}).
doi:\doiurl{10.1111/j.1365-2966.2008.14020.x}.
\burl{http://doi.wiley.com/10.1111/j.1365-2966.2008.14020.x}
\end{barticle}
\endbibitem

\bibitem[\protect\citeauthoryear{Bate et~al.}{2011}]{bib5:Bate2011}
\begin{barticle}
\bauthor{\binits{N.F.} \bsnm{Bate}},
\bauthor{\binits{D.J.E.} \bsnm{Floyd}},
\bauthor{\binits{R.L.} \bsnm{Webster}},
\bauthor{\binits{J.S.B.} \bsnm{Wyithe}},
\batitle{A microlensing measurement of dark matter fractions in three lensing
  galaxies}.
\bjtitle{The Astrophysical Journal}
\bvolume{731},
\bfpage{71}
(\byear{2011}).
doi:\doiurl{10.1088/0004-637X/731/1/71}.
\burl{http://stacks.iop.org/0004-637X/731/i=1/a=71?key=crossref.7becc3e19823d48e952c57c17e390513}
\end{barticle}
\endbibitem

\bibitem[\protect\citeauthoryear{Bate et~al.}{2018}]{bib5:Bate2018}
\begin{barticle}
\bauthor{\binits{N.F.} \bsnm{Bate}},
\bauthor{\binits{G.} \bsnm{Vernardos}},
\bauthor{\binits{M.J.} \bsnm{O'Dowd}},
\bauthor{\binits{M.D.} \bsnm{Neri-Larios}},
\batitle{Hst imaging of four lensed quasars}.
\bjtitle{Monthly Notices of the Royal Astronomical Society}
\bvolume{479},
\bfpage{4796}--\blpage{4814}
(\byear{2018})
\end{barticle}
\endbibitem

\bibitem[\protect\citeauthoryear{Belle and Lewis}{2000}]{bib5:Belle2000}
\begin{barticle}
\bauthor{\binits{K.E.} \bsnm{Belle}},
\bauthor{\binits{G.F.} \bsnm{Lewis}},
\batitle{Microlensing of broad absorption line quasars : Polarization
  variability}.
\bjtitle{Publications of the Astronomical Society of the Pacific}
\bvolume{112},
\bfpage{320}
(\byear{2000})
\end{barticle}
\endbibitem

\bibitem[\protect\citeauthoryear{Bennert et~al.}{2002}]{bib5:Bennert2002}
\begin{barticle}
\bauthor{\binits{N.} \bsnm{Bennert}},
\bauthor{\binits{H.} \bsnm{Falcke}},
\bauthor{\binits{H.} \bsnm{Schulz}},
\bauthor{\binits{A.S.} \bsnm{Wilson}},
\bauthor{\binits{B.J.} \bsnm{Wills}},
\batitle{Size and structure of the narrow-line region of quasars 1}.
\bjtitle{The Astrophysical Journal}
\bvolume{574},
\bfpage{105}--\blpage{109}
(\byear{2002})
\end{barticle}
\endbibitem

\bibitem[\protect\citeauthoryear{{Benning} and
  {Burger}}{2018}]{bib5:Benning2018}
\begin{botherref}
\oauthor{\binits{M.} \bsnm{{Benning}}},
\oauthor{\binits{M.} \bsnm{{Burger}}},
Modern regularization methods for inverse problems.
arXiv e-prints,
1801--09922
(2018).
doi:\doiurl{10.48550/arXiv.1801.09922}
\end{botherref}
\endbibitem

\bibitem[\protect\citeauthoryear{Bentz and Katz}{2015}]{bib5:Bentz2015}
\begin{barticle}
\bauthor{\binits{M.C.} \bsnm{Bentz}},
\bauthor{\binits{S.} \bsnm{Katz}},
\batitle{The agn black hole mass database}.
\bjtitle{Publications of the Astronomical Society of the Pacific}
\bvolume{127},
\bfpage{67}
(\byear{2015})
\end{barticle}
\endbibitem

\bibitem[\protect\citeauthoryear{Bentz et~al.}{2013}]{bib5:Bentz2013}
\begin{botherref}
\oauthor{\binits{M.C.} \bsnm{Bentz}},
\oauthor{\binits{K.D.} \bsnm{Denney}},
\oauthor{\binits{C.J.} \bsnm{Grier}},
\oauthor{\binits{A.J.} \bsnm{Barth}},
\oauthor{\binits{B.M.} \bsnm{Peterson}},
\oauthor{\binits{M.} \bsnm{Vestergaard}},
\oauthor{\binits{V.N.} \bsnm{Bennert}},
\oauthor{\binits{G.} \bsnm{Canalizo}},
\oauthor{\binits{G.D.} \bsnm{Rosa}},
\oauthor{\binits{A.V.} \bsnm{Filippenko}},
\oauthor{\binits{E.L.} \bsnm{Gates}},
\oauthor{\binits{J.E.} \bsnm{Greene}},
\oauthor{\binits{W.} \bsnm{Li}},
\oauthor{\binits{M.A.} \bsnm{Malkan}},
\oauthor{\binits{R.W.} \bsnm{Pogge}},
\oauthor{\binits{D.} \bsnm{Stern}},
\oauthor{\binits{T.} \bsnm{Treu}},
\oauthor{\binits{J.H.} \bsnm{Woo}},
The low-luminosity end of the radius-luminosity relationship for active
  galactic nuclei.
Astrophysical Journal
\textbf{767}
(2013).
doi:\doiurl{10.1088/0004-637X/767/2/149}
\end{botherref}
\endbibitem

\bibitem[\protect\citeauthoryear{Bentz et~al.}{2021}]{bib5:Bentz2021}
\begin{barticle}
\bauthor{\binits{M.C.} \bsnm{Bentz}},
\bauthor{\binits{P.R.} \bsnm{Williams}},
\bauthor{\binits{R.} \bsnm{Street}},
\bauthor{\binits{C.A.} \bsnm{Onken}},
\bauthor{\binits{M.} \bsnm{Valluri}},
\bauthor{\binits{T.} \bsnm{Treu}},
\batitle{A detailed view of the broad-line region in ngc 3783 from
  velocity-resolved reverberation mapping}.
\bjtitle{The Astrophysical Journal}
\bvolume{920},
\bfpage{112}
(\byear{2021}).
doi:\doiurl{10.3847/1538-4357/ac19af}
\end{barticle}
\endbibitem

\bibitem[\protect\citeauthoryear{Best et~al.}{2022}]{bib5:Best2022}
\begin{botherref}
\oauthor{\binits{H.} \bsnm{Best}},
\oauthor{\binits{J.} \bsnm{Fagin}},
\oauthor{\binits{G.} \bsnm{Vernardos}},
\oauthor{\binits{M.} \bsnm{O'Dowd}},
Resolving the vicinity of supermassive black holes with gravitational
  microlensing
(2022).
\url{http://arxiv.org/abs/2210.10500}
\end{botherref}
\endbibitem

\bibitem[\protect\citeauthoryear{Bhatiani et~al.}{2019}]{bib5:Bhatiani2019}
\begin{barticle}
\bauthor{\binits{S.} \bsnm{Bhatiani}},
\bauthor{\binits{X.} \bsnm{Dai}},
\bauthor{\binits{E.} \bsnm{Guerras}},
\batitle{Confirmation of planet-mass objects in extragalactic systems}.
\bjtitle{The Astrophysical Journal}
\bvolume{885},
\bfpage{77}
(\byear{2019}).
doi:\doiurl{10.3847/1538-4357/ab46ac}.
\burl{http://dx.doi.org/10.3847/1538-4357/ab46ac}
\end{barticle}
\endbibitem

\bibitem[\protect\citeauthoryear{Biggs}{2023}]{bib5:Biggs2023}
\begin{barticle}
\bauthor{\binits{A.D.} \bsnm{Biggs}},
\batitle{A vla monitoring study of jvas b1422+231: Investigation of time delays
  and detection of extrinsic variability}.
\bjtitle{Monthly Notices of the Royal Astronomical Society}
(\byear{2023}).
doi:\doiurl{10.1093/mnras/stad870}
\end{barticle}
\endbibitem

\bibitem[\protect\citeauthoryear{Biggs and Browne}{2018}]{bib5:Biggs2018}
\begin{barticle}
\bauthor{\binits{A.D.} \bsnm{Biggs}},
\bauthor{\binits{I.W.A.} \bsnm{Browne}},
\batitle{A revised lens time delay for jvas b0218+357 from a reanalysis of vla
  monitoring data}.
\bjtitle{Monthly Notices of the Royal Astronomical Society}
\bvolume{476},
\bfpage{5393}--\blpage{5407}
(\byear{2018}).
doi:\doiurl{10.1093/MNRAS/STY565}
\end{barticle}
\endbibitem

\bibitem[\protect\citeauthoryear{Blackburne et~al.}{2011}]{bib5:Blackburne2011}
\begin{barticle}
\bauthor{\binits{J.A.} \bsnm{Blackburne}},
\bauthor{\binits{D.} \bsnm{Pooley}},
\bauthor{\binits{S.} \bsnm{Rappaport}},
\bauthor{\binits{P.L.} \bsnm{Schechter}},
\batitle{Sizes and temperature profiles of quasar accretion disks from
  chromatic microlensing}.
\bjtitle{The Astrophysical Journal}
\bvolume{729},
\bfpage{34}
(\byear{2011}).
doi:\doiurl{10.1088/0004-637X/729/1/34}.
\burl{http://stacks.iop.org/0004-637X/729/i=1/a=34?key=crossref.b8d368904b30adf291e182145cf26520}
\end{barticle}
\endbibitem

\bibitem[\protect\citeauthoryear{Blackburne et~al.}{2014}]{bib5:Blackburne2014}
\begin{barticle}
\bauthor{\binits{J.A.} \bsnm{Blackburne}},
\bauthor{\binits{C.S.} \bsnm{Kochanek}},
\bauthor{\binits{B.} \bsnm{Chen}},
\bauthor{\binits{X.} \bsnm{Dai}},
\bauthor{\binits{G.} \bsnm{Chartas}},
\batitle{the optical, ultraviolet, and x-ray structure of the quasar he
  0435$-$1223}.
\bjtitle{The Astrophysical Journal}
\bvolume{789},
\bfpage{125}
(\byear{2014}).
doi:\doiurl{10.1088/0004-637X/789/2/125}.
\burl{http://stacks.iop.org/0004-637X/789/i=2/a=125?key=crossref.6cd0b035991ffbc29b0403cff7ba1e19}
\end{barticle}
\endbibitem

\bibitem[\protect\citeauthoryear{Blackburne and
  Kochanek}{2010}]{bib5:Blackburne2010}
\begin{barticle}
\bauthor{\binits{J.A.} \bsnm{Blackburne}},
\bauthor{\binits{C.S.} \bsnm{Kochanek}},
\batitle{The effect of a time-varying accretion disk size on quasar
  microlensing light curves}.
\bjtitle{Astrophysical Journal}
\bvolume{718},
\bfpage{1079}--\blpage{1084}
(\byear{2010}).
doi:\doiurl{10.1088/0004-637X/718/2/1079}
\end{barticle}
\endbibitem

\bibitem[\protect\citeauthoryear{Blandford and
  Mckee}{1982}]{bib5:Blandford1982}
\begin{barticle}
\bauthor{\binits{R.D.} \bsnm{Blandford}},
\bauthor{\binits{C.F.} \bsnm{Mckee}},
\batitle{Reverberation mapping of the emission line regions of seyfert galaxies
  and quasars}.
\bjtitle{The Astrophysical Journal}
\bvolume{255},
\bfpage{419}--\blpage{439}
(\byear{1982})
\end{barticle}
\endbibitem

\bibitem[\protect\citeauthoryear{Bogdanov and
  Cherepashchuk}{2002}]{bib5:Bogdanov2002}
\begin{barticle}
\bauthor{\binits{M.B.} \bsnm{Bogdanov}},
\bauthor{\binits{A.M.} \bsnm{Cherepashchuk}},
\batitle{Reconstruction of the strip brightness distribution in a quasar
  accretion disk from gravitational microlensing data}.
\bjtitle{Astronomy Reports}
\bvolume{46},
\bfpage{626}--\blpage{633}
(\byear{2002})
\end{barticle}
\endbibitem

\bibitem[\protect\citeauthoryear{Borguet and
  Hutsem{\'e}kers}{2010}]{bib5:Borguet2010}
\begin{botherref}
\oauthor{\binits{B.} \bsnm{Borguet}},
\oauthor{\binits{D.} \bsnm{Hutsem{\'e}kers}},
A polar+equatorial wind model for broad absorption line quasars.
Astronomy and Astrophysics
\textbf{515}
(2010).
doi:\doiurl{10.1051/0004-6361/200913255}
\end{botherref}
\endbibitem

\bibitem[\protect\citeauthoryear{Bourassa and
  Kantowski}{1975}]{bib5:Bourassa1975}
\begin{barticle}
\bauthor{\binits{R.R.} \bsnm{Bourassa}},
\bauthor{\binits{R.} \bsnm{Kantowski}},
\batitle{The theory of transparent gravitational lenses}.
\bjtitle{The Astrophysical Journal}
\bvolume{195},
\bfpage{13}--\blpage{21}
(\byear{1975})
\end{barticle}
\endbibitem

\bibitem[\protect\citeauthoryear{Bourassa et~al.}{1973}]{bib5:Bourassa1973}
\begin{barticle}
\bauthor{\binits{R.R.} \bsnm{Bourassa}},
\bauthor{\binits{R.} \bsnm{Kantowski}},
\bauthor{\binits{T.D.} \bsnm{Norton}},
\batitle{The spheroidal gravitational lens}.
\bjtitle{The Astrophysical Journal}
\bvolume{185},
\bfpage{747}--\blpage{756}
(\byear{1973})
\end{barticle}
\endbibitem

\bibitem[\protect\citeauthoryear{Bozza}{2010}]{bib5:Bozza2010}
\begin{barticle}
\bauthor{\binits{V.} \bsnm{Bozza}},
\batitle{Microlensing with an advanced contour integration algorithm: Green's
  theorem to third order, error control, optimal sampling and limb darkening}.
\bjtitle{Monthly Notices of the Royal Astronomical Society}
\bvolume{408},
\bfpage{2188}--\blpage{2200}
(\byear{2010}).
doi:\doiurl{10.1111/j.1365-2966.2010.17265.x}
\end{barticle}
\endbibitem

\bibitem[\protect\citeauthoryear{Braibant et~al.}{2014}]{bib5:Braibant2014}
\begin{botherref}
\oauthor{\binits{L.} \bsnm{Braibant}},
\oauthor{\binits{D.} \bsnm{Hutsem{\'e}kers}},
\oauthor{\binits{D.} \bsnm{Sluse}},
\oauthor{\binits{T.} \bsnm{Anguita}},
\oauthor{\binits{C.J.} \bsnm{Garc{\'i}a-Vergara}},
Microlensing of the broad-line region in the quadruply imaged quasar
  he0435$-$1223.
Astronomy and Astrophysics
\textbf{565}
(2014).
doi:\doiurl{10.1051/0004-6361/201423633}
\end{botherref}
\endbibitem

\bibitem[\protect\citeauthoryear{Braibant et~al.}{2016}]{bib5:Braibant2016}
\begin{botherref}
\oauthor{\binits{L.} \bsnm{Braibant}},
\oauthor{\binits{D.} \bsnm{Hutsem{\'e}kers}},
\oauthor{\binits{D.} \bsnm{Sluse}},
\oauthor{\binits{T.} \bsnm{Anguita}},
The different origins of high- and low-ionization broad emission lines revealed
  by gravitational microlensing in the einstein cross.
Astronomy and Astrophysics
\textbf{592}
(2016).
doi:\doiurl{10.1051/0004-6361/201628594}
\end{botherref}
\endbibitem

\bibitem[\protect\citeauthoryear{Braibant et~al.}{2017}]{bib5:Braibant2017}
\begin{botherref}
\oauthor{\binits{L.} \bsnm{Braibant}},
\oauthor{\binits{D.} \bsnm{Hutsem{\'e}kers}},
\oauthor{\binits{D.} \bsnm{Sluse}},
\oauthor{\binits{R.} \bsnm{Goosmann}},
Constraining the geometry and kinematics of the quasar broad emission line
  region using gravitational microlensing: I. models and simulations.
Astronomy and Astrophysics
\textbf{607}
(2017).
doi:\doiurl{10.1051/0004-6361/201731086}
\end{botherref}
\endbibitem

\bibitem[\protect\citeauthoryear{{Brenneman} and
  {Reynolds}}{2006}]{bib5:Brenneman2006}
\begin{barticle}
\bauthor{\binits{L.W.} \bsnm{{Brenneman}}},
\bauthor{\binits{C.S.} \bsnm{{Reynolds}}},
\batitle{Constraining black hole spin via x-ray spectroscopy}.
\bjtitle{\apj}
\bvolume{652}(\bissue{2}),
\bfpage{1028}--\blpage{1043}
(\byear{2006}).
doi:\doiurl{10.1086/508146}
\end{barticle}
\endbibitem

\bibitem[\protect\citeauthoryear{Bruce et~al.}{2017}]{bib5:Bruce2017}
\begin{barticle}
\bauthor{\binits{A.} \bsnm{Bruce}},
\bauthor{\binits{A.} \bsnm{Lawrence}},
\bauthor{\binits{C.} \bsnm{MacLeod}},
\bauthor{\binits{M.} \bsnm{Elvis}},
\bauthor{\binits{M.J.} \bsnm{Ward}},
\bauthor{\binits{J.S.} \bsnm{Collinson}},
\bauthor{\binits{S.} \bsnm{Gezari}},
\bauthor{\binits{P.J.} \bsnm{Marshall}},
\bauthor{\binits{M.C.} \bsnm{Lam}},
\bauthor{\binits{R.} \bsnm{Kotak}},
\bauthor{\binits{C.} \bsnm{Inserra}},
\bauthor{\binits{J.} \bsnm{Polshaw}},
\bauthor{\binits{N.} \bsnm{Kaiser}},
\bauthor{\binits{R.P.} \bsnm{Kudritzki}},
\bauthor{\binits{E.A.} \bsnm{Magnier}},
\bauthor{\binits{C.} \bsnm{Waters}},
\batitle{Spectral analysis of four 'hypervariable' agn: A microneedle in the
  haystack?}
\bjtitle{Monthly Notices of the Royal Astronomical Society}
\bvolume{467},
\bfpage{1259}--\blpage{1280}
(\byear{2017}).
doi:\doiurl{10.1093/mnras/stx168}
\end{barticle}
\endbibitem

\bibitem[\protect\citeauthoryear{{Cackett} et~al.}{2021}]{bib5:Cackett2021}
\begin{barticle}
\bauthor{\binits{E.M.} \bsnm{{Cackett}}},
\bauthor{\binits{M.C.} \bsnm{{Bentz}}},
\bauthor{\binits{E.} \bsnm{{Kara}}},
\batitle{Reverberation mapping of active galactic nuclei: from x-ray corona to
  dusty torus}.
\bjtitle{iScience}
\bvolume{24}(\bissue{6}),
\bfpage{102557}
(\byear{2021}).
doi:\doiurl{10.1016/j.isci.2021.102557}
\end{barticle}
\endbibitem

\bibitem[\protect\citeauthoryear{Cackett et~al.}{2020}]{bib5:Cackett2020}
\begin{barticle}
\bauthor{\binits{E.M.} \bsnm{Cackett}},
\bauthor{\binits{J.} \bsnm{Gelbord}},
\bauthor{\binits{Y.-R.} \bsnm{Li}},
\bauthor{\binits{K.} \bsnm{Horne}},
\bauthor{\binits{J.-M.} \bsnm{Wang}},
\bauthor{\binits{A.J.} \bsnm{Barth}},
\bauthor{\binits{J.-M.} \bsnm{Bai}},
\bauthor{\binits{W.-H.} \bsnm{Bian}},
\bauthor{\binits{R.W.} \bsnm{Carroll}},
\bauthor{\binits{P.} \bsnm{Du}},
\bauthor{\binits{R.} \bsnm{Edelson}},
\bauthor{\binits{M.R.} \bsnm{Goad}},
\bauthor{\binits{L.C.} \bsnm{Ho}},
\bauthor{\binits{C.} \bsnm{Hu}},
\bauthor{\binits{V.C.} \bsnm{Khatu}},
\bauthor{\binits{B.} \bsnm{Luo}},
\bauthor{\binits{J.} \bsnm{Miller}},
\bauthor{\binits{Y.-F.} \bsnm{Yuan}},
\batitle{Supermassive black holes with high accretion rates in active galactic
  nuclei. xi. accretion disk reverberation mapping of mrk 142}.
\bjtitle{The Astrophysical Journal}
\bvolume{896},
\bfpage{1}
(\byear{2020}).
doi:\doiurl{10.3847/1538-4357/ab91b5}
\end{barticle}
\endbibitem

\bibitem[\protect\citeauthoryear{Canizares}{1982}]{bib5:Canizares1982}
\begin{barticle}
\bauthor{\binits{C.R.} \bsnm{Canizares}},
\batitle{Manifestations of a cosmological density of compact objects in quasar
  light}.
\bjtitle{The Astrophysical Journal}
\bvolume{263},
\bfpage{508}
(\byear{1982})
\end{barticle}
\endbibitem

\bibitem[\protect\citeauthoryear{Cappellari}{2016}]{bib5:Cappellari2016}
\begin{barticle}
\bauthor{\binits{M.} \bsnm{Cappellari}},
\batitle{Structure and kinematics of early-type galaxies from integral field
  spectroscopy}.
\bjtitle{Annual Review of Astronomy and Astrophysics}
\bvolume{54},
\bfpage{597}--\blpage{665}
(\byear{2016}).
doi:\doiurl{10.1146/annurev-astro-082214-122432}
\end{barticle}
\endbibitem

\bibitem[\protect\citeauthoryear{Cardelli and Savage}{1988}]{bib5:Cardelli1988}
\begin{barticle}
\bauthor{\binits{J.A.} \bsnm{Cardelli}},
\bauthor{\binits{B.D.} \bsnm{Savage}},
\batitle{Two lines of sight with exceedingly anomalous ultraviolet interstellar
  extinction}.
\bjtitle{The Astrophysical Journal}
\bvolume{325},
\bfpage{864}--\blpage{879}
(\byear{1988})
\end{barticle}
\endbibitem

\bibitem[\protect\citeauthoryear{Chae et~al.}{2001}]{bib5:Chae2001}
\begin{barticle}
\bauthor{\binits{K.-H.} \bsnm{Chae}},
\bauthor{\binits{D.A.} \bsnm{Turnshek}},
\bauthor{\binits{R.E.} \bsnm{Schulte-Ladbeck}},
\bauthor{\binits{S.M.} \bsnm{Rao}},
\bauthor{\binits{O.L.} \bsnm{Lupie4}},
\batitle{Hubble space t el escope observations of the gravitationally lensed
  cloverleaf broad absorption line qso h1413]1143 : Imaging polarimetry and
  evidence for microlensing of a scattering region1}.
\bjtitle{The Astrophysical Journal}
\bvolume{561},
\bfpage{653}--\blpage{659}
(\byear{2001})
\end{barticle}
\endbibitem

\bibitem[\protect\citeauthoryear{Chan et~al.}{2021}]{bib5:Chan2021}
\begin{botherref}
\oauthor{\binits{J.H.H.} \bsnm{Chan}},
\oauthor{\binits{K.} \bsnm{Rojas}},
\oauthor{\binits{M.} \bsnm{Millon}},
\oauthor{\binits{F.} \bsnm{Courbin}},
\oauthor{\binits{V.} \bsnm{Bonvin}},
\oauthor{\binits{G.} \bsnm{Jauffret}},
Measuring accretion disk sizes of lensed quasars with microlensing time delay
  in multi-band light curves.
Astronomy and Astrophysics
\textbf{647}
(2021).
doi:\doiurl{10.1051/0004-6361/202038971}
\end{botherref}
\endbibitem

\bibitem[\protect\citeauthoryear{Chang and Refsdal}{1979}]{bib5:Chang1979}
\begin{barticle}
\bauthor{\binits{K.} \bsnm{Chang}},
\bauthor{\binits{S.} \bsnm{Refsdal}},
\batitle{Flux variations of qso 0957$+$561 a, b and image splitting by stars
  near the light path}.
\bjtitle{Nature}
\bvolume{282},
\bfpage{561}
(\byear{1979})
\end{barticle}
\endbibitem

\bibitem[\protect\citeauthoryear{Chang and Refsdal}{1984}]{bib5:Chang1984}
\begin{barticle}
\bauthor{\binits{K.} \bsnm{Chang}},
\bauthor{\binits{S.} \bsnm{Refsdal}},
\batitle{Star disturbances in gravitational lens galaxies}.
\bjtitle{Astronomy \& Astrophysics}
\bvolume{132},
\bfpage{168}
(\byear{1984})
\end{barticle}
\endbibitem

\bibitem[\protect\citeauthoryear{{Chartas} et~al.}{2002}]{bib5:Chartas2002}
\begin{barticle}
\bauthor{\binits{G.} \bsnm{{Chartas}}},
\bauthor{\binits{E.} \bsnm{{Agol}}},
\bauthor{\binits{M.} \bsnm{{Eracleous}}},
\bauthor{\binits{G.} \bsnm{{Garmire}}},
\bauthor{\binits{M.W.} \bsnm{{Bautz}}},
\bauthor{\binits{N.D.} \bsnm{{Morgan}}},
\batitle{Caught in the act: Chandra observations of microlensing of the
  radio-loud quasar mg j0414+0534}.
\bjtitle{\apj}
\bvolume{568}(\bissue{2}),
\bfpage{509}--\blpage{521}
(\byear{2002}).
doi:\doiurl{10.1086/339162}
\end{barticle}
\endbibitem

\bibitem[\protect\citeauthoryear{{Chartas} et~al.}{2004}]{bib5:Chartas2004}
\begin{barticle}
\bauthor{\binits{G.} \bsnm{{Chartas}}},
\bauthor{\binits{M.} \bsnm{{Eracleous}}},
\bauthor{\binits{E.} \bsnm{{Agol}}},
\bauthor{\binits{S.C.} \bsnm{{Gallagher}}},
\batitle{Chandra observations of the cloverleaf quasar h1413+117: A unique
  laboratory for microlensing studies of a lobal quasar}.
\bjtitle{\apj}
\bvolume{606}(\bissue{1}),
\bfpage{78}--\blpage{84}
(\byear{2004}).
doi:\doiurl{10.1086/382743}
\end{barticle}
\endbibitem

\bibitem[\protect\citeauthoryear{{Chartas} et~al.}{2007}]{bib5:Chartas2007}
\begin{barticle}
\bauthor{\binits{G.} \bsnm{{Chartas}}},
\bauthor{\binits{M.} \bsnm{{Eracleous}}},
\bauthor{\binits{X.} \bsnm{{Dai}}},
\bauthor{\binits{E.} \bsnm{{Agol}}},
\bauthor{\binits{S.} \bsnm{{Gallagher}}},
\batitle{Discovery of probable relativistic fe emission and absorption in the
  cloverleaf quasar h 1413+117}.
\bjtitle{\apj}
\bvolume{661}(\bissue{2}),
\bfpage{678}--\blpage{692}
(\byear{2007}).
doi:\doiurl{10.1086/516816}
\end{barticle}
\endbibitem

\bibitem[\protect\citeauthoryear{Chartas et~al.}{2009}]{bib5:Chartas2009}
\begin{barticle}
\bauthor{\binits{G.} \bsnm{Chartas}},
\bauthor{\binits{C.S.} \bsnm{Kochanek}},
\bauthor{\binits{X.} \bsnm{Dai}},
\bauthor{\binits{S.} \bsnm{Poindexter}},
\bauthor{\binits{G.} \bsnm{Garmire}},
\batitle{X-ray microlensing in rxj1131-1231 and he1104-1805}.
\bjtitle{The Astrophysical Journal}
\bvolume{693},
\bfpage{174}--\blpage{185}
(\byear{2009}).
doi:\doiurl{10.1088/0004-637X/693/1/174}.
\burl{http://stacks.iop.org/0004-637X/693/i=1/a=174?key=crossref.0cdea05ef9165a8c0c6d5c38ca0cfa1d}
\end{barticle}
\endbibitem

\bibitem[\protect\citeauthoryear{{Chartas} et~al.}{2012}]{bib5:Chartas2012}
\begin{barticle}
\bauthor{\binits{G.} \bsnm{{Chartas}}},
\bauthor{\binits{C.S.} \bsnm{{Kochanek}}},
\bauthor{\binits{X.} \bsnm{{Dai}}},
\bauthor{\binits{D.} \bsnm{{Moore}}},
\bauthor{\binits{A.M.} \bsnm{{Mosquera}}},
\bauthor{\binits{J.A.} \bsnm{{Blackburne}}},
\batitle{Revealing the structure of an accretion disk through energy-dependent
  x-ray microlensing}.
\bjtitle{\apj}
\bvolume{757}(\bissue{2}),
\bfpage{137}
(\byear{2012}).
doi:\doiurl{10.1088/0004-637X/757/2/137}
\end{barticle}
\endbibitem

\bibitem[\protect\citeauthoryear{Chartas et~al.}{2017}]{bib5:Chartas2017}
\begin{barticle}
\bauthor{\binits{G.} \bsnm{Chartas}},
\bauthor{\binits{H.} \bsnm{Krawczynski}},
\bauthor{\binits{L.} \bsnm{Zalesky}},
\bauthor{\binits{C.S.} \bsnm{Kochanek}},
\bauthor{\binits{X.} \bsnm{Dai}},
\bauthor{\binits{C.W.} \bsnm{Morgan}},
\bauthor{\binits{A.} \bsnm{Mosquera}},
\batitle{Measuring the innermost stable circular orbits of supermassive black
  holes}.
\bjtitle{The Astrophysical Journal}
\bvolume{837},
\bfpage{26}
(\byear{2017}).
doi:\doiurl{10.3847/1538-4357/aa5d50}
\end{barticle}
\endbibitem

\bibitem[\protect\citeauthoryear{Chartas et~al.}{2021}]{bib5:Chartas2021}
\begin{barticle}
\bauthor{\binits{G.} \bsnm{Chartas}},
\bauthor{\binits{M.} \bsnm{Cappi}},
\bauthor{\binits{C.} \bsnm{Vignali}},
\bauthor{\binits{M.} \bsnm{Dadina}},
\bauthor{\binits{V.} \bsnm{James}},
\bauthor{\binits{G.} \bsnm{Lanzuisi}},
\bauthor{\binits{M.} \bsnm{Giustini}},
\bauthor{\binits{M.} \bsnm{Gaspari}},
\bauthor{\binits{S.} \bsnm{Strickland}},
\bauthor{\binits{E.} \bsnm{Bertola}},
\batitle{Multiphase powerful outflows detected in high-z quasars}.
\bjtitle{The Astrophysical Journal}
\bvolume{920},
\bfpage{24}
(\byear{2021}).
doi:\doiurl{10.3847/1538-4357/ac0ef2}
\end{barticle}
\endbibitem

\bibitem[\protect\citeauthoryear{Chelouche}{2005}]{bib5:Chelouche2005}
\begin{barticle}
\bauthor{\binits{D.} \bsnm{Chelouche}},
\batitle{Gravitational microlensing and the structure of quasar outflows}.
\bjtitle{The Astrophysical Journal}
\bvolume{629},
\bfpage{667}
(\byear{2005})
\end{barticle}
\endbibitem

\bibitem[\protect\citeauthoryear{{Chen} et~al.}{2012}]{bib5:Chen2012}
\begin{barticle}
\bauthor{\binits{B.} \bsnm{{Chen}}},
\bauthor{\binits{X.} \bsnm{{Dai}}},
\bauthor{\binits{C.S.} \bsnm{{Kochanek}}},
\bauthor{\binits{G.} \bsnm{{Chartas}}},
\bauthor{\binits{J.A.} \bsnm{{Blackburne}}},
\bauthor{\binits{C.W.} \bsnm{{Morgan}}},
\batitle{X-ray monitoring of gravitational lenses with chandra}.
\bjtitle{\apj}
\bvolume{755}(\bissue{1}),
\bfpage{24}
(\byear{2012}).
doi:\doiurl{10.1088/0004-637X/755/1/24}
\end{barticle}
\endbibitem

\bibitem[\protect\citeauthoryear{Chen et~al.}{2017}]{bib5:Chen2017}
\begin{botherref}
\oauthor{\binits{B.} \bsnm{Chen}},
\oauthor{\binits{R.} \bsnm{Kantowski}},
\oauthor{\binits{X.} \bsnm{Dai}},
\oauthor{\binits{E.} \bsnm{Baron}},
\oauthor{\binits{P.V.} \bparticle{der} \bsnm{Mark}},
Accelerating gravitational microlensing simulations using the xeon phi
  coprocessor
(2017).
\url{http://arxiv.org/abs/1703.09707}
\end{botherref}
\endbibitem

\bibitem[\protect\citeauthoryear{Chen et~al.}{2016}]{bib5:Chen2016}
\begin{barticle}
\bauthor{\binits{G.C.F.} \bsnm{Chen}},
\bauthor{\binits{S.H.} \bsnm{Suyu}},
\bauthor{\binits{K.C.} \bsnm{Wong}},
\bauthor{\binits{C.D.} \bsnm{Fassnacht}},
\bauthor{\binits{T.} \bsnm{Chiueh}},
\bauthor{\binits{A.} \bsnm{Halkola}},
\bauthor{\binits{I.S.} \bsnm{Hu}},
\bauthor{\binits{M.W.} \bsnm{Auger}},
\bauthor{\binits{L.V.E.} \bsnm{Koopmans}},
\bauthor{\binits{D.J.} \bsnm{Lagattuta}},
\bauthor{\binits{J.P.} \bsnm{McKean}},
\bauthor{\binits{S.} \bsnm{Vegetti}},
\batitle{Sharp - iii. first use of adaptive-optics imaging to constrain
  cosmology with gravitational lens time delays}.
\bjtitle{Monthly Notices of the Royal Astronomical Society}
\bvolume{462},
\bfpage{3457}--\blpage{3475}
(\byear{2016}).
doi:\doiurl{10.1093/mnras/stw991}
\end{barticle}
\endbibitem

\bibitem[\protect\citeauthoryear{Chiba et~al.}{2005}]{bib5:Chiba2005}
\begin{barticle}
\bauthor{\binits{M.} \bsnm{Chiba}},
\bauthor{\binits{T.} \bsnm{Minezaki}},
\bauthor{\binits{N.} \bsnm{Kashikawa}},
\bauthor{\binits{H.} \bsnm{Kataza}},
\bauthor{\binits{K.T.} \bsnm{Inoue}},
\batitle{Subaru mid-infrared imaging of the quadruple lenses pg 1115+080 and
  b1422+231: Limits on substructure lensing 1}.
\bjtitle{The Astrophysical Journal}
\bvolume{627},
\bfpage{53}
(\byear{2005}).
\burl{http://cfa-www.harvard.edu}
\end{barticle}
\endbibitem

\bibitem[\protect\citeauthoryear{Collin et~al.}{2002}]{bib5:Collin2002}
\begin{barticle}
\bauthor{\binits{S.} \bsnm{Collin}},
\bauthor{\binits{C.} \bsnm{Boisson}},
\bauthor{\binits{M.} \bsnm{Mouchet}},
\bauthor{\binits{A.M.} \bsnm{Dumont}},
\bauthor{\binits{S.} \bsnm{Coup{\'e}}},
\bauthor{\binits{D.} \bsnm{Porquet}},
\bauthor{\binits{E.} \bsnm{Rokaki}},
\batitle{Are quasars accreting at super-eddington rates?}
\bjtitle{Astronomy and Astrophysics}
\bvolume{388},
\bfpage{771}--\blpage{786}
(\byear{2002}).
doi:\doiurl{10.1051/0004-6361:20020550}
\end{barticle}
\endbibitem

\bibitem[\protect\citeauthoryear{Congdon et~al.}{2007}]{bib5:Congdon2007}
\begin{barticle}
\bauthor{\binits{A.B.} \bsnm{Congdon}},
\bauthor{\binits{C.R.} \bsnm{Keeton}},
\bauthor{\binits{S.J.} \bsnm{Osmer}},
\batitle{Microlensing of an extended source by a power-law mass distribution}.
\bjtitle{Monthly Notices of the Royal Astronomical Society}
\bvolume{376},
\bfpage{263}--\blpage{272}
(\byear{2007}).
doi:\doiurl{10.1111/j.1365-2966.2007.11426.x}
\end{barticle}
\endbibitem

\bibitem[\protect\citeauthoryear{Cornachione and
  Morgan}{2020}]{bib5:Cornachione2020a}
\begin{barticle}
\bauthor{\binits{M.A.} \bsnm{Cornachione}},
\bauthor{\binits{C.W.} \bsnm{Morgan}},
\batitle{Quasar microlensing variability studies favor shallow accretion disk
  temperature profiles}.
\bjtitle{The Astrophysical Journal}
\bvolume{895},
\bfpage{93}
(\byear{2020}).
doi:\doiurl{10.3847/1538-4357/ab8aed}.
\burl{http://dx.doi.org/10.3847/1538-4357/ab8aed}
\end{barticle}
\endbibitem

\bibitem[\protect\citeauthoryear{Cornachione
  et~al.}{2020a}]{bib5:Cornachione2020b}
\begin{barticle}
\bauthor{\binits{M.A.} \bsnm{Cornachione}},
\bauthor{\binits{C.W.} \bsnm{Morgan}},
\bauthor{\binits{M.} \bsnm{Millon}},
\bauthor{\binits{M.C.} \bsnm{Bentz}},
\bauthor{\binits{F.} \bsnm{Courbin}},
\bauthor{\binits{V.} \bsnm{Bonvin}},
\bauthor{\binits{E.E.} \bsnm{Falco}},
\batitle{A microlensing accretion disk size measurement in the lensed quasar
  wfi 2026$-$4536}.
\bjtitle{The Astrophysical Journal}
\bvolume{895},
\bfpage{125}
(\byear{2020}a).
doi:\doiurl{10.3847/1538-4357/ab557a}.
\burl{http://dx.doi.org/10.3847/1538-4357/ab557a}
\end{barticle}
\endbibitem

\bibitem[\protect\citeauthoryear{Cornachione
  et~al.}{2020b}]{bib5:Cornachione2020c}
\begin{barticle}
\bauthor{\binits{M.A.} \bsnm{Cornachione}},
\bauthor{\binits{C.W.} \bsnm{Morgan}},
\bauthor{\binits{H.R.} \bsnm{Burger}},
\bauthor{\binits{V.N.} \bsnm{Shalyapin}},
\bauthor{\binits{L.J.} \bsnm{Goicoechea}},
\bauthor{\binits{F.J.} \bsnm{Vrba}},
\bauthor{\binits{S.E.} \bsnm{Dahm}},
\bauthor{\binits{T.M.} \bsnm{Tilleman}},
\batitle{Near-infrared and optical continuum emission region size measurements
  in the gravitationally lensed quasars q0957$+$561 and sbs0909$+$532}.
\bjtitle{The Astrophysical Journal}
\bvolume{905},
\bfpage{7}
(\byear{2020}b).
doi:\doiurl{10.3847/1538-4357/abc25d}.
\burl{http://dx.doi.org/10.3847/1538-4357/abc25d}
\end{barticle}
\endbibitem

\bibitem[\protect\citeauthoryear{Czerny et~al.}{2017}]{bib5:Czerny2017}
\begin{barticle}
\bauthor{\binits{B.} \bsnm{Czerny}},
\bauthor{\binits{Y.-R.} \bsnm{Li}},
\bauthor{\binits{K.} \bsnm{Hryniewicz}},
\bauthor{\binits{S.} \bsnm{Panda}},
\bauthor{\binits{C.} \bsnm{Wildy}},
\bauthor{\binits{M.} \bsnm{Sniegowska}},
\bauthor{\binits{J.-M.} \bsnm{Wang}},
\bauthor{\binits{J.} \bsnm{Sredzinska}},
\bauthor{\binits{V.} \bsnm{Karas}},
\batitle{Failed radiatively accelerated dusty outflow model of the broad line
  region in active galactic nuclei. i. analytical solution}.
\bjtitle{The Astrophysical Journal}
\bvolume{846},
\bfpage{154}
(\byear{2017}).
doi:\doiurl{10.3847/1538-4357/aa8810}
\end{barticle}
\endbibitem

\bibitem[\protect\citeauthoryear{Dai and Pascale}{2021}]{bib5:Dai2021}
\begin{botherref}
\oauthor{\binits{L.} \bsnm{Dai}},
\oauthor{\binits{M.} \bsnm{Pascale}},
New approximation of magnification statistics for random microlensing of
  magnified sources
(2021).
\url{http://arxiv.org/abs/2104.12009}
\end{botherref}
\endbibitem

\bibitem[\protect\citeauthoryear{{Dai} et~al.}{2003}]{bib5:Dai2003}
\begin{barticle}
\bauthor{\binits{X.} \bsnm{{Dai}}},
\bauthor{\binits{G.} \bsnm{{Chartas}}},
\bauthor{\binits{E.} \bsnm{{Agol}}},
\bauthor{\binits{M.W.} \bsnm{{Bautz}}},
\bauthor{\binits{G.P.} \bsnm{{Garmire}}},
\batitle{Chandra observations of qso 2237+0305}.
\bjtitle{\apj}
\bvolume{589}(\bissue{1}),
\bfpage{100}--\blpage{110}
(\byear{2003}).
doi:\doiurl{10.1086/374548}
\end{barticle}
\endbibitem

\bibitem[\protect\citeauthoryear{Dai et~al.}{2010}]{bib5:Dai2010}
\begin{barticle}
\bauthor{\binits{X.} \bsnm{Dai}},
\bauthor{\binits{C.S.} \bsnm{Kochanek}},
\bauthor{\binits{G.} \bsnm{Chartas}},
\bauthor{\binits{S.} \bsnm{Koz{\l}owski}},
\bauthor{\binits{C.W.} \bsnm{Morgan}},
\bauthor{\binits{G.} \bsnm{Garmire}},
\bauthor{\binits{E.} \bsnm{Agol}},
\batitle{The sizes of the x-ray and optical emission regions of rxj 1131-1231}.
\bjtitle{The Astrophysical Journal}
\bvolume{709},
\bfpage{278}--\blpage{285}
(\byear{2010}).
doi:\doiurl{10.1088/0004-637X/709/1/278}.
\burl{http://stacks.iop.org/0004-637X/709/i=1/a=278?key=crossref.50b2f0dd99b063734fa9f0e6a5448833}
\end{barticle}
\endbibitem

\bibitem[\protect\citeauthoryear{Dai and Guerras}{2018}]{bib5:Dai2018}
\begin{barticle}
\bauthor{\binits{X.} \bsnm{Dai}},
\bauthor{\binits{E.} \bsnm{Guerras}},
\batitle{Probing extragalactic planets using quasar microlensing}.
\bjtitle{The Astrophysical Journal}
\bvolume{853},
\bfpage{27}
(\byear{2018}).
doi:\doiurl{10.3847/2041-8213/aaa5fb}.
\burl{http://dx.doi.org/10.3847/2041-8213/aaa5fb}
\end{barticle}
\endbibitem

\bibitem[\protect\citeauthoryear{Dalal and Kochanek}{2002}]{bib5:Dalal2002}
\begin{barticle}
\bauthor{\binits{N.} \bsnm{Dalal}},
\bauthor{\binits{C.S.} \bsnm{Kochanek}},
\batitle{Direct detection of cold dark matter substructure}.
\bjtitle{The Astrophysical Journal}
\bvolume{572},
\bfpage{25}--\blpage{33}
(\byear{2002})
\end{barticle}
\endbibitem

\bibitem[\protect\citeauthoryear{Davis and Laor}{2011}]{bib5:Davis2011}
\begin{botherref}
\oauthor{\binits{S.W.} \bsnm{Davis}},
\oauthor{\binits{A.} \bsnm{Laor}},
The radiative efficiency of accretion flows in individual active galactic
  nuclei.
Astrophysical Journal
\textbf{728}
(2011).
doi:\doiurl{10.1088/0004-637X/728/2/98}
\end{botherref}
\endbibitem

\bibitem[\protect\citeauthoryear{Dempsey and Zakamska}{2018}]{bib5:Dempsey2018}
\begin{barticle}
\bauthor{\binits{R.} \bsnm{Dempsey}},
\bauthor{\binits{N.L.} \bsnm{Zakamska}},
\batitle{The size-luminosity relationship of quasar narrow-line regions}.
\bjtitle{Monthly Notices of the Royal Astronomical Society}
\bvolume{477},
\bfpage{4615}--\blpage{4626}
(\byear{2018}).
doi:\doiurl{10.1093/mnras/sty941}
\end{barticle}
\endbibitem

\bibitem[\protect\citeauthoryear{Dexter and Agol}{2011}]{bib5:Dexter2011}
\begin{barticle}
\bauthor{\binits{J.} \bsnm{Dexter}},
\bauthor{\binits{E.} \bsnm{Agol}},
\batitle{Quasar accretion disks are strongly inhomogeneous}.
\bjtitle{The Astrophysical Journal Letters}
\bvolume{727},
\bfpage{24}
(\byear{2011}).
doi:\doiurl{10.1088/2041-8205/727/1/L24}.
\burl{http://stacks.iop.org/2041-8205/727/i=1/a=L24?key=crossref.001391153ce1c00f07fe87dc43360477}
\end{barticle}
\endbibitem

\bibitem[\protect\citeauthoryear{{Di Matteo}}{1998}]{bib5:Dimatteo1998}
\begin{barticle}
\bauthor{\binits{T.} \bsnm{{Di Matteo}}},
\batitle{Magnetic reconnection: flares and coronal heating in active galactic
  nuclei}.
\bjtitle{\mnras}
\bvolume{299}(\bissue{1}),
\bfpage{15}--\blpage{20}
(\byear{1998}).
doi:\doiurl{10.1046/j.1365-8711.1998.01950.x}
\end{barticle}
\endbibitem

\bibitem[\protect\citeauthoryear{{Di Valentino}
  et~al.}{2021}]{bib5:diValentino2021}
\begin{barticle}
\bauthor{\binits{E.} \bsnm{{Di Valentino}}},
\bauthor{\binits{L.A.} \bsnm{{Anchordoqui}}},
\bauthor{\binits{{\"O}.} \bsnm{{Akarsu}}},
\bauthor{\binits{Y.} \bsnm{{Ali-Haimoud}}},
\bauthor{\binits{L.} \bsnm{{Amendola}}},
\bauthor{\binits{N.} \bsnm{{Arendse}}},
\bauthor{\binits{M.} \bsnm{{Asgari}}},
\bauthor{\binits{M.} \bsnm{{Ballardini}}},
\bauthor{\binits{S.} \bsnm{{Basilakos}}},
\bauthor{\binits{E.} \bsnm{{Battistelli}}},
\bauthor{\binits{M.} \bsnm{{Benetti}}},
\bauthor{\binits{S.} \bsnm{{Birrer}}},
\bauthor{\binits{F.R.} \bsnm{{Bouchet}}},
\bauthor{\binits{M.} \bsnm{{Bruni}}},
\bauthor{\binits{E.} \bsnm{{Calabrese}}},
\bauthor{\binits{D.} \bsnm{{Camarena}}},
\bauthor{\binits{S.} \bsnm{{Capozziello}}},
\bauthor{\binits{A.} \bsnm{{Chen}}},
\bauthor{\binits{J.} \bsnm{{Chluba}}},
\bauthor{\binits{A.} \bsnm{{Chudaykin}}},
\bauthor{\binits{E.{\'O}.} \bsnm{{Colg{\'a}in}}},
\bauthor{\binits{F.-Y.} \bsnm{{Cyr-Racine}}},
\bauthor{\binits{P.} \bsnm{{de Bernardis}}},
\bauthor{\binits{J.} \bsnm{{de Cruz P{\'e}rez}}},
\bauthor{\binits{J.} \bsnm{{Delabrouille}}},
\bauthor{\binits{J.} \bsnm{{Dunkley}}},
\bauthor{\binits{C.} \bsnm{{Escamilla-Rivera}}},
\bauthor{\binits{A.} \bsnm{{Fert{\'e}}}},
\bauthor{\binits{F.} \bsnm{{Finelli}}},
\bauthor{\binits{W.} \bsnm{{Freedman}}},
\bauthor{\binits{N.} \bsnm{{Frusciante}}},
\bauthor{\binits{E.} \bsnm{{Giusarma}}},
\bauthor{\binits{A.} \bsnm{{G{\'o}mez-Valent}}},
\bauthor{\binits{W.} \bsnm{{Handley}}},
\bauthor{\binits{I.} \bsnm{{Harrison}}},
\bauthor{\binits{L.} \bsnm{{Hart}}},
\bauthor{\binits{A.} \bsnm{{Heavens}}},
\bauthor{\binits{H.} \bsnm{{Hildebrandt}}},
\bauthor{\binits{D.} \bsnm{{Holz}}},
\bauthor{\binits{D.} \bsnm{{Huterer}}},
\bauthor{\binits{M.M.} \bsnm{{Ivanov}}},
\bauthor{\binits{S.} \bsnm{{Joudaki}}},
\bauthor{\binits{M.} \bsnm{{Kamionkowski}}},
\bauthor{\binits{T.} \bsnm{{Karwal}}},
\bauthor{\binits{L.} \bsnm{{Knox}}},
\bauthor{\binits{S.} \bsnm{{Kumar}}},
\bauthor{\binits{L.} \bsnm{{Lamagna}}},
\bauthor{\binits{J.} \bsnm{{Lesgourgues}}},
\bauthor{\binits{M.} \bsnm{{Lucca}}},
\bauthor{\binits{V.} \bsnm{{Marra}}},
\bauthor{\binits{S.} \bsnm{{Masi}}},
\bauthor{\binits{S.} \bsnm{{Matarrese}}},
\bauthor{\binits{A.} \bsnm{{Mazumdar}}},
\bauthor{\binits{A.} \bsnm{{Melchiorri}}},
\bauthor{\binits{O.} \bsnm{{Mena}}},
\bauthor{\binits{L.} \bsnm{{Mersini-Houghton}}},
\bauthor{\binits{V.} \bsnm{{Miranda}}},
\bauthor{\binits{C.} \bsnm{{Moreno-Pulido}}},
\bauthor{\binits{D.F.} \bsnm{{Mota}}},
\bauthor{\binits{J.} \bsnm{{Muir}}},
\bauthor{\binits{A.} \bsnm{{Mukherjee}}},
\bauthor{\binits{F.} \bsnm{{Niedermann}}},
\bauthor{\binits{A.} \bsnm{{Notari}}},
\bauthor{\binits{R.C.} \bsnm{{Nunes}}},
\bauthor{\binits{F.} \bsnm{{Pace}}},
\bauthor{\binits{A.} \bsnm{{Paliathanasis}}},
\bauthor{\binits{A.} \bsnm{{Palmese}}},
\bauthor{\binits{S.} \bsnm{{Pan}}},
\bauthor{\binits{D.} \bsnm{{Paoletti}}},
\bauthor{\binits{V.} \bsnm{{Pettorino}}},
\bauthor{\binits{F.} \bsnm{{Piacentini}}},
\bauthor{\binits{V.} \bsnm{{Poulin}}},
\bauthor{\binits{M.} \bsnm{{Raveri}}},
\bauthor{\binits{A.G.} \bsnm{{Riess}}},
\bauthor{\binits{V.} \bsnm{{Salzano}}},
\bauthor{\binits{E.N.} \bsnm{{Saridakis}}},
\bauthor{\binits{A.A.} \bsnm{{Sen}}},
\bauthor{\binits{A.} \bsnm{{Shafieloo}}},
\bauthor{\binits{A.J.} \bsnm{{Shajib}}},
\bauthor{\binits{J.} \bsnm{{Silk}}},
\bauthor{\binits{A.} \bsnm{{Silvestri}}},
\bauthor{\binits{M.S.} \bsnm{{Sloth}}},
\bauthor{\binits{T.L.} \bsnm{{Smith}}},
\bauthor{\binits{J.} \bsnm{{Sol{\`a} Peracaula}}},
\bauthor{\binits{C.} \bsnm{{van de Bruck}}},
\bauthor{\binits{L.} \bsnm{{Verde}}},
\bauthor{\binits{L.} \bsnm{{Visinelli}}},
\bauthor{\binits{B.D.} \bsnm{{Wandelt}}},
\bauthor{\binits{D.} \bsnm{{Wang}}},
\bauthor{\binits{J.-M.} \bsnm{{Wang}}},
\bauthor{\binits{A.K.} \bsnm{{Yadav}}},
\bauthor{\binits{W.} \bsnm{{Yang}}},
\batitle{Cosmology intertwined iii: f{\ensuremath{\sigma}}$_{8}$ and s$_{8}$}.
\bjtitle{Astroparticle Physics}
\bvolume{131},
\bfpage{102604}
(\byear{2021}).
doi:\doiurl{10.1016/j.astropartphys.2021.102604}
\end{barticle}
\endbibitem

\bibitem[\protect\citeauthoryear{DIego et~al.}{2019}]{bib5:Diego2019}
\begin{botherref}
\oauthor{\binits{J.M.} \bsnm{DIego}},
\oauthor{\binits{O.A.} \bsnm{Hannuksela}},
\oauthor{\binits{P.L.} \bsnm{Kelly}},
\oauthor{\binits{G.} \bsnm{Pagano}},
\oauthor{\binits{T.} \bsnm{Broadhurst}},
\oauthor{\binits{K.} \bsnm{Kim}},
\oauthor{\binits{T.G.F.} \bsnm{Li}},
\oauthor{\binits{G.F.} \bsnm{Smoot}},
Observational signatures of microlensing in gravitational waves at ligo/virgo
  frequencies.
Astronomy and Astrophysics
\textbf{627}
(2019).
doi:\doiurl{10.1051/0004-6361/201935490}
\end{botherref}
\endbibitem

\bibitem[\protect\citeauthoryear{Dobler et~al.}{2007}]{bib5:Dobler2007}
\begin{barticle}
\bauthor{\binits{G.} \bsnm{Dobler}},
\bauthor{\binits{C.R.} \bsnm{Keeton}},
\bauthor{\binits{J.} \bsnm{Wambsganss}},
\batitle{Microlensing of central images in strong gravitational lens systems}.
\bjtitle{Monthly Notices of the Royal Astronomical Society}
\bvolume{377},
\bfpage{977}--\blpage{986}
(\byear{2007}).
doi:\doiurl{10.1111/j.1365-2966.2007.11695.x}
\end{barticle}
\endbibitem

\bibitem[\protect\citeauthoryear{Edelson et~al.}{2015}]{bib5:Edelson2015}
\begin{botherref}
\oauthor{\binits{R.} \bsnm{Edelson}},
\oauthor{\binits{J.M.} \bsnm{Gelbord}},
\oauthor{\binits{K.} \bsnm{Horne}},
\oauthor{\binits{I.M.} \bsnm{McHardy}},
\oauthor{\binits{B.M.} \bsnm{Peterson}},
\oauthor{\binits{P.} \bsnm{Ar{\'e}valo}},
\oauthor{\binits{A.A.} \bsnm{Breeveld}},
\oauthor{\binits{G.D.} \bsnm{Rosa}},
\oauthor{\binits{P.A.} \bsnm{Evans}},
\oauthor{\binits{M.R.} \bsnm{Goad}},
\oauthor{\binits{G.A.} \bsnm{Kriss}},
\oauthor{\binits{W.N.} \bsnm{Brandt}},
\oauthor{\binits{N.} \bsnm{Gehrels}},
\oauthor{\binits{D.} \bsnm{Grupe}},
\oauthor{\binits{J.A.} \bsnm{Kennea}},
\oauthor{\binits{C.S.} \bsnm{Kochanek}},
\oauthor{\binits{J.A.} \bsnm{Nousek}},
\oauthor{\binits{I.} \bsnm{Papadakis}},
\oauthor{\binits{M.} \bsnm{Siegel}},
\oauthor{\binits{D.} \bsnm{Starkey}},
\oauthor{\binits{P.} \bsnm{Uttley}},
\oauthor{\binits{S.} \bsnm{Vaughan}},
\oauthor{\binits{S.} \bsnm{Young}},
\oauthor{\binits{A.J.} \bsnm{Barth}},
\oauthor{\binits{M.C.} \bsnm{Bentz}},
\oauthor{\binits{B.J.} \bsnm{Brewer}},
\oauthor{\binits{D.M.} \bsnm{Crenshaw}},
\oauthor{\binits{E.D.B.A.D.L.} \bsnm{C{\'a}ceres}},
\oauthor{\binits{K.D.} \bsnm{Denney}},
\oauthor{\binits{M.} \bsnm{Dietrich}},
\oauthor{\binits{J.} \bsnm{Ely}},
\oauthor{\binits{M.M.} \bsnm{Fausnaugh}},
\oauthor{\binits{C.J.} \bsnm{Grier}},
\oauthor{\binits{P.B.} \bsnm{Hall}},
\oauthor{\binits{J.} \bsnm{Kaastra}},
\oauthor{\binits{B.C.} \bsnm{Kelly}},
\oauthor{\binits{K.T.} \bsnm{Korista}},
\oauthor{\binits{P.} \bsnm{Lira}},
\oauthor{\binits{S.} \bsnm{Mathur}},
\oauthor{\binits{H.} \bsnm{Netzer}},
\oauthor{\binits{A.} \bsnm{Pancoast}},
\oauthor{\binits{L.} \bsnm{Pei}},
\oauthor{\binits{R.W.} \bsnm{Pogge}},
\oauthor{\binits{J.S.} \bsnm{Schimoia}},
\oauthor{\binits{T.} \bsnm{Treu}},
\oauthor{\binits{M.} \bsnm{Vestergaard}},
\oauthor{\binits{C.} \bsnm{Villforth}},
\oauthor{\binits{H.} \bsnm{Yan}},
\oauthor{\binits{Y.} \bsnm{Zu}},
Space telescope and optical reverberation mapping project. ii. swift and hst
  reverberation mapping of the accretion disk of ngc 5548.
Astrophysical Journal
\textbf{806}
(2015).
doi:\doiurl{10.1088/0004-637X/806/1/129}
\end{botherref}
\endbibitem

\bibitem[\protect\citeauthoryear{Eigenbrod et~al.}{2008a}]{bib5:Eigenbrod2008b}
\begin{barticle}
\bauthor{\binits{A.} \bsnm{Eigenbrod}},
\bauthor{\binits{F.} \bsnm{Courbin}},
\bauthor{\binits{D.} \bsnm{Sluse}},
\bauthor{\binits{G.} \bsnm{Meylan}},
\bauthor{\binits{E.} \bsnm{Agol}},
\batitle{Microlensing variability in the gravitationally lensed quasar qso
  2237$+$0305 - the einstein cross}.
\bjtitle{Astronomy and Astrophysics}
\bvolume{480},
\bfpage{647}--\blpage{661}
(\byear{2008}a).
doi:\doiurl{10.1051/0004-6361:20078703}
\end{barticle}
\endbibitem

\bibitem[\protect\citeauthoryear{Eigenbrod et~al.}{2008b}]{bib5:Eigenbrod2008a}
\begin{barticle}
\bauthor{\binits{A.} \bsnm{Eigenbrod}},
\bauthor{\binits{F.} \bsnm{Courbin}},
\bauthor{\binits{G.} \bsnm{Meylan}},
\bauthor{\binits{E.} \bsnm{Agol}},
\bauthor{\binits{T.} \bsnm{Anguita}},
\bauthor{\binits{R.W.} \bsnm{Schmidt}},
\bauthor{\binits{J.} \bsnm{Wambsganss}},
\batitle{Microlensing variability in the gravitationally lensed quasar qso
  2237$+$0305 - the einstein cross ii. energy profile of the accretion disk}.
\bjtitle{Astronomy \& Astrophysics}
\bvolume{490},
\bfpage{933}--\blpage{943}
(\byear{2008}b).
doi:\doiurl{10.1051/0004-6361}
\end{barticle}
\endbibitem

\bibitem[\protect\citeauthoryear{Elitzur et~al.}{2014}]{bib5:Elitzur2014}
\begin{barticle}
\bauthor{\binits{M.} \bsnm{Elitzur}},
\bauthor{\binits{L.C.} \bsnm{Ho}},
\bauthor{\binits{J.R.} \bsnm{Trump}},
\batitle{Evolution of broad-line emission from active galactic nuclei}.
\bjtitle{Monthly Notices of the Royal Astronomical Society}
\bvolume{438},
\bfpage{3340}--\blpage{3351}
(\byear{2014}).
doi:\doiurl{10.1093/mnras/stt2445}
\end{barticle}
\endbibitem

\bibitem[\protect\citeauthoryear{Esteban-Guti{\'e}rrez
  et~al.}{2020}]{bib5:Esteban2020}
\begin{barticle}
\bauthor{\binits{A.} \bsnm{Esteban-Guti{\'e}rrez}},
\bauthor{\binits{N.} \bsnm{Ag{\"u}es-Paszkowsky}},
\bauthor{\binits{E.} \bsnm{Mediavilla}},
\bauthor{\binits{J.} \bsnm{Jim{\'e}nez-Vicente}},
\bauthor{\binits{J.A.} \bsnm{Mu{\~n}oz}},
\bauthor{\binits{S.} \bsnm{Heydenreich}},
\batitle{The impact of the mass spectrum of lenses in quasar microlensing
  studies. constraints on a mixed population of primordial black holes and
  stars}.
\bjtitle{The Astrophysical Journal}
\bvolume{904},
\bfpage{176}
(\byear{2020}).
doi:\doiurl{10.3847/1538-4357/abbdf7}.
\burl{http://dx.doi.org/10.3847/1538-4357/abbdf7}
\end{barticle}
\endbibitem

\bibitem[\protect\citeauthoryear{Esteban-Guti{\'e}rrez
  et~al.}{2022}]{bib5:Esteban2022}
\begin{barticle}
\bauthor{\binits{A.} \bsnm{Esteban-Guti{\'e}rrez}},
\bauthor{\binits{N.} \bsnm{Ag{\"u}es-Paszkowsky}},
\bauthor{\binits{E.} \bsnm{Mediavilla}},
\bauthor{\binits{J.} \bsnm{Jim{\'e}nez-Vicente}},
\bauthor{\binits{J.A.} \bsnm{Mu{\~n}oz}},
\bauthor{\binits{S.} \bsnm{Heydenreich}},
\batitle{Abundance of ligo/virgo black holes from microlensing observations of
  quasars with reverberation mapping size estimates}.
\bjtitle{The Astrophysical Journal}
\bvolume{929},
\bfpage{123}
(\byear{2022}).
doi:\doiurl{10.3847/1538-4357/ac57c5}
\end{barticle}
\endbibitem

\bibitem[\protect\citeauthoryear{{Event Horizon Telescope Collaboration}
  et~al.}{2019}]{bib5:EHT2019}
\begin{botherref}
\oauthor{\bsnm{{Event Horizon Telescope Collaboration}}},
\oauthor{\binits{K.} \bsnm{{Akiyama}}},
\oauthor{\binits{A.} \bsnm{{Alberdi}}},
\oauthor{\binits{W.} \bsnm{{Alef}}},
\oauthor{\binits{K.} \bsnm{{Asada}}},
\oauthor{\binits{R.} \bsnm{{Azulay}}},
\oauthor{\binits{A.-K.} \bsnm{{Baczko}}},
\oauthor{\binits{D.} \bsnm{{Ball}}},
\oauthor{\binits{M.} \bsnm{{Balokovi{\'c}}}},
\oauthor{\binits{J.} \bsnm{{Barrett}}},
\oauthor{\binits{D.} \bsnm{{Bintley}}},
\oauthor{\binits{L.} \bsnm{{Blackburn}}},
\oauthor{\binits{W.} \bsnm{{Boland}}},
\oauthor{\binits{K.L.} \bsnm{{Bouman}}},
\oauthor{\binits{G.C.} \bsnm{{Bower}}},
\oauthor{\binits{M.} \bsnm{{Bremer}}},
\oauthor{\binits{C.D.} \bsnm{{Brinkerink}}},
\oauthor{\binits{R.} \bsnm{{Brissenden}}},
\oauthor{\binits{S.} \bsnm{{Britzen}}},
\oauthor{\binits{A.E.} \bsnm{{Broderick}}},
\oauthor{\binits{D.} \bsnm{{Broguiere}}},
\oauthor{\binits{T.} \bsnm{{Bronzwaer}}},
\oauthor{\binits{D.-Y.} \bsnm{{Byun}}},
\oauthor{\binits{J.E.} \bsnm{{Carlstrom}}},
\oauthor{\binits{A.} \bsnm{{Chael}}},
\oauthor{\binits{C.-k.} \bsnm{{Chan}}},
\oauthor{\binits{S.} \bsnm{{Chatterjee}}},
\oauthor{\binits{K.} \bsnm{{Chatterjee}}},
\oauthor{\binits{M.-T.} \bsnm{{Chen}}},
\oauthor{\binits{Y.} \bsnm{{Chen}}},
\oauthor{\binits{I.} \bsnm{{Cho}}},
\oauthor{\binits{P.} \bsnm{{Christian}}},
\oauthor{\binits{J.E.} \bsnm{{Conway}}},
\oauthor{\binits{J.M.} \bsnm{{Cordes}}},
\oauthor{\binits{G.B.} \bsnm{{Crew}}},
\oauthor{\binits{Y.} \bsnm{{Cui}}},
\oauthor{\binits{J.} \bsnm{{Davelaar}}},
\oauthor{\binits{M.} \bsnm{{De Laurentis}}},
\oauthor{\binits{R.} \bsnm{{Deane}}},
\oauthor{\binits{J.} \bsnm{{Dempsey}}},
\oauthor{\binits{G.} \bsnm{{Desvignes}}},
\oauthor{\binits{J.} \bsnm{{Dexter}}},
\oauthor{\binits{S.S.} \bsnm{{Doeleman}}},
\oauthor{\binits{R.P.} \bsnm{{Eatough}}},
\oauthor{\binits{H.} \bsnm{{Falcke}}},
\oauthor{\binits{V.L.} \bsnm{{Fish}}},
\oauthor{\binits{E.} \bsnm{{Fomalont}}},
\oauthor{\binits{R.} \bsnm{{Fraga-Encinas}}},
\oauthor{\binits{P.} \bsnm{{Friberg}}},
\oauthor{\binits{C.M.} \bsnm{{Fromm}}},
\oauthor{\binits{J.L.} \bsnm{{G{\'o}mez}}},
\oauthor{\binits{P.} \bsnm{{Galison}}},
\oauthor{\binits{C.F.} \bsnm{{Gammie}}},
\oauthor{\binits{R.} \bsnm{{Garc{\'\i}a}}},
\oauthor{\binits{O.} \bsnm{{Gentaz}}},
\oauthor{\binits{B.} \bsnm{{Georgiev}}},
\oauthor{\binits{C.} \bsnm{{Goddi}}},
\oauthor{\binits{R.} \bsnm{{Gold}}},
\oauthor{\binits{M.} \bsnm{{Gu}}},
\oauthor{\binits{M.} \bsnm{{Gurwell}}},
\oauthor{\binits{K.} \bsnm{{Hada}}},
\oauthor{\binits{M.H.} \bsnm{{Hecht}}},
\oauthor{\binits{R.} \bsnm{{Hesper}}},
\oauthor{\binits{L.C.} \bsnm{{Ho}}},
\oauthor{\binits{P.} \bsnm{{Ho}}},
\oauthor{\binits{M.} \bsnm{{Honma}}},
\oauthor{\binits{C.-W.L.} \bsnm{{Huang}}},
\oauthor{\binits{L.} \bsnm{{Huang}}},
\oauthor{\binits{D.H.} \bsnm{{Hughes}}},
\oauthor{\binits{S.} \bsnm{{Ikeda}}},
\oauthor{\binits{M.} \bsnm{{Inoue}}},
\oauthor{\binits{S.} \bsnm{{Issaoun}}},
\oauthor{\binits{D.J.} \bsnm{{James}}},
\oauthor{\binits{B.T.} \bsnm{{Jannuzi}}},
\oauthor{\binits{M.} \bsnm{{Janssen}}},
\oauthor{\binits{B.} \bsnm{{Jeter}}},
\oauthor{\binits{W.} \bsnm{{Jiang}}},
\oauthor{\binits{M.D.} \bsnm{{Johnson}}},
\oauthor{\binits{S.} \bsnm{{Jorstad}}},
\oauthor{\binits{T.} \bsnm{{Jung}}},
\oauthor{\binits{M.} \bsnm{{Karami}}},
\oauthor{\binits{R.} \bsnm{{Karuppusamy}}},
\oauthor{\binits{T.} \bsnm{{Kawashima}}},
\oauthor{\binits{G.K.} \bsnm{{Keating}}},
\oauthor{\binits{M.} \bsnm{{Kettenis}}},
\oauthor{\binits{J.-Y.} \bsnm{{Kim}}},
\oauthor{\binits{J.} \bsnm{{Kim}}},
\oauthor{\binits{J.} \bsnm{{Kim}}},
\oauthor{\binits{M.} \bsnm{{Kino}}},
\oauthor{\binits{J.Y.} \bsnm{{Koay}}},
\oauthor{\binits{P.M.} \bsnm{{Koch}}},
\oauthor{\binits{S.} \bsnm{{Koyama}}},
\oauthor{\binits{M.} \bsnm{{Kramer}}},
\oauthor{\binits{C.} \bsnm{{Kramer}}},
\oauthor{\binits{T.P.} \bsnm{{Krichbaum}}},
\oauthor{\binits{C.-Y.} \bsnm{{Kuo}}},
\oauthor{\binits{T.R.} \bsnm{{Lauer}}},
\oauthor{\binits{S.-S.} \bsnm{{Lee}}},
\oauthor{\binits{Y.-R.} \bsnm{{Li}}},
\oauthor{\binits{Z.} \bsnm{{Li}}},
\oauthor{\binits{M.} \bsnm{{Lindqvist}}},
\oauthor{\binits{K.} \bsnm{{Liu}}},
\oauthor{\binits{E.} \bsnm{{Liuzzo}}},
\oauthor{\binits{W.-P.} \bsnm{{Lo}}},
\oauthor{\binits{A.P.} \bsnm{{Lobanov}}},
\oauthor{\binits{L.} \bsnm{{Loinard}}},
\oauthor{\binits{C.} \bsnm{{Lonsdale}}},
\oauthor{\binits{R.-S.} \bsnm{{Lu}}},
\oauthor{\binits{N.R.} \bsnm{{MacDonald}}},
\oauthor{\binits{J.} \bsnm{{Mao}}},
\oauthor{\binits{S.} \bsnm{{Markoff}}},
\oauthor{\binits{D.P.} \bsnm{{Marrone}}},
\oauthor{\binits{A.P.} \bsnm{{Marscher}}},
\oauthor{\binits{I.} \bsnm{{Mart{\'\i}-Vidal}}},
\oauthor{\binits{S.} \bsnm{{Matsushita}}},
\oauthor{\binits{L.D.} \bsnm{{Matthews}}},
\oauthor{\binits{L.} \bsnm{{Medeiros}}},
\oauthor{\binits{K.M.} \bsnm{{Menten}}},
\oauthor{\binits{Y.} \bsnm{{Mizuno}}},
\oauthor{\binits{I.} \bsnm{{Mizuno}}},
\oauthor{\binits{J.M.} \bsnm{{Moran}}},
\oauthor{\binits{K.} \bsnm{{Moriyama}}},
\oauthor{\binits{M.} \bsnm{{Moscibrodzka}}},
\oauthor{\binits{C.} \bsnm{{M{\"u}ller}}},
\oauthor{\binits{H.} \bsnm{{Nagai}}},
\oauthor{\binits{N.M.} \bsnm{{Nagar}}},
\oauthor{\binits{M.} \bsnm{{Nakamura}}},
\oauthor{\binits{R.} \bsnm{{Narayan}}},
\oauthor{\binits{G.} \bsnm{{Narayanan}}},
\oauthor{\binits{I.} \bsnm{{Natarajan}}},
\oauthor{\binits{R.} \bsnm{{Neri}}},
\oauthor{\binits{C.} \bsnm{{Ni}}},
\oauthor{\binits{A.} \bsnm{{Noutsos}}},
\oauthor{\binits{H.} \bsnm{{Okino}}},
\oauthor{\binits{H.} \bsnm{{Olivares}}},
\oauthor{\binits{T.} \bsnm{{Oyama}}},
\oauthor{\binits{F.} \bsnm{{{\"O}zel}}},
\oauthor{\binits{D.C.M.} \bsnm{{Palumbo}}},
\oauthor{\binits{N.} \bsnm{{Patel}}},
\oauthor{\binits{U.-L.} \bsnm{{Pen}}},
\oauthor{\binits{D.W.} \bsnm{{Pesce}}},
\oauthor{\binits{V.} \bsnm{{Pi{\'e}tu}}},
\oauthor{\binits{R.} \bsnm{{Plambeck}}},
\oauthor{\binits{A.} \bsnm{{PopStefanija}}},
\oauthor{\binits{O.} \bsnm{{Porth}}},
\oauthor{\binits{B.} \bsnm{{Prather}}},
\oauthor{\binits{J.A.} \bsnm{{Preciado-L{\'o}pez}}},
\oauthor{\binits{D.} \bsnm{{Psaltis}}},
\oauthor{\binits{H.-Y.} \bsnm{{Pu}}},
\oauthor{\binits{V.} \bsnm{{Ramakrishnan}}},
\oauthor{\binits{R.} \bsnm{{Rao}}},
\oauthor{\binits{M.G.} \bsnm{{Rawlings}}},
\oauthor{\binits{A.W.} \bsnm{{Raymond}}},
\oauthor{\binits{L.} \bsnm{{Rezzolla}}},
\oauthor{\binits{B.} \bsnm{{Ripperda}}},
\oauthor{\binits{F.} \bsnm{{Roelofs}}},
\oauthor{\binits{A.} \bsnm{{Rogers}}},
\oauthor{\binits{E.} \bsnm{{Ros}}},
\oauthor{\binits{M.} \bsnm{{Rose}}},
\oauthor{\binits{A.} \bsnm{{Roshanineshat}}},
\oauthor{\binits{H.} \bsnm{{Rottmann}}},
\oauthor{\binits{A.L.} \bsnm{{Roy}}},
\oauthor{\binits{C.} \bsnm{{Ruszczyk}}},
\oauthor{\binits{B.R.} \bsnm{{Ryan}}},
\oauthor{\binits{K.L.J.} \bsnm{{Rygl}}},
\oauthor{\binits{S.} \bsnm{{S{\'a}nchez}}},
\oauthor{\binits{D.} \bsnm{{S{\'a}nchez-Arguelles}}},
\oauthor{\binits{M.} \bsnm{{Sasada}}},
\oauthor{\binits{T.} \bsnm{{Savolainen}}},
\oauthor{\binits{F.P.} \bsnm{{Schloerb}}},
\oauthor{\binits{K.-F.} \bsnm{{Schuster}}},
\oauthor{\binits{L.} \bsnm{{Shao}}},
\oauthor{\binits{Z.} \bsnm{{Shen}}},
\oauthor{\binits{D.} \bsnm{{Small}}},
\oauthor{\binits{B.W.} \bsnm{{Sohn}}},
\oauthor{\binits{J.} \bsnm{{SooHoo}}},
\oauthor{\binits{F.} \bsnm{{Tazaki}}},
\oauthor{\binits{P.} \bsnm{{Tiede}}},
\oauthor{\binits{R.P.J.} \bsnm{{Tilanus}}},
\oauthor{\binits{M.} \bsnm{{Titus}}},
\oauthor{\binits{K.} \bsnm{{Toma}}},
\oauthor{\binits{P.} \bsnm{{Torne}}},
\oauthor{\binits{T.} \bsnm{{Trent}}},
\oauthor{\binits{S.} \bsnm{{Trippe}}},
\oauthor{\binits{S.} \bsnm{{Tsuda}}},
\oauthor{\binits{I.} \bsnm{{van Bemmel}}},
\oauthor{\binits{H.J.} \bsnm{{van Langevelde}}},
\oauthor{\binits{D.R.} \bsnm{{van Rossum}}},
\oauthor{\binits{J.} \bsnm{{Wagner}}},
\oauthor{\binits{J.} \bsnm{{Wardle}}},
\oauthor{\binits{J.} \bsnm{{Weintroub}}},
\oauthor{\binits{N.} \bsnm{{Wex}}},
\oauthor{\binits{R.} \bsnm{{Wharton}}},
\oauthor{\binits{M.} \bsnm{{Wielgus}}},
\oauthor{\binits{G.N.} \bsnm{{Wong}}},
\oauthor{\binits{Q.} \bsnm{{Wu}}},
\oauthor{\binits{A.} \bsnm{{Young}}},
\oauthor{\binits{K.} \bsnm{{Young}}},
\oauthor{\binits{Z.} \bsnm{{Younsi}}},
\oauthor{\binits{F.} \bsnm{{Yuan}}},
\oauthor{\binits{Y.-F.} \bsnm{{Yuan}}},
\oauthor{\binits{J.A.} \bsnm{{Zensus}}},
\oauthor{\binits{G.} \bsnm{{Zhao}}},
\oauthor{\binits{S.-S.} \bsnm{{Zhao}}},
\oauthor{\binits{Z.} \bsnm{{Zhu}}},
\oauthor{\binits{J.R.} \bsnm{{Farah}}},
\oauthor{\binits{Z.} \bsnm{{Meyer-Zhao}}},
\oauthor{\binits{D.} \bsnm{{Michalik}}},
\oauthor{\binits{A.} \bsnm{{Nadolski}}},
\oauthor{\binits{H.} \bsnm{{Nishioka}}},
\oauthor{\binits{N.} \bsnm{{Pradel}}},
\oauthor{\binits{R.A.} \bsnm{{Primiani}}},
\oauthor{\binits{K.} \bsnm{{Souccar}}},
\oauthor{\binits{L.} \bsnm{{Vertatschitsch}}},
\oauthor{\binits{P.} \bsnm{{Yamaguchi}}},
First m87 event horizon telescope results. iv. imaging the central supermassive
  black hole
\textbf{875},
4
(2019).
doi:\doiurl{10.3847/2041-8213/ab0e85}
\end{botherref}
\endbibitem

\bibitem[\protect\citeauthoryear{{Event Horizon Telescope Collaboration}
  et~al.}{2022}]{bib5:EHT2022}
\begin{botherref}
\oauthor{\bsnm{{Event Horizon Telescope Collaboration}}},
\oauthor{\binits{K.} \bsnm{{Akiyama}}},
\oauthor{\binits{A.} \bsnm{{Alberdi}}},
\oauthor{\binits{W.} \bsnm{{Alef}}},
\oauthor{\binits{J.C.} \bsnm{{Algaba}}},
\oauthor{\binits{R.} \bsnm{{Anantua}}},
\oauthor{\binits{K.} \bsnm{{Asada}}},
\oauthor{\binits{R.} \bsnm{{Azulay}}},
\oauthor{\binits{U.} \bsnm{{Bach}}},
\oauthor{\binits{A.-K.} \bsnm{{Baczko}}},
\oauthor{\binits{D.} \bsnm{{Ball}}},
\oauthor{\binits{M.} \bsnm{{Balokovi{\'c}}}},
\oauthor{\binits{J.} \bsnm{{Barrett}}},
\oauthor{\binits{M.} \bsnm{{Baub{\"o}ck}}},
\oauthor{\binits{B.A.} \bsnm{{Benson}}},
\oauthor{\binits{D.} \bsnm{{Bintley}}},
\oauthor{\binits{L.} \bsnm{{Blackburn}}},
\oauthor{\binits{R.} \bsnm{{Blundell}}},
\oauthor{\binits{K.L.} \bsnm{{Bouman}}},
\oauthor{\binits{G.C.} \bsnm{{Bower}}},
\oauthor{\binits{H.} \bsnm{{Boyce}}},
\oauthor{\binits{M.} \bsnm{{Bremer}}},
\oauthor{\binits{C.D.} \bsnm{{Brinkerink}}},
\oauthor{\binits{R.} \bsnm{{Brissenden}}},
\oauthor{\binits{S.} \bsnm{{Britzen}}},
\oauthor{\binits{A.E.} \bsnm{{Broderick}}},
\oauthor{\binits{D.} \bsnm{{Broguiere}}},
\oauthor{\binits{T.} \bsnm{{Bronzwaer}}},
\oauthor{\binits{S.} \bsnm{{Bustamante}}},
\oauthor{\binits{D.-Y.} \bsnm{{Byun}}},
\oauthor{\binits{J.E.} \bsnm{{Carlstrom}}},
\oauthor{\binits{C.} \bsnm{{Ceccobello}}},
\oauthor{\binits{A.} \bsnm{{Chael}}},
\oauthor{\binits{C.-k.} \bsnm{{Chan}}},
\oauthor{\binits{K.} \bsnm{{Chatterjee}}},
\oauthor{\binits{S.} \bsnm{{Chatterjee}}},
\oauthor{\binits{M.-T.} \bsnm{{Chen}}},
\oauthor{\binits{Y.} \bsnm{{Chen}}},
\oauthor{\binits{X.} \bsnm{{Cheng}}},
\oauthor{\binits{I.} \bsnm{{Cho}}},
\oauthor{\binits{P.} \bsnm{{Christian}}},
\oauthor{\binits{N.S.} \bsnm{{Conroy}}},
\oauthor{\binits{J.E.} \bsnm{{Conway}}},
\oauthor{\binits{J.M.} \bsnm{{Cordes}}},
\oauthor{\binits{T.M.} \bsnm{{Crawford}}},
\oauthor{\binits{G.B.} \bsnm{{Crew}}},
\oauthor{\binits{A.} \bsnm{{Cruz-Osorio}}},
\oauthor{\binits{Y.} \bsnm{{Cui}}},
\oauthor{\binits{J.} \bsnm{{Davelaar}}},
\oauthor{\binits{M.} \bsnm{{De Laurentis}}},
\oauthor{\binits{R.} \bsnm{{Deane}}},
\oauthor{\binits{J.} \bsnm{{Dempsey}}},
\oauthor{\binits{G.} \bsnm{{Desvignes}}},
\oauthor{\binits{J.} \bsnm{{Dexter}}},
\oauthor{\binits{V.} \bsnm{{Dhruv}}},
\oauthor{\binits{S.S.} \bsnm{{Doeleman}}},
\oauthor{\binits{S.} \bsnm{{Dougal}}},
\oauthor{\binits{S.A.} \bsnm{{Dzib}}},
\oauthor{\binits{R.P.} \bsnm{{Eatough}}},
\oauthor{\binits{R.} \bsnm{{Emami}}},
\oauthor{\binits{H.} \bsnm{{Falcke}}},
\oauthor{\binits{J.} \bsnm{{Farah}}},
\oauthor{\binits{V.L.} \bsnm{{Fish}}},
\oauthor{\binits{E.} \bsnm{{Fomalont}}},
\oauthor{\binits{H.A.} \bsnm{{Ford}}},
\oauthor{\binits{R.} \bsnm{{Fraga-Encinas}}},
\oauthor{\binits{W.T.} \bsnm{{Freeman}}},
\oauthor{\binits{P.} \bsnm{{Friberg}}},
\oauthor{\binits{C.M.} \bsnm{{Fromm}}},
\oauthor{\binits{A.} \bsnm{{Fuentes}}},
\oauthor{\binits{P.} \bsnm{{Galison}}},
\oauthor{\binits{C.F.} \bsnm{{Gammie}}},
\oauthor{\binits{R.} \bsnm{{Garc{\'\i}a}}},
\oauthor{\binits{O.} \bsnm{{Gentaz}}},
\oauthor{\binits{B.} \bsnm{{Georgiev}}},
\oauthor{\binits{C.} \bsnm{{Goddi}}},
\oauthor{\binits{R.} \bsnm{{Gold}}},
\oauthor{\binits{A.I.} \bsnm{{G{\'o}mez-Ruiz}}},
\oauthor{\binits{J.L.} \bsnm{{G{\'o}mez}}},
\oauthor{\binits{M.} \bsnm{{Gu}}},
\oauthor{\binits{M.} \bsnm{{Gurwell}}},
\oauthor{\binits{K.} \bsnm{{Hada}}},
\oauthor{\binits{D.} \bsnm{{Haggard}}},
\oauthor{\binits{K.} \bsnm{{Haworth}}},
\oauthor{\binits{M.H.} \bsnm{{Hecht}}},
\oauthor{\binits{R.} \bsnm{{Hesper}}},
\oauthor{\binits{D.} \bsnm{{Heumann}}},
\oauthor{\binits{L.C.} \bsnm{{Ho}}},
\oauthor{\binits{P.} \bsnm{{Ho}}},
\oauthor{\binits{M.} \bsnm{{Honma}}},
\oauthor{\binits{C.-W.L.} \bsnm{{Huang}}},
\oauthor{\binits{L.} \bsnm{{Huang}}},
\oauthor{\binits{D.H.} \bsnm{{Hughes}}},
\oauthor{\binits{S.} \bsnm{{Ikeda}}},
\oauthor{\binits{C.M.V.} \bsnm{{Impellizzeri}}},
\oauthor{\binits{M.} \bsnm{{Inoue}}},
\oauthor{\binits{S.} \bsnm{{Issaoun}}},
\oauthor{\binits{D.J.} \bsnm{{James}}},
\oauthor{\binits{B.T.} \bsnm{{Jannuzi}}},
\oauthor{\binits{M.} \bsnm{{Janssen}}},
\oauthor{\binits{B.} \bsnm{{Jeter}}},
\oauthor{\binits{W.} \bsnm{{Jiang}}},
\oauthor{\binits{A.} \bsnm{{Jim{\'e}nez-Rosales}}},
\oauthor{\binits{M.D.} \bsnm{{Johnson}}},
\oauthor{\binits{S.} \bsnm{{Jorstad}}},
\oauthor{\binits{A.V.} \bsnm{{Joshi}}},
\oauthor{\binits{T.} \bsnm{{Jung}}},
\oauthor{\binits{M.} \bsnm{{Karami}}},
\oauthor{\binits{R.} \bsnm{{Karuppusamy}}},
\oauthor{\binits{T.} \bsnm{{Kawashima}}},
\oauthor{\binits{G.K.} \bsnm{{Keating}}},
\oauthor{\binits{M.} \bsnm{{Kettenis}}},
\oauthor{\binits{D.-J.} \bsnm{{Kim}}},
\oauthor{\binits{J.-Y.} \bsnm{{Kim}}},
\oauthor{\binits{J.} \bsnm{{Kim}}},
\oauthor{\binits{J.} \bsnm{{Kim}}},
\oauthor{\binits{M.} \bsnm{{Kino}}},
\oauthor{\binits{J.Y.} \bsnm{{Koay}}},
\oauthor{\binits{P.} \bsnm{{Kocherlakota}}},
\oauthor{\binits{Y.} \bsnm{{Kofuji}}},
\oauthor{\binits{P.M.} \bsnm{{Koch}}},
\oauthor{\binits{S.} \bsnm{{Koyama}}},
\oauthor{\binits{C.} \bsnm{{Kramer}}},
\oauthor{\binits{M.} \bsnm{{Kramer}}},
\oauthor{\binits{T.P.} \bsnm{{Krichbaum}}},
\oauthor{\binits{C.-Y.} \bsnm{{Kuo}}},
\oauthor{\binits{N.} \bsnm{{La Bella}}},
\oauthor{\binits{T.R.} \bsnm{{Lauer}}},
\oauthor{\binits{D.} \bsnm{{Lee}}},
\oauthor{\binits{S.-S.} \bsnm{{Lee}}},
\oauthor{\binits{P.K.} \bsnm{{Leung}}},
\oauthor{\binits{A.} \bsnm{{Levis}}},
\oauthor{\binits{Z.} \bsnm{{Li}}},
\oauthor{\binits{R.} \bsnm{{Lico}}},
\oauthor{\binits{G.} \bsnm{{Lindahl}}},
\oauthor{\binits{M.} \bsnm{{Lindqvist}}},
\oauthor{\binits{M.} \bsnm{{Lisakov}}},
\oauthor{\binits{J.} \bsnm{{Liu}}},
\oauthor{\binits{K.} \bsnm{{Liu}}},
\oauthor{\binits{E.} \bsnm{{Liuzzo}}},
\oauthor{\binits{W.-P.} \bsnm{{Lo}}},
\oauthor{\binits{A.P.} \bsnm{{Lobanov}}},
\oauthor{\binits{L.} \bsnm{{Loinard}}},
\oauthor{\binits{C.J.} \bsnm{{Lonsdale}}},
\oauthor{\binits{R.-S.} \bsnm{{Lu}}},
\oauthor{\binits{J.} \bsnm{{Mao}}},
\oauthor{\binits{N.} \bsnm{{Marchili}}},
\oauthor{\binits{S.} \bsnm{{Markoff}}},
\oauthor{\binits{D.P.} \bsnm{{Marrone}}},
\oauthor{\binits{A.P.} \bsnm{{Marscher}}},
\oauthor{\binits{I.} \bsnm{{Mart{\'\i}-Vidal}}},
\oauthor{\binits{S.} \bsnm{{Matsushita}}},
\oauthor{\binits{L.D.} \bsnm{{Matthews}}},
\oauthor{\binits{L.} \bsnm{{Medeiros}}},
\oauthor{\binits{K.M.} \bsnm{{Menten}}},
\oauthor{\binits{D.} \bsnm{{Michalik}}},
\oauthor{\binits{I.} \bsnm{{Mizuno}}},
\oauthor{\binits{Y.} \bsnm{{Mizuno}}},
\oauthor{\binits{J.M.} \bsnm{{Moran}}},
\oauthor{\binits{K.} \bsnm{{Moriyama}}},
\oauthor{\binits{M.} \bsnm{{Moscibrodzka}}},
\oauthor{\binits{C.} \bsnm{{M{\"u}ller}}},
\oauthor{\binits{A.} \bsnm{{Mus}}},
\oauthor{\binits{G.} \bsnm{{Musoke}}},
\oauthor{\binits{I.} \bsnm{{Myserlis}}},
\oauthor{\binits{A.} \bsnm{{Nadolski}}},
\oauthor{\binits{H.} \bsnm{{Nagai}}},
\oauthor{\binits{N.M.} \bsnm{{Nagar}}},
\oauthor{\binits{M.} \bsnm{{Nakamura}}},
\oauthor{\binits{R.} \bsnm{{Narayan}}},
\oauthor{\binits{G.} \bsnm{{Narayanan}}},
\oauthor{\binits{I.} \bsnm{{Natarajan}}},
\oauthor{\binits{A.} \bsnm{{Nathanail}}},
\oauthor{\binits{S.N.} \bsnm{{Fuentes}}},
\oauthor{\binits{J.} \bsnm{{Neilsen}}},
\oauthor{\binits{R.} \bsnm{{Neri}}},
\oauthor{\binits{C.} \bsnm{{Ni}}},
\oauthor{\binits{A.} \bsnm{{Noutsos}}},
\oauthor{\binits{M.A.} \bsnm{{Nowak}}},
\oauthor{\binits{J.} \bsnm{{Oh}}},
\oauthor{\binits{H.} \bsnm{{Okino}}},
\oauthor{\binits{H.} \bsnm{{Olivares}}},
\oauthor{\binits{G.N.} \bsnm{{Ortiz-Le{\'o}n}}},
\oauthor{\binits{T.} \bsnm{{Oyama}}},
\oauthor{\binits{F.} \bsnm{{{\"O}zel}}},
\oauthor{\binits{D.C.M.} \bsnm{{Palumbo}}},
\oauthor{\binits{G.F.} \bsnm{{Paraschos}}},
\oauthor{\binits{J.} \bsnm{{Park}}},
\oauthor{\binits{H.} \bsnm{{Parsons}}},
\oauthor{\binits{N.} \bsnm{{Patel}}},
\oauthor{\binits{U.-L.} \bsnm{{Pen}}},
\oauthor{\binits{D.W.} \bsnm{{Pesce}}},
\oauthor{\binits{V.} \bsnm{{Pi{\'e}tu}}},
\oauthor{\binits{R.} \bsnm{{Plambeck}}},
\oauthor{\binits{A.} \bsnm{{PopStefanija}}},
\oauthor{\binits{O.} \bsnm{{Porth}}},
\oauthor{\binits{F.M.} \bsnm{{P{\"o}tzl}}},
\oauthor{\binits{B.} \bsnm{{Prather}}},
\oauthor{\binits{J.A.} \bsnm{{Preciado-L{\'o}pez}}},
\oauthor{\binits{D.} \bsnm{{Psaltis}}},
\oauthor{\binits{H.-Y.} \bsnm{{Pu}}},
\oauthor{\binits{V.} \bsnm{{Ramakrishnan}}},
\oauthor{\binits{R.} \bsnm{{Rao}}},
\oauthor{\binits{M.G.} \bsnm{{Rawlings}}},
\oauthor{\binits{A.W.} \bsnm{{Raymond}}},
\oauthor{\binits{L.} \bsnm{{Rezzolla}}},
\oauthor{\binits{A.} \bsnm{{Ricarte}}},
\oauthor{\binits{B.} \bsnm{{Ripperda}}},
\oauthor{\binits{F.} \bsnm{{Roelofs}}},
\oauthor{\binits{A.} \bsnm{{Rogers}}},
\oauthor{\binits{E.} \bsnm{{Ros}}},
\oauthor{\binits{C.} \bsnm{{Romero-Ca{\~n}izales}}},
\oauthor{\binits{A.} \bsnm{{Roshanineshat}}},
\oauthor{\binits{H.} \bsnm{{Rottmann}}},
\oauthor{\binits{A.L.} \bsnm{{Roy}}},
\oauthor{\binits{I.} \bsnm{{Ruiz}}},
\oauthor{\binits{C.} \bsnm{{Ruszczyk}}},
\oauthor{\binits{K.L.J.} \bsnm{{Rygl}}},
\oauthor{\binits{S.} \bsnm{{S{\'a}nchez}}},
\oauthor{\binits{D.} \bsnm{{S{\'a}nchez-Arg{\"u}elles}}},
\oauthor{\binits{M.} \bsnm{{S{\'a}nchez-Portal}}},
\oauthor{\binits{M.} \bsnm{{Sasada}}},
\oauthor{\binits{K.} \bsnm{{Satapathy}}},
\oauthor{\binits{T.} \bsnm{{Savolainen}}},
\oauthor{\binits{F.P.} \bsnm{{Schloerb}}},
\oauthor{\binits{J.} \bsnm{{Schonfeld}}},
\oauthor{\binits{K.-F.} \bsnm{{Schuster}}},
\oauthor{\binits{L.} \bsnm{{Shao}}},
\oauthor{\binits{Z.} \bsnm{{Shen}}},
\oauthor{\binits{D.} \bsnm{{Small}}},
\oauthor{\binits{B.W.} \bsnm{{Sohn}}},
\oauthor{\binits{J.} \bsnm{{SooHoo}}},
\oauthor{\binits{K.} \bsnm{{Souccar}}},
\oauthor{\binits{H.} \bsnm{{Sun}}},
\oauthor{\binits{F.} \bsnm{{Tazaki}}},
\oauthor{\binits{A.J.} \bsnm{{Tetarenko}}},
\oauthor{\binits{P.} \bsnm{{Tiede}}},
\oauthor{\binits{R.P.J.} \bsnm{{Tilanus}}},
\oauthor{\binits{M.} \bsnm{{Titus}}},
\oauthor{\binits{P.} \bsnm{{Torne}}},
\oauthor{\binits{E.} \bsnm{{Traianou}}},
\oauthor{\binits{T.} \bsnm{{Trent}}},
\oauthor{\binits{S.} \bsnm{{Trippe}}},
\oauthor{\binits{M.} \bsnm{{Turk}}},
\oauthor{\binits{I.} \bsnm{{van Bemmel}}},
\oauthor{\binits{H.J.} \bsnm{{van Langevelde}}},
\oauthor{\binits{D.R.} \bsnm{{van Rossum}}},
\oauthor{\binits{J.} \bsnm{{Vos}}},
\oauthor{\binits{J.} \bsnm{{Wagner}}},
\oauthor{\binits{D.} \bsnm{{Ward-Thompson}}},
\oauthor{\binits{J.} \bsnm{{Wardle}}},
\oauthor{\binits{J.} \bsnm{{Weintroub}}},
\oauthor{\binits{N.} \bsnm{{Wex}}},
\oauthor{\binits{R.} \bsnm{{Wharton}}},
\oauthor{\binits{M.} \bsnm{{Wielgus}}},
\oauthor{\binits{K.} \bsnm{{Wiik}}},
\oauthor{\binits{G.} \bsnm{{Witzel}}},
\oauthor{\binits{M.F.} \bsnm{{Wondrak}}},
\oauthor{\binits{G.N.} \bsnm{{Wong}}},
\oauthor{\binits{Q.} \bsnm{{Wu}}},
\oauthor{\binits{P.} \bsnm{{Yamaguchi}}},
\oauthor{\binits{D.} \bsnm{{Yoon}}},
\oauthor{\binits{A.} \bsnm{{Young}}},
\oauthor{\binits{K.} \bsnm{{Young}}},
\oauthor{\binits{Z.} \bsnm{{Younsi}}},
\oauthor{\binits{F.} \bsnm{{Yuan}}},
\oauthor{\binits{Y.-F.} \bsnm{{Yuan}}},
\oauthor{\binits{J.A.} \bsnm{{Zensus}}},
\oauthor{\binits{S.} \bsnm{{Zhang}}},
\oauthor{\binits{G.-Y.} \bsnm{{Zhao}}},
\oauthor{\binits{S.-S.} \bsnm{{Zhao}}},
First sagittarius a event horizon telescope results. iii. imaging of the
  galactic center supermassive black hole
\textbf{930},
14
(2022).
doi:\doiurl{10.3847/2041-8213/ac6429}
\end{botherref}
\endbibitem

\bibitem[\protect\citeauthoryear{{Fabian} et~al.}{1989}]{bib5:Fabian1989}
\begin{barticle}
\bauthor{\binits{A.C.} \bsnm{{Fabian}}},
\bauthor{\binits{M.J.} \bsnm{{Rees}}},
\bauthor{\binits{L.} \bsnm{{Stella}}},
\bauthor{\binits{N.E.} \bsnm{{White}}},
\batitle{X-ray fluorescence from the inner disc in cygnus x-1.}
\bjtitle{\mnras}
\bvolume{238},
\bfpage{729}--\blpage{736}
(\byear{1989}).
doi:\doiurl{10.1093/mnras/238.3.729}
\end{barticle}
\endbibitem

\bibitem[\protect\citeauthoryear{{Fabian} et~al.}{2009}]{bib5:Fabian2009}
\begin{barticle}
\bauthor{\binits{A.C.} \bsnm{{Fabian}}},
\bauthor{\binits{A.} \bsnm{{Zoghbi}}},
\bauthor{\binits{R.R.} \bsnm{{Ross}}},
\bauthor{\binits{P.} \bsnm{{Uttley}}},
\bauthor{\binits{L.C.} \bsnm{{Gallo}}},
\bauthor{\binits{W.N.} \bsnm{{Brandt}}},
\bauthor{\binits{A.J.} \bsnm{{Blustin}}},
\bauthor{\binits{T.} \bsnm{{Boller}}},
\bauthor{\binits{M.D.} \bsnm{{Caballero-Garcia}}},
\bauthor{\binits{J.} \bsnm{{Larsson}}},
\bauthor{\binits{J.M.} \bsnm{{Miller}}},
\bauthor{\binits{G.} \bsnm{{Miniutti}}},
\bauthor{\binits{G.} \bsnm{{Ponti}}},
\bauthor{\binits{R.C.} \bsnm{{Reis}}},
\bauthor{\binits{C.S.} \bsnm{{Reynolds}}},
\bauthor{\binits{Y.} \bsnm{{Tanaka}}},
\bauthor{\binits{A.J.} \bsnm{{Young}}},
\batitle{Broad line emission from iron k- and l-shell transitions in the active
  galaxy 1h0707-495}.
\bjtitle{\nat}
\bvolume{459}(\bissue{7246}),
\bfpage{540}--\blpage{542}
(\byear{2009}).
doi:\doiurl{10.1038/nature08007}
\end{barticle}
\endbibitem

\bibitem[\protect\citeauthoryear{{Fabian} et~al.}{2013}]{bib5:Fabian2013}
\begin{barticle}
\bauthor{\binits{A.C.} \bsnm{{Fabian}}},
\bauthor{\binits{E.} \bsnm{{Kara}}},
\bauthor{\binits{D.J.} \bsnm{{Walton}}},
\bauthor{\binits{D.R.} \bsnm{{Wilkins}}},
\bauthor{\binits{R.R.} \bsnm{{Ross}}},
\bauthor{\binits{K.} \bsnm{{Lozanov}}},
\bauthor{\binits{P.} \bsnm{{Uttley}}},
\bauthor{\binits{L.C.} \bsnm{{Gallo}}},
\bauthor{\binits{A.} \bsnm{{Zoghbi}}},
\bauthor{\binits{G.} \bsnm{{Miniutti}}},
\bauthor{\binits{T.} \bsnm{{Boller}}},
\bauthor{\binits{W.N.} \bsnm{{Brandt}}},
\bauthor{\binits{E.M.} \bsnm{{Cackett}}},
\bauthor{\binits{C.-Y.} \bsnm{{Chiang}}},
\bauthor{\binits{T.} \bsnm{{Dwelly}}},
\bauthor{\binits{J.} \bsnm{{Malzac}}},
\bauthor{\binits{J.M.} \bsnm{{Miller}}},
\bauthor{\binits{E.} \bsnm{{Nardini}}},
\bauthor{\binits{G.} \bsnm{{Ponti}}},
\bauthor{\binits{R.C.} \bsnm{{Reis}}},
\bauthor{\binits{C.S.} \bsnm{{Reynolds}}},
\bauthor{\binits{J.F.} \bsnm{{Steiner}}},
\bauthor{\binits{Y.} \bsnm{{Tanaka}}},
\bauthor{\binits{A.J.} \bsnm{{Young}}},
\batitle{Long xmm observation of the narrow-line seyfert 1 galaxy iras
  13224-3809: rapid variability, high spin and a soft lag}.
\bjtitle{\mnras}
\bvolume{429}(\bissue{4}),
\bfpage{2917}--\blpage{2923}
(\byear{2013}).
doi:\doiurl{10.1093/mnras/sts504}
\end{barticle}
\endbibitem

\bibitem[\protect\citeauthoryear{Fadely and Keeton}{2011}]{bib5:Fadely2011}
\begin{botherref}
\oauthor{\binits{R.} \bsnm{Fadely}},
\oauthor{\binits{C.R.} \bsnm{Keeton}},
Near-infrared k and l' flux ratios in six lensed quasars.
Astronomical Journal
\textbf{141}
(2011).
doi:\doiurl{10.1088/0004-6256/141/3/101}
\end{botherref}
\endbibitem

\bibitem[\protect\citeauthoryear{Falco et~al.}{1999}]{bib5:Falco1999}
\begin{barticle}
\bauthor{\binits{E.E.} \bsnm{Falco}},
\bauthor{\binits{C.D.} \bsnm{Impey}},
\bauthor{\binits{C.S.} \bsnm{Kochanek}},
\bauthor{\binits{J.} \bsnm{Leha}},
\bauthor{\binits{B.A.} \bsnm{Mcleod}},
\bauthor{\binits{H.-W.} \bsnm{Rix}},
\bauthor{\binits{C.R.} \bsnm{Keeton}},
\bauthor{\binits{J.A.} \bsnm{Mun}},
\bauthor{\binits{C.Y.} \bsnm{Peng}},
\batitle{Dust and extinction curves in galaxies with z [ 0 : The interstellar
  medium of gravitational lens galaxies1}.
\bjtitle{The Astrophysical Journal}
\bvolume{523},
\bfpage{617}--\blpage{632}
(\byear{1999})
\end{barticle}
\endbibitem

\bibitem[\protect\citeauthoryear{Falco et~al.}{1985}]{bib5:Falco1985}
\begin{barticle}
\bauthor{\binits{E.E.} \bsnm{Falco}},
\bauthor{\binits{V.M.} \bsnm{Gorenstein}},
\bauthor{\binits{I.I.} \bsnm{Shapiro}},
\batitle{On model-dependent bounds on h0 from gravitational images: application
  to q0957$+$561a,b}.
\bjtitle{The Astrophysical Journal}
\bvolume{289},
\bfpage{1}
(\byear{1985})
\end{barticle}
\endbibitem

\bibitem[\protect\citeauthoryear{Ferland et~al.}{2013}]{bib5:Ferland2013}
\begin{barticle}
\bauthor{\binits{G.J.} \bsnm{Ferland}},
\bauthor{\binits{R.L.} \bsnm{Porter}},
\bauthor{\binits{P.A.M.} \bparticle{van} \bsnm{Hoof}},
\bauthor{\binits{R.J.R.} \bsnm{Williams}},
\bauthor{\binits{N.P.} \bsnm{Abel}},
\bauthor{\binits{M.L.} \bsnm{Lykins}},
\bauthor{\binits{G.} \bsnm{Shaw}},
\bauthor{\binits{W.J.} \bsnm{Henney}},
\bauthor{\binits{P.C.} \bsnm{Stancil}},
\batitle{The 2013 release of cloudy}.
\bjtitle{Revista Mexicana de Astronom{\'i}a y Astrof{\'i}sica}
\bvolume{49},
\bfpage{137}--\blpage{163}
(\byear{2013}).
doi:\doiurl{10.48550/arXiv.1302.4485}
\end{barticle}
\endbibitem

\bibitem[\protect\citeauthoryear{Ferland et~al.}{2009}]{bib5:Ferland2009}
\begin{botherref}
\oauthor{\binits{G.J.} \bsnm{Ferland}},
\oauthor{\binits{C.} \bsnm{Hu}},
\oauthor{\binits{J.M.} \bsnm{Wang}},
\oauthor{\binits{J.A.} \bsnm{Baldwin}},
\oauthor{\binits{R.L.} \bsnm{Porter}},
\oauthor{\binits{P.A.M.V.} \bsnm{Hoof}},
\oauthor{\binits{R.J.R.} \bsnm{Williams}},
Implications of infalling fe ii-emitting clouds in active galactic nuclei:
  Anisotropic properties.
Astrophysical Journal
\textbf{707}
(2009).
doi:\doiurl{10.1088/0004-637X/707/1/L82}
\end{botherref}
\endbibitem

\bibitem[\protect\citeauthoryear{{Fian} et~al.}{2016}]{bib5:Fian2016}
\begin{barticle}
\bauthor{\binits{C.} \bsnm{{Fian}}},
\bauthor{\binits{E.} \bsnm{{Mediavilla}}},
\bauthor{\binits{A.} \bsnm{{Hanslmeier}}},
\bauthor{\binits{A.} \bsnm{{Oscoz}}},
\bauthor{\binits{M.} \bsnm{{Serra-Ricart}}},
\bauthor{\binits{J.A.} \bsnm{{Mu{\~n}oz}}},
\bauthor{\binits{J.} \bsnm{{Jim{\'e}nez-Vicente}}},
\batitle{Size of the accretion disk in the graviationally lensed quasar sdss
  j1004$+$4112 from the statistics of microlensing magnifications}.
\bjtitle{\apj}
\bvolume{830}(\bissue{2}),
\bfpage{149}
(\byear{2016}).
doi:\doiurl{10.3847/0004-637X/830/2/149}.
\burl{http://dx.doi.org/10.3847/0004-637X/830/2/149}
\end{barticle}
\endbibitem

\bibitem[\protect\citeauthoryear{Fian et~al.}{2018}]{bib5:Fian2018}
\begin{barticle}
\bauthor{\binits{C.} \bsnm{Fian}},
\bauthor{\binits{E.} \bsnm{Mediavilla}},
\bauthor{\binits{J.} \bsnm{Jim{\'e}nez-Vicente}},
\bauthor{\binits{J.A.} \bsnm{Mu{\~n}oz}},
\bauthor{\binits{A.} \bsnm{Hanslmeier}},
\batitle{Estimate of the accretion disk size in the gravitationally lensed
  quasar he 0435$-$1223 using microlensing magnification statistics}.
\bjtitle{The Astrophysical Journal}
\bvolume{869},
\bfpage{132}
(\byear{2018}).
doi:\doiurl{10.3847/1538-4357/aaeed5}.
\burl{http://dx.doi.org/10.3847/1538-4357/aaeed5}
\end{barticle}
\endbibitem

\bibitem[\protect\citeauthoryear{Fian et~al.}{2021a}]{bib5:Fian2021b}
\begin{botherref}
\oauthor{\binits{C.} \bsnm{Fian}},
\oauthor{\binits{E.} \bsnm{Mediavilla}},
\oauthor{\binits{V.} \bsnm{Motta}},
\oauthor{\binits{J.} \bsnm{Jim{\'e}nez-Vicente}},
\oauthor{\binits{J.A.} \bsnm{Mu{\~n}oz}},
\oauthor{\binits{D.} \bsnm{Chelouche}},
\oauthor{\binits{A.} \bsnm{Hanslmeier}},
Microlensing of the broad emission lines in 27 gravitationally lensed quasars:
  Broad line region structure and kinematics.
Astronomy and Astrophysics
\textbf{653}
(2021a).
doi:\doiurl{10.1051/0004-6361/202039829}
\end{botherref}
\endbibitem

\bibitem[\protect\citeauthoryear{Fian et~al.}{2021b}]{bib5:Fian2021a}
\begin{botherref}
\oauthor{\binits{C.} \bsnm{Fian}},
\oauthor{\binits{E.} \bsnm{Mediavilla}},
\oauthor{\binits{J.} \bsnm{Jim{\'e}nez-Vicente}},
\oauthor{\binits{V.} \bsnm{Motta}},
\oauthor{\binits{J.A.} \bsnm{Mu{\~n}oz}},
\oauthor{\binits{D.} \bsnm{Chelouche}},
\oauthor{\binits{P.} \bsnm{Gom{\'e}z-Alvarez}},
\oauthor{\binits{K.} \bsnm{Rojas}},
\oauthor{\binits{A.} \bsnm{Hanslmeier}},
Revealing the structure of the lensed quasar q 0957$+$561: I. accretion disk
  size.
Astronomy and Astrophysics
\textbf{654}
(2021b).
doi:\doiurl{10.1051/0004-6361/202039854}
\end{botherref}
\endbibitem

\bibitem[\protect\citeauthoryear{Floyd et~al.}{2009}]{bib5:Floyd2009}
\begin{barticle}
\bauthor{\binits{D.J.E.} \bsnm{Floyd}},
\bauthor{\binits{N.F.} \bsnm{Bate}},
\bauthor{\binits{R.L.} \bsnm{Webster}},
\batitle{The accretion disc in the quasar sdss j0924$+$0219}.
\bjtitle{Monthly Notices of the Royal Astronomical Society}
\bvolume{398},
\bfpage{233}--\blpage{239}
(\byear{2009}).
doi:\doiurl{10.1111/j.1365-2966.2009.15045.x}.
\burl{http://doi.wiley.com/10.1111/j.1365-2966.2009.15045.x}
\end{barticle}
\endbibitem

\bibitem[\protect\citeauthoryear{Fluke and Webster}{1999}]{bib5:Fluke1999}
\begin{barticle}
\bauthor{\binits{C.J.} \bsnm{Fluke}},
\bauthor{\binits{R.L.} \bsnm{Webster}},
\batitle{Investigating the geometry of quasars with microlensing}.
\bjtitle{Monthly Notices of the Royal Astronomical Society}
\bvolume{302},
\bfpage{68}--\blpage{74}
(\byear{1999}).
doi:\doiurl{10.1046/j.1365-8711.1999.02109.x}
\end{barticle}
\endbibitem

\bibitem[\protect\citeauthoryear{Fluke et~al.}{2011}]{bib5:Fluke2011}
\begin{barticle}
\bauthor{\binits{C.J.} \bsnm{Fluke}},
\bauthor{\binits{D.G.} \bsnm{Barnes}},
\bauthor{\binits{B.R.} \bsnm{Barsdell}},
\bauthor{\binits{A.H.} \bsnm{Hassan}},
\batitle{Astrophysical supercomputing with gpus : Critical decisions for early
  adopters}.
\bjtitle{Publications of the Astron. Soc. of Australia}
\bvolume{28},
\bfpage{15}--\blpage{27}
(\byear{2011})
\end{barticle}
\endbibitem

\bibitem[\protect\citeauthoryear{Foxley-Marrable
  et~al.}{2018}]{bib5:FoxleyMarrable2018}
\begin{barticle}
\bauthor{\binits{M.} \bsnm{Foxley-Marrable}},
\bauthor{\binits{T.E.} \bsnm{Collett}},
\bauthor{\binits{G.} \bsnm{Vernardos}},
\bauthor{\binits{D.A.} \bsnm{Goldstein}},
\bauthor{\binits{D.} \bsnm{Bacon}},
\batitle{The impact of microlensing on the standardisation of strongly lensed
  type ia supernovae}.
\bjtitle{Monthly Notices of the Royal Astronomical Society}
\bvolume{478},
\bfpage{5081}--\blpage{5090}
(\byear{2018}).
\burl{http://arxiv.org/abs/1802.07738}
\end{barticle}
\endbibitem

\bibitem[\protect\citeauthoryear{{Galeev} et~al.}{1979}]{bib5:Galeev1979}
\begin{barticle}
\bauthor{\binits{A.A.} \bsnm{{Galeev}}},
\bauthor{\binits{R.} \bsnm{{Rosner}}},
\bauthor{\binits{G.S.} \bsnm{{Vaiana}}},
\batitle{Structured coronae of accretion disks.}
\bjtitle{\apj}
\bvolume{229},
\bfpage{318}--\blpage{326}
(\byear{1979}).
doi:\doiurl{10.1086/156957}
\end{barticle}
\endbibitem

\bibitem[\protect\citeauthoryear{{Gallagher} et~al.}{2006}]{bib5:Gallagher2006}
\begin{barticle}
\bauthor{\binits{S.C.} \bsnm{{Gallagher}}},
\bauthor{\binits{W.N.} \bsnm{{Brandt}}},
\bauthor{\binits{G.} \bsnm{{Chartas}}},
\bauthor{\binits{R.} \bsnm{{Priddey}}},
\bauthor{\binits{G.P.} \bsnm{{Garmire}}},
\bauthor{\binits{R.M.} \bsnm{{Sambruna}}},
\batitle{An exploratory chandra survey of a well-defined sample of 35 large
  bright quasar survey broad absorption line quasars}.
\bjtitle{\apj}
\bvolume{644}(\bissue{2}),
\bfpage{709}--\blpage{724}
(\byear{2006}).
doi:\doiurl{10.1086/503762}
\end{barticle}
\endbibitem

\bibitem[\protect\citeauthoryear{{Gallo} et~al.}{2015}]{bib5:Gallo2015}
\begin{barticle}
\bauthor{\binits{L.C.} \bsnm{{Gallo}}},
\bauthor{\binits{D.R.} \bsnm{{Wilkins}}},
\bauthor{\binits{K.} \bsnm{{Bonson}}},
\bauthor{\binits{C.-Y.} \bsnm{{Chiang}}},
\bauthor{\binits{D.} \bsnm{{Grupe}}},
\bauthor{\binits{M.L.} \bsnm{{Parker}}},
\bauthor{\binits{A.} \bsnm{{Zoghbi}}},
\bauthor{\binits{A.C.} \bsnm{{Fabian}}},
\bauthor{\binits{S.} \bsnm{{Komossa}}},
\bauthor{\binits{A.L.} \bsnm{{Longinotti}}},
\batitle{Suzaku observations of mrk 335: confronting partial covering and
  relativistic reflection}.
\bjtitle{\mnras}
\bvolume{446}(\bissue{1}),
\bfpage{633}--\blpage{650}
(\byear{2015}).
doi:\doiurl{10.1093/mnras/stu2108}
\end{barticle}
\endbibitem

\bibitem[\protect\citeauthoryear{Gardner and Done}{2017}]{bib5:Gardner2017}
\begin{barticle}
\bauthor{\binits{E.} \bsnm{Gardner}},
\bauthor{\binits{C.} \bsnm{Done}},
\batitle{The origin of the uv/optical lags in ngc 5548}.
\bjtitle{Monthly Notices of the Royal Astronomical Society}
\bvolume{470},
\bfpage{3591}--\blpage{3605}
(\byear{2017}).
doi:\doiurl{10.1093/mnras/stx946}
\end{barticle}
\endbibitem

\bibitem[\protect\citeauthoryear{Garsden and Lewis}{2010}]{bib5:Garsden2010}
\begin{botherref}
\oauthor{\binits{H.} \bsnm{Garsden}},
\oauthor{\binits{G.F.} \bsnm{Lewis}},
Gravitational microlensing: A parallel, large-data implementation 1.
New Astronomy,
181
(2010)
\end{botherref}
\endbibitem

\bibitem[\protect\citeauthoryear{{Garsden} et~al.}{2011}]{bib5:Garsden2011}
\begin{barticle}
\bauthor{\binits{H.} \bsnm{{Garsden}}},
\bauthor{\binits{N.F.} \bsnm{{Bate}}},
\bauthor{\binits{G.F.} \bsnm{{Lewis}}},
\batitle{Gravitational microlensing of a reverberating quasar broad-line region
  - i. method and qualitative results}.
\bjtitle{\mnras}
\bvolume{418}(\bissue{2}),
\bfpage{1012}--\blpage{1027}
(\byear{2011}).
doi:\doiurl{10.1111/j.1365-2966.2011.19552.x}
\end{barticle}
\endbibitem

\bibitem[\protect\citeauthoryear{Gaudi and Petters}{2002a}]{bib5:Gaudi2002a}
\begin{barticle}
\bauthor{\binits{B.S.} \bsnm{Gaudi}},
\bauthor{\binits{A.O.} \bsnm{Petters}},
\batitle{Gravitational microlensing near caustics. i. folds}.
\bjtitle{The Astrophysical Journal}
\bvolume{574},
\bfpage{970}--\blpage{984}
(\byear{2002}a).
doi:\doiurl{10.1086/341063}.
\burl{http://iopscience.iop.org/0004-637X/574/2/970/fulltext/}
\end{barticle}
\endbibitem

\bibitem[\protect\citeauthoryear{Gaudi and Petters}{2002b}]{bib5:Gaudi2002b}
\begin{barticle}
\bauthor{\binits{B.S.} \bsnm{Gaudi}},
\bauthor{\binits{A.O.} \bsnm{Petters}},
\batitle{Gravitational microlensing near caustics. ii. cusps}.
\bjtitle{The Astrophysical Journal}
\bvolume{580},
\bfpage{468}
(\byear{2002}b)
\end{barticle}
\endbibitem

\bibitem[\protect\citeauthoryear{{Gil{-}Merino}
  et~al.}{2018}]{bib5:GilMerino2018}
\begin{botherref}
\oauthor{\binits{R.} \bsnm{{Gil{-}Merino}}},
\oauthor{\binits{L.J.} \bsnm{Goicoechea}},
\oauthor{\binits{V.N.} \bsnm{Shalyapin}},
\oauthor{\binits{A.} \bsnm{Oscoz}},
New database for a sample of optically bright lensed quasars in the northern
  hemisphere.
Astronomy and Astrophysics
\textbf{616}
(2018).
doi:\doiurl{10.1051/0004-6361/201832737}
\end{botherref}
\endbibitem

\bibitem[\protect\citeauthoryear{Goicoechea et~al.}{2020}]{bib5:Goicoechea2020}
\begin{botherref}
\oauthor{\binits{L.J.} \bsnm{Goicoechea}},
\oauthor{\binits{B.P.} \bsnm{Artamonov}},
\oauthor{\binits{V.N.} \bsnm{Shalyapin}},
\oauthor{\binits{A.V.} \bsnm{Sergeyev}},
\oauthor{\binits{O.A.} \bsnm{Burkhonov}},
\oauthor{\binits{T.A.} \bsnm{Akhunov}},
\oauthor{\binits{I.M.} \bsnm{Asfandiyarov}},
\oauthor{\binits{V.V.} \bsnm{Bruevich}},
\oauthor{\binits{S.A.} \bsnm{Ehgamberdiev}},
\oauthor{\binits{E.V.} \bsnm{Shimanovskaya}},
\oauthor{\binits{A.P.} \bsnm{Zheleznyak}},
Liverpool-maidanak monitoring of the einstein cross in 2006-2019.
Astronomy and Astrophysics
\textbf{637}
(2020).
doi:\doiurl{10.1051/0004-6361/202037902}
\end{botherref}
\endbibitem

\bibitem[\protect\citeauthoryear{Gorenstein et~al.}{1988}]{bib5:Gorenstein1988}
\begin{barticle}
\bauthor{\binits{V.M.} \bsnm{Gorenstein}},
\bauthor{\binits{E.E.} \bsnm{Falco}},
\bauthor{\binits{I.I.} \bsnm{Shapiro}},
\batitle{Degeneracies in parameter estimates for models of gravitational lens
  systems}.
\bjtitle{The Astrophysical Journal}
\bvolume{327},
\bfpage{693}
(\byear{1988})
\end{barticle}
\endbibitem

\bibitem[\protect\citeauthoryear{Gott}{1981}]{bib5:Gott1981}
\begin{barticle}
\bauthor{\binits{J.R.} \bsnm{Gott}},
\batitle{Are heavy halos made of low mass stars? a gravitational lens test}.
\bjtitle{The Astrophysical Journal}
\bvolume{243},
\bfpage{140}--\blpage{146}
(\byear{1981})
\end{barticle}
\endbibitem

\bibitem[\protect\citeauthoryear{Gould and
  Miralda-Escude}{1997}]{bib5:Gould1997}
\begin{barticle}
\bauthor{\binits{A.} \bsnm{Gould}},
\bauthor{\binits{J.} \bsnm{Miralda-Escude}},
\batitle{Signatures of accretion disks in quasar microlensing}.
\bjtitle{The Astrophysical Journal}
\bvolume{483},
\bfpage{13}
(\byear{1997})
\end{barticle}
\endbibitem

\bibitem[\protect\citeauthoryear{Graham}{2007}]{bib5:Graham2007}
\begin{barticle}
\bauthor{\binits{A.W.} \bsnm{Graham}},
\batitle{The black hole mass - spheroid luminosity relation}.
\bjtitle{Monthly Notices of the Royal Astronomical Society}
\bvolume{379},
\bfpage{711}--\blpage{722}
(\byear{2007}).
doi:\doiurl{10.1111/j.1365-2966.2007.11950.x}
\end{barticle}
\endbibitem

\bibitem[\protect\citeauthoryear{Graham}{2016}]{bib5:Graham2016}
\begin{botherref}
\oauthor{\binits{A.W.} \bsnm{Graham}},
Galaxy Bulges and Their Massive Black Holes: A Review,
ed. by E. Laurikainen, R. Peletier, D. Gadotti,
2016,
p. 263.
doi:\doiurl{10.1007/978-3-319-19378-6-11}
\end{botherref}
\endbibitem

\bibitem[\protect\citeauthoryear{Granot et~al.}{2003}]{bib5:Granot2003}
\begin{barticle}
\bauthor{\binits{J.} \bsnm{Granot}},
\bauthor{\binits{P.L.} \bsnm{Schechter}},
\bauthor{\binits{J.} \bsnm{Wambsganss}},
\batitle{The mean number of extra microimage pairs for macrolensed quasars}.
\bjtitle{The Astrophysical Journal}
\bvolume{583},
\bfpage{575}--\blpage{583}
(\byear{2003})
\end{barticle}
\endbibitem

\bibitem[\protect\citeauthoryear{{GRAVITY Collaboration}
  et~al.}{2020}]{bib5:gravity2020}
\begin{botherref}
\oauthor{\bsnm{{GRAVITY Collaboration}}},
\oauthor{\binits{O.} \bsnm{Pfuhl}},
\oauthor{\binits{R.} \bsnm{Davies}},
\oauthor{\binits{J.} \bsnm{Dexter}},
\oauthor{\binits{H.} \bsnm{Netzer}},
\oauthor{\binits{S.} \bsnm{H{\"o}nig}},
\oauthor{\binits{D.} \bsnm{Lutz}},
\oauthor{\binits{M.} \bsnm{Schartmann}},
\oauthor{\binits{E.} \bsnm{Sturm}},
\oauthor{\binits{A.} \bsnm{Amorim}},
\oauthor{\binits{W.} \bsnm{Brandner}},
\oauthor{\binits{Y.} \bsnm{Cl{\'e}net}},
\oauthor{\binits{P.T.D.} \bsnm{Zeeuw}},
\oauthor{\binits{A.} \bsnm{Eckart}},
\oauthor{\binits{F.} \bsnm{Eisenhauer}},
\oauthor{\binits{N.M.F.} \bsnm{Schreiber}},
\oauthor{\binits{F.} \bsnm{Gao}},
\oauthor{\binits{P.J.V.} \bsnm{Garcia}},
\oauthor{\binits{R.} \bsnm{Genzel}},
\oauthor{\binits{S.} \bsnm{Gillessen}},
\oauthor{\binits{D.} \bsnm{Gratadour}},
\oauthor{\binits{M.} \bsnm{Kishimoto}},
\oauthor{\binits{S.} \bsnm{Lacour}},
\oauthor{\binits{F.} \bsnm{Millour}},
\oauthor{\binits{T.} \bsnm{Ott}},
\oauthor{\binits{T.} \bsnm{Paumard}},
\oauthor{\binits{K.} \bsnm{Perraut}},
\oauthor{\binits{G.} \bsnm{Perrin}},
\oauthor{\binits{B.M.} \bsnm{Peterson}},
\oauthor{\binits{P.O.} \bsnm{Petrucci}},
\oauthor{\binits{M.A.} \bsnm{Prieto}},
\oauthor{\binits{D.} \bsnm{Rouan}},
\oauthor{\binits{J.} \bsnm{Shangguan}},
\oauthor{\binits{T.} \bsnm{Shimizu}},
\oauthor{\binits{A.} \bsnm{Sternberg}},
\oauthor{\binits{O.} \bsnm{Straub}},
\oauthor{\binits{C.} \bsnm{Straubmeier}},
\oauthor{\binits{L.J.} \bsnm{Tacconi}},
\oauthor{\binits{K.R.W.} \bsnm{Tristram}},
\oauthor{\binits{P.} \bsnm{Vermot}},
\oauthor{\binits{I.} \bsnm{Waisberg}},
\oauthor{\binits{F.} \bsnm{Widmann}},
\oauthor{\binits{J.} \bsnm{Woillez}},
An image of the dust sublimation region in the nucleus of ngc 1068.
Astronomy and Astrophysics
\textbf{634}
(2020).
doi:\doiurl{10.1051/0004-6361/201936255}
\end{botherref}
\endbibitem

\bibitem[\protect\citeauthoryear{{GRAVITY Collaboration}
  et~al.}{2021}]{bib5:gravity2021}
\begin{botherref}
\oauthor{\bsnm{{GRAVITY Collaboration}}},
\oauthor{\binits{A.} \bsnm{{Amorim}}},
\oauthor{\binits{M.} \bsnm{{Baub{\"o}ck}}},
\oauthor{\binits{M.C.} \bsnm{{Bentz}}},
\oauthor{\binits{W.} \bsnm{{Brandner}}},
\oauthor{\binits{M.} \bsnm{{Bolzer}}},
\oauthor{\binits{Y.} \bsnm{{Cl{\'e}net}}},
\oauthor{\binits{R.} \bsnm{{Davies}}},
\oauthor{\binits{P.T.} \bsnm{{de Zeeuw}}},
\oauthor{\binits{J.} \bsnm{{Dexter}}},
\oauthor{\binits{A.} \bsnm{{Drescher}}},
\oauthor{\binits{A.} \bsnm{{Eckart}}},
\oauthor{\binits{F.} \bsnm{{Eisenhauer}}},
\oauthor{\binits{N.M.} \bsnm{{F{\"o}rster Schreiber}}},
\oauthor{\binits{P.J.V.} \bsnm{{Garcia}}},
\oauthor{\binits{R.} \bsnm{{Genzel}}},
\oauthor{\binits{S.} \bsnm{{Gillessen}}},
\oauthor{\binits{D.} \bsnm{{Gratadour}}},
\oauthor{\binits{S.} \bsnm{{H{\"o}nig}}},
\oauthor{\binits{D.} \bsnm{{Kaltenbrunner}}},
\oauthor{\binits{M.} \bsnm{{Kishimoto}}},
\oauthor{\binits{S.} \bsnm{{Lacour}}},
\oauthor{\binits{D.} \bsnm{{Lutz}}},
\oauthor{\binits{F.} \bsnm{{Millour}}},
\oauthor{\binits{H.} \bsnm{{Netzer}}},
\oauthor{\binits{C.A.} \bsnm{{Onken}}},
\oauthor{\binits{T.} \bsnm{{Ott}}},
\oauthor{\binits{T.} \bsnm{{Paumard}}},
\oauthor{\binits{K.} \bsnm{{Perraut}}},
\oauthor{\binits{G.} \bsnm{{Perrin}}},
\oauthor{\binits{P.O.} \bsnm{{Petrucci}}},
\oauthor{\binits{O.} \bsnm{{Pfuhl}}},
\oauthor{\binits{M.A.} \bsnm{{Prieto}}},
\oauthor{\binits{D.} \bsnm{{Rouan}}},
\oauthor{\binits{J.} \bsnm{{Shangguan}}},
\oauthor{\binits{T.} \bsnm{{Shimizu}}},
\oauthor{\binits{J.} \bsnm{{Stadler}}},
\oauthor{\binits{A.} \bsnm{{Sternberg}}},
\oauthor{\binits{O.} \bsnm{{Straub}}},
\oauthor{\binits{C.} \bsnm{{Straubmeier}}},
\oauthor{\binits{R.} \bsnm{{Street}}},
\oauthor{\binits{E.} \bsnm{{Sturm}}},
\oauthor{\binits{L.J.} \bsnm{{Tacconi}}},
\oauthor{\binits{K.R.W.} \bsnm{{Tristram}}},
\oauthor{\binits{P.} \bsnm{{Vermot}}},
\oauthor{\binits{S.} \bsnm{{von Fellenberg}}},
\oauthor{\binits{F.} \bsnm{{Widmann}}},
\oauthor{\binits{J.} \bsnm{{Woillez}}},
The central parsec of ngc 3783: A rotating broad emission line region,
  asymmetric hot dust structure, and compact coronal line region.
Astronomy and Astrophysics
\textbf{648}
(2021).
doi:\doiurl{10.1051/0004-6361/202040061}
\end{botherref}
\endbibitem

\bibitem[\protect\citeauthoryear{Green}{2006}]{bib5:Green2006}
\begin{barticle}
\bauthor{\binits{P.J.} \bsnm{Green}},
\batitle{Lens-aided multi-angle spectroscopy (lamas) reveals small-scale
  outflow structure in quasars}.
\bjtitle{The Astrophysical Journal}
\bvolume{644},
\bfpage{733}
(\byear{2006})
\end{barticle}
\endbibitem

\bibitem[\protect\citeauthoryear{Greencard and
  Rokhlin}{1987}]{bib5:Greengard1987}
\begin{barticle}
\bauthor{\binits{L.} \bsnm{Greencard}},
\bauthor{\binits{V.} \bsnm{Rokhlin}},
\batitle{A fast algorithm for particle simulations}.
\bjtitle{JOURNAL OF COMPUTATIONAL PHYSICS}
\bvolume{73},
\bfpage{315}--\blpage{348}
(\byear{1987})
\end{barticle}
\endbibitem

\bibitem[\protect\citeauthoryear{Grieger et~al.}{1988}]{bib5:Grieger1988}
\begin{barticle}
\bauthor{\binits{B.} \bsnm{Grieger}},
\bauthor{\binits{R.} \bsnm{Kayser}},
\bauthor{\binits{S.} \bsnm{Refsdal}},
\batitle{Gravitational micro-lensing as a clue to quasar structure}.
\bjtitle{Astronomy \& Astrophysics}
\bvolume{194},
\bfpage{54}
(\byear{1988})
\end{barticle}
\endbibitem

\bibitem[\protect\citeauthoryear{Grieger et~al.}{1991}]{bib5:Grieger1991}
\begin{barticle}
\bauthor{\binits{B.} \bsnm{Grieger}},
\bauthor{\binits{R.} \bsnm{Kayser}},
\bauthor{\binits{T.} \bsnm{Schramm}},
\batitle{The deconvolution of the quasar structure from microlensing light
  curves}.
\bjtitle{Astronomy \& Astrophysics}
\bvolume{252},
\bfpage{508}
(\byear{1991})
\end{barticle}
\endbibitem

\bibitem[\protect\citeauthoryear{Grier et~al.}{2017}]{bib5:Grier2017}
\begin{barticle}
\bauthor{\binits{C.J.} \bsnm{Grier}},
\bauthor{\binits{A.} \bsnm{Pancoast}},
\bauthor{\binits{A.J.} \bsnm{Barth}},
\bauthor{\binits{M.M.} \bsnm{Fausnaugh}},
\bauthor{\binits{B.J.} \bsnm{Brewer}},
\bauthor{\binits{T.} \bsnm{Treu}},
\bauthor{\binits{B.M.} \bsnm{Peterson}},
\batitle{The structure of the broad-line region in active galactic nuclei. ii.
  dynamical modeling of data from the agn10 reverberation mapping campaign}.
\bjtitle{The Astrophysical Journal}
\bvolume{849},
\bfpage{146}
(\byear{2017}).
doi:\doiurl{10.3847/1538-4357/aa901b}
\end{barticle}
\endbibitem

\bibitem[\protect\citeauthoryear{Grier et~al.}{2019}]{bib5:Grier2019}
\begin{barticle}
\bauthor{\binits{C.J.} \bsnm{Grier}},
\bauthor{\binits{Y.} \bsnm{Shen}},
\bauthor{\binits{K.} \bsnm{Horne}},
\bauthor{\binits{W.N.} \bsnm{Brandt}},
\bauthor{\binits{J.R.} \bsnm{Trump}},
\bauthor{\binits{P.B.} \bsnm{Hall}},
\bauthor{\binits{K.} \bsnm{Kinemuchi}},
\bauthor{\binits{D.} \bsnm{Starkey}},
\bauthor{\binits{D.P.} \bsnm{Schneider}},
\bauthor{\binits{L.C.} \bsnm{Ho}},
\bauthor{\binits{Y.} \bsnm{Homayouni}},
\bauthor{\binits{J.I.-H.} \bsnm{Li}},
\bauthor{\binits{I.D.} \bsnm{McGreer}},
\bauthor{\binits{B.M.} \bsnm{Peterson}},
\bauthor{\binits{D.} \bsnm{Bizyaev}},
\bauthor{\binits{Y.} \bsnm{Chen}},
\bauthor{\binits{K.S.} \bsnm{Dawson}},
\bauthor{\binits{S.} \bsnm{Eftekharzadeh}},
\bauthor{\binits{Y.} \bsnm{Guo}},
\bauthor{\binits{S.} \bsnm{Jia}},
\bauthor{\binits{L.} \bsnm{Jiang}},
\bauthor{\binits{J.-P.} \bsnm{Kneib}},
\bauthor{\binits{F.} \bsnm{Li}},
\bauthor{\binits{Z.} \bsnm{Li}},
\bauthor{\binits{J.} \bsnm{Nie}},
\bauthor{\binits{A.} \bsnm{Oravetz}},
\bauthor{\binits{D.} \bsnm{Oravetz}},
\bauthor{\binits{K.} \bsnm{Pan}},
\bauthor{\binits{P.} \bsnm{Petitjean}},
\bauthor{\binits{K.A.} \bsnm{Ponder}},
\bauthor{\binits{J.} \bsnm{Rogerson}},
\bauthor{\binits{M.} \bsnm{Vivek}},
\bauthor{\binits{T.} \bsnm{Zhang}},
\bauthor{\binits{H.} \bsnm{Zou}},
\batitle{The sloan digital sky survey reverberation mapping project: Initial c
  iv lag results from four years of data}.
\bjtitle{The Astrophysical Journal}
\bvolume{887},
\bfpage{38}
(\byear{2019}).
doi:\doiurl{10.3847/1538-4357/ab4ea5}
\end{barticle}
\endbibitem

\bibitem[\protect\citeauthoryear{{Grz{\k{e}}dzielski}
  et~al.}{2017}]{bib5:Grzedzielski2017}
\begin{botherref}
\oauthor{\binits{M.} \bsnm{{Grz{\k{e}}dzielski}}},
\oauthor{\binits{A.} \bsnm{{Janiuk}}},
\oauthor{\binits{B.} \bsnm{{Czerny}}},
\oauthor{\binits{Q.} \bsnm{{Wu}}},
Modified viscosity in accretion disks: Application to galactic black hole
  binaries, intermediate mass black holes, and active galactic nuclei.
Astronomy and Astrophysics
\textbf{603}
(2017).
doi:\doiurl{10.1051/0004-6361/201629672}
\end{botherref}
\endbibitem

\bibitem[\protect\citeauthoryear{Guerras et~al.}{2013a}]{bib5:Guerras2013a}
\begin{barticle}
\bauthor{\binits{E.} \bsnm{Guerras}},
\bauthor{\binits{E.} \bsnm{Mediavilla}},
\bauthor{\binits{J.} \bsnm{Jim{\'e}nez-Vicente}},
\bauthor{\binits{C.S.} \bsnm{Kochanek}},
\bauthor{\binits{J.} \bparticle{a.} \bsnm{Mu{\~n}oz}},
\bauthor{\binits{E.} \bsnm{Falco}},
\bauthor{\binits{V.} \bsnm{Motta}},
\batitle{Microlensing of quasar broad emission lines: Constraints on broad line
  region size}.
\bjtitle{The Astrophysical Journal}
\bvolume{764},
\bfpage{160}
(\byear{2013}a).
doi:\doiurl{10.1088/0004-637X/764/2/160}.
\burl{http://stacks.iop.org/0004-637X/764/i=2/a=160?key=crossref.96a988c6d640e715b6839ba0fb168be2}
\end{barticle}
\endbibitem

\bibitem[\protect\citeauthoryear{Guerras et~al.}{2013b}]{bib5:Guerras2013b}
\begin{botherref}
\oauthor{\binits{E.} \bsnm{Guerras}},
\oauthor{\binits{E.} \bsnm{Mediavilla}},
\oauthor{\binits{J.} \bsnm{Jimenez-Vicente}},
\oauthor{\binits{C.S.} \bsnm{Kochanek}},
\oauthor{\binits{J.A.} \bsnm{Mu{\~n}oz}},
\oauthor{\binits{E.} \bsnm{Falco}},
\oauthor{\binits{V.} \bsnm{Motta}},
\oauthor{\binits{K.} \bsnm{Rojas}},
Microlensing of quasar ultraviolet iron emission.
Astrophysical Journal
\textbf{778}
(2013b).
doi:\doiurl{10.1088/0004-637X/778/2/123}
\end{botherref}
\endbibitem

\bibitem[\protect\citeauthoryear{Guerras et~al.}{2020}]{bib5:Guerras2020}
\begin{barticle}
\bauthor{\binits{E.} \bsnm{Guerras}},
\bauthor{\binits{X.} \bsnm{Dai}},
\bauthor{\binits{E.} \bsnm{Mediavilla}},
\batitle{A second-order moment of microlensing variability as a novel tool to
  constrain source emission size or discrete lens demographics in extragalactic
  research}.
\bjtitle{The Astrophysical Journal}
\bvolume{896},
\bfpage{111}
(\byear{2020}).
doi:\doiurl{10.3847/1538-4357/ab76b9}.
\burl{http://dx.doi.org/10.3847/1538-4357/ab76b9}
\end{barticle}
\endbibitem

\bibitem[\protect\citeauthoryear{{Guilbert} and
  {Rees}}{1988}]{bib5:Guilbert1988}
\begin{barticle}
\bauthor{\binits{P.W.} \bsnm{{Guilbert}}},
\bauthor{\binits{M.J.} \bsnm{{Rees}}},
\batitle{'cold' material in non-thermal sources.}
\bjtitle{\mnras}
\bvolume{233},
\bfpage{475}--\blpage{484}
(\byear{1988}).
doi:\doiurl{10.1093/mnras/233.2.475}
\end{barticle}
\endbibitem

\bibitem[\protect\citeauthoryear{Guo et~al.}{2022}]{bib5:Guo2022}
\begin{barticle}
\bauthor{\binits{W.-J.} \bsnm{Guo}},
\bauthor{\binits{Y.-R.} \bsnm{Li}},
\bauthor{\binits{Z.-X.} \bsnm{Zhang}},
\bauthor{\binits{L.C.} \bsnm{Ho}},
\bauthor{\binits{J.-M.} \bsnm{Wang}},
\batitle{Accretion disk size measurements of active galactic nuclei monitored
  by the zwicky transient facility}.
\bjtitle{The Astrophysical Journal}
\bvolume{929},
\bfpage{19}
(\byear{2022}).
doi:\doiurl{10.3847/1538-4357/ac4e84}
\end{barticle}
\endbibitem

\bibitem[\protect\citeauthoryear{{Haardt} and
  {Maraschi}}{1991}]{bib5:Haardt1991}
\begin{barticle}
\bauthor{\binits{F.} \bsnm{{Haardt}}},
\bauthor{\binits{L.} \bsnm{{Maraschi}}},
\batitle{A two-phase model for the x-ray emission from seyfert galaxies}.
\bjtitle{\apjl}
\bvolume{380},
\bfpage{51}
(\byear{1991}).
doi:\doiurl{10.1086/186171}
\end{barticle}
\endbibitem

\bibitem[\protect\citeauthoryear{{Haardt} and
  {Maraschi}}{1993}]{bib5:Haardt1993}
\begin{barticle}
\bauthor{\binits{F.} \bsnm{{Haardt}}},
\bauthor{\binits{L.} \bsnm{{Maraschi}}},
\batitle{X-ray spectra from two-phase accretion disks}.
\bjtitle{\apj}
\bvolume{413},
\bfpage{507}
(\byear{1993}).
doi:\doiurl{10.1086/173020}
\end{barticle}
\endbibitem

\bibitem[\protect\citeauthoryear{{Haardt} et~al.}{1994}]{bib5:Haardt1994}
\begin{barticle}
\bauthor{\binits{F.} \bsnm{{Haardt}}},
\bauthor{\binits{L.} \bsnm{{Maraschi}}},
\bauthor{\binits{G.} \bsnm{{Ghisellini}}},
\batitle{A model for the x-ray and ultraviolet emission from seyfert galaxies
  and galactic black holes}.
\bjtitle{\apjl}
\bvolume{432},
\bfpage{95}
(\byear{1994}).
doi:\doiurl{10.1086/187520}
\end{barticle}
\endbibitem

\bibitem[\protect\citeauthoryear{Hainline et~al.}{2012}]{bib5:Hainline2012}
\begin{barticle}
\bauthor{\binits{L.J.} \bsnm{Hainline}},
\bauthor{\binits{C.W.} \bsnm{Morgan}},
\bauthor{\binits{J.N.} \bsnm{Beach}},
\bauthor{\binits{C.S.} \bsnm{Kochanek}},
\bauthor{\binits{H.C.} \bsnm{Harris}},
\bauthor{\binits{T.} \bsnm{Tilleman}},
\bauthor{\binits{R.} \bsnm{Fadely}},
\bauthor{\binits{E.E.} \bsnm{Falco}},
\bauthor{\binits{T.X.} \bsnm{Le}},
\batitle{a new microlensing event in the doubly imaged quasar q 0957$+$561}.
\bjtitle{The Astrophysical Journal}
\bvolume{744},
\bfpage{104}
(\byear{2012}).
doi:\doiurl{10.1088/0004-637X/744/2/104}.
\burl{http://stacks.iop.org/0004-637X/744/i=2/a=104?key=crossref.5aac13a2114ae420a1502308595718ae}
\end{barticle}
\endbibitem

\bibitem[\protect\citeauthoryear{Hainline et~al.}{2013}]{bib5:Hainline2013}
\begin{botherref}
\oauthor{\binits{L.J.} \bsnm{Hainline}},
\oauthor{\binits{C.W.} \bsnm{Morgan}},
\oauthor{\binits{C.L.} \bsnm{Macleod}},
\oauthor{\binits{Z.D.} \bsnm{Landaal}},
\oauthor{\binits{C.S.} \bsnm{Kochanek}},
\oauthor{\binits{H.C.} \bsnm{Harris}},
\oauthor{\binits{T.} \bsnm{Tilleman}},
\oauthor{\binits{L.J.} \bsnm{Goicoechea}},
\oauthor{\binits{V.N.} \bsnm{Shalyapin}},
\oauthor{\binits{E.E.} \bsnm{Falco}},
Time delay and accretion disk size measurements in the lensed quasar sbs
  0909$+$532 from multiwavelength microlensing analysis.
Astrophysical Journal
\textbf{774}
(2013).
doi:\doiurl{10.1088/0004-637X/774/1/69}
\end{botherref}
\endbibitem

\bibitem[\protect\citeauthoryear{Hales and Lewis}{2007}]{bib5:Hales2007}
\begin{barticle}
\bauthor{\binits{C.A.} \bsnm{Hales}},
\bauthor{\binits{G.F.} \bsnm{Lewis}},
\batitle{Resolving the structure at the heart of bal quasars through
  microlensing induced polarisation variability}.
\bjtitle{Publications of the Astronomical Society of Australia}
\bvolume{24},
\bfpage{30}--\blpage{40}
(\byear{2007}).
doi:\doiurl{10.1071/AS07002}
\end{barticle}
\endbibitem

\bibitem[\protect\citeauthoryear{Hall et~al.}{2014}]{bib5:Hall2014}
\begin{barticle}
\bauthor{\binits{P.B.} \bsnm{Hall}},
\bauthor{\binits{E.S.} \bsnm{Noordeh}},
\bauthor{\binits{L.S.} \bsnm{Chajet}},
\bauthor{\binits{E.} \bsnm{Weiss}},
\bauthor{\binits{C.J.} \bsnm{Nixon}},
\batitle{Modelling spikes in quasar accretion disc temperature}.
\bjtitle{Monthly Notices of the Royal Astronomical Society}
\bvolume{442},
\bfpage{1090}--\blpage{1109}
(\byear{2014}).
doi:\doiurl{10.1093/mnras/stu890}
\end{barticle}
\endbibitem

\bibitem[\protect\citeauthoryear{Hall et~al.}{2018}]{bib5:Hall2018}
\begin{barticle}
\bauthor{\binits{P.B.} \bsnm{Hall}},
\bauthor{\binits{G.T.} \bsnm{Sarrouh}},
\bauthor{\binits{K.} \bsnm{Horne}},
\batitle{Non-blackbody disks can help explain inferred agn accretion disk
  sizes}.
\bjtitle{The Astrophysical Journal}
\bvolume{854},
\bfpage{93}
(\byear{2018}).
doi:\doiurl{10.3847/1538-4357/aaa768}
\end{barticle}
\endbibitem

\bibitem[\protect\citeauthoryear{Hamann et~al.}{2013}]{bib5:Hamann2013}
\begin{barticle}
\bauthor{\binits{F.} \bsnm{Hamann}},
\bauthor{\binits{G.} \bsnm{Chartas}},
\bauthor{\binits{S.} \bsnm{McGraw}},
\bauthor{\binits{P.R.} \bsnm{Hidalgo}},
\bauthor{\binits{J.} \bsnm{Shields}},
\bauthor{\binits{D.} \bsnm{Capellupo}},
\bauthor{\binits{J.} \bsnm{Charlton}},
\bauthor{\binits{M.} \bsnm{Eracleous}},
\batitle{Extreme-velocity quasar outflows and the role of x-ray shielding}.
\bjtitle{Monthly Notices of the Royal Astronomical Society}
\bvolume{435},
\bfpage{133}--\blpage{148}
(\byear{2013}).
doi:\doiurl{10.1093/mnras/stt1231}
\end{barticle}
\endbibitem

\bibitem[\protect\citeauthoryear{{Hartley} et~al.}{2019}]{bib5:Hartley2019}
\begin{barticle}
\bauthor{\binits{P.} \bsnm{{Hartley}}},
\bauthor{\binits{N.} \bsnm{{Jackson}}},
\bauthor{\binits{D.} \bsnm{{Sluse}}},
\bauthor{\binits{H.R.} \bsnm{{Stacey}}},
\bauthor{\binits{H.} \bsnm{{Vives-Arias}}},
\batitle{Strong lensing reveals jets in a sub-microjy radio-quiet quasar}.
\bjtitle{\mnras}
\bvolume{485}(\bissue{3}),
\bfpage{3009}--\blpage{3023}
(\byear{2019}).
doi:\doiurl{10.1093/mnras/stz510}
\end{barticle}
\endbibitem

\bibitem[\protect\citeauthoryear{Hartley et~al.}{2021}]{bib5:Hartley2021}
\begin{barticle}
\bauthor{\binits{P.} \bsnm{Hartley}},
\bauthor{\binits{N.} \bsnm{Jackson}},
\bauthor{\binits{S.} \bsnm{Badole}},
\bauthor{\binits{J.P.} \bsnm{McKean}},
\bauthor{\binits{D.} \bsnm{Sluse}},
\bauthor{\binits{H.} \bsnm{Vives-Arias}},
\batitle{Using strong lensing to understand the microjy radio emission in two
  radio quiet quasars at redshift 1.7}.
\bjtitle{Monthly Notices of the Royal Astronomical Society}
\bvolume{508},
\bfpage{4625}--\blpage{4638}
(\byear{2021}).
doi:\doiurl{10.1093/mnras/stab2758}
\end{barticle}
\endbibitem

\bibitem[\protect\citeauthoryear{Hawkins}{2020}]{bib5:Hawkins2020}
\begin{barticle}
\bauthor{\binits{M.R.S.} \bsnm{Hawkins}},
\batitle{The signature of primordial black holes in the dark matter halos of
  galaxies}.
\bjtitle{Astronomy \& Astrophysics}
\bvolume{633},
\bfpage{107}
(\byear{2020}).
doi:\doiurl{10.1051/0004-6361/201936462}
\end{barticle}
\endbibitem

\bibitem[\protect\citeauthoryear{Hawkins}{2022}]{bib5:Hawkins2022}
\begin{barticle}
\bauthor{\binits{M.R.S.} \bsnm{Hawkins}},
\batitle{New evidence for a cosmological distribution of stellar mass
  primordial black holes}.
\bjtitle{Monthly Notices of the Royal Astronomical Society}
\bvolume{512},
\bfpage{5706}--\blpage{5714}
(\byear{2022}).
doi:\doiurl{10.1093/mnras/stac863}
\end{barticle}
\endbibitem

\bibitem[\protect\citeauthoryear{Hawkins}{2015}]{bib5:Hawkins2015}
\begin{botherref}
\oauthor{\binits{M.R.S.} \bsnm{Hawkins}},
A new look at microlensing limits on dark matter in the galactic halo.
Astronomy and Astrophysics
\textbf{575}
(2015).
doi:\doiurl{10.1051/0004-6361/201425400}
\end{botherref}
\endbibitem

\bibitem[\protect\citeauthoryear{{Heyrovsk{\'y}} and
  {Loeb}}{1997}]{bib5:Heyrovsky1997}
\begin{barticle}
\bauthor{\binits{D.} \bsnm{{Heyrovsk{\'y}}}},
\bauthor{\binits{A.} \bsnm{{Loeb}}},
\batitle{Microlensing of an elliptical source by a point mass}.
\bjtitle{\apj}
\bvolume{490}(\bissue{1}),
\bfpage{38}--\blpage{50}
(\byear{1997}).
doi:\doiurl{10.1086/304855}
\end{barticle}
\endbibitem

\bibitem[\protect\citeauthoryear{Higginbottom
  et~al.}{2014}]{bib5:Higginbottom2014}
\begin{botherref}
\oauthor{\binits{N.} \bsnm{Higginbottom}},
\oauthor{\binits{D.} \bsnm{Proga}},
\oauthor{\binits{C.} \bsnm{Knigge}},
\oauthor{\binits{K.S.} \bsnm{Long}},
\oauthor{\binits{J.H.} \bsnm{Matthews}},
\oauthor{\binits{S.A.} \bsnm{Sim}},
Line-driven disk winds in active galactic nuclei: The critical importance of
  ionization and radiative transfer.
Astrophysical Journal
\textbf{789}
(2014).
doi:\doiurl{10.1088/0004-637X/789/1/19}
\end{botherref}
\endbibitem

\bibitem[\protect\citeauthoryear{Hjorth et~al.}{2002}]{bib5:Hjorth2002}
\begin{barticle}
\bauthor{\binits{J.} \bsnm{Hjorth}},
\bauthor{\binits{I.} \bsnm{Burud}},
\bauthor{\binits{A.O.} \bsnm{Jaunsen}},
\bauthor{\binits{P.L.} \bsnm{Schechter}},
\bauthor{\binits{J.-P.} \bsnm{Kneib}},
\bauthor{\binits{M.I.} \bsnm{Andersen}},
\bauthor{\binits{H.} \bsnm{Korhonen}},
\bauthor{\binits{J.W.} \bsnm{Clasen}},
\bauthor{\binits{A.A.} \bsnm{Kaas}},
\bauthor{\binits{R.} \bsnm{Ostensen}},
\bauthor{\binits{J.} \bsnm{Pelt}},
\bauthor{\binits{F.P.} \bsnm{Pijpers}},
\batitle{The time delay of the quadruple quasar rx j0911.40551 1}.
\bjtitle{The Astrophysical Journal}
\bvolume{572},
\bfpage{11}--\blpage{14}
(\byear{2002})
\end{barticle}
\endbibitem

\bibitem[\protect\citeauthoryear{Homayouni et~al.}{2020}]{bib5:Homayouni2020}
\begin{barticle}
\bauthor{\binits{Y.} \bsnm{Homayouni}},
\bauthor{\binits{J.R.} \bsnm{Trump}},
\bauthor{\binits{C.J.} \bsnm{Grier}},
\bauthor{\binits{K.} \bsnm{Horne}},
\bauthor{\binits{Y.} \bsnm{Shen}},
\bauthor{\binits{W.N.} \bsnm{Brandt}},
\bauthor{\binits{K.S.} \bsnm{Dawson}},
\bauthor{\binits{G.F.} \bsnm{Alvarez}},
\bauthor{\binits{P.J.} \bsnm{Green}},
\bauthor{\binits{P.B.} \bsnm{Hall}},
\bauthor{\binits{J.V.H.} \bsnm{Santisteban}},
\bauthor{\binits{L.C.} \bsnm{Ho}},
\bauthor{\binits{K.} \bsnm{Kinemuchi}},
\bauthor{\binits{C.S.} \bsnm{Kochanek}},
\bauthor{\binits{J.I.-H.} \bsnm{Li}},
\bauthor{\binits{B.M.} \bsnm{Peterson}},
\bauthor{\binits{D.P.} \bsnm{Schneider}},
\bauthor{\binits{D.A.} \bsnm{Starkey}},
\bauthor{\binits{D.} \bsnm{Bizyaev}},
\bauthor{\binits{K.} \bsnm{Pan}},
\bauthor{\binits{D.} \bsnm{Oravetz}},
\bauthor{\binits{A.} \bsnm{Simmons}},
\batitle{The sloan digital sky survey reverberation mapping project: Mg ii lag
  results from four years of monitoring}.
\bjtitle{The Astrophysical Journal}
\bvolume{901},
\bfpage{55}
(\byear{2020}).
doi:\doiurl{10.3847/1538-4357/ababa9}
\end{barticle}
\endbibitem

\bibitem[\protect\citeauthoryear{Horne et~al.}{1991}]{bib5:Horne1991}
\begin{barticle}
\bauthor{\binits{K.} \bsnm{Horne}},
\bauthor{\binits{W.F.} \bsnm{Welsh}},
\bauthor{\binits{B.M.} \bsnm{Peterson}},
\batitle{Echo mapping of broad h$\beta$ emission in ngc 5548}.
\bjtitle{The Astrophysical Journal}
\bvolume{367},
\bfpage{5}
(\byear{1991})
\end{barticle}
\endbibitem

\bibitem[\protect\citeauthoryear{Hu et~al.}{2015}]{bib5:Hu2015}
\begin{botherref}
\oauthor{\binits{C.} \bsnm{Hu}},
\oauthor{\binits{P.} \bsnm{Du}},
\oauthor{\binits{K.X.} \bsnm{Lu}},
\oauthor{\binits{Y.R.} \bsnm{Li}},
\oauthor{\binits{F.} \bsnm{Wang}},
\oauthor{\binits{J.} \bsnm{Qiu}},
\oauthor{\binits{J.M.} \bsnm{Bai}},
\oauthor{\binits{S.} \bsnm{Kaspi}},
\oauthor{\binits{L.C.} \bsnm{Ho}},
\oauthor{\binits{H.} \bsnm{Netzer}},
\oauthor{\binits{J.M.} \bsnm{Wang}},
Supermassive black holes with high accretion rates in active galactic nuclei.
  iii. detection of fe ii reverberation in nine narrow-line seyfert 1 galaxies.
Astrophysical Journal
\textbf{804}
(2015).
doi:\doiurl{10.1088/0004-637X/804/2/138}
\end{botherref}
\endbibitem

\bibitem[\protect\citeauthoryear{Hubeny et~al.}{2000}]{bib5:Hubeny2000}
\begin{barticle}
\bauthor{\binits{I.} \bsnm{Hubeny}},
\bauthor{\binits{E.} \bsnm{Agol}},
\bauthor{\binits{O.} \bsnm{Blaes}},
\bauthor{\binits{J.H.} \bsnm{Krolik}},
\batitle{Non-lte models and theoretical spectra of accretion disks in active
  galactic nuclei. iii. integrated spectra for hydrogen-helium disks}.
\bjtitle{The Astrophysical Journal}
\bvolume{533},
\bfpage{710}--\blpage{728}
(\byear{2000})
\end{barticle}
\endbibitem

\bibitem[\protect\citeauthoryear{Huchra et~al.}{1985}]{bib5:Huchra1985}
\begin{barticle}
\bauthor{\binits{J.} \bsnm{Huchra}},
\bauthor{\binits{M.} \bsnm{Gorenstein}},
\bauthor{\binits{S.} \bsnm{Kent}},
\bauthor{\binits{I.} \bsnm{Shapiro}},
\bauthor{\binits{G.} \bsnm{Smith}},
\bauthor{\binits{E.} \bsnm{Horine}},
\bauthor{\binits{F.L.} \bsnm{Whipple}},
\bauthor{\binits{R.} \bsnm{Perley}},
\batitle{2237$+$0305: A new and unusual gravitational lens}.
\bjtitle{Astronomical Journal}
\bvolume{90},
\bfpage{691}
(\byear{1985})
\end{barticle}
\endbibitem

\bibitem[\protect\citeauthoryear{Hutsem{\'e}kers}{1993}]{bib5:Hutsemekers1993}
\begin{barticle}
\bauthor{\binits{D.} \bsnm{Hutsem{\'e}kers}},
\batitle{Selective gravitational microlensing and line profile variations in
  the bal quasar h1413+117}.
\bjtitle{Astronomy and Astrophysics}
\bvolume{280},
\bfpage{435}
(\byear{1993})
\end{barticle}
\endbibitem

\bibitem[\protect\citeauthoryear{Hutsem{\'e}kers}{1994}]{bib5:Hutsemekers1994}
\begin{barticle}
\bauthor{\binits{D.} \bsnm{Hutsem{\'e}kers}},
\batitle{The use of gravitational microlensing to scan the structure of bal
  qsos}.
\bjtitle{Astrophysics and Space Science}
\bvolume{216},
\bfpage{361}
(\byear{1994})
\end{barticle}
\endbibitem

\bibitem[\protect\citeauthoryear{Hutsem{\'e}kers and
  Sluse}{2021}]{bib5:Hutsemekers2021}
\begin{botherref}
\oauthor{\binits{D.} \bsnm{Hutsem{\'e}kers}},
\oauthor{\binits{D.} \bsnm{Sluse}},
Geometry and kinematics of the broad emission line region in the lensed quasar
  q2237+0305.
Astronomy and Astrophysics
\textbf{654}
(2021).
doi:\doiurl{10.1051/0004-6361/202141820}
\end{botherref}
\endbibitem

\bibitem[\protect\citeauthoryear{Hutsem{\'e}kers
  et~al.}{2020}]{bib5:Hutsemekers2020}
\begin{botherref}
\oauthor{\binits{D.} \bsnm{Hutsem{\'e}kers}},
\oauthor{\binits{D.} \bsnm{Sluse}},
\oauthor{\binits{P.} \bsnm{Kumar}},
Spatially separated continuum sources revealed by microlensing in the
  gravitationally lensed broad absorption line quasar sdss j081830.46+060138.0.
Astronomy and Astrophysics
\textbf{633}
(2020).
doi:\doiurl{10.1051/0004-6361/201936973}
\end{botherref}
\endbibitem

\bibitem[\protect\citeauthoryear{Hutsem{\'e}kers
  et~al.}{2010}]{bib5:Hutsemekers2010}
\begin{botherref}
\oauthor{\binits{D.} \bsnm{Hutsem{\'e}kers}},
\oauthor{\binits{B.} \bsnm{Borguet}},
\oauthor{\binits{D.} \bsnm{Sluse}},
\oauthor{\binits{P.} \bsnm{Riaud}},
\oauthor{\binits{T.} \bsnm{Anguita}},
Microlensing in h1413+117: Disentangling line profile emission and absorption
  in a broad absorption line quasar.
Astronomy and Astrophysics
\textbf{519}
(2010).
doi:\doiurl{10.1051/0004-6361/200913247}
\end{botherref}
\endbibitem

\bibitem[\protect\citeauthoryear{Hutsem{\'e}kers
  et~al.}{2015}]{bib5:Hutsemekers2015}
\begin{botherref}
\oauthor{\binits{D.} \bsnm{Hutsem{\'e}kers}},
\oauthor{\binits{D.} \bsnm{Sluse}},
\oauthor{\binits{L.} \bsnm{Braibant}},
\oauthor{\binits{T.} \bsnm{Anguita}},
Polarization microlensing in the quadruply imaged broad absorption line quasar
  h1413+117.
Astronomy and Astrophysics
\textbf{584}
(2015).
doi:\doiurl{10.1051/0004-6361/201527243}
\end{botherref}
\endbibitem

\bibitem[\protect\citeauthoryear{Hutsemekers
  et~al.}{2019}]{bib5:Hutsemekers2019}
\begin{botherref}
\oauthor{\binits{D.} \bsnm{Hutsemekers}},
\oauthor{\binits{L.} \bsnm{Braibant}},
\oauthor{\binits{D.} \bsnm{Sluse}},
\oauthor{\binits{R.} \bsnm{Goosmann}},
Constraining the geometry and kinematics of the quasar broad emission line
  region using gravitational microlensing: Ii. comparing models with
  observations in the lensed quasar he0435$-$1223.
Astronomy and Astrophysics
\textbf{629}
(2019).
doi:\doiurl{10.1051/0004-6361/201731087}
\end{botherref}
\endbibitem

\bibitem[\protect\citeauthoryear{Hutsem{\'e}kers
  et~al.}{2023}]{bib5:Hutsemekers2023}
\begin{barticle}
\bauthor{\binits{D.} \bsnm{Hutsem{\'e}kers}},
\bauthor{\binits{D.} \bsnm{Sluse}},
\bauthor{\binits{D.} \bsnm{Savic}},
\bauthor{\binits{G.T.} \bsnm{Richards}},
\batitle{Microlensing of the broad emission line region in the lensed quasar
  j1004+4112}.
\bjtitle{Astronomy \& Astrophysics}
(\byear{2023}).
doi:\doiurl{10.1051/0004-6361/202245490}
\end{barticle}
\endbibitem

\bibitem[\protect\citeauthoryear{{Hutsem{\'e}kers}
  et~al.}{2017}]{bib5:Hutsemekers2017}
\begin{barticle}
\bauthor{\binits{D.} \bsnm{{Hutsem{\'e}kers}}},
\bauthor{\binits{L.} \bsnm{{Braibant}}},
\bauthor{\binits{D.} \bsnm{{Sluse}}},
\bauthor{\binits{T.} \bsnm{{Anguita}}},
\bauthor{\binits{R.} \bsnm{{Goosmann}}},
\batitle{New constraints on quasar broad absorption and emission line regions
  from gravitational microlensing}.
\bjtitle{Frontiers in Astronomy and Space Sciences}
\bvolume{4},
\bfpage{18}
(\byear{2017}).
doi:\doiurl{10.3389/fspas.2017.00018}
\end{barticle}
\endbibitem

\bibitem[\protect\citeauthoryear{Hyde and Bernardi}{2009}]{bib5:Hyde2009}
\begin{barticle}
\bauthor{\binits{J.B.} \bsnm{Hyde}},
\bauthor{\binits{M.} \bsnm{Bernardi}},
\batitle{The luminosity and stellar mass fundamental plane of early-type
  galaxies}.
\bjtitle{Monthly Notices of the Royal Astronomical Society}
\bvolume{396},
\bfpage{1171}--\blpage{1185}
(\byear{2009}).
doi:\doiurl{10.1111/j.1365-2966.2009.14783.x}
\end{barticle}
\endbibitem

\bibitem[\protect\citeauthoryear{Ichimaru}{1977}]{bib5:Ichimaru1977}
\begin{botherref}
\oauthor{\binits{S.} \bsnm{Ichimaru}},
Bimodal behavior of accretion disks: theory and application to cygnus x-1
  transitions.
\textbf{214},
840--855
(1977).
doi:\doiurl{10.1086/155314}
\end{botherref}
\endbibitem

\bibitem[\protect\citeauthoryear{Inoue}{2016}]{bib5:Inoue2016}
\begin{barticle}
\bauthor{\binits{K.T.} \bsnm{Inoue}},
\batitle{On the origin of the flux ratio anomaly in quadruple lens systems}.
\bjtitle{Monthly Notices of the Royal Astronomical Society}
\bvolume{461},
\bfpage{164}--\blpage{175}
(\byear{2016}).
doi:\doiurl{10.1093/mnras/stw1270}
\end{barticle}
\endbibitem

\bibitem[\protect\citeauthoryear{Irwin et~al.}{1989}]{bib5:Irwin1989}
\begin{barticle}
\bauthor{\binits{M.J.} \bsnm{Irwin}},
\bauthor{\binits{R.L.} \bsnm{Webster}},
\bauthor{\binits{P.C.} \bsnm{Hewett}},
\bauthor{\binits{R.T.} \bsnm{Corrigan}},
\bauthor{\binits{R.I.} \bsnm{Jedrzejewski}},
\batitle{Photometric variations in the q2237$+$0305 system: first detection of
  a microlensing event}.
\bjtitle{The Astronomical Journal}
\bvolume{98},
\bfpage{1989}
(\byear{1989})
\end{barticle}
\endbibitem

\bibitem[\protect\citeauthoryear{Jiang et~al.}{2016}]{bib5:Jiang2016}
\begin{barticle}
\bauthor{\binits{Y.-F.} \bsnm{Jiang}},
\bauthor{\binits{S.W.} \bsnm{Davis}},
\bauthor{\binits{J.M.} \bsnm{Stone}},
\batitle{Iron opacity bump changes the stability and structure of accretion
  disks in active galactic nuclei}.
\bjtitle{The Astrophysical Journal}
\bvolume{827},
\bfpage{10}
(\byear{2016}).
doi:\doiurl{10.3847/0004-637x/827/1/10}
\end{barticle}
\endbibitem

\bibitem[\protect\citeauthoryear{Jim{\'e}nez-Vicente and
  Mediavilla}{2019}]{bib5:Jimenez2019}
\begin{barticle}
\bauthor{\binits{J.} \bsnm{Jim{\'e}nez-Vicente}},
\bauthor{\binits{E.} \bsnm{Mediavilla}},
\batitle{The initial mass function of lens galaxies from quasar microlensing}.
\bjtitle{The Astrophysical Journal}
\bvolume{885},
\bfpage{75}
(\byear{2019}).
doi:\doiurl{10.3847/1538-4357/ab46b8}.
\burl{http://dx.doi.org/10.3847/1538-4357/ab46b8}
\end{barticle}
\endbibitem

\bibitem[\protect\citeauthoryear{Jim{\'e}nez-Vicente and
  Mediavilla}{2022}]{bib5:JimenezVicente2022}
\begin{barticle}
\bauthor{\binits{J.} \bsnm{Jim{\'e}nez-Vicente}},
\bauthor{\binits{E.} \bsnm{Mediavilla}},
\batitle{Fast multipole method for gravitational lensing: Application to
  high-magnification quasar microlensing}.
\bjtitle{The Astrophysical Journal}
\bvolume{941},
\bfpage{80}
(\byear{2022}).
doi:\doiurl{10.3847/1538-4357/ac9e59}
\end{barticle}
\endbibitem

\bibitem[\protect\citeauthoryear{Jim{\'e}nez-Vicente
  et~al.}{2012}]{bib5:Jimenez2012}
\begin{botherref}
\oauthor{\binits{J.} \bsnm{Jim{\'e}nez-Vicente}},
\oauthor{\binits{E.} \bsnm{Mediavilla}},
\oauthor{\binits{J.A.} \bsnm{Mu{\~n}oz}},
\oauthor{\binits{C.S.} \bsnm{Kochanek}},
A robust determination of the size of quasar accretion disks using
  gravitational microlensing.
Astrophysical Journal
\textbf{751}
(2012).
doi:\doiurl{10.1088/0004-637X/751/2/106}
\end{botherref}
\endbibitem

\bibitem[\protect\citeauthoryear{Jim{\'e}nez-Vicente
  et~al.}{2014}]{bib5:Jimenez2014}
\begin{barticle}
\bauthor{\binits{J.} \bsnm{Jim{\'e}nez-Vicente}},
\bauthor{\binits{E.} \bsnm{Mediavilla}},
\bauthor{\binits{C.S.} \bsnm{Kochanek}},
\bauthor{\binits{J.A.} \bsnm{Mu{\~n}oz}},
\bauthor{\binits{V.} \bsnm{Motta}},
\bauthor{\binits{E.} \bsnm{Falco}},
\bauthor{\binits{A.M.} \bsnm{Mosquera}},
\batitle{The average size and temperature profile of quasar accretion disks}.
\bjtitle{The Astrophysical Journal}
\bvolume{783},
\bfpage{47}
(\byear{2014})
\end{barticle}
\endbibitem

\bibitem[\protect\citeauthoryear{Jim{\'e}nez-Vicente
  et~al.}{2015a}]{bib5:Jimenez2015a}
\begin{barticle}
\bauthor{\binits{J.} \bsnm{Jim{\'e}nez-Vicente}},
\bauthor{\binits{E.} \bsnm{Mediavilla}},
\bauthor{\binits{C.S.} \bsnm{Kochanek}},
\bauthor{\binits{J.A.} \bsnm{Mu{\~n}oz}},
\batitle{Dark matter mass fraction in lens galaxies: New estimates from
  microlensing}.
\bjtitle{The Astrophysical Journal}
\bvolume{799},
\bfpage{149}
(\byear{2015}a).
doi:\doiurl{10.1088/0004-637X/783/1/47}.
\burl{http://stacks.iop.org/0004-637X/783/i=1/a=47?key=crossref.00fa7daf9eb49afaa7c164a3cbbf828b}
\end{barticle}
\endbibitem

\bibitem[\protect\citeauthoryear{Jim{\'e}nez-Vicente
  et~al.}{2015b}]{bib5:Jimenez2015b}
\begin{barticle}
\bauthor{\binits{J.} \bsnm{Jim{\'e}nez-Vicente}},
\bauthor{\binits{E.} \bsnm{Mediavilla}},
\bauthor{\binits{C.S.} \bsnm{Kochanek}},
\bauthor{\binits{J.A.} \bsnm{Mu{\~n}oz}},
\batitle{Probing the dark matter radial profile in lens galaxies and the size
  of x-ray emitting region in quasars with microlensing}.
\bjtitle{The Astrophysical Journal}
\bvolume{806},
\bfpage{251}
(\byear{2015}b).
doi:\doiurl{10.1088/0004-637X/799/2/149}.
\burl{http://stacks.iop.org/0004-637X/799/i=2/a=149?key=crossref.5f37efa6dd6da11e6b57838c5451075a}
\end{barticle}
\endbibitem

\bibitem[\protect\citeauthoryear{{Jorstad} et~al.}{2023}]{bib5:Jorstad2023}
\begin{botherref}
\oauthor{\binits{S.} \bsnm{{Jorstad}}},
\oauthor{\binits{M.} \bsnm{{Wielgus}}},
\oauthor{\binits{R.} \bsnm{{Lico}}},
\oauthor{\binits{S.} \bsnm{{Issaoun}}},
\oauthor{\binits{A.E.} \bsnm{{Broderick}}},
\oauthor{\binits{D.W.} \bsnm{{Pesce}}},
\oauthor{\binits{J.} \bsnm{{Liu}}},
\oauthor{\binits{G.-Y.} \bsnm{{Zhao}}},
\oauthor{\binits{T.P.} \bsnm{{Krichbaum}}},
\oauthor{\binits{L.} \bsnm{{Blackburn}}},
\oauthor{\binits{C.-k.} \bsnm{{Chan}}},
\oauthor{\binits{M.} \bsnm{{Janssen}}},
\oauthor{\binits{V.} \bsnm{{Ramakrishnan}}},
\oauthor{\binits{K.} \bsnm{{Akiyama}}},
\oauthor{\binits{A.} \bsnm{{Alberdi}}},
\oauthor{\binits{J.C.} \bsnm{{Algaba}}},
\oauthor{\binits{K.L.} \bsnm{{Bouman}}},
\oauthor{\binits{I.} \bsnm{{Cho}}},
\oauthor{\binits{A.} \bsnm{{Fuentes}}},
\oauthor{\binits{J.L.} \bsnm{{G{\'o}mez}}},
\oauthor{\binits{M.} \bsnm{{Gurwell}}},
\oauthor{\binits{M.D.} \bsnm{{Johnson}}},
\oauthor{\binits{J.-Y.} \bsnm{{Kim}}},
\oauthor{\binits{R.-S.} \bsnm{{Lu}}},
\oauthor{\binits{I.} \bsnm{{Mart{\'\i}-Vidal}}},
\oauthor{\binits{M.} \bsnm{{Moscibrodzka}}},
\oauthor{\binits{F.M.} \bsnm{{P{\"o}tzl}}},
\oauthor{\binits{E.} \bsnm{{Traianou}}},
\oauthor{\binits{I.} \bsnm{{van Bemmel}}},
\oauthor{\binits{W.} \bsnm{{Alef}}},
\oauthor{\binits{R.} \bsnm{{Anantua}}},
\oauthor{\binits{K.} \bsnm{{Asada}}},
\oauthor{\binits{R.} \bsnm{{Azulay}}},
\oauthor{\binits{U.} \bsnm{{Bach}}},
\oauthor{\binits{A.-K.} \bsnm{{Baczko}}},
\oauthor{\binits{D.} \bsnm{{Ball}}},
\oauthor{\binits{M.} \bsnm{{Balokovi{\'c}}}},
\oauthor{\binits{J.} \bsnm{{Barrett}}},
\oauthor{\binits{M.} \bsnm{{Baub{\"o}ck}}},
\oauthor{\binits{B.A.} \bsnm{{Benson}}},
\oauthor{\binits{D.} \bsnm{{Bintley}}},
\oauthor{\binits{R.} \bsnm{{Blundell}}},
\oauthor{\binits{G.C.} \bsnm{{Bower}}},
\oauthor{\binits{H.} \bsnm{{Boyce}}},
\oauthor{\binits{M.} \bsnm{{Bremer}}},
\oauthor{\binits{C.D.} \bsnm{{Brinkerink}}},
\oauthor{\binits{R.} \bsnm{{Brissenden}}},
\oauthor{\binits{S.} \bsnm{{Britzen}}},
\oauthor{\binits{D.} \bsnm{{Broguiere}}},
\oauthor{\binits{T.} \bsnm{{Bronzwaer}}},
\oauthor{\binits{S.} \bsnm{{Bustamante}}},
\oauthor{\binits{D.-Y.} \bsnm{{Byun}}},
\oauthor{\binits{J.E.} \bsnm{{Carlstrom}}},
\oauthor{\binits{C.} \bsnm{{Ceccobello}}},
\oauthor{\binits{A.} \bsnm{{Chael}}},
\oauthor{\binits{K.} \bsnm{{Chatterjee}}},
\oauthor{\binits{S.} \bsnm{{Chatterjee}}},
\oauthor{\binits{M.-T.} \bsnm{{Chen}}},
\oauthor{\binits{Y.} \bsnm{{Chen}}},
\oauthor{\binits{X.} \bsnm{{Cheng}}},
\oauthor{\binits{P.} \bsnm{{Christian}}},
\oauthor{\binits{N.S.} \bsnm{{Conroy}}},
\oauthor{\binits{J.E.} \bsnm{{Conway}}},
\oauthor{\binits{J.M.} \bsnm{{Cordes}}},
\oauthor{\binits{T.M.} \bsnm{{Crawford}}},
\oauthor{\binits{G.B.} \bsnm{{Crew}}},
\oauthor{\binits{A.} \bsnm{{Cruz-Osorio}}},
\oauthor{\binits{Y.} \bsnm{{Cui}}},
\oauthor{\binits{J.} \bsnm{{Davelaar}}},
\oauthor{\binits{M.} \bsnm{{De Laurentis}}},
\oauthor{\binits{R.} \bsnm{{Deane}}},
\oauthor{\binits{J.} \bsnm{{Dempsey}}},
\oauthor{\binits{G.} \bsnm{{Desvignes}}},
\oauthor{\binits{J.} \bsnm{{Dexter}}},
\oauthor{\binits{V.} \bsnm{{Dhruv}}},
\oauthor{\binits{S.S.} \bsnm{{Doeleman}}},
\oauthor{\binits{S.} \bsnm{{Dougal}}},
\oauthor{\binits{S.A.} \bsnm{{Dzib}}},
\oauthor{\binits{R.P.} \bsnm{{Eatough}}},
\oauthor{\binits{R.} \bsnm{{Emami}}},
\oauthor{\binits{H.} \bsnm{{Falcke}}},
\oauthor{\binits{J.} \bsnm{{Farah}}},
\oauthor{\binits{V.L.} \bsnm{{Fish}}},
\oauthor{\binits{E.} \bsnm{{Fomalont}}},
\oauthor{\binits{H.A.} \bsnm{{Ford}}},
\oauthor{\binits{R.} \bsnm{{Fraga-Encinas}}},
\oauthor{\binits{W.T.} \bsnm{{Freeman}}},
\oauthor{\binits{P.} \bsnm{{Friberg}}},
\oauthor{\binits{C.M.} \bsnm{{Fromm}}},
\oauthor{\binits{P.} \bsnm{{Galison}}},
\oauthor{\binits{C.F.} \bsnm{{Gammie}}},
\oauthor{\binits{R.} \bsnm{{Garc{\'\i}a}}},
\oauthor{\binits{O.} \bsnm{{Gentaz}}},
\oauthor{\binits{B.} \bsnm{{Georgiev}}},
\oauthor{\binits{C.} \bsnm{{Goddi}}},
\oauthor{\binits{R.} \bsnm{{Gold}}},
\oauthor{\binits{A.I.} \bsnm{{G{\'o}mez-Ruiz}}},
\oauthor{\binits{M.} \bsnm{{Gu}}},
\oauthor{\binits{K.} \bsnm{{Hada}}},
\oauthor{\binits{D.} \bsnm{{Haggard}}},
\oauthor{\binits{K.} \bsnm{{Haworth}}},
\oauthor{\binits{M.H.} \bsnm{{Hecht}}},
\oauthor{\binits{R.} \bsnm{{Hesper}}},
\oauthor{\binits{D.} \bsnm{{Heumann}}},
\oauthor{\binits{L.C.} \bsnm{{Ho}}},
\oauthor{\binits{P.} \bsnm{{Ho}}},
\oauthor{\binits{M.} \bsnm{{Honma}}},
\oauthor{\binits{C.-W.L.} \bsnm{{Huang}}},
\oauthor{\binits{L.} \bsnm{{Huang}}},
\oauthor{\binits{D.H.} \bsnm{{Hughes}}},
\oauthor{\binits{S.} \bsnm{{Ikeda}}},
\oauthor{\binits{C.M.V.} \bsnm{{Impellizzeri}}},
\oauthor{\binits{M.} \bsnm{{Inoue}}},
\oauthor{\binits{D.J.} \bsnm{{James}}},
\oauthor{\binits{B.T.} \bsnm{{Jannuzi}}},
\oauthor{\binits{B.} \bsnm{{Jeter}}},
\oauthor{\binits{W.} \bsnm{{Jiang}}},
\oauthor{\binits{A.} \bsnm{{Jim{\'e}nez-Rosales}}},
\oauthor{\binits{A.V.} \bsnm{{Joshi}}},
\oauthor{\binits{T.} \bsnm{{Jung}}},
\oauthor{\binits{M.} \bsnm{{Karami}}},
\oauthor{\binits{R.} \bsnm{{Karuppusamy}}},
\oauthor{\binits{T.} \bsnm{{Kawashima}}},
\oauthor{\binits{G.K.} \bsnm{{Keating}}},
\oauthor{\binits{M.} \bsnm{{Kettenis}}},
\oauthor{\binits{D.-J.} \bsnm{{Kim}}},
\oauthor{\binits{J.} \bsnm{{Kim}}},
\oauthor{\binits{J.} \bsnm{{Kim}}},
\oauthor{\binits{M.} \bsnm{{Kino}}},
\oauthor{\binits{J.Y.} \bsnm{{Koay}}},
\oauthor{\binits{P.} \bsnm{{Kocherlakota}}},
\oauthor{\binits{Y.} \bsnm{{Kofuji}}},
\oauthor{\binits{S.} \bsnm{{Koyama}}},
\oauthor{\binits{C.} \bsnm{{Kramer}}},
\oauthor{\binits{M.} \bsnm{{Kramer}}},
\oauthor{\binits{C.-Y.} \bsnm{{Kuo}}},
\oauthor{\binits{N.} \bsnm{{La Bella}}},
\oauthor{\binits{T.R.} \bsnm{{Lauer}}},
\oauthor{\binits{D.} \bsnm{{Lee}}},
\oauthor{\binits{S.-S.} \bsnm{{Lee}}},
\oauthor{\binits{P.K.} \bsnm{{Leung}}},
\oauthor{\binits{A.} \bsnm{{Levis}}},
\oauthor{\binits{Z.} \bsnm{{Li}}},
\oauthor{\binits{G.} \bsnm{{Lindahl}}},
\oauthor{\binits{M.} \bsnm{{Lindqvist}}},
\oauthor{\binits{M.} \bsnm{{Lisakov}}},
\oauthor{\binits{K.} \bsnm{{Liu}}},
\oauthor{\binits{E.} \bsnm{{Liuzzo}}},
\oauthor{\binits{W.-P.} \bsnm{{Lo}}},
\oauthor{\binits{A.P.} \bsnm{{Lobanov}}},
\oauthor{\binits{L.} \bsnm{{Loinard}}},
\oauthor{\binits{C.J.} \bsnm{{Lonsdale}}},
\oauthor{\binits{N.R.} \bsnm{{MacDonald}}},
\oauthor{\binits{J.} \bsnm{{Mao}}},
\oauthor{\binits{N.} \bsnm{{Marchili}}},
\oauthor{\binits{S.} \bsnm{{Markoff}}},
\oauthor{\binits{D.P.} \bsnm{{Marrone}}},
\oauthor{\binits{A.P.} \bsnm{{Marscher}}},
\oauthor{\binits{S.} \bsnm{{Matsushita}}},
\oauthor{\binits{L.D.} \bsnm{{Matthews}}},
\oauthor{\binits{L.} \bsnm{{Medeiros}}},
\oauthor{\binits{K.M.} \bsnm{{Menten}}},
\oauthor{\binits{D.} \bsnm{{Michalik}}},
\oauthor{\binits{I.} \bsnm{{Mizuno}}},
\oauthor{\binits{Y.} \bsnm{{Mizuno}}},
\oauthor{\binits{J.M.} \bsnm{{Moran}}},
\oauthor{\binits{K.} \bsnm{{Moriyama}}},
\oauthor{\binits{C.} \bsnm{{M{\"u}ller}}},
\oauthor{\binits{A.} \bsnm{{Mus}}},
\oauthor{\binits{G.} \bsnm{{Musoke}}},
\oauthor{\binits{I.} \bsnm{{Myserlis}}},
\oauthor{\binits{A.} \bsnm{{Nadolski}}},
\oauthor{\binits{H.} \bsnm{{Nagai}}},
\oauthor{\binits{N.M.} \bsnm{{Nagar}}},
\oauthor{\binits{M.} \bsnm{{Nakamura}}},
\oauthor{\binits{R.} \bsnm{{Narayan}}},
\oauthor{\binits{G.} \bsnm{{Narayanan}}},
\oauthor{\binits{I.} \bsnm{{Natarajan}}},
\oauthor{\binits{A.} \bsnm{{Nathanail}}},
\oauthor{\binits{S.N.} \bsnm{{Fuentes}}},
\oauthor{\binits{J.} \bsnm{{Neilsen}}},
\oauthor{\binits{R.} \bsnm{{Neri}}},
\oauthor{\binits{C.} \bsnm{{Ni}}},
\oauthor{\binits{A.} \bsnm{{Noutsos}}},
\oauthor{\binits{M.A.} \bsnm{{Nowak}}},
\oauthor{\binits{J.} \bsnm{{Oh}}},
\oauthor{\binits{H.} \bsnm{{Okino}}},
\oauthor{\binits{H.} \bsnm{{Olivares}}},
\oauthor{\binits{G.N.} \bsnm{{Ortiz-Le{\'o}n}}},
\oauthor{\binits{T.} \bsnm{{Oyama}}},
\oauthor{\binits{F.} \bsnm{{{\"O}zel}}},
\oauthor{\binits{D.C.M.} \bsnm{{Palumbo}}},
\oauthor{\binits{G.F.} \bsnm{{Paraschos}}},
\oauthor{\binits{J.} \bsnm{{Park}}},
\oauthor{\binits{H.} \bsnm{{Parsons}}},
\oauthor{\binits{N.} \bsnm{{Patel}}},
\oauthor{\binits{U.-L.} \bsnm{{Pen}}},
\oauthor{\binits{V.} \bsnm{{Pi{\'e}tu}}},
\oauthor{\binits{R.} \bsnm{{Plambeck}}},
\oauthor{\binits{A.} \bsnm{{PopStefanija}}},
\oauthor{\binits{O.} \bsnm{{Porth}}},
\oauthor{\binits{B.} \bsnm{{Prather}}},
\oauthor{\binits{J.A.} \bsnm{{Preciado-L{\'o}pez}}},
\oauthor{\binits{D.} \bsnm{{Psaltis}}},
\oauthor{\binits{H.-Y.} \bsnm{{Pu}}},
\oauthor{\binits{R.} \bsnm{{Rao}}},
\oauthor{\binits{M.G.} \bsnm{{Rawlings}}},
\oauthor{\binits{A.W.} \bsnm{{Raymond}}},
\oauthor{\binits{L.} \bsnm{{Rezzolla}}},
\oauthor{\binits{A.} \bsnm{{Ricarte}}},
\oauthor{\binits{B.} \bsnm{{Ripperda}}},
\oauthor{\binits{F.} \bsnm{{Roelofs}}},
\oauthor{\binits{A.} \bsnm{{Rogers}}},
\oauthor{\binits{E.} \bsnm{{Ros}}},
\oauthor{\binits{C.} \bsnm{{Romero-Ca{\~n}izales}}},
\oauthor{\binits{A.} \bsnm{{Roshanineshat}}},
\oauthor{\binits{H.} \bsnm{{Rottmann}}},
\oauthor{\binits{A.L.} \bsnm{{Roy}}},
\oauthor{\binits{I.} \bsnm{{Ruiz}}},
\oauthor{\binits{C.} \bsnm{{Ruszczyk}}},
\oauthor{\binits{K.L.J.} \bsnm{{Rygl}}},
\oauthor{\binits{S.} \bsnm{{S{\'a}nchez}}},
\oauthor{\binits{D.} \bsnm{{S{\'a}nchez-Arg{\"u}elles}}},
\oauthor{\binits{M.} \bsnm{{S{\'a}nchez-Portal}}},
\oauthor{\binits{M.} \bsnm{{Sasada}}},
\oauthor{\binits{K.} \bsnm{{Satapathy}}},
\oauthor{\binits{T.} \bsnm{{Savolainen}}},
\oauthor{\binits{F.P.} \bsnm{{Schloerb}}},
\oauthor{\binits{J.} \bsnm{{Schonfeld}}},
\oauthor{\binits{K.-F.} \bsnm{{Schuster}}},
\oauthor{\binits{L.} \bsnm{{Shao}}},
\oauthor{\binits{Z.} \bsnm{{Shen}}},
\oauthor{\binits{D.} \bsnm{{Small}}},
\oauthor{\binits{B.W.} \bsnm{{Sohn}}},
\oauthor{\binits{J.} \bsnm{{SooHoo}}},
\oauthor{\binits{K.} \bsnm{{Souccar}}},
\oauthor{\binits{H.} \bsnm{{Sun}}},
\oauthor{\binits{F.} \bsnm{{Tazaki}}},
\oauthor{\binits{A.J.} \bsnm{{Tetarenko}}},
\oauthor{\binits{P.} \bsnm{{Tiede}}},
\oauthor{\binits{R.P.J.} \bsnm{{Tilanus}}},
\oauthor{\binits{M.} \bsnm{{Titus}}},
\oauthor{\binits{P.} \bsnm{{Torne}}},
\oauthor{\binits{T.} \bsnm{{Trent}}},
\oauthor{\binits{S.} \bsnm{{Trippe}}},
\oauthor{\binits{M.} \bsnm{{Turk}}},
\oauthor{\binits{H.J.} \bsnm{{van Langevelde}}},
\oauthor{\binits{D.R.} \bsnm{{van Rossum}}},
\oauthor{\binits{J.} \bsnm{{Vos}}},
\oauthor{\binits{J.} \bsnm{{Wagner}}},
\oauthor{\binits{D.} \bsnm{{Ward-Thompson}}},
\oauthor{\binits{J.} \bsnm{{Wardle}}},
\oauthor{\binits{J.} \bsnm{{Weintroub}}},
\oauthor{\binits{N.} \bsnm{{Wex}}},
\oauthor{\binits{R.} \bsnm{{Wharton}}},
\oauthor{\binits{K.} \bsnm{{Wiik}}},
\oauthor{\binits{G.} \bsnm{{Witzel}}},
\oauthor{\binits{M.F.} \bsnm{{Wondrak}}},
\oauthor{\binits{G.N.} \bsnm{{Wong}}},
\oauthor{\binits{Q.} \bsnm{{Wu}}},
\oauthor{\binits{P.} \bsnm{{Yamaguchi}}},
\oauthor{\binits{D.} \bsnm{{Yoon}}},
\oauthor{\binits{A.} \bsnm{{Young}}},
\oauthor{\binits{K.} \bsnm{{Young}}},
\oauthor{\binits{Z.} \bsnm{{Younsi}}},
\oauthor{\binits{F.} \bsnm{{Yuan}}},
\oauthor{\binits{Y.-F.} \bsnm{{Yuan}}},
\oauthor{\binits{J.A.} \bsnm{{Zensus}}},
\oauthor{\binits{S.} \bsnm{{Zhang}}},
\oauthor{\binits{S.-S.} \bsnm{{Zhao}}},
The event horizon telescope image of the quasar nrao 530
\textbf{943},
170
(2023).
doi:\doiurl{10.3847/1538-4357/acaea8}
\end{botherref}
\endbibitem

\bibitem[\protect\citeauthoryear{{Jovanovi{\'c}}
  et~al.}{2009}]{bib5:Jovanovic2009}
\begin{barticle}
\bauthor{\binits{P.} \bsnm{{Jovanovi{\'c}}}},
\bauthor{\binits{L.{\v{C}}.} \bsnm{{Popovi{\'c}}}},
\bauthor{\binits{S.} \bsnm{{Simi{\'c}}}},
\batitle{Influence of gravitational microlensing on broad absorption lines of
  qsos: The case of the fe k {\ensuremath{\alpha}} line}.
\bjtitle{New Astronomy Reviews}
\bvolume{53}(\bissue{7-10}),
\bfpage{156}--\blpage{161}
(\byear{2009}).
doi:\doiurl{10.1016/j.newar.2009.07.008}
\end{barticle}
\endbibitem

\bibitem[\protect\citeauthoryear{Kaspi et~al.}{2000}]{bib5:Kaspi2000}
\begin{barticle}
\bauthor{\binits{S.} \bsnm{Kaspi}},
\bauthor{\binits{P.S.} \bsnm{Smith}},
\bauthor{\binits{H.} \bsnm{Netzer}},
\bauthor{\binits{D.} \bsnm{Maoz}},
\bauthor{\binits{B.T.} \bsnm{Jannuzi}},
\bauthor{\binits{U.} \bsnm{Giveon1}},
\batitle{Reverberation measurements for 17 quasars and the size-mass-luminosity
  relations in active galactic nuclei}.
\bjtitle{The Astrophysical Journal}
\bvolume{533},
\bfpage{631}--\blpage{649}
(\byear{2000})
\end{barticle}
\endbibitem

\bibitem[\protect\citeauthoryear{Kaspi et~al.}{2007}]{bib5:Kaspi2007}
\begin{barticle}
\bauthor{\binits{S.} \bsnm{Kaspi}},
\bauthor{\binits{W.N.} \bsnm{Brandt}},
\bauthor{\binits{D.} \bsnm{Maoz}},
\bauthor{\binits{H.} \bsnm{Netzer}},
\bauthor{\binits{D.P.} \bsnm{Schneider}},
\bauthor{\binits{O.} \bsnm{Shemmer}},
\batitle{Reverberation mapping of high-luminosity quasars: First results}.
\bjtitle{The Astrophysical Journal}
\bvolume{659},
\bfpage{997}
(\byear{2007})
\end{barticle}
\endbibitem

\bibitem[\protect\citeauthoryear{{Katz}}{1976}]{bib5:Katz1976}
\begin{barticle}
\bauthor{\binits{J.I.} \bsnm{{Katz}}},
\batitle{Nonrelativistic compton scattering and models of quasars.}
\bjtitle{\apj}
\bvolume{206},
\bfpage{910}--\blpage{916}
(\byear{1976}).
doi:\doiurl{10.1086/154455}
\end{barticle}
\endbibitem

\bibitem[\protect\citeauthoryear{Katz et~al.}{1986}]{bib5:Katz1986}
\begin{barticle}
\bauthor{\binits{N.} \bsnm{Katz}},
\bauthor{\binits{S.} \bsnm{Balbus}},
\bauthor{\binits{B.} \bsnm{Paczynski}},
\batitle{Random scattering approach to gravitational microlensing}.
\bjtitle{Astrophysical Journal}
\bvolume{306},
\bfpage{2}
(\byear{1986})
\end{barticle}
\endbibitem

\bibitem[\protect\citeauthoryear{Kauffmann et~al.}{2003}]{bib5:Kauffmann2003}
\begin{barticle}
\bauthor{\binits{G.} \bsnm{Kauffmann}},
\bauthor{\binits{T.M.} \bsnm{Heckman}},
\bauthor{\binits{S.D.M.} \bsnm{White}},
\bauthor{\binits{S.} \bsnm{Charlot}},
\bauthor{\binits{C.} \bsnm{Tremonti}},
\bauthor{\binits{J.} \bsnm{Brinchmann}},
\bauthor{\binits{G.} \bsnm{Bruzual}},
\bauthor{\binits{E.W.} \bsnm{Peng}},
\bauthor{\binits{M.} \bsnm{Seibert}},
\bauthor{\binits{M.} \bsnm{Bernardi}},
\bauthor{\binits{M.} \bsnm{Blanton}},
\bauthor{\binits{J.} \bsnm{Brinkmann}},
\bauthor{\binits{F.} \bsnm{Castander}},
\bauthor{\binits{I.} \bsnm{Cs{\'a}bai}},
\bauthor{\binits{M.} \bsnm{Fukugita}},
\bauthor{\binits{Z.} \bsnm{Ivezic}},
\bauthor{\binits{J.A.} \bsnm{Munn}},
\bauthor{\binits{R.C.} \bsnm{Nichol}},
\bauthor{\binits{N.} \bsnm{Padmanabhan}},
\bauthor{\binits{A.R.} \bsnm{Thakar}},
\bauthor{\binits{D.H.} \bsnm{Weinberg}},
\bauthor{\binits{D.} \bsnm{York}},
\batitle{Stellar masses and star formation histories for 10 5 galaxies from the
  sloan digital sky survey}.
\bjtitle{Mon. Not. R. Astron. Soc}
\bvolume{341},
\bfpage{33}--\blpage{53}
(\byear{2003})
\end{barticle}
\endbibitem

\bibitem[\protect\citeauthoryear{{Kayser}}{1992}]{bib5:Kayser92}
\begin{bchapter}
\bauthor{\binits{R.} \bsnm{{Kayser}}},
\bctitle{Gravitational microlensing},
in \bbtitle{Gravitational Lenses},
ed. by \beditor{\binits{R.} \bsnm{Kayser}},
\beditor{\binits{T.} \bsnm{Schramm}},
\beditor{\binits{L.} \bsnm{Nieser}}
\bsertitle{Lecture Notes in Physics},
vol. \bseriesno{406}
(\bpublisher{Springer},
\blocation{Heidelberg}, \byear{1992}),
pp. \bfpage{143}--\blpage{155}
\end{bchapter}
\endbibitem

\bibitem[\protect\citeauthoryear{Kayser et~al.}{1986}]{bib5:Kayser1986}
\begin{barticle}
\bauthor{\binits{R.} \bsnm{Kayser}},
\bauthor{\binits{S.} \bsnm{Refsdal}},
\bauthor{\binits{R.} \bsnm{Stabell}},
\batitle{Astrophysical aplications of gravitational micro-lensing}.
\bjtitle{Astronomy and Astrophysics}
\bvolume{166},
\bfpage{36}
(\byear{1986})
\end{barticle}
\endbibitem

\bibitem[\protect\citeauthoryear{Kedziora et~al.}{2011}]{bib5:Kedziora2011}
\begin{barticle}
\bauthor{\binits{D.J.} \bsnm{Kedziora}},
\bauthor{\binits{H.} \bsnm{Garsden}},
\bauthor{\binits{G.F.} \bsnm{Lewis}},
\batitle{Gravitational microlensing as a probe of the electron-scattering
  region in q2237+0305}.
\bjtitle{Monthly Notices of the Royal Astronomical Society}
\bvolume{415},
\bfpage{1409}--\blpage{1418}
(\byear{2011}).
doi:\doiurl{10.1111/j.1365-2966.2011.18787.x}
\end{barticle}
\endbibitem

\bibitem[\protect\citeauthoryear{Keeton et~al.}{2006}]{bib5:Keeton2006}
\begin{barticle}
\bauthor{\binits{C.R.} \bsnm{Keeton}},
\bauthor{\binits{S.} \bsnm{Burles}},
\bauthor{\binits{P.L.} \bsnm{Schechter}},
\bauthor{\binits{J.} \bsnm{Wambsganss}},
\batitle{Differential microlensing of the continuum and broad emission lines in
  sdss j0924$+$0219 , the most anomalous lensed quasar 1}.
\bjtitle{The Astrophysical Journal}
\bvolume{639},
\bfpage{1}--\blpage{6}
(\byear{2006})
\end{barticle}
\endbibitem

\bibitem[\protect\citeauthoryear{{Khavinson} and
  {Neumann}}{2004}]{bib5:Khavinson2004}
\begin{botherref}
\oauthor{\binits{D.} \bsnm{{Khavinson}}},
\oauthor{\binits{G.} \bsnm{{Neumann}}},
On the number of zeros of certain rational harmonic functions.
arXiv Mathematics e-prints,
0401188
(2004).
doi:\doiurl{10.48550/arXiv.math/0401188}
\end{botherref}
\endbibitem

\bibitem[\protect\citeauthoryear{Kishimoto et~al.}{2007}]{bib5:Kishimoto2007}
\begin{barticle}
\bauthor{\binits{M.} \bsnm{Kishimoto}},
\bauthor{\binits{S.F.} \bsnm{H{\"o}nig}},
\bauthor{\binits{T.} \bsnm{Beckert}},
\bauthor{\binits{G.} \bsnm{Weigelt}},
\batitle{The innermost region of agn tori: Implications from the hst/nicmos
  type 1 point sources and near-ir reverberation}.
\bjtitle{Astronomy and Astrophysics}
\bvolume{476},
\bfpage{713}--\blpage{721}
(\byear{2007}).
doi:\doiurl{10.1051/0004-6361:20077911}
\end{barticle}
\endbibitem

\bibitem[\protect\citeauthoryear{Kishimoto et~al.}{2011a}]{bib5:Kishimoto2011a}
\begin{botherref}
\oauthor{\binits{M.} \bsnm{Kishimoto}},
\oauthor{\binits{S.F.} \bsnm{H{\"o}nig}},
\oauthor{\binits{R.} \bsnm{Antonucci}},
\oauthor{\binits{R.} \bsnm{Barvainis}},
\oauthor{\binits{T.} \bsnm{Kotani}},
\oauthor{\binits{K.R.W.} \bsnm{Tristram}},
\oauthor{\binits{G.} \bsnm{Weigelt}},
\oauthor{\binits{K.} \bsnm{Levin}},
The innermost dusty structure in active galactic nuclei as probed by the keck
  interferometer.
Astronomy and Astrophysics
\textbf{527}
(2011a).
doi:\doiurl{10.1051/0004-6361/201016054}
\end{botherref}
\endbibitem

\bibitem[\protect\citeauthoryear{Kishimoto et~al.}{2011b}]{bib5:Kishimoto2011b}
\begin{botherref}
\oauthor{\binits{M.} \bsnm{Kishimoto}},
\oauthor{\binits{S.F.} \bsnm{H{\"o}nig}},
\oauthor{\binits{R.} \bsnm{Antonucci}},
\oauthor{\binits{F.} \bsnm{Millour}},
\oauthor{\binits{K.R.W.} \bsnm{Tristram}},
\oauthor{\binits{G.} \bsnm{Weigelt}},
Mapping the radial structure of agn tori.
Astronomy and Astrophysics
\textbf{536}
(2011b).
doi:\doiurl{10.1051/0004-6361/201117367}
\end{botherref}
\endbibitem

\bibitem[\protect\citeauthoryear{Kochanek}{2004}]{bib5:Kochanek2004}
\begin{barticle}
\bauthor{\binits{C.S.} \bsnm{Kochanek}},
\batitle{Quantitative interpretation of quasar microlensing light curves}.
\bjtitle{The Astrophysical Journal}
\bvolume{605},
\bfpage{58}--\blpage{77}
(\byear{2004}).
doi:\doiurl{10.1086/382180}.
\burl{http://stacks.iop.org/0004-637X/605/i=1/a=58}
\end{barticle}
\endbibitem

\bibitem[\protect\citeauthoryear{{Koda} et~al.}{2014}]{bib5:Koda2014}
\begin{barticle}
\bauthor{\binits{J.} \bsnm{{Koda}}},
\bauthor{\binits{C.} \bsnm{{Blake}}},
\bauthor{\binits{T.} \bsnm{{Davis}}},
\bauthor{\binits{C.} \bsnm{{Magoulas}}},
\bauthor{\binits{C.M.} \bsnm{{Springob}}},
\bauthor{\binits{M.} \bsnm{{Scrimgeour}}},
\bauthor{\binits{A.} \bsnm{{Johnson}}},
\bauthor{\binits{G.B.} \bsnm{{Poole}}},
\bauthor{\binits{L.} \bsnm{{Staveley-Smith}}},
\batitle{Are peculiar velocity surveys competitive as a cosmological probe?}
\bjtitle{\mnras}
\bvolume{445}(\bissue{4}),
\bfpage{4267}--\blpage{4286}
(\byear{2014}).
doi:\doiurl{10.1093/mnras/stu1610}
\end{barticle}
\endbibitem

\bibitem[\protect\citeauthoryear{Kollmeier et~al.}{2006}]{bib5:Kollmeier2006}
\begin{barticle}
\bauthor{\binits{J.A.} \bsnm{Kollmeier}},
\bauthor{\binits{C.A.} \bsnm{Onken}},
\bauthor{\binits{C.S.} \bsnm{Kochanek}},
\bauthor{\binits{A.} \bsnm{Gould}},
\bauthor{\binits{D.H.} \bsnm{Weinberg}},
\bauthor{\binits{M.} \bsnm{Dietrich}},
\bauthor{\binits{R.} \bsnm{Cool}},
\bauthor{\binits{A.} \bsnm{Dey}},
\bauthor{\binits{D.J.} \bsnm{Eisenstein}},
\bauthor{\binits{B.T.} \bsnm{Jannuzi}},
\bauthor{\binits{E.L.} \bsnm{Floc'}},
\bauthor{\binits{D.} \bsnm{Stern}},
\batitle{Black hole masses and eddington ratios at 0.3 < z < 4 1}.
\bjtitle{The Astrophysical Journal}
\bvolume{648},
\bfpage{128}
(\byear{2006}).
\burl{http://www.noao.edu/noao/noaodeep/.}
\end{barticle}
\endbibitem

\bibitem[\protect\citeauthoryear{Koopmans and Bruyn}{2000}]{bib5:Koopmans2000}
\begin{barticle}
\bauthor{\binits{L.V.E.} \bsnm{Koopmans}},
\bauthor{\binits{A.G.D.} \bsnm{Bruyn}},
\batitle{Microlensing of multiply-imaged compact radio sources evidence for
  compact halo objects in the disk galaxy of b1600+434}.
\bjtitle{Astron. Astrophys}
\bvolume{358},
\bfpage{793}--\blpage{811}
(\byear{2000})
\end{barticle}
\endbibitem

\bibitem[\protect\citeauthoryear{Koptelova et~al.}{2007}]{bib5:Koptelova2007}
\begin{barticle}
\bauthor{\binits{E.} \bsnm{Koptelova}},
\bauthor{\binits{E.} \bsnm{Shimanovskaya}},
\bauthor{\binits{B.} \bsnm{Artamonov}},
\bauthor{\binits{A.} \bsnm{Yagola}},
\batitle{Analysis of the q2237+0305 light-curve variability with regularization
  technique}.
\bjtitle{Monthly Notices of the Royal Astronomical Society}
\bvolume{381},
\bfpage{1655}--\blpage{1662}
(\byear{2007}).
doi:\doiurl{10.1111/j.1365-2966.2007.12335.x}
\end{barticle}
\endbibitem

\bibitem[\protect\citeauthoryear{{Korista} et~al.}{1995}]{bib5:Korista1995}
\begin{barticle}
\bauthor{\binits{K.T.} \bsnm{{Korista}}},
\bauthor{\binits{D.} \bsnm{{Alloin}}},
\bauthor{\binits{P.} \bsnm{{Barr}}},
\bauthor{\binits{J.} \bsnm{{Clavel}}},
\bauthor{\binits{R.D.} \bsnm{{Cohen}}},
\bauthor{\binits{D.M.} \bsnm{{Crenshaw}}},
\bauthor{\binits{I.N.} \bsnm{{Evans}}},
\bauthor{\binits{K.} \bsnm{{Horne}}},
\bauthor{\binits{A.P.} \bsnm{{Koratkar}}},
\bauthor{\binits{G.A.} \bsnm{{Kriss}}},
\bauthor{\binits{J.H.} \bsnm{{Krolik}}},
\bauthor{\binits{M.A.} \bsnm{{Malkan}}},
\bauthor{\binits{S.L.} \bsnm{{Morris}}},
\bauthor{\binits{H.} \bsnm{{Netzer}}},
\bauthor{\binits{P.T.} \bsnm{{O'Brien}}},
\bauthor{\binits{B.M.} \bsnm{{Peterson}}},
\bauthor{\binits{G.A.} \bsnm{{Reichert}}},
\bauthor{\binits{P.M.} \bsnm{{Rodriguez-Pascual}}},
\bauthor{\binits{W.} \bsnm{{Wamsteker}}},
\bauthor{\binits{K.S.J.} \bsnm{{Anderson}}},
\bauthor{\binits{D.J.} \bsnm{{Axon}}},
\bauthor{\binits{E.} \bsnm{{Benitez}}},
\bauthor{\binits{P.} \bsnm{{Berlind}}},
\bauthor{\binits{R.} \bsnm{{Bertram}}},
\bauthor{\binits{J.} \bsnm{{Blackwell}} \bsuffix{J.~H.}},
\bauthor{\binits{N.G.} \bsnm{{Bochkarev}}},
\bauthor{\binits{C.} \bsnm{{Boisson}}},
\bauthor{\binits{M.} \bsnm{{Carini}}},
\bauthor{\binits{R.} \bsnm{{Carrillo}}},
\bauthor{\binits{T.E.} \bsnm{{Carone}}},
\bauthor{\binits{F.-Z.} \bsnm{{Cheng}}},
\bauthor{\binits{J.A.} \bsnm{{Christensen}}},
\bauthor{\binits{K.K.} \bsnm{{Chuvaev}}},
\bauthor{\binits{M.} \bsnm{{Dietrich}}},
\bauthor{\binits{J.J.} \bsnm{{Dokter}}},
\bauthor{\binits{V.} \bsnm{{Doroshenko}}},
\bauthor{\binits{D.} \bsnm{{Dultzin-Hacyan}}},
\bauthor{\binits{M.N.} \bsnm{{England}}},
\bauthor{\binits{B.R.} \bsnm{{Espey}}},
\bauthor{\binits{A.V.} \bsnm{{Filippenko}}},
\bauthor{\binits{C.M.} \bsnm{{Gaskell}}},
\bauthor{\binits{M.R.} \bsnm{{Goad}}},
\bauthor{\binits{L.C.} \bsnm{{Ho}}},
\bauthor{\binits{J.P.} \bsnm{{Huchra}}},
\bauthor{\binits{X.J.} \bsnm{{Jiang}}},
\bauthor{\binits{S.} \bsnm{{Kaspi}}},
\bauthor{\binits{W.} \bsnm{{Kollatschny}}},
\bauthor{\binits{A.} \bsnm{{Laor}}},
\bauthor{\binits{J.-P.} \bsnm{{Luminet}}},
\bauthor{\binits{G.M.} \bsnm{{MacAlpine}}},
\bauthor{\binits{J.W.} \bsnm{{MacKenty}}},
\bauthor{\binits{Y.F.} \bsnm{{Malkov}}},
\bauthor{\binits{D.} \bsnm{{Maoz}}},
\bauthor{\binits{P.G.} \bsnm{{Martin}}},
\bauthor{\binits{T.} \bsnm{{Matheson}}},
\bauthor{\binits{B.} \bsnm{{McCollum}}},
\bauthor{\binits{N.} \bsnm{{Merkulova}}},
\bauthor{\binits{L.} \bsnm{{Metik}}},
\bauthor{\binits{M.} \bsnm{{Mignoli}}},
\bauthor{\binits{H.R.} \bsnm{{Miller}}},
\bauthor{\binits{M.G.} \bsnm{{Pastoriza}}},
\bauthor{\binits{D.} \bsnm{{Pelat}}},
\bauthor{\binits{J.} \bsnm{{Penfold}}},
\bauthor{\binits{M.} \bsnm{{Perez}}},
\bauthor{\binits{G.C.} \bsnm{{Perola}}},
\bauthor{\binits{J.L.} \bsnm{{Persaud}}},
\bauthor{\binits{J.} \bsnm{{Peters}}},
\bauthor{\binits{R.} \bsnm{{Pitts}}},
\bauthor{\binits{R.W.} \bsnm{{Pogge}}},
\bauthor{\binits{I.} \bsnm{{Pronik}}},
\bauthor{\binits{V.I.} \bsnm{{Pronik}}},
\bauthor{\binits{R.L.} \bsnm{{Ptak}}},
\bauthor{\binits{L.} \bsnm{{Rawley}}},
\bauthor{\binits{M.C.} \bsnm{{Recondo-Gonzalez}}},
\bauthor{\binits{J.M.} \bsnm{{Rodriguez-Espinosa}}},
\bauthor{\binits{W.} \bsnm{{Romanishin}}},
\bauthor{\binits{A.C.} \bsnm{{Sadun}}},
\bauthor{\binits{I.} \bsnm{{Salamanca}}},
\bauthor{\binits{M.} \bsnm{{Santos-Lleo}}},
\bauthor{\binits{K.} \bsnm{{Sekiguchi}}},
\bauthor{\binits{S.G.} \bsnm{{Sergeev}}},
\bauthor{\binits{A.I.} \bsnm{{Shapovalova}}},
\bauthor{\binits{J.C.} \bsnm{{Shields}}},
\bauthor{\binits{C.} \bsnm{{Shrader}}},
\bauthor{\binits{J.M.} \bsnm{{Shull}}},
\bauthor{\binits{N.A.} \bsnm{{Silbermann}}},
\bauthor{\binits{M.L.} \bsnm{{Sitko}}},
\bauthor{\binits{D.R.} \bsnm{{Skillman}}},
\bauthor{\binits{H.A.} \bsnm{{Smith}}},
\bauthor{\binits{S.M.} \bsnm{{Smith}}},
\bauthor{\binits{M.A.J.} \bsnm{{Snijders}}},
\bauthor{\binits{L.S.} \bsnm{{Sparke}}},
\bauthor{\binits{G.M.} \bsnm{{Stirpe}}},
\bauthor{\binits{R.E.} \bsnm{{Stoner}}},
\bauthor{\binits{W.-H.} \bsnm{{Sun}}},
\bauthor{\binits{U.} \bsnm{{Thiele}}},
\bauthor{\binits{S.} \bsnm{{Tokarz}}},
\bauthor{\binits{Z.I.} \bsnm{{Tsvetanov}}},
\bauthor{\binits{D.A.} \bsnm{{Turnshek}}},
\bauthor{\binits{S.} \bsnm{{Veilleux}}},
\bauthor{\binits{R.M.} \bsnm{{Wagner}}},
\bauthor{\binits{S.J.} \bsnm{{Wagner}}},
\bauthor{\binits{I.} \bsnm{{Wanders}}},
\bauthor{\binits{T.} \bsnm{{Wang}}},
\bauthor{\binits{W.F.} \bsnm{{Welsh}}},
\bauthor{\binits{R.J.} \bsnm{{Weymann}}},
\bauthor{\binits{R.J.} \bsnm{{White}}},
\bauthor{\binits{B.J.} \bsnm{{Wilkes}}},
\bauthor{\binits{B.J.} \bsnm{{Wills}}},
\bauthor{\binits{C.} \bsnm{{Winge}}},
\bauthor{\binits{H.} \bsnm{{Wu}}},
\bauthor{\binits{Z.L.} \bsnm{{Zou}}},
\batitle{Steps toward determination of the size and structure of the broad-line
  region in active galactic nuclei. viii. an intensive hst, iue, and
  ground-based study of ngc 5548}.
\bjtitle{The Astrophysical Journal Supplement Series}
\bvolume{97},
\bfpage{13}
(\byear{1995})
\end{barticle}
\endbibitem

\bibitem[\protect\citeauthoryear{Korista and Goad}{2001}]{bib5:Korista2001}
\begin{barticle}
\bauthor{\binits{K.T.} \bsnm{Korista}},
\bauthor{\binits{M.R.} \bsnm{Goad}},
\batitle{The variable diffuse continuum emission of broad-line clouds}.
\bjtitle{The Astrophysical Journal}
\bvolume{553},
\bfpage{695}--\blpage{708}
(\byear{2001})
\end{barticle}
\endbibitem

\bibitem[\protect\citeauthoryear{Koshida et~al.}{2014}]{bib5:Koshida2014}
\begin{botherref}
\oauthor{\binits{S.} \bsnm{Koshida}},
\oauthor{\binits{T.} \bsnm{Minezaki}},
\oauthor{\binits{Y.} \bsnm{Yoshii}},
\oauthor{\binits{Y.} \bsnm{Kobayashi}},
\oauthor{\binits{Y.} \bsnm{Sakata}},
\oauthor{\binits{S.} \bsnm{Sugawara}},
\oauthor{\binits{K.} \bsnm{Enya}},
\oauthor{\binits{M.} \bsnm{Suganuma}},
\oauthor{\binits{H.} \bsnm{Tomita}},
\oauthor{\binits{T.} \bsnm{Aoki}},
\oauthor{\binits{B.A.} \bsnm{Peterson}},
Reverberation measurements of the inner radius of the dust torus in 17 seyfert
  galaxies.
Astrophysical Journal
\textbf{788}
(2014).
doi:\doiurl{10.1088/0004-637X/788/2/159}
\end{botherref}
\endbibitem

\bibitem[\protect\citeauthoryear{{Krawczynski} and
  {Chartas}}{2017}]{bib5:Krawczynski2017}
\begin{barticle}
\bauthor{\binits{H.} \bsnm{{Krawczynski}}},
\bauthor{\binits{G.} \bsnm{{Chartas}}},
\batitle{Simulations of the fe k{\ensuremath{\alpha}} energy spectra from
  gravitationally microlensed quasars}.
\bjtitle{\apj}
\bvolume{843}(\bissue{2}),
\bfpage{118}
(\byear{2017}).
doi:\doiurl{10.3847/1538-4357/aa7896}
\end{barticle}
\endbibitem

\bibitem[\protect\citeauthoryear{Kundic and Wambsganss}{1993}]{bib5:Kundic1993}
\begin{barticle}
\bauthor{\binits{T.} \bsnm{Kundic}},
\bauthor{\binits{J.} \bsnm{Wambsganss}},
\batitle{Gravitational microlensing: the effect of random motion of individual
  stars in the lensing galaxy}.
\bjtitle{The Astrophysical Journal}
\bvolume{404},
\bfpage{455}
(\byear{1993}).
doi:\doiurl{10.1192/bjp.111.479.1009-a}
\end{barticle}
\endbibitem

\bibitem[\protect\citeauthoryear{Lamer et~al.}{2006}]{bib5:Lamer2006}
\begin{barticle}
\bauthor{\binits{G.} \bsnm{Lamer}},
\bauthor{\binits{A.} \bsnm{Schwope}},
\bauthor{\binits{L.} \bsnm{Wisotzki}},
\bauthor{\binits{L.} \bsnm{Christensen}},
\batitle{Strange magnification pattern in the large separation lens sdss
  j1004+4112 from optical to x-rays}.
\bjtitle{Astronomy and Astrophysics}
\bvolume{454},
\bfpage{493}--\blpage{501}
(\byear{2006}).
doi:\doiurl{10.1051/0004-6361:20064934}
\end{barticle}
\endbibitem

\bibitem[\protect\citeauthoryear{{Lan{\v{c}}ov{\'a}}
  et~al.}{2019}]{bib5:Lancova2019}
\begin{barticle}
\bauthor{\binits{D.} \bsnm{{Lan{\v{c}}ov{\'a}}}},
\bauthor{\binits{D.} \bsnm{{Abarca}}},
\bauthor{\binits{W.} \bsnm{{Klu{\'z}niak}}},
\bauthor{\binits{M.} \bsnm{{Wielgus}}},
\bauthor{\binits{A.} \bsnm{{S{\k{a}}dowski}}},
\bauthor{\binits{R.} \bsnm{{Narayan}}},
\bauthor{\binits{J.} \bsnm{{Schee}}},
\bauthor{\binits{G.} \bsnm{{T{\"o}r{\"o}k}}},
\bauthor{\binits{M.} \bsnm{{Abramowicz}}},
\batitle{Puffy accretion disks: Sub-eddington, optically thick, and stable}.
\bjtitle{The Astrophysical Journal}
\bvolume{884},
\bfpage{37}
(\byear{2019}).
doi:\doiurl{10.3847/2041-8213/ab48f5}
\end{barticle}
\endbibitem

\bibitem[\protect\citeauthoryear{{Laor}}{1991}]{bib5:Laor1991}
\begin{barticle}
\bauthor{\binits{A.} \bsnm{{Laor}}},
\batitle{Line profiles from a disk around a rotating black hole}.
\bjtitle{\apj}
\bvolume{376},
\bfpage{90}
(\byear{1991}).
doi:\doiurl{10.1086/170257}
\end{barticle}
\endbibitem

\bibitem[\protect\citeauthoryear{Laor and Behar}{2008}]{bib5:Laor2008}
\begin{barticle}
\bauthor{\binits{A.} \bsnm{Laor}},
\bauthor{\binits{E.} \bsnm{Behar}},
\batitle{On the origin of radio emission in radio-quiet quasars}.
\bjtitle{Monthly Notices of the Royal Astronomical Society}
\bvolume{390},
\bfpage{847}--\blpage{862}
(\byear{2008}).
doi:\doiurl{10.1111/j.1365-2966.2008.13806.x}
\end{barticle}
\endbibitem

\bibitem[\protect\citeauthoryear{Lasota}{2023}]{bib5:Lasota2023}
\begin{bbook}
\bauthor{\binits{J.P.} \bsnm{Lasota}},
\bbtitle{Active galactic nuclei}
(\bpublisher{wiley}, \blocation{???}, \byear{2023}),
pp. \bfpage{1}--\blpage{320}.
\bisbn{9781394163724}.
doi:\doiurl{10.1002/9781394163724}
\end{bbook}
\endbibitem

\bibitem[\protect\citeauthoryear{Lasota et~al.}{2015}]{bib5:Lasota2015}
\begin{botherref}
\oauthor{\binits{J.P.} \bsnm{Lasota}},
\oauthor{\binits{A.R.} \bsnm{King}},
\oauthor{\binits{G.} \bsnm{Dubus}},
X-ray transients: Hyper- or hypo-luminous?
Astrophysical Journal Letters
\textbf{801}
(2015).
doi:\doiurl{10.1088/2041-8205/801/1/L4}
\end{botherref}
\endbibitem

\bibitem[\protect\citeauthoryear{Lawrence}{2012}]{bib5:Lawrence2012}
\begin{barticle}
\bauthor{\binits{A.} \bsnm{Lawrence}},
\batitle{The uv peak in active galactic nuclei: A false continuum from blurred
  reflection?}
\bjtitle{Monthly Notices of the Royal Astronomical Society}
\bvolume{423},
\bfpage{451}--\blpage{463}
(\byear{2012}).
doi:\doiurl{10.1111/j.1365-2966.2012.20889.x}
\end{barticle}
\endbibitem

\bibitem[\protect\citeauthoryear{Lawther et~al.}{2023}]{bib5:Lawther2023}
\begin{barticle}
\bauthor{\binits{D.} \bsnm{Lawther}},
\bauthor{\binits{M.} \bsnm{Vestergaard}},
\bauthor{\binits{S.} \bsnm{Raimundo}},
\bauthor{\binits{J.Y.} \bsnm{Koay}},
\bauthor{\binits{B.M.} \bsnm{Peterson}},
\bauthor{\binits{X.} \bsnm{Fan}},
\bauthor{\binits{D.} \bsnm{Grupe}},
\bauthor{\binits{S.} \bsnm{Mathur}},
\batitle{Flares in the changing look agn mrk 590 - i. the uv response to x-ray
  outbursts suggests a more complex reprocessing geometry than a standard
  disc}.
\bjtitle{Monthly Notices of the Royal Astronomical Society}
\bvolume{519},
\bfpage{3903}--\blpage{3922}
(\byear{2023}).
doi:\doiurl{10.1093/mnras/stac3515}
\end{barticle}
\endbibitem

\bibitem[\protect\citeauthoryear{{Ledvina} et~al.}{2018}]{bib5:Ledvina2018}
\begin{barticle}
\bauthor{\binits{L.} \bsnm{{Ledvina}}},
\bauthor{\binits{D.} \bsnm{{Heyrovsk{\'y}}}},
\bauthor{\binits{M.} \bsnm{{Dov{\v{c}}iak}}},
\batitle{X-ray line profile variations during quasar microlensing}.
\bjtitle{\apj}
\bvolume{863}(\bissue{1}),
\bfpage{66}
(\byear{2018}).
doi:\doiurl{10.3847/1538-4357/aad0f3}
\end{barticle}
\endbibitem

\bibitem[\protect\citeauthoryear{Lehar et~al.}{2000}]{bib5:Lehar2000}
\begin{barticle}
\bauthor{\binits{J.} \bsnm{Lehar}},
\bauthor{\binits{E.E.} \bsnm{Falco}},
\bauthor{\binits{C.S.} \bsnm{Kochanek}},
\bauthor{\binits{B.A.} \bsnm{McLeod}},
\bauthor{\binits{J.A.} \bsnm{Mu{\~n}oz}},
\bauthor{\binits{C.D.} \bsnm{Impey}},
\bauthor{\binits{H.-W.} \bsnm{Rix}},
\bauthor{\binits{C.R.} \bsnm{Keeton}},
\bauthor{\binits{C.Y.} \bsnm{Peng}},
\batitle{Hubble space telescope observations of 10 two-image gravitational
  lenses}.
\bjtitle{The Astrophysical Journal}
\bvolume{536},
\bfpage{584}
(\byear{2000})
\end{barticle}
\endbibitem

\bibitem[\protect\citeauthoryear{Lewis and Belle}{1998}]{bib5:Lewis1998b}
\begin{barticle}
\bauthor{\binits{G.F.} \bsnm{Lewis}},
\bauthor{\binits{K.E.} \bsnm{Belle}},
\batitle{Microlensing of broad absorption line quasars}.
\bjtitle{Monthly Notices of the Royal Astronomical Society}
\bvolume{297},
\bfpage{69}--\blpage{76}
(\byear{1998})
\end{barticle}
\endbibitem

\bibitem[\protect\citeauthoryear{Lewis and Irwin}{1995}]{bib5:Lewis1995}
\begin{barticle}
\bauthor{\binits{G.F.} \bsnm{Lewis}},
\bauthor{\binits{M.J.} \bsnm{Irwin}},
\batitle{The statistics of micolensing light curves - i. amplification
  probability distributions}.
\bjtitle{Monthly Notices of the Royal Astronomical Society}
\bvolume{276},
\bfpage{103}
(\byear{1995})
\end{barticle}
\endbibitem

\bibitem[\protect\citeauthoryear{Lewis et~al.}{1993}]{bib5:Lewis1993}
\begin{barticle}
\bauthor{\binits{G.F.} \bsnm{Lewis}},
\bauthor{\binits{J.} \bsnm{Miralda-Escud{\'e}}},
\bauthor{\binits{D.C.} \bsnm{Richardson}},
\bauthor{\binits{J.} \bsnm{Wambsganss}},
\batitle{Microlensing light curves: a new and efficient numerical method}.
\bjtitle{Monthly Notices of the Royal Astronomical Society}
\bvolume{261},
\bfpage{647}
(\byear{1993})
\end{barticle}
\endbibitem

\bibitem[\protect\citeauthoryear{Lewis et~al.}{1998}]{bib5:Lewis1998a}
\begin{barticle}
\bauthor{\binits{G.F.} \bsnm{Lewis}},
\bauthor{\binits{M.J.} \bsnm{Irwin}},
\bauthor{\binits{P.C.} \bsnm{Hewett}},
\bauthor{\binits{C.B.} \bsnm{Foltz}},
\batitle{Microlensing-induced spectral variability in q2237$+$0305}.
\bjtitle{Monthly Notices of the Royal Astronomical Society}
\bvolume{295},
\bfpage{573}--\blpage{586}
(\byear{1998})
\end{barticle}
\endbibitem

\bibitem[\protect\citeauthoryear{Lewis}{2020}]{bib5:Lewis2020}
\begin{barticle}
\bauthor{\binits{G.F.} \bsnm{Lewis}},
\batitle{Gravitational microlensing time delays at high optical depth: Image
  parities and the temporal properties of fast radio bursts}.
\bjtitle{Monthly Notices of the Royal Astronomical Society}
\bvolume{497},
\bfpage{1583}--\blpage{1589}
(\byear{2020}).
doi:\doiurl{10.1093/mnras/staa2044}
\end{barticle}
\endbibitem

\bibitem[\protect\citeauthoryear{{Lewis} and {Ibata}}{2004}]{Lewis2004}
\begin{barticle}
\bauthor{\binits{G.F.} \bsnm{{Lewis}}},
\bauthor{\binits{R.A.} \bsnm{{Ibata}}},
\batitle{Gravitational microlensing of quasar broad-line regions at large
  optical depths}.
\bjtitle{Monthly Notices of the Royal Astronomical Society}
\bvolume{348}(\bissue{1}),
\bfpage{24}--\blpage{33}
(\byear{2004}).
doi:\doiurl{10.1111/j.1365-2966.2004.07349.x}
\end{barticle}
\endbibitem

\bibitem[\protect\citeauthoryear{{Lewis} and {Ibata}}{1998}]{bib5:Lewis1998c}
\begin{barticle}
\bauthor{\binits{G.F.} \bsnm{{Lewis}}},
\bauthor{\binits{R.A.} \bsnm{{Ibata}}},
\batitle{Quasar image shifts resulting from gravitational microlensing}.
\bjtitle{\apj}
\bvolume{501}(\bissue{2}),
\bfpage{478}--\blpage{485}
(\byear{1998}).
doi:\doiurl{10.1086/305860}
\end{barticle}
\endbibitem

\bibitem[\protect\citeauthoryear{Li et~al.}{2019}]{bib5:Li2019}
\begin{barticle}
\bauthor{\binits{Y.P.} \bsnm{Li}},
\bauthor{\binits{F.} \bsnm{Yuan}},
\bauthor{\binits{X.} \bsnm{Dai}},
\batitle{Reconciling the quasar microlensing disc size problem with a wind
  model of active galactic nucleus}.
\bjtitle{Monthly Notices of the Royal Astronomical Society}
\bvolume{483},
\bfpage{2275}--\blpage{2281}
(\byear{2019}).
doi:\doiurl{10.1093/mnras/sty3245}
\end{barticle}
\endbibitem

\bibitem[\protect\citeauthoryear{Li et~al.}{2013}]{bib5:Li2013}
\begin{botherref}
\oauthor{\binits{Y.R.} \bsnm{Li}},
\oauthor{\binits{J.M.} \bsnm{Wang}},
\oauthor{\binits{L.C.} \bsnm{Ho}},
\oauthor{\binits{P.} \bsnm{Du}},
\oauthor{\binits{J.M.} \bsnm{Bai}},
A bayesian approach to estimate the size and structure of the broad-line region
  in active galactic nuclei using reverberation mapping data.
Astrophysical Journal
\textbf{779}
(2013).
doi:\doiurl{10.1088/0004-637X/779/2/110}
\end{botherref}
\endbibitem

\bibitem[\protect\citeauthoryear{{Lightman} and
  {White}}{1988}]{bib5:Lightman1988}
\begin{barticle}
\bauthor{\binits{A.P.} \bsnm{{Lightman}}},
\bauthor{\binits{T.R.} \bsnm{{White}}},
\batitle{Effects of cold matter in active galactic nuclei: A broad hump in the
  x-ray spectra}.
\bjtitle{\apj}
\bvolume{335},
\bfpage{57}
(\byear{1988}).
doi:\doiurl{10.1086/166905}
\end{barticle}
\endbibitem

\bibitem[\protect\citeauthoryear{Luhtaru et~al.}{2021}]{bib5:Luhtaru2021}
\begin{barticle}
\bauthor{\binits{R.} \bsnm{Luhtaru}},
\bauthor{\binits{P.L.} \bsnm{Schechter}},
\bauthor{\binits{K.M.} \bparticle{de} \bsnm{Soto}},
\batitle{What makes quadruply lensed quasars quadruple?}
\bjtitle{The Astrophysical Journal}
\bvolume{915},
\bfpage{4}
(\byear{2021}).
doi:\doiurl{10.3847/1538-4357/abfda1}
\end{barticle}
\endbibitem

\bibitem[\protect\citeauthoryear{Lynden-Bell}{1969}]{bib5:LyndenBell1969}
\begin{barticle}
\bauthor{\binits{D.} \bsnm{Lynden-Bell}},
\batitle{Galactic nuclei as collapsed old quasars}.
\bjtitle{Nature}
\bvolume{223},
\bfpage{690}--\blpage{694}
(\byear{1969}).
doi:\doiurl{10.1038/223690a0}
\end{barticle}
\endbibitem

\bibitem[\protect\citeauthoryear{MacLeod et~al.}{2015}]{bib5:MacLeod2015}
\begin{barticle}
\bauthor{\binits{C.L.} \bsnm{MacLeod}},
\bauthor{\binits{C.W.} \bsnm{Morgan}},
\bauthor{\binits{A.M.} \bsnm{Mosquera}},
\bauthor{\binits{C.S.} \bsnm{Kochanek}},
\bauthor{\binits{M.} \bsnm{Tewes}},
\bauthor{\binits{F.} \bsnm{Courbin}},
\bauthor{\binits{G.} \bsnm{Meylan}},
\bauthor{\binits{B.} \bsnm{Chen}},
\bauthor{\binits{X.} \bsnm{Dai}},
\bauthor{\binits{G.} \bsnm{Chartas}},
\batitle{A consistent picture emerges: A compact x-ray continuum emission
  region in the gravitationally lensed quasar sdss j0924$+$0219}.
\bjtitle{Astrophysical Journal}
\bvolume{806},
\bfpage{258}
(\byear{2015}).
doi:\doiurl{10.1088/0004-637X/806/2/258}.
\burl{http://dx.doi.org/10.1088/0004-637X/806/2/258}
\end{barticle}
\endbibitem

\bibitem[\protect\citeauthoryear{{Makarov} and
  {Secrest}}{2022}]{bib5:Makarov2022}
\begin{barticle}
\bauthor{\binits{V.V.} \bsnm{{Makarov}}},
\bauthor{\binits{N.J.} \bsnm{{Secrest}}},
\batitle{Quasars with proper motions and the link to double and multiple agns}.
\bjtitle{\apj}
\bvolume{933}(\bissue{1}),
\bfpage{28}
(\byear{2022}).
doi:\doiurl{10.3847/1538-4357/ac7047}
\end{barticle}
\endbibitem

\bibitem[\protect\citeauthoryear{Mao}{1993}]{bib5:Mao1993}
\begin{barticle}
\bauthor{\binits{S.} \bsnm{Mao}},
\batitle{Gravitational microlensing of gamma-ray bursts}.
\bjtitle{The Astrophysical Journal}
\bvolume{402},
\bfpage{382}--\blpage{386}
(\byear{1993})
\end{barticle}
\endbibitem

\bibitem[\protect\citeauthoryear{Mao and Schneider}{1998}]{bib5:Mao1998}
\begin{barticle}
\bauthor{\binits{S.} \bsnm{Mao}},
\bauthor{\binits{P.} \bsnm{Schneider}},
\batitle{Evidence for substructure in lens galaxies?}
\bjtitle{Monthly Notices of the Royal Astronomical Society}
\bvolume{295},
\bfpage{587}--\blpage{594}
(\byear{1998})
\end{barticle}
\endbibitem

\bibitem[\protect\citeauthoryear{Marconi and Hunt}{2003}]{bib5:Marconi2003}
\begin{barticle}
\bauthor{\binits{A.} \bsnm{Marconi}},
\bauthor{\binits{L.K.} \bsnm{Hunt}},
\batitle{The relation between black hole mass, bulge mass, and near-infrared
  luminosity}.
\bjtitle{The Astrophysical Journal}
\bvolume{589},
\bfpage{21}--\blpage{24}
(\byear{2003}).
\burl{http://leda.univ-lyon1.fr.}
\end{barticle}
\endbibitem

\bibitem[\protect\citeauthoryear{Marziani et~al.}{2001}]{bib5:Marziani2001}
\begin{barticle}
\bauthor{\binits{P.} \bsnm{Marziani}},
\bauthor{\binits{J.W.} \bsnm{Sulentic}},
\bauthor{\binits{T.} \bsnm{Zwitter}},
\bauthor{\binits{D.} \bsnm{Dultzin-Hacyan}},
\bauthor{},
\bauthor{\binits{M.} \bsnm{Calvani1}},
\batitle{Searching for the physical drivers of the eigenvector 1 correlation
  space}.
\bjtitle{The Astrophysical Journal}
\bvolume{558},
\bfpage{553}--\blpage{560}
(\byear{2001})
\end{barticle}
\endbibitem

\bibitem[\protect\citeauthoryear{{McGill} et~al.}{2023}]{bib5:McGill2023}
\begin{barticle}
\bauthor{\binits{P.} \bsnm{{McGill}}},
\bauthor{\binits{J.} \bsnm{{Anderson}}},
\bauthor{\binits{S.} \bsnm{{Casertano}}},
\bauthor{\binits{K.C.} \bsnm{{Sahu}}},
\bauthor{\binits{P.} \bsnm{{Bergeron}}},
\bauthor{\binits{S.} \bsnm{{Blouin}}},
\bauthor{\binits{P.} \bsnm{{Dufour}}},
\bauthor{\binits{L.C.} \bsnm{{Smith}}},
\bauthor{\binits{N.W.} \bsnm{{Evans}}},
\bauthor{\binits{V.} \bsnm{{Belokurov}}},
\bauthor{\binits{R.L.} \bsnm{{Smart}}},
\bauthor{\binits{A.} \bsnm{{Bellini}}},
\bauthor{\binits{A.} \bsnm{{Calamida}}},
\bauthor{\binits{M.} \bsnm{{Dominik}}},
\bauthor{\binits{N.} \bsnm{{Kains}}},
\bauthor{\binits{J.} \bsnm{{Kl{\"u}ter}}},
\bauthor{\binits{M.B.} \bsnm{{Nielsen}}},
\bauthor{\binits{J.} \bsnm{{Wambsganss}}},
\batitle{First semi-empirical test of the white dwarf mass-radius relationship
  using a single white dwarf via astrometric microlensing}.
\bjtitle{\mnras}
\bvolume{520}(\bissue{1}),
\bfpage{259}--\blpage{280}
(\byear{2023}).
doi:\doiurl{10.1093/mnras/stac3532}
\end{barticle}
\endbibitem

\bibitem[\protect\citeauthoryear{Mediavilla et~al.}{2005}]{bib5:Mediavilla2005}
\begin{barticle}
\bauthor{\binits{E.} \bsnm{Mediavilla}},
\bauthor{\binits{J.A.} \bsnm{Mu{\~n}oz}},
\bauthor{\binits{C.S.} \bsnm{Kochanek}},
\bauthor{\binits{E.E.} \bsnm{Falco}},
\bauthor{\binits{S.} \bsnm{Arribas}},
\bauthor{\binits{V.} \bsnm{Motta}},
\batitle{The first precise determination of an optical-far-ultraviolet
  extinction curve beyond the local group (z ¼ 0:83)}.
\bjtitle{The Astrophysical Journal}
\bvolume{619},
\bfpage{749}
(\byear{2005})
\end{barticle}
\endbibitem

\bibitem[\protect\citeauthoryear{Mediavilla et~al.}{2006}]{bib5:Mediavilla2006}
\begin{barticle}
\bauthor{\binits{E.} \bsnm{Mediavilla}},
\bauthor{\binits{J.A.} \bsnm{Munoz}},
\bauthor{\binits{P.} \bsnm{Lopez}},
\bauthor{\binits{T.} \bsnm{Mediavilla}},
\bauthor{\binits{C.} \bsnm{Abajas}},
\bauthor{\binits{C.} \bsnm{Gonzalez{-}Morcillo}},
\bauthor{\binits{R.} \bsnm{{Gil{-}Merino}}},
\batitle{A fast and very accurate approach to the computation of microlensing
  magnification patterns based on inverse polygon mapping}.
\bjtitle{The Astrophysical Journal}
\bvolume{653},
\bfpage{942}--\blpage{953}
(\byear{2006}).
doi:\doiurl{10.1086/508796}.
\burl{http://stacks.iop.org/0004-637X/653/i=2/a=942}
\end{barticle}
\endbibitem

\bibitem[\protect\citeauthoryear{Mediavilla et~al.}{2009}]{bib5:Mediavilla2009}
\begin{barticle}
\bauthor{\binits{E.} \bsnm{Mediavilla}},
\bauthor{\binits{J.A.} \bsnm{Mu{\~n}oz}},
\bauthor{\binits{E.} \bsnm{Falco}},
\bauthor{\binits{V.} \bsnm{Motta}},
\bauthor{\binits{E.} \bsnm{Guerras}},
\bauthor{\binits{H.} \bsnm{Canovas}},
\bauthor{\binits{C.} \bsnm{Jean}},
\bauthor{\binits{A.} \bsnm{Oscoz}},
\bauthor{\binits{A.M.} \bsnm{Mosquera}},
\batitle{Microlensing-based estimate of the mass fraction in compact objects in
  lens galaxies}.
\bjtitle{The Astrophysical Journal}
\bvolume{706},
\bfpage{1451}--\blpage{1462}
(\byear{2009}).
doi:\doiurl{10.1088/0004-637X/706/2/1451}.
\burl{http://stacks.iop.org/0004-637X/706/i=2/a=1451?key=crossref.bf03148af2851be46375bd6def511382}
\end{barticle}
\endbibitem

\bibitem[\protect\citeauthoryear{Mediavilla
  et~al.}{2011}]{bib5:Mediavilla2011b}
\begin{barticle}
\bauthor{\binits{E.} \bsnm{Mediavilla}},
\bauthor{\binits{T.} \bsnm{Mediavilla}},
\bauthor{\binits{J.A.} \bsnm{Mu{\~n}oz}},
\bauthor{\binits{O.} \bsnm{Ariza}},
\bauthor{\binits{P.} \bsnm{Lopez}},
\bauthor{\binits{C.} \bsnm{Gonzalez{-}Morcillo}},
\bauthor{\binits{J.} \bsnm{Jim{\'e}nez-Vicente}},
\batitle{New developments on inverse polygon mapping to calculate gravitational
  lensing magnification maps: Optimized computations}.
\bjtitle{The Astrophysical Journal}
\bvolume{741},
\bfpage{42}
(\byear{2011}).
doi:\doiurl{10.1088/0004-637X/741/1/42}.
\burl{http://stacks.iop.org/0004-637X/741/i=1/a=42?key=crossref.957486f84599f12bc6b40d9fbe3d715d}
\end{barticle}
\endbibitem

\bibitem[\protect\citeauthoryear{Mediavilla et~al.}{2015}]{bib5:Mediavilla2015}
\begin{botherref}
\oauthor{\binits{E.} \bsnm{Mediavilla}},
\oauthor{\binits{J.} \bsnm{Jim{\'e}nez-Vicente}},
\oauthor{\binits{J.A.} \bsnm{Mu{\~n}oz}},
\oauthor{\binits{T.} \bsnm{Mediavilla}},
Resolving the innermost region of the accretion disk of the lensed quasar
  q2237$+$0305 through gravitational microlensing.
Astrophysical Journal Letters
\textbf{814}
(2015).
doi:\doiurl{10.1088/2041-8205/814/2/L26}
\end{botherref}
\endbibitem

\bibitem[\protect\citeauthoryear{{Mediavilla}
  et~al.}{2016}]{bib5:Mediavilla2016}
\begin{barticle}
\bauthor{\binits{E.} \bsnm{{Mediavilla}}},
\bauthor{\binits{J.} \bsnm{{Jim{\'e}nez-Vicente}}},
\bauthor{\binits{J.A.} \bsnm{{Mu{\~n}oz}}},
\bauthor{\binits{E.} \bsnm{{Battaner}}},
\batitle{Peculiar transverse velocities of galaxies from quasar microlensing.
  tentative estimate of the peculiar velocity dispersion at z
  \raisebox{-0.5ex}\textasciitilde 0.5}.
\bjtitle{\apj}
\bvolume{832}(\bissue{1}),
\bfpage{46}
(\byear{2016}).
doi:\doiurl{10.3847/0004-637X/832/1/46}
\end{barticle}
\endbibitem

\bibitem[\protect\citeauthoryear{Mediavilla et~al.}{2017}]{bib5:Mediavilla2017}
\begin{barticle}
\bauthor{\binits{E.} \bsnm{Mediavilla}},
\bauthor{\binits{J.} \bsnm{Jim{\'e}nez-Vicente}},
\bauthor{\binits{J.A.} \bsnm{Mu{\~n}oz}},
\bauthor{\binits{H.} \bsnm{Vives-Arias}},
\bauthor{\binits{J.} \bsnm{Calder{\'o}n-Infante}},
\batitle{Limits on the mass and abundance of primordial black holes from quasar
  gravitational microlensing}.
\bjtitle{The Astrophysical Journal}
\bvolume{836},
\bfpage{18}
(\byear{2017}).
doi:\doiurl{10.3847/2041-8213/aa5dab}.
\burl{http://dx.doi.org/10.3847/2041-8213/aa5dab}
\end{barticle}
\endbibitem

\bibitem[\protect\citeauthoryear{Mediavilla et~al.}{2020}]{bib5:Mediavilla2020}
\begin{barticle}
\bauthor{\binits{E.} \bsnm{Mediavilla}},
\bauthor{\binits{J.} \bsnm{Jim{\'e}nez-vicente}},
\bauthor{\binits{J.} \bsnm{Mej{\'i}a-restrepo}},
\bauthor{\binits{V.} \bsnm{Motta}},
\bauthor{\binits{E.} \bsnm{Falco}},
\bauthor{\binits{J.A.} \bsnm{Mu{\~n}oz}},
\bauthor{\binits{C.} \bsnm{Fian}},
\bauthor{\binits{E.} \bsnm{Guerras}},
\batitle{Individual estimates of the virial factor in 10 quasars: Implications
  on the kinematics of the broad-line region}.
\bjtitle{The Astrophysical Journal}
\bvolume{895},
\bfpage{111}
(\byear{2020}).
doi:\doiurl{10.3847/1538-4357/ab8ae0}
\end{barticle}
\endbibitem

\bibitem[\protect\citeauthoryear{Melo et~al.}{2021}]{bib5:Melo2021}
\begin{botherref}
\oauthor{\binits{A.} \bsnm{Melo}},
\oauthor{\binits{V.} \bsnm{Motta}},
\oauthor{\binits{N.} \bsnm{Godoy}},
\oauthor{\binits{J.} \bsnm{Mejia-Restrepo}},
\oauthor{\binits{R.J.} \bsnm{Assef}},
\oauthor{\binits{E.} \bsnm{Mediavilla}},
\oauthor{\binits{E.} \bsnm{Falco}},
\oauthor{\binits{F.} \bsnm{{\'a}vila-Vera}},
\oauthor{\binits{R.} \bsnm{Jerez}},
First black hole mass estimation for the quadruple lensed system wgd2038-4008.
Astronomy and Astrophysics
\textbf{656}
(2021).
doi:\doiurl{10.1051/0004-6361/202141869}
\end{botherref}
\endbibitem

\bibitem[\protect\citeauthoryear{{Merloni} and
  {Fabian}}{2001}]{bib5:Merloni2001}
\begin{barticle}
\bauthor{\binits{A.} \bsnm{{Merloni}}},
\bauthor{\binits{A.C.} \bsnm{{Fabian}}},
\batitle{Thunderclouds and accretion discs: a model for the spectral and
  temporal variability of seyfert 1 galaxies}.
\bjtitle{\mnras}
\bvolume{328}(\bissue{3}),
\bfpage{958}--\blpage{968}
(\byear{2001}).
doi:\doiurl{10.1046/j.1365-8711.2001.04925.x}
\end{barticle}
\endbibitem

\bibitem[\protect\citeauthoryear{Metcalf and Zhao}{2002}]{bib5:Metcalf2002}
\begin{barticle}
\bauthor{\binits{R.B.} \bsnm{Metcalf}},
\bauthor{\binits{H.} \bsnm{Zhao}},
\batitle{Flux ratios as a probe of dark substructures in quadruple-image
  gravitational lenses}.
\bjtitle{The Astrophysical Journal Letters}
\bvolume{567},
\bfpage{5}--\blpage{8}
(\byear{2002})
\end{barticle}
\endbibitem

\bibitem[\protect\citeauthoryear{Middleton et~al.}{2015}]{bib5:Middleton2015}
\begin{barticle}
\bauthor{\binits{M.J.} \bsnm{Middleton}},
\bauthor{\binits{L.} \bsnm{Heil}},
\bauthor{\binits{F.} \bsnm{Pintore}},
\bauthor{\binits{D.J.} \bsnm{Walton}},
\bauthor{\binits{T.P.} \bsnm{Roberts}},
\batitle{A spectral-timing model for ulxs in the supercritical regime}.
\bjtitle{Monthly Notices of the Royal Astronomical Society}
\bvolume{447},
\bfpage{3243}--\blpage{3263}
(\byear{2015}).
doi:\doiurl{10.1093/mnras/stu2644}
\end{barticle}
\endbibitem

\bibitem[\protect\citeauthoryear{Millon et~al.}{2020a}]{bib5:Millon2020a}
\begin{barticle}
\bauthor{\binits{M.} \bsnm{Millon}},
\bauthor{\binits{F.} \bsnm{Courbin}},
\bauthor{\binits{V.} \bsnm{Bonvin}},
\bauthor{\binits{E.} \bsnm{Paic}},
\bauthor{\binits{G.} \bsnm{Meylan}},
\bauthor{\binits{M.} \bsnm{Tewes}},
\bauthor{\binits{D.} \bsnm{Sluse}},
\bauthor{\binits{P.} \bsnm{Magain}},
\bauthor{\binits{J.H.H.} \bsnm{Chan}},
\bauthor{\binits{A.} \bsnm{Galan}},
\bauthor{\binits{R.} \bsnm{Joseph}},
\bauthor{\binits{C.} \bsnm{Lemon}},
\bauthor{\binits{O.} \bsnm{Tihhonova}},
\bauthor{\binits{R.I.} \bsnm{Anderson}},
\bauthor{\binits{M.} \bsnm{Marmier}},
\bauthor{\binits{B.} \bsnm{Chazelas}},
\bauthor{\binits{M.} \bsnm{Lendl}},
\bauthor{\binits{A.H.M.J.} \bsnm{Triaud}},
\bauthor{\binits{A.} \bsnm{Wyttenbach}},
\batitle{Cosmograil. xix. time delays in 18 strongly lensed quasars from 15
  years of optical monitoring.}
\bjtitle{Astronomy \& Astrophysics}
\bvolume{640},
\bfpage{105}
(\byear{2020}a).
doi:\doiurl{10.1051/0004-6361/202037740}
\end{barticle}
\endbibitem

\bibitem[\protect\citeauthoryear{Millon et~al.}{2020b}]{bib5:Millon2020b}
\begin{barticle}
\bauthor{\binits{M.} \bsnm{Millon}},
\bauthor{\binits{A.} \bsnm{Galan}},
\bauthor{\binits{F.} \bsnm{Courbin}},
\bauthor{\binits{T.} \bsnm{Treu}},
\bauthor{\binits{S.H.} \bsnm{Suyu}},
\bauthor{\binits{X.} \bsnm{Ding}},
\bauthor{\binits{S.} \bsnm{Birrer}},
\bauthor{\binits{G.C.F.} \bsnm{Chen}},
\bauthor{\binits{A.J.} \bsnm{Shajib}},
\bauthor{\binits{D.} \bsnm{Sluse}},
\bauthor{\binits{K.C.} \bsnm{Wong}},
\bauthor{\binits{A.} \bsnm{Agnello}},
\bauthor{\binits{M.W.} \bsnm{Auger}},
\bauthor{\binits{E.J.} \bsnm{Buckley-Geer}},
\bauthor{\binits{J.H.H.} \bsnm{Chan}},
\bauthor{\binits{T.} \bsnm{Collett}},
\batitle{Tdcosmo: I. an exploration of systematic uncertainties in the
  inference of h0 from time-delay cosmography}.
\bjtitle{Astronomy and Astrophysics}
\bvolume{639},
\bfpage{1}--\blpage{19}
(\byear{2020}b).
doi:\doiurl{10.1051/0004-6361/201937351}
\end{barticle}
\endbibitem

\bibitem[\protect\citeauthoryear{Millon et~al.}{2022}]{bib5:Millon2022}
\begin{botherref}
\oauthor{\binits{M.} \bsnm{Millon}},
\oauthor{\binits{C.} \bsnm{Dalang}},
\oauthor{\binits{C.} \bsnm{Lemon}},
\oauthor{\binits{D.} \bsnm{Sluse}},
\oauthor{\binits{E.} \bsnm{Paic}},
\oauthor{\binits{J.H.H.} \bsnm{Chan}},
\oauthor{\binits{F.} \bsnm{Courbin}},
Evidence for a milliparsec-separation supermassive binary black hole with
  quasar microlensing.
Astronomy and Astrophysics
\textbf{668}
(2022).
doi:\doiurl{10.1051/0004-6361/202244440}
\end{botherref}
\endbibitem

\bibitem[\protect\citeauthoryear{Mineshige and
  Yonehara}{1999}]{bib5:Mineshige1999}
\begin{barticle}
\bauthor{\binits{S.} \bsnm{Mineshige}},
\bauthor{\binits{A.} \bsnm{Yonehara}},
\batitle{Gravitational microlens mapping of a quasar accretion disk}.
\bjtitle{Publications of the Astronomical Society of Japan}
\bvolume{51},
\bfpage{497}
(\byear{1999})
\end{barticle}
\endbibitem

\bibitem[\protect\citeauthoryear{Minezaki et~al.}{2009}]{bib5:Minezaki2009}
\begin{barticle}
\bauthor{\binits{T.} \bsnm{Minezaki}},
\bauthor{\binits{M.} \bsnm{Chiba}},
\bauthor{\binits{N.} \bsnm{Kashikawa}},
\bauthor{\binits{K.T.} \bsnm{Inoue}},
\bauthor{\binits{H.} \bsnm{Kataza}},
\batitle{Subaru mid-infrared imaging of the quadruple lenses. ii. unveiling
  lens structure of mg0414+0534 and q2237+030}.
\bjtitle{Astrophysical Journal}
\bvolume{697},
\bfpage{610}--\blpage{618}
(\byear{2009}).
doi:\doiurl{10.1088/0004-637X/697/1/610}
\end{barticle}
\endbibitem

\bibitem[\protect\citeauthoryear{Moore}{1965}]{bib5:Moore1965}
\begin{barticle}
\bauthor{\binits{G.E.} \bsnm{Moore}},
\batitle{Cramming more components onto integrated circuits}.
\bjtitle{Electronics Magazine}
\bvolume{38},
\bfpage{4}
(\byear{1965})
\end{barticle}
\endbibitem

\bibitem[\protect\citeauthoryear{Morgan et~al.}{2006}]{bib5:Morgan2006}
\begin{barticle}
\bauthor{\binits{C.W.} \bsnm{Morgan}},
\bauthor{\binits{C.S.} \bsnm{Kochanek}},
\bauthor{\binits{N.D.} \bsnm{Morgan}},
\bauthor{\binits{E.E.} \bsnm{Falco}},
\batitle{Microlensing of the lensed quasar sdss 0924$+$0219}.
\bjtitle{The Astrophysical Journal}
\bvolume{647},
\bfpage{874}--\blpage{885}
(\byear{2006}).
doi:\doiurl{10.1086/505569}.
\burl{http://stacks.iop.org/0004-637X/647/i=2/a=874}
\end{barticle}
\endbibitem

\bibitem[\protect\citeauthoryear{Morgan et~al.}{2008}]{bib5:Morgan2008b}
\begin{barticle}
\bauthor{\binits{C.W.} \bsnm{Morgan}},
\bauthor{\binits{C.S.} \bsnm{Kochanek}},
\bauthor{\binits{X.} \bsnm{Dai}},
\bauthor{\binits{N.D.} \bsnm{Morgan}},
\bauthor{\binits{E.E.} \bsnm{Falco}},
\bauthor{\binits{M.E.T.} \bsnm{Al}},
\batitle{X-ray and optical microlensing in the lensed quasar pg 1115 $+$ 080
  1}.
\bjtitle{The Astrophysical Journal}
\bvolume{689},
\bfpage{755}
(\byear{2008})
\end{barticle}
\endbibitem

\bibitem[\protect\citeauthoryear{Morgan et~al.}{2010}]{bib5:Morgan2010}
\begin{barticle}
\bauthor{\binits{C.W.} \bsnm{Morgan}},
\bauthor{\binits{C.S.} \bsnm{Kochanek}},
\bauthor{\binits{N.D.} \bsnm{Morgan}},
\bauthor{\binits{E.E.} \bsnm{Falco}},
\batitle{The quasar accretion disk size-black hole mass relation}.
\bjtitle{The Astrophysical Journal}
\bvolume{712},
\bfpage{1129}--\blpage{1136}
(\byear{2010}).
doi:\doiurl{10.1088/0004-637X/712/2/1129}.
\burl{http://stacks.iop.org/0004-637X/712/i=2/a=1129?key=crossref.355ea7287be512fda3fc2983bd7421c3}
\end{barticle}
\endbibitem

\bibitem[\protect\citeauthoryear{Morgan et~al.}{2012}]{bib5:Morgan2012}
\begin{barticle}
\bauthor{\binits{C.W.} \bsnm{Morgan}},
\bauthor{\binits{L.J.} \bsnm{Hainline}},
\bauthor{\binits{B.} \bsnm{Chen}},
\bauthor{\binits{M.} \bsnm{Tewes}},
\bauthor{\binits{C.S.} \bsnm{Kochanek}},
\bauthor{\binits{X.} \bsnm{Dai}},
\bauthor{\binits{S.} \bsnm{Kozlowski}},
\bauthor{\binits{J.A.} \bsnm{Blackburne}},
\bauthor{\binits{A.M.} \bsnm{Mosquera}},
\bauthor{\binits{G.} \bsnm{Chartas}},
\bauthor{\binits{F.} \bsnm{Courbin}},
\bauthor{\binits{G.} \bsnm{Meylan}},
\batitle{Further evidence that quasar x-ray emitting regions are compact: X-ray
  and optical microlensing in the lensed quasar q j0158-4325}.
\bjtitle{The Astrophysical Journal}
\bvolume{756},
\bfpage{52}
(\byear{2012}).
doi:\doiurl{10.1088/0004-637X/756/1/52}
\end{barticle}
\endbibitem

\bibitem[\protect\citeauthoryear{Morgan et~al.}{2018}]{bib5:Morgan2018}
\begin{barticle}
\bauthor{\binits{C.W.} \bsnm{Morgan}},
\bauthor{\binits{G.E.} \bsnm{Hyer}},
\bauthor{\binits{V.} \bsnm{Bonvin}},
\bauthor{\binits{A.M.} \bsnm{Mosquera}},
\bauthor{\binits{M.} \bsnm{Cornachione}},
\bauthor{\binits{F.} \bsnm{Courbin}},
\bauthor{\binits{C.S.} \bsnm{Kochanek}},
\bauthor{\binits{E.E.} \bsnm{Falco}},
\batitle{Accretion disk size measurement and time delays in the lensed quasar
  wfi 2033$-$4723}.
\bjtitle{The Astrophysical Journal}
\bvolume{869},
\bfpage{106}
(\byear{2018}).
doi:\doiurl{10.3847/1538-4357/aaed3e}.
\burl{http://dx.doi.org/10.3847/1538-4357/aaed3e}
\end{barticle}
\endbibitem

\bibitem[\protect\citeauthoryear{Mortonson et~al.}{2005}]{bib5:Mortonson2005}
\begin{barticle}
\bauthor{\binits{M.J.} \bsnm{Mortonson}},
\bauthor{\binits{P.L.} \bsnm{Schechter}},
\bauthor{\binits{J.} \bsnm{Wambsganss}},
\batitle{Size is everything: Universal features of quasar microlensing with
  extended sources}.
\bjtitle{The Astrophysical Journal}
\bvolume{628},
\bfpage{594}--\blpage{603}
(\byear{2005}).
doi:\doiurl{10.1086/431195}.
\burl{http://stacks.iop.org/0004-637X/628/i=2/a=594}
\end{barticle}
\endbibitem

\bibitem[\protect\citeauthoryear{Mosquera and
  Kochanek}{2011}]{bib5:Mosquera2011b}
\begin{barticle}
\bauthor{\binits{A.M.} \bsnm{Mosquera}},
\bauthor{\binits{C.S.} \bsnm{Kochanek}},
\batitle{The microlensing properties of a sample of 87 lensed quasars}.
\bjtitle{The Astrophysical Journal}
\bvolume{738},
\bfpage{96}
(\byear{2011}).
doi:\doiurl{10.1088/0004-637X/738/1/96}.
\burl{http://stacks.iop.org/0004-637X/738/i=1/a=96?key=crossref.d2f854b93c9a36dbfbcd550ed85d0320}
\end{barticle}
\endbibitem

\bibitem[\protect\citeauthoryear{Mosquera et~al.}{2009}]{bib5:Mosquera2009}
\begin{barticle}
\bauthor{\binits{A.M.} \bsnm{Mosquera}},
\bauthor{\binits{J.A.} \bsnm{Mu{\~n}oz}},
\bauthor{\binits{E.} \bsnm{Mediavilla}},
\batitle{Detection of chromatic microlensing in q 2237 $+$ 0305 a}.
\bjtitle{The Astrophysical Journal}
\bvolume{691},
\bfpage{1292}--\blpage{1299}
(\byear{2009}).
doi:\doiurl{10.1088/0004-637X/691/2/1292}.
\burl{http://stacks.iop.org/0004-637X/691/i=2/a=1292?key=crossref.76a5efcf2ecd4471342c364e7df63046}
\end{barticle}
\endbibitem

\bibitem[\protect\citeauthoryear{Mosquera et~al.}{2011}]{bib5:Mosquera2011a}
\begin{barticle}
\bauthor{\binits{A.M.} \bsnm{Mosquera}},
\bauthor{\binits{J.A.} \bsnm{Mu{\~n}oz}},
\bauthor{\binits{E.} \bsnm{Mediavilla}},
\bauthor{\binits{C.S.} \bsnm{Kochanek}},
\batitle{A study of gravitational lens chromaticity using ground-based
  narrowband photometry}.
\bjtitle{The Astrophysical Journal}
\bvolume{728},
\bfpage{145}
(\byear{2011}).
doi:\doiurl{10.1088/0004-637X/728/2/145}.
\burl{http://stacks.iop.org/0004-637X/728/i=2/a=145?key=crossref.e42408f766a7a6cfac6e7a9cd453a2e8}
\end{barticle}
\endbibitem

\bibitem[\protect\citeauthoryear{Mosquera et~al.}{2013}]{bib5:Mosquera2013}
\begin{barticle}
\bauthor{\binits{A.M.} \bsnm{Mosquera}},
\bauthor{\binits{C.S.} \bsnm{Kochanek}},
\bauthor{\binits{B.} \bsnm{Chen}},
\bauthor{\binits{X.} \bsnm{Dai}},
\bauthor{\binits{J.A.} \bsnm{Blackburne}},
\bauthor{\binits{G.} \bsnm{Chartas}},
\batitle{The structure of the x-ray and optical emitting regions of the lensed
  quasar q 2237$+$0305}.
\bjtitle{The Astrophysical Journal}
\bvolume{769},
\bfpage{53}
(\byear{2013})
\end{barticle}
\endbibitem

\bibitem[\protect\citeauthoryear{Motta et~al.}{2002}]{bib5:Motta2002}
\begin{barticle}
\bauthor{\binits{V.} \bsnm{Motta}},
\bauthor{\binits{E.} \bsnm{Mediavilla}},
\bauthor{\binits{J.A.} \bsnm{Mu{\~n}oz}},
\bauthor{\binits{E.} \bsnm{Falco}},
\bauthor{\binits{C.S.} \bsnm{Kochanek}},
\bauthor{\binits{S.} \bsnm{Arribas}},
\bauthor{\binits{B.} \bsnm{{García{-}Lorenzo}}},
\bauthor{\binits{A.} \bsnm{Oscoz}},
\bauthor{\binits{M.} \bsnm{{Serra{-}Ricart}}},
\batitle{Detection of the 2175 \aa extinction feature at z = 0.83}.
\bjtitle{The Astrophysical Journal}
\bvolume{574},
\bfpage{719}
(\byear{2002})
\end{barticle}
\endbibitem

\bibitem[\protect\citeauthoryear{Motta et~al.}{2012}]{bib5:Motta2012}
\begin{barticle}
\bauthor{\binits{V.} \bsnm{Motta}},
\bauthor{\binits{E.} \bsnm{Mediavilla}},
\bauthor{\binits{E.} \bsnm{Falco}},
\bauthor{\binits{J.A.} \bsnm{Munoz}},
\batitle{Measuring microlensing using spectra of multiply lensed quasars}.
\bjtitle{The Astrophysical Journal}
\bvolume{755},
\bfpage{82}
(\byear{2012}).
doi:\doiurl{10.1088/0004-637X/755/1/82}.
\burl{http://stacks.iop.org/0004-637X/755/i=1/a=82?key=crossref.8a462a69eac5b5dfad6ad5bb86c6c89b}
\end{barticle}
\endbibitem

\bibitem[\protect\citeauthoryear{{Moustakas} et~al.}{2019}]{bib5:Moustakas2019}
\begin{botherref}
\oauthor{\binits{L.} \bsnm{{Moustakas}}},
\oauthor{\binits{M.} \bsnm{{O'Dowd}}},
\oauthor{\binits{T.} \bsnm{{Anguita}}},
\oauthor{\binits{R.} \bsnm{{Webster}}},
\oauthor{\binits{G.} \bsnm{{Chartas}}},
\oauthor{\binits{M.} \bsnm{{Cornachione}}},
\oauthor{\binits{X.} \bsnm{{Dai}}},
\oauthor{\binits{C.} \bsnm{{Fian}}},
\oauthor{\binits{D.} \bsnm{{Hutsemekers}}},
\oauthor{\binits{J.} \bsnm{{Jimenez-Vicente}}},
\oauthor{\binits{K.} \bsnm{{Labrie}}},
\oauthor{\binits{G.} \bsnm{{Lewis}}},
\oauthor{\binits{C.} \bsnm{{Macleod}}},
\oauthor{\binits{E.} \bsnm{{Mediavilla}}},
\oauthor{\binits{C.W.} \bsnm{{Morgan}}},
\oauthor{\binits{V.} \bsnm{{Motta}}},
\oauthor{\binits{A.} \bsnm{{Nierenberg}}},
\oauthor{\binits{D.} \bsnm{{Pooley}}},
\oauthor{\binits{K.} \bsnm{{Rojas}}},
\oauthor{\binits{D.} \bsnm{{Sluse}}},
\oauthor{\binits{G.} \bsnm{{Vernardos}}},
\oauthor{\binits{J.} \bsnm{{Wambsganss}}},
\oauthor{\binits{S.Y.} \bsnm{{Yong}}},
Astro2020 science white paper - quasar microlensing: Revolutionizing our
  understanding of quasar structure and dynamics.
arXiv e-prints,
1904--12967
(2019).
doi:\doiurl{10.48550/arXiv.1904.12967}
\end{botherref}
\endbibitem

\bibitem[\protect\citeauthoryear{Mu{\~n}oz et~al.}{2011}]{bib5:Munoz2011}
\begin{barticle}
\bauthor{\binits{J.A.} \bsnm{Mu{\~n}oz}},
\bauthor{\binits{E.} \bsnm{Mediavilla}},
\bauthor{\binits{C.S.} \bsnm{Kochanek}},
\bauthor{\binits{E.E.} \bsnm{Falco}},
\bauthor{\binits{A.M.} \bsnm{Mosquera}},
\batitle{a study of gravitational lens chromaticity with the hubble space
  telescope}.
\bjtitle{The Astrophysical Journal}
\bvolume{742},
\bfpage{67}
(\byear{2011}).
doi:\doiurl{10.1088/0004-637X/742/2/67}.
\burl{http://stacks.iop.org/0004-637X/742/i=2/a=67?key=crossref.064a3e4ce395c9ee4c2fe688e95a0b12}
\end{barticle}
\endbibitem

\bibitem[\protect\citeauthoryear{Mu{\~n}oz et~al.}{2016}]{bib5:Munoz2016}
\begin{barticle}
\bauthor{\binits{J.A.} \bsnm{Mu{\~n}oz}},
\bauthor{\binits{H.} \bsnm{Vives-Arias}},
\bauthor{\binits{A.M.} \bsnm{Mosquera}},
\bauthor{\binits{J.} \bsnm{Jim{\'e}nez-Vicente}},
\bauthor{\binits{C.S.} \bsnm{Kochanek}},
\bauthor{\binits{E.} \bsnm{Mediavilla}},
\batitle{Structure of the accretion disk in the lensed quasar q2237+0305 from
  multi-epoch and multi-wavelength narrowband photometry}.
\bjtitle{The Astrophysical Journal}
\bvolume{817},
\bfpage{155}
(\byear{2016}).
doi:\doiurl{10.3847/0004-637x/817/2/155}
\end{barticle}
\endbibitem

\bibitem[\protect\citeauthoryear{Murray and Chiang}{1997}]{bib5:Murray1997}
\begin{barticle}
\bauthor{\binits{N.} \bsnm{Murray}},
\bauthor{\binits{J.} \bsnm{Chiang}},
\batitle{Disk winds and disk emission lines}.
\bjtitle{The Astrophysical Journal}
\bvolume{474},
\bfpage{91}--\blpage{103}
(\byear{1997})
\end{barticle}
\endbibitem

\bibitem[\protect\citeauthoryear{Murray et~al.}{1995}]{bib5:Murray1995}
\begin{barticle}
\bauthor{\binits{N.} \bsnm{Murray}},
\bauthor{\binits{J.} \bsnm{Chiang}},
\bauthor{\binits{S.A.} \bsnm{Grossman}},
\bauthor{\binits{G.M.} \bsnm{Voit}},
\batitle{Accretion disk winds from active galactic nuclei}.
\bjtitle{The Astrophysical Journal}
\bvolume{451},
\bfpage{498}
(\byear{1995})
\end{barticle}
\endbibitem

\bibitem[\protect\citeauthoryear{{Mushotzky}}{1984}]{bib5:Mushotzky1984}
\begin{barticle}
\bauthor{\binits{R.F.} \bsnm{{Mushotzky}}},
\batitle{X-ray spectra and time variability of active galactic nuclei}.
\bjtitle{Advances in Space Research}
\bvolume{3}(\bissue{10-12}),
\bfpage{157}--\blpage{165}
(\byear{1984}).
doi:\doiurl{10.1016/0273-1177(84)90081-4}
\end{barticle}
\endbibitem

\bibitem[\protect\citeauthoryear{Nagao et~al.}{2006}]{bib5:Nagao2006}
\begin{barticle}
\bauthor{\binits{T.} \bsnm{Nagao}},
\bauthor{\binits{A.} \bsnm{Marconi}},
\bauthor{\binits{R.} \bsnm{Maiolino}},
\batitle{The evolution of the broad-line region among sdss quasars}.
\bjtitle{Astronomy and Astrophysics}
\bvolume{447},
\bfpage{157}--\blpage{172}
(\byear{2006}).
doi:\doiurl{10.1051/0004-6361:20054024}
\end{barticle}
\endbibitem

\bibitem[\protect\citeauthoryear{Narayan and Yi}{1994}]{bib5:Narayan1994}
\begin{barticle}
\bauthor{\binits{R.} \bsnm{Narayan}},
\bauthor{\binits{I.} \bsnm{Yi}},
\batitle{Advection-dominated accretion: A self-similar solution}.
\bjtitle{The Astrophysical Journal Letters}
\bvolume{428},
\bfpage{13}--\blpage{16}
(\byear{1994})
\end{barticle}
\endbibitem

\bibitem[\protect\citeauthoryear{Neira et~al.}{2020}]{bib5:Neira2020}
\begin{barticle}
\bauthor{\binits{F.} \bsnm{Neira}},
\bauthor{\binits{T.} \bsnm{Anguita}},
\bauthor{\binits{G.} \bsnm{Vernardos}},
\batitle{A quasar microlensing light-curve generator for lsst}.
\bjtitle{Monthly Notices of the Royal Astronomical Society}
\bvolume{495},
\bfpage{544}--\blpage{553}
(\byear{2020}).
doi:\doiurl{10.1093/mnras/staa1208}
\end{barticle}
\endbibitem

\bibitem[\protect\citeauthoryear{{Neronov} and {Vovk}}{2016}]{bib5:Neronov2016}
\begin{barticle}
\bauthor{\binits{A.} \bsnm{{Neronov}}},
\bauthor{\binits{I.} \bsnm{{Vovk}}},
\batitle{Test of relativistic gravity using microlensing of relativistically
  broadened lines in gravitationally lensed quasars}.
\bjtitle{\prd}
\bvolume{93}(\bissue{2}),
\bfpage{023006}
(\byear{2016}).
doi:\doiurl{10.1103/PhysRevD.93.023006}
\end{barticle}
\endbibitem

\bibitem[\protect\citeauthoryear{Netzer}{2015}]{bib5:Netzer2015}
\begin{barticle}
\bauthor{\binits{H.} \bsnm{Netzer}},
\batitle{Revisiting the unified model of active galactic nuclei}.
\bjtitle{Annual Review of Astronomy and Astrophysics}
\bvolume{53},
\bfpage{365}--\blpage{408}
(\byear{2015}).
doi:\doiurl{10.1146/annurev-astro-082214-122302}
\end{barticle}
\endbibitem

\bibitem[\protect\citeauthoryear{Nixon et~al.}{2012}]{bib5:Nixon2012}
\begin{botherref}
\oauthor{\binits{C.} \bsnm{Nixon}},
\oauthor{\binits{A.} \bsnm{King}},
\oauthor{\binits{D.} \bsnm{Price}},
\oauthor{\binits{J.} \bsnm{Frank}},
Tearing up the disk: How black holes accrete.
Astrophysical Journal Letters
\textbf{757}
(2012).
doi:\doiurl{10.1088/2041-8205/757/2/L24}
\end{botherref}
\endbibitem

\bibitem[\protect\citeauthoryear{Novikov and Thorne}{1973}]{bib5:Novikov1973}
\begin{botherref}
\oauthor{\binits{I.D.} \bsnm{Novikov}},
\oauthor{\binits{K.S.} \bsnm{Thorne}},
Astrophysics of Black Holes,
ed. by C. DeWitt, B.S. DeWitt
(Gordon and Breach, 1973),
pp. 343--450
\end{botherref}
\endbibitem

\bibitem[\protect\citeauthoryear{{O'Dowd} et~al.}{2011}]{bib5:ODowd2011}
\begin{barticle}
\bauthor{\binits{M.} \bsnm{{O'Dowd}}},
\bauthor{\binits{N.F.} \bsnm{{Bate}}},
\bauthor{\binits{R.L.} \bsnm{{Webster}}},
\bauthor{\binits{R.} \bsnm{{Wayth}}},
\bauthor{\binits{K.} \bsnm{{Labrie}}},
\batitle{Differential microlensing measurements of quasar broad-line kinematics
  in q2237$+$0305}.
\bjtitle{Monthly Notices of the Royal Astronomical Society}
\bvolume{415},
\bfpage{1985}--\blpage{1998}
(\byear{2011}).
doi:\doiurl{10.1111/j.1365-2966.2010.18119.x}.
\burl{http://doi.wiley.com/10.1111/j.1365-2966.2010.18119.x}
\end{barticle}
\endbibitem

\bibitem[\protect\citeauthoryear{{O'Dowd} et~al.}{2015}]{bib5:ODowd2015}
\begin{barticle}
\bauthor{\binits{M.J.} \bsnm{{O'Dowd}}},
\bauthor{\binits{N.F.} \bsnm{{Bate}}},
\bauthor{\binits{R.L.} \bsnm{{Webster}}},
\bauthor{\binits{K.} \bsnm{{Labrie}}},
\bauthor{\binits{J.} \bsnm{{Rogers}}},
\batitle{Microlensing constraints on broad absorption and emission line flows
  in the quasar h1413$+$117}.
\bjtitle{The Astrophysical Journal}
\bvolume{813},
\bfpage{62}
(\byear{2015}).
doi:\doiurl{10.1088/0004-637X/813/1/62}.
\burl{http://stacks.iop.org/0004-637X/813/i=1/a=62?key=crossref.7458f31e10f08e8bd8ceb32cd5324f8c}
\end{barticle}
\endbibitem

\bibitem[\protect\citeauthoryear{Oguri and Marshall}{2010}]{bib5:Oguri2010}
\begin{barticle}
\bauthor{\binits{M.} \bsnm{Oguri}},
\bauthor{\binits{P.J.} \bsnm{Marshall}},
\batitle{Gravitationally lensed quasars and supernovae in future wide-field
  optical imaging surveys}.
\bjtitle{Monthly Notices of the Royal Astronomical Society}
\bvolume{405},
\bfpage{2579}
(\byear{2010}).
doi:\doiurl{10.1111/j.1365-2966.2010.16639.x}.
\burl{http://doi.wiley.com/10.1111/j.1365-2966.2010.16639.x}
\end{barticle}
\endbibitem

\bibitem[\protect\citeauthoryear{Oguri et~al.}{2014}]{bib5:Oguri2014}
\begin{barticle}
\bauthor{\binits{M.} \bsnm{Oguri}},
\bauthor{\binits{C.E.} \bsnm{Rusu}},
\bauthor{\binits{E.E.} \bsnm{Falco}},
\batitle{The stellar and dark matter distributions in elliptical galaxies from
  the ensemble of strong gravitational lenses}.
\bjtitle{Monthly Notices of the Royal Astronomical Society}
\bvolume{439},
\bfpage{2494}--\blpage{2504}
(\byear{2014}).
doi:\doiurl{10.1093/mnras/stu106}
\end{barticle}
\endbibitem

\bibitem[\protect\citeauthoryear{{Oshima} et~al.}{2001}]{bib5:Oshima2001}
\begin{barticle}
\bauthor{\binits{T.} \bsnm{{Oshima}}},
\bauthor{\binits{K.} \bsnm{{Mitsuda}}},
\bauthor{\binits{R.} \bsnm{{Fujimoto}}},
\bauthor{\binits{N.} \bsnm{{Iyomoto}}},
\bauthor{\binits{K.} \bsnm{{Futamoto}}},
\bauthor{\binits{M.} \bsnm{{Hattori}}},
\bauthor{\binits{N.} \bsnm{{Ota}}},
\bauthor{\binits{K.} \bsnm{{Mori}}},
\bauthor{\binits{Y.} \bsnm{{Ikebe}}},
\bauthor{\binits{J.M.} \bsnm{{Miralles}}},
\bauthor{\binits{J.-P.} \bsnm{{Kneib}}},
\batitle{Detection of an iron emission feature from the lensed broad absorption
  line qso h1413+117 at z = 2.56}.
\bjtitle{\apjl}
\bvolume{563}(\bissue{2}),
\bfpage{103}--\blpage{106}
(\byear{2001}).
doi:\doiurl{10.1086/338653}
\end{barticle}
\endbibitem

\bibitem[\protect\citeauthoryear{Ostensen et~al.}{1996}]{bib5:Ostensen1996}
\begin{barticle}
\bauthor{\binits{R.} \bsnm{Ostensen}},
\bauthor{\binits{S.} \bsnm{Refsdal}},
\bauthor{\binits{R.} \bsnm{Stabell}},
\bauthor{\binits{J.} \bsnm{Teuber}},
\bauthor{\binits{P.I.} \bsnm{Emanuelsen}},
\bauthor{\binits{L.} \bsnm{Festin}},
\bauthor{\binits{R.} \bsnm{Florentin-Nielsen}},
\bauthor{\binits{G.} \bsnm{Gahm}},
\batitle{1996a\&a...309..}
\bjtitle{Astronomy \& Astrophysics}
\bvolume{309},
\bfpage{59}
(\byear{1996})
\end{barticle}
\endbibitem

\bibitem[\protect\citeauthoryear{{Ota} et~al.}{2006}]{bib5:Ota2006}
\begin{barticle}
\bauthor{\binits{N.} \bsnm{{Ota}}},
\bauthor{\binits{N.} \bsnm{{Inada}}},
\bauthor{\binits{M.} \bsnm{{Oguri}}},
\bauthor{\binits{K.} \bsnm{{Mitsuda}}},
\bauthor{\binits{G.T.} \bsnm{{Richards}}},
\bauthor{\binits{Y.} \bsnm{{Suto}}},
\bauthor{\binits{W.N.} \bsnm{{Brandt}}},
\bauthor{\binits{F.J.} \bsnm{{Castander}}},
\bauthor{\binits{R.} \bsnm{{Fujimoto}}},
\bauthor{\binits{P.B.} \bsnm{{Hall}}},
\bauthor{\binits{C.R.} \bsnm{{Keeton}}},
\bauthor{\binits{R.C.} \bsnm{{Nichol}}},
\bauthor{\binits{D.P.} \bsnm{{Schneider}}},
\bauthor{\binits{D.E.} \bsnm{{Eisenstein}}},
\bauthor{\binits{J.A.} \bsnm{{Frieman}}},
\bauthor{\binits{E.L.} \bsnm{{Turner}}},
\bauthor{\binits{T.} \bsnm{{Minezaki}}},
\bauthor{\binits{Y.} \bsnm{{Yoshii}}},
\batitle{Chandra observations of sdss j1004+4112: Constraints on the lensing
  cluster and anomalous x-ray flux ratios of the quadruply imaged quasar}.
\bjtitle{\apj}
\bvolume{647}(\bissue{1}),
\bfpage{215}--\blpage{221}
(\byear{2006}).
doi:\doiurl{10.1086/505385}
\end{barticle}
\endbibitem

\bibitem[\protect\citeauthoryear{Paczynski}{1986}]{bib5:Paczynski1986}
\begin{barticle}
\bauthor{\binits{B.} \bsnm{Paczynski}},
\batitle{Gravitational microlensing at large optical depth}.
\bjtitle{The Astrophysical Journal}
\bvolume{301},
\bfpage{503}
(\byear{1986})
\end{barticle}
\endbibitem

\bibitem[\protect\citeauthoryear{Padovani et~al.}{2017}]{bib5:Padovani2017}
\begin{barticle}
\bauthor{\binits{P.} \bsnm{Padovani}},
\bauthor{\binits{D.M.} \bsnm{Alexander}},
\bauthor{\binits{R.J.} \bsnm{Assef}},
\bauthor{\binits{B.D.} \bsnm{Marco}},
\bauthor{\binits{P.} \bsnm{Giommi}},
\bauthor{\binits{R.C.} \bsnm{Hickox}},
\bauthor{\binits{G.T.} \bsnm{Richards}},
\bauthor{\binits{V.} \bsnm{Smol{\'C}i{\'c}}},
\bauthor{\binits{E.} \bsnm{Hatziminaoglou}},
\bauthor{\binits{V.} \bsnm{Mainieri}},
\bauthor{\binits{M.} \bsnm{Salvato}},
\batitle{Active galactic nuclei: what's in a name?}
\bjtitle{Astronomy and Astrophysics Review}
\bvolume{25},
\bfpage{2}
(\byear{2017}).
doi:\doiurl{10.1007/s00159-017-0102-9}
\end{barticle}
\endbibitem

\bibitem[\protect\citeauthoryear{Paic et~al.}{2022}]{bib5:Paic2022}
\begin{botherref}
\oauthor{\binits{E.} \bsnm{Paic}},
\oauthor{\binits{G.} \bsnm{Vernardos}},
\oauthor{\binits{D.} \bsnm{Sluse}},
\oauthor{\binits{M.} \bsnm{Millon}},
\oauthor{\binits{F.} \bsnm{Courbin}},
\oauthor{\binits{J.H.} \bsnm{Chan}},
\oauthor{\binits{V.} \bsnm{Bonvin}},
Constraining quasar structure using high-frequency microlensing variations and
  continuum reverberation.
Astronomy and Astrophysics
\textbf{659}
(2022).
doi:\doiurl{10.1051/0004-6361/202141808}
\end{botherref}
\endbibitem

\bibitem[\protect\citeauthoryear{Pancoast et~al.}{2011}]{bib5:Pancoast2011}
\begin{botherref}
\oauthor{\binits{A.} \bsnm{Pancoast}},
\oauthor{\binits{B.J.} \bsnm{Brewer}},
\oauthor{\binits{T.} \bsnm{Treu}},
Geometric and dynamical models of reverberation mapping data.
Astrophysical Journal
\textbf{730}
(2011).
doi:\doiurl{10.1088/0004-637X/730/2/139}
\end{botherref}
\endbibitem

\bibitem[\protect\citeauthoryear{Pancoast et~al.}{2014}]{bib5:Pancoast2014}
\begin{barticle}
\bauthor{\binits{A.} \bsnm{Pancoast}},
\bauthor{\binits{B.J.} \bsnm{Brewer}},
\bauthor{\binits{T.} \bsnm{Treu}},
\batitle{Modelling reverberation mapping data - i. improved geometric and
  dynamical models and comparison with cross-correlation results}.
\bjtitle{Monthly Notices of the Royal Astronomical Society}
\bvolume{445},
\bfpage{3055}--\blpage{3072}
(\byear{2014}).
doi:\doiurl{10.1093/mnras/stu1809}
\end{barticle}
\endbibitem

\bibitem[\protect\citeauthoryear{Papadakis et~al.}{2022}]{bib5:Papadakis2022}
\begin{botherref}
\oauthor{\binits{I.E.} \bsnm{Papadakis}},
\oauthor{\binits{M.} \bsnm{Dov{\'C}iak}},
\oauthor{\binits{E.S.} \bsnm{Kammoun}},
X-ray illuminated accretion discs and quasar microlensing disc sizes.
Astronomy and Astrophysics
\textbf{666}
(2022).
doi:\doiurl{10.1051/0004-6361/202142962}
\end{botherref}
\endbibitem

\bibitem[\protect\citeauthoryear{{Parker} et~al.}{2015}]{bib5:Parker2015}
\begin{barticle}
\bauthor{\binits{M.L.} \bsnm{{Parker}}},
\bauthor{\binits{J.A.} \bsnm{{Tomsick}}},
\bauthor{\binits{J.M.} \bsnm{{Miller}}},
\bauthor{\binits{K.} \bsnm{{Yamaoka}}},
\bauthor{\binits{A.} \bsnm{{Lohfink}}},
\bauthor{\binits{M.} \bsnm{{Nowak}}},
\bauthor{\binits{A.C.} \bsnm{{Fabian}}},
\bauthor{\binits{W.N.} \bsnm{{Alston}}},
\bauthor{\binits{S.E.} \bsnm{{Boggs}}},
\bauthor{\binits{F.E.} \bsnm{{Christensen}}},
\bauthor{\binits{W.W.} \bsnm{{Craig}}},
\bauthor{\binits{F.} \bsnm{{F{\"u}rst}}},
\bauthor{\binits{P.} \bsnm{{Gandhi}}},
\bauthor{\binits{B.W.} \bsnm{{Grefenstette}}},
\bauthor{\binits{V.} \bsnm{{Grinberg}}},
\bauthor{\binits{C.J.} \bsnm{{Hailey}}},
\bauthor{\binits{F.A.} \bsnm{{Harrison}}},
\bauthor{\binits{E.} \bsnm{{Kara}}},
\bauthor{\binits{A.L.} \bsnm{{King}}},
\bauthor{\binits{D.} \bsnm{{Stern}}},
\bauthor{\binits{D.J.} \bsnm{{Walton}}},
\bauthor{\binits{J.} \bsnm{{Wilms}}},
\bauthor{\binits{W.W.} \bsnm{{Zhang}}},
\batitle{Nustar and suzaku observations of the hard state in cygnus x-1:
  Locating the inner accretion disk}.
\bjtitle{\apj}
\bvolume{808}(\bissue{1}),
\bfpage{9}
(\byear{2015}).
doi:\doiurl{10.1088/0004-637X/808/1/9}
\end{barticle}
\endbibitem

\bibitem[\protect\citeauthoryear{{Pashchenko}
  et~al.}{2020}]{bib5:Paschenko2020}
\begin{barticle}
\bauthor{\binits{I.N.} \bsnm{{Pashchenko}}},
\bauthor{\binits{A.V.} \bsnm{{Plavin}}},
\bauthor{\binits{A.M.} \bsnm{{Kutkin}}},
\bauthor{\binits{Y.Y.} \bsnm{{Kovalev}}},
\batitle{A bias in vlbi measurements of the core shift effect in agn jets}.
\bjtitle{\mnras}
\bvolume{499}(\bissue{3}),
\bfpage{4515}--\blpage{4525}
(\byear{2020}).
doi:\doiurl{10.1093/mnras/staa3140}
\end{barticle}
\endbibitem

\bibitem[\protect\citeauthoryear{Peterson and Wandel}{1999}]{bib5:Peterson1999}
\begin{barticle}
\bauthor{\binits{B.M.} \bsnm{Peterson}},
\bauthor{\binits{A.} \bsnm{Wandel}},
\batitle{Keplerian motion of broad-line region gas as evidence for supermassive
  black holes in active galactic nuclei}.
\bjtitle{The Astrophysical Journal}
\bvolume{521},
\bfpage{95}--\blpage{98}
(\byear{1999}).
\burl{http://www.astronomy.ohio-state.edu/}
\end{barticle}
\endbibitem

\bibitem[\protect\citeauthoryear{Peterson et~al.}{1985}]{bib5:Peterson1985}
\begin{barticle}
\bauthor{\binits{B.M.} \bsnm{Peterson}},
\bauthor{\binits{D.M.} \bsnm{Crenshaw}},
\bauthor{\binits{K.A.} \bsnm{Meyers}},
\batitle{Variability of the emission-line spectra and optical continua of
  seyfert galaxies. iii. results for a homogeneous sample}.
\bjtitle{The Astrophysical Journal}
\bvolume{298},
\bfpage{283}--\blpage{291}
(\byear{1985})
\end{barticle}
\endbibitem

\bibitem[\protect\citeauthoryear{{Peterson} et~al.}{2005}]{bib5:Peterson2005}
\begin{barticle}
\bauthor{\binits{B.M.} \bsnm{{Peterson}}},
\bauthor{\binits{M.C.} \bsnm{{Bentz}}},
\bauthor{\binits{L.-B.} \bsnm{{Desroches}}},
\bauthor{\binits{A.V.} \bsnm{{Filippenko}}},
\bauthor{\binits{L.C.} \bsnm{{Ho}}},
\bauthor{\binits{S.} \bsnm{{Kaspi}}},
\bauthor{\binits{A.} \bsnm{{Laor}}},
\bauthor{\binits{D.} \bsnm{{Maoz}}},
\bauthor{\binits{E.C.} \bsnm{{Moran}}},
\bauthor{\binits{R.W.} \bsnm{{Pogge}}},
\bauthor{\binits{A.C.} \bsnm{{Quillen}}},
\batitle{Multiwavelength monitoring of the dwarf seyfert 1 galaxy ngc 4395. i.
  a reverberation-based measurement of the black hole mass}.
\bjtitle{\apj}
\bvolume{632}(\bissue{2}),
\bfpage{799}--\blpage{808}
(\byear{2005}).
doi:\doiurl{10.1086/444494}
\end{barticle}
\endbibitem

\bibitem[\protect\citeauthoryear{Petters}{1992}]{bib5:Petters1992}
\begin{barticle}
\bauthor{\binits{A.O.} \bsnm{Petters}},
\batitle{Morse theory and gravitational microlensing}.
\bjtitle{Journal of Mathematical Physics}
\bvolume{33},
\bfpage{1915}--\blpage{1931}
(\byear{1992}).
doi:\doiurl{10.1063/1.529667}
\end{barticle}
\endbibitem

\bibitem[\protect\citeauthoryear{Petters et~al.}{2001}]{bib5:Petters2001}
\begin{bbook}
\bauthor{\binits{A.O.} \bsnm{Petters}},
\bauthor{\binits{H.} \bsnm{Levine}},
\bauthor{\binits{J.} \bsnm{Wambsganss}},
\bbtitle{Singularity theory and gravitational lensing}
\byear{2001}
\end{bbook}
\endbibitem

\bibitem[\protect\citeauthoryear{Petters et~al.}{2009a}]{bib5:Petters2009a}
\begin{barticle}
\bauthor{\binits{A.O.} \bsnm{Petters}},
\bauthor{\binits{B.} \bsnm{Rider}},
\bauthor{\binits{A.M.} \bsnm{Teguia}},
\batitle{A mathematical theory of stochastic microlensing i. random time-delay
  functions and lensing maps}.
\bjtitle{Journal of Mathematical Physics}
\bvolume{50},
\bfpage{072503}
(\byear{2009}a).
doi:\doiurl{10.1063/1.3158854}.
\burl{http://arxiv.org/abs/0807.0232 http://dx.doi.org/10.1063/1.3158854}
\end{barticle}
\endbibitem

\bibitem[\protect\citeauthoryear{Petters et~al.}{2009b}]{bib5:Petters2009b}
\begin{botherref}
\oauthor{\binits{A.O.} \bsnm{Petters}},
\oauthor{\binits{B.} \bsnm{Rider}},
\oauthor{\binits{A.M.} \bsnm{Teguia}},
A mathematical theory of stochastic microlensing ii. random images, shear, and
  the kac-rice formula.
Journal of Mathematical Physics
\textbf{50}
(2009b).
doi:\doiurl{10.1063/1.3267859}.
\url{http://arxiv.org/abs/0807.4984 http://dx.doi.org/10.1063/1.3267859}
\end{botherref}
\endbibitem

\bibitem[\protect\citeauthoryear{Poindexter and
  Kochanek}{2010}]{bib5:Poindexter2010a}
\begin{barticle}
\bauthor{\binits{S.} \bsnm{Poindexter}},
\bauthor{\binits{C.S.} \bsnm{Kochanek}},
\batitle{The transverse peculiar velocity of the q2237$+$0305 lens galaxy and
  the mean mass of its stars}.
\bjtitle{The Astrophysical Journal}
\bvolume{712},
\bfpage{658}--\blpage{667}
(\byear{2010}).
doi:\doiurl{10.1088/0004-637X/712/1/658}.
\burl{http://stacks.iop.org/0004-637X/712/i=1/a=658?key=crossref.142f4fb59c205cd9e630b021192c44dd}
\end{barticle}
\endbibitem

\bibitem[\protect\citeauthoryear{Poindexter et~al.}{2008}]{bib5:Poindexter2008}
\begin{barticle}
\bauthor{\binits{S.} \bsnm{Poindexter}},
\bauthor{\binits{N.D.} \bsnm{Morgan}},
\bauthor{\binits{C.S.} \bsnm{Kochanek}},
\batitle{The spatial structure of an accretion disk}.
\bjtitle{The Astrophysical Journal}
\bvolume{673},
\bfpage{34}--\blpage{38}
(\byear{2008}).
doi:\doiurl{10.1086/524190}.
\burl{http://stacks.iop.org/0004-637X/673/i=1/a=34}
\end{barticle}
\endbibitem

\bibitem[\protect\citeauthoryear{Pooley et~al.}{2007}]{bib5:Pooley2007}
\begin{barticle}
\bauthor{\binits{D.} \bsnm{Pooley}},
\bauthor{\binits{J.A.} \bsnm{Blackburne}},
\bauthor{\binits{S.} \bsnm{Rappaport}},
\bauthor{\binits{P.L.} \bsnm{Schechter}},
\batitle{X-ray and optical flux ratio anomalies in quadruply lensed quasars . i
  . zooming in on quasar emission regions}.
\bjtitle{The Astrophysical Journal}
\bvolume{661},
\bfpage{19}--\blpage{29}
(\byear{2007})
\end{barticle}
\endbibitem

\bibitem[\protect\citeauthoryear{Pooley et~al.}{2012}]{bib5:Pooley2012}
\begin{barticle}
\bauthor{\binits{D.} \bsnm{Pooley}},
\bauthor{\binits{S.} \bsnm{Rappaport}},
\bauthor{\binits{J.A.} \bsnm{Blackburne}},
\bauthor{\binits{P.L.} \bsnm{Schechter}},
\bauthor{\binits{J.} \bsnm{Wambsganss}},
\batitle{X-ray and optical flux ratio anomalies in quadruply lensed quasars.
  ii. mapping the dark matter content in elliptical galaxies}.
\bjtitle{The Astrophysical Journal}
\bvolume{744},
\bfpage{111}
(\byear{2012}).
doi:\doiurl{10.1088/0004-637X/744/2/111}.
\burl{http://stacks.iop.org/0004-637X/744/i=2/a=111?key=crossref.3e1962113f6b1509a77ff18d68f2c8ce}
\end{barticle}
\endbibitem

\bibitem[\protect\citeauthoryear{Popovi{\'c} et~al.}{2020}]{bib5:Popovic2020}
\begin{botherref}
\oauthor{\binits{L.A.} \bsnm{Popovi{\'c}}},
\oauthor{\binits{V.L.} \bsnm{Afanasiev}},
\oauthor{\binits{A.} \bsnm{Moiseev}},
\oauthor{\binits{A.} \bsnm{Smirnova}},
\oauthor{\binits{S.} \bsnm{Simi{\'c}}},
\oauthor{\binits{D.} \bsnm{Savi{\'c}}},
\oauthor{\binits{E.G.} \bsnm{Mediavilla}},
\oauthor{\binits{C.} \bsnm{Fian}},
Spectroscopy and polarimetry of the gravitationally lensed quasar sdss
  j1004+4112 with the 6m sao ras telescope.
Astronomy and Astrophysics
\textbf{634}
(2020).
doi:\doiurl{10.1051/0004-6361/201936088}
\end{botherref}
\endbibitem

\bibitem[\protect\citeauthoryear{Popovi{\'c} and
  Chartas}{2005}]{bib5:Popovic2005}
\begin{barticle}
\bauthor{\binits{L.{\'C}.} \bsnm{Popovi{\'c}}},
\bauthor{\binits{G.} \bsnm{Chartas}},
\batitle{The influence of gravitational lensing on the spectra of lensed
  quasi-stellar objects}.
\bjtitle{Monthly Notices of the Royal Astronomical Society}
\bvolume{357},
\bfpage{135}--\blpage{144}
(\byear{2005}).
doi:\doiurl{10.1111/j.1365-2966.2004.08619.x}
\end{barticle}
\endbibitem

\bibitem[\protect\citeauthoryear{{Popovi{\'c}} et~al.}{2006}]{bib5:Popovic2006}
\begin{barticle}
\bauthor{\binits{L.{\v{C}}.} \bsnm{{Popovi{\'c}}}},
\bauthor{\binits{P.} \bsnm{{Jovanovi{\'c}}}},
\bauthor{\binits{E.} \bsnm{{Mediavilla}}},
\bauthor{\binits{A.F.} \bsnm{{Zakharov}}},
\bauthor{\binits{C.} \bsnm{{Abajas}}},
\bauthor{\binits{J.A.} \bsnm{{Mu{\~n}oz}}},
\bauthor{\binits{G.} \bsnm{{Chartas}}},
\batitle{A study of the correlation between the amplification of the fe
  k{\ensuremath{\alpha}} line and the x-ray continuum of quasars due to
  microlensing}.
\bjtitle{\apj}
\bvolume{637}(\bissue{2}),
\bfpage{620}--\blpage{630}
(\byear{2006}).
doi:\doiurl{10.1086/498558}
\end{barticle}
\endbibitem

\bibitem[\protect\citeauthoryear{Popovi{\'c} et~al.}{2021}]{bib5:Popovic2021}
\begin{botherref}
\oauthor{\binits{L.C.} \bsnm{Popovi{\'c}}},
\oauthor{\binits{V.L.} \bsnm{Afanasiev}},
\oauthor{\binits{E.S.} \bsnm{Shablovinskaya}},
\oauthor{\binits{V.I.} \bsnm{Ardilanov}},
\oauthor{\binits{D.J.} \bsnm{Savi{\'c}}},
Spectroscopy and polarimetry of the gravitationally lensed quasar q0957+561.
Astronomy and Astrophysics
\textbf{647}
(2021).
doi:\doiurl{10.1051/0004-6361/202039914}
\end{botherref}
\endbibitem

\bibitem[\protect\citeauthoryear{{Pozdnyakov}
  et~al.}{1976}]{bib5:Pozdnyakov1976}
\begin{barticle}
\bauthor{\binits{L.A.} \bsnm{{Pozdnyakov}}},
\bauthor{\binits{I.M.} \bsnm{{Sobol}}},
\bauthor{\binits{R.A.} \bsnm{{Syunyaev}}},
\batitle{Multiple compton scattering by relativistic electrons - monte carlo
  calculations of the emission spectrum}.
\bjtitle{Soviet Astronomy Letters}
\bvolume{2},
\bfpage{55}--\blpage{57}
(\byear{1976})
\end{barticle}
\endbibitem

\bibitem[\protect\citeauthoryear{Press and Gunn}{1973}]{bib5:PressGunn1973}
\begin{barticle}
\bauthor{\binits{W.H.} \bsnm{Press}},
\bauthor{\binits{J.E.} \bsnm{Gunn}},
\batitle{Method for detecting a cosmological density of condensed objects}.
\bjtitle{The Astrophysical Journal}
\bvolume{185},
\bfpage{397}--\blpage{412}
(\byear{1973})
\end{barticle}
\endbibitem

\bibitem[\protect\citeauthoryear{Rauch et~al.}{1992}]{bib5:Rauch1992}
\begin{barticle}
\bauthor{\binits{K.P.} \bsnm{Rauch}},
\bauthor{\binits{S.} \bsnm{Mao}},
\bauthor{\binits{J.} \bsnm{Wambsganss}},
\bauthor{\binits{B.} \bsnm{Paczynski}},
\batitle{Caustic-induced features in microlensing magnification probability
  distributions}.
\bjtitle{The Astrophysical Journal}
\bvolume{386},
\bfpage{30}
(\byear{1992})
\end{barticle}
\endbibitem

\bibitem[\protect\citeauthoryear{Rauch and Blandford}{1991}]{bib5:Rauch1991}
\begin{barticle}
\bauthor{\binits{K.P.} \bsnm{Rauch}},
\bauthor{\binits{R.D.} \bsnm{Blandford}},
\batitle{Microlensing and the structure of active galactic nucleus accretion
  discs}.
\bjtitle{The Astrophysical Journal}
\bvolume{381},
\bfpage{39}
(\byear{1991})
\end{barticle}
\endbibitem

\bibitem[\protect\citeauthoryear{Refsdal and Stabell}{1997}]{bib5:Refsdal1997}
\begin{barticle}
\bauthor{\binits{S.} \bsnm{Refsdal}},
\bauthor{\binits{R.} \bsnm{Stabell}},
\batitle{Gravitational microlensing of large sources including shear term
  effects}.
\bjtitle{ASTRONOMY AND ASTROPHYSICS}
\bvolume{325},
\bfpage{877}--\blpage{880}
(\byear{1997})
\end{barticle}
\endbibitem

\bibitem[\protect\citeauthoryear{Reynolds and Nowak}{2003}]{bib5:Reynolds2003}
\begin{barticle}
\bauthor{\binits{C.S.} \bsnm{Reynolds}},
\bauthor{\binits{M.A.} \bsnm{Nowak}},
\batitle{Fluorescent iron lines as a probe of astrophysical black hole
  systems}.
\bjtitle{Physics Reports}
\bvolume{377},
\bfpage{389}
(\byear{2003})
\end{barticle}
\endbibitem

\bibitem[\protect\citeauthoryear{Richards et~al.}{2004}]{bib5:Richards2004}
\begin{barticle}
\bauthor{\binits{G.T.} \bsnm{Richards}},
\bauthor{\binits{C.R.} \bsnm{Keeton}},
\bauthor{\binits{B.} \bsnm{Pindor}},
\bauthor{\binits{J.F.} \bsnm{Hennawi}},
\bauthor{\binits{P.B.} \bsnm{Hall}},
\bauthor{\binits{E.L.} \bsnm{Turner}},
\bauthor{\binits{N.} \bsnm{Inada}},
\bauthor{\binits{M.} \bsnm{Oguri}},
\bauthor{\binits{S.-I.} \bsnm{Ichikawa}},
\bauthor{\binits{R.H.} \bsnm{Becker}},
\bauthor{\binits{M.D.} \bsnm{Gregg}},
\bauthor{\binits{R.L.} \bsnm{White}},
\bauthor{\binits{J.} \bsnm{Stuart}},
\bauthor{\binits{B.} \bsnm{Wyithe}},
\bauthor{\binits{D.P.} \bsnm{Schneider}},
\bauthor{\binits{D.E.} \bsnm{Johnston}},
\bauthor{\binits{J.A.} \bsnm{Frieman}},
\bauthor{\binits{J.} \bsnm{Brinkmann}},
\batitle{Microlensing of the broad emission line region in the quadruple lens
  sdss j1004+4112}.
\bjtitle{The Astrophysical Journal}
\bvolume{610},
\bfpage{679}
(\byear{2004})
\end{barticle}
\endbibitem

\bibitem[\protect\citeauthoryear{{Risaliti} and
  {Lusso}}{2019}]{bib5:Risaliti2019}
\begin{barticle}
\bauthor{\binits{G.} \bsnm{{Risaliti}}},
\bauthor{\binits{E.} \bsnm{{Lusso}}},
\batitle{Cosmological constraints from the hubble diagram of quasars at high
  redshifts}.
\bjtitle{Nature Astronomy}
\bvolume{3},
\bfpage{272}--\blpage{277}
(\byear{2019}).
doi:\doiurl{10.1038/s41550-018-0657-z}
\end{barticle}
\endbibitem

\bibitem[\protect\citeauthoryear{Rojas et~al.}{2014}]{bib5:Rojas2014}
\begin{botherref}
\oauthor{\binits{K.} \bsnm{Rojas}},
\oauthor{\binits{V.} \bsnm{Motta}},
\oauthor{\binits{E.} \bsnm{Mediavilla}},
\oauthor{\binits{E.} \bsnm{Falco}},
\oauthor{\binits{J.} \bsnm{Jim{\'e}nez-Vicente}},
\oauthor{\binits{J.A.} \bsnm{Mu{\~n}oz}},
Strong chromatic microlensing in he0047-1756 and sdss1155$+$6346.
The Astrophysical Journal
\textbf{accepted}
(2014)
\end{botherref}
\endbibitem

\bibitem[\protect\citeauthoryear{{Rothschild}
  et~al.}{1983}]{bib5:Rothschild1983}
\begin{barticle}
\bauthor{\binits{R.E.} \bsnm{{Rothschild}}},
\bauthor{\binits{F.R.} \bsnm{{Mushotzky}}},
\bauthor{\binits{W.A.} \bsnm{{Baity}}},
\bauthor{\binits{D.E.} \bsnm{{Gruber}}},
\bauthor{\binits{J.L.} \bsnm{{Matteson}}},
\bauthor{\binits{L.E.} \bsnm{{Peterson}}},
\batitle{2-165 kev observations of active galaxies and the diffuse background.}
\bjtitle{\apj}
\bvolume{269},
\bfpage{423}--\blpage{437}
(\byear{1983}).
doi:\doiurl{10.1086/161053}
\end{barticle}
\endbibitem

\bibitem[\protect\citeauthoryear{Rusu et~al.}{2020}]{bib5:Rusu2020}
\begin{botherref}
\oauthor{\binits{C.E.} \bsnm{Rusu}},
\oauthor{\binits{K.C.} \bsnm{Wong}},
\oauthor{\binits{V.} \bsnm{Bonvin}},
\oauthor{\binits{D.} \bsnm{Sluse}},
\oauthor{\binits{S.H.} \bsnm{Suyu}},
\oauthor{\binits{C.D.} \bsnm{Fassnacht}},
\oauthor{\binits{J.H.H.} \bsnm{Chan}},
\oauthor{\binits{S.} \bsnm{Hilbert}},
\oauthor{\binits{M.W.} \bsnm{Auger}},
\oauthor{\binits{A.} \bsnm{Sonnenfeld}},
\oauthor{\binits{S.} \bsnm{Birrer}},
\oauthor{\binits{F.} \bsnm{Courbin}},
\oauthor{\binits{T.} \bsnm{Treu}},
\oauthor{\binits{G.C.-F.} \bsnm{Chen}},
\oauthor{\binits{A.} \bsnm{Halkola}},
\oauthor{\binits{L.V.E.} \bsnm{Koopmans}},
\oauthor{\binits{P.J.} \bsnm{Marshall}},
\oauthor{\binits{A.J.} \bsnm{Shajib}},
H0licow xii. lens mass model of wfi2033$-$4723 and blind measurement of its
  time-delay distance and $h_0$
\textbf{498},
1440--1468
(2020).
doi:\doiurl{10.1093/mnras/stz3451}
\end{botherref}
\endbibitem

\bibitem[\protect\citeauthoryear{Saha}{2000}]{bib5:Saha2000}
\begin{barticle}
\bauthor{\binits{P.} \bsnm{Saha}},
\batitle{Lensing degeneracies revisited}.
\bjtitle{The Astronomical Journal}
\bvolume{120},
\bfpage{1654}--\blpage{1659}
(\byear{2000})
\end{barticle}
\endbibitem

\bibitem[\protect\citeauthoryear{Saha and Williams}{2011}]{bib5:Saha2011}
\begin{barticle}
\bauthor{\binits{P.} \bsnm{Saha}},
\bauthor{\binits{L.L.R.} \bsnm{Williams}},
\batitle{Understanding micro-image configurations in quasar microlensing}.
\bjtitle{Monthly Notices of the Royal Astronomical Society}
\bvolume{411},
\bfpage{1671}--\blpage{1677}
(\byear{2011}).
doi:\doiurl{10.1111/j.1365-2966.2010.17797.x}
\end{barticle}
\endbibitem

\bibitem[\protect\citeauthoryear{Sarkar et~al.}{2021}]{bib5:Sarkar2021}
\begin{barticle}
\bauthor{\binits{A.} \bsnm{Sarkar}},
\bauthor{\binits{G.J.} \bsnm{Ferland}},
\bauthor{\binits{M.} \bsnm{Chatzikos}},
\bauthor{\binits{F.} \bsnm{Guzm{\'a}n}},
\bauthor{\binits{P.A.M.} \bparticle{van} \bsnm{Hoof}},
\bauthor{\binits{R.T.} \bsnm{Smyth}},
\bauthor{\binits{C.A.} \bsnm{Ramsbottom}},
\bauthor{\binits{F.P.} \bsnm{Keenan}},
\bauthor{\binits{C.P.} \bsnm{Ballance}},
\batitle{Improved fe ii emission-line models for agns using new atomic data
  sets}.
\bjtitle{The Astrophysical Journal}
\bvolume{907},
\bfpage{12}
(\byear{2021}).
doi:\doiurl{10.3847/1538-4357/abcaa6}
\end{barticle}
\endbibitem

\bibitem[\protect\citeauthoryear{Schechter and
  Wambsganss}{2002}]{bib5:Schechter2002}
\begin{barticle}
\bauthor{\binits{P.L.} \bsnm{Schechter}},
\bauthor{\binits{J.} \bsnm{Wambsganss}},
\batitle{Quasar microlensing at high magnification and the role of dark matter:
  Enhanced fluctuations and suppressed saddle points}.
\bjtitle{The Astrophysical Journal}
\bvolume{580},
\bfpage{685}--\blpage{695}
(\byear{2002}).
doi:\doiurl{10.1086/343856}.
\burl{http://stacks.iop.org/0004-637X/580/i=2/a=685}
\end{barticle}
\endbibitem

\bibitem[\protect\citeauthoryear{Schechter et~al.}{2004}]{bib5:Schechter2004}
\begin{barticle}
\bauthor{\binits{P.L.} \bsnm{Schechter}},
\bauthor{\binits{J.} \bsnm{Wambsganss}},
\bauthor{\binits{G.F.} \bsnm{Lewis}},
\batitle{Qualitative aspects of quasar microlensing with two mass components :
  Magnification patterns and probability distributions}.
\bjtitle{The Astrophysical Journal}
\bvolume{613},
\bfpage{77}--\blpage{85}
(\byear{2004})
\end{barticle}
\endbibitem

\bibitem[\protect\citeauthoryear{{Schechter} et~al.}{2003}]{bib5:Schechter2003}
\begin{barticle}
\bauthor{\binits{P.L.} \bsnm{{Schechter}}},
\bauthor{\binits{A.} \bsnm{{Udalski}}},
\bauthor{\binits{M.} \bsnm{{Szyma{\'n}ski}}},
\bauthor{\binits{M.} \bsnm{{Kubiak}}},
\bauthor{\binits{G.} \bsnm{{Pietrzy{\'n}ski}}},
\bauthor{\binits{I.} \bsnm{{Soszy{\'n}ski}}},
\bauthor{\binits{P.} \bsnm{{Wo{\'z}niak}}},
\bauthor{\binits{K.} \bsnm{{{\.Z}ebru{\'n}}}},
\bauthor{\binits{O.} \bsnm{{Szewczyk}}},
\bauthor{\binits{{\L}.} \bsnm{{Wyrzykowski}}},
\batitle{Microlensing of relativistic knots in the quasar he 1104-1805 ab}.
\bjtitle{The Astrophysical Journal}
\bvolume{584},
\bfpage{657}
(\byear{2003})
\end{barticle}
\endbibitem

\bibitem[\protect\citeauthoryear{Schechter et~al.}{2014}]{bib5:Schechter2014}
\begin{botherref}
\oauthor{\binits{P.L.} \bsnm{Schechter}},
\oauthor{\binits{D.} \bsnm{Pooley}},
\oauthor{\binits{J.A.} \bsnm{Blackburne}},
\oauthor{\binits{J.} \bsnm{Wambsganss}},
A calibration of the stellar mass fundamental plane at z $\sim$ 0.5 using the
  micro-lensing-induced flux ratio anomalies of macro-lensed quasars.
Astrophysical Journal
\textbf{793}
(2014).
doi:\doiurl{10.1088/0004-637X/793/2/96}
\end{botherref}
\endbibitem

\bibitem[\protect\citeauthoryear{Schild}{1996}]{bib5:Schild1996}
\begin{barticle}
\bauthor{\binits{R.E.} \bsnm{Schild}},
\batitle{Microlensing variability of the gravitationally lensed quasar
  q0957+561 a,b}.
\bjtitle{The Astrophysical Journal}
\bvolume{464},
\bfpage{125}
(\byear{1996})
\end{barticle}
\endbibitem

\bibitem[\protect\citeauthoryear{Schmidt and
  Wambsganss}{2010}]{bib5:Schmidt2010}
\begin{barticle}
\bauthor{\binits{R.W.} \bsnm{Schmidt}},
\bauthor{\binits{J.} \bsnm{Wambsganss}},
\batitle{Quasar microlensing}.
\bjtitle{Gen. Relativ. and Gravit.}
\bvolume{42},
\bfpage{2127}--\blpage{2150}
(\byear{2010}).
\bisbn{1071401009}.
doi:\doiurl{10.1007/s10714-010-0956-x}.
\burl{http://www.springerlink.com/index/10.1007/s10714-010-0956-x}
\end{barticle}
\endbibitem

\bibitem[\protect\citeauthoryear{Schneider and
  Wambsganss}{1990}]{bib5:Schneider1990}
\begin{barticle}
\bauthor{\binits{P.} \bsnm{Schneider}},
\bauthor{\binits{J.} \bsnm{Wambsganss}},
\batitle{Are the broad emission lines of quasars affected by gravitational
  microlensing?}
\bjtitle{Astronomy \& Astrophysics}
\bvolume{237},
\bfpage{42}--\blpage{53}
(\byear{1990})
\end{barticle}
\endbibitem

\bibitem[\protect\citeauthoryear{Schneider and
  Weiss}{1987}]{bib5:Schneider1987a}
\begin{barticle}
\bauthor{\binits{P.} \bsnm{Schneider}},
\bauthor{\binits{A.} \bsnm{Weiss}},
\batitle{A gravitational lens origin for agn-variability? consequences of
  micro-lensing}.
\bjtitle{Astronomy \& Astrophysics}
\bvolume{171},
\bfpage{49}
(\byear{1987})
\end{barticle}
\endbibitem

\bibitem[\protect\citeauthoryear{Schneider et~al.}{1992}]{bib5:Schneider1992}
\begin{bbook}
\bauthor{\binits{P.} \bsnm{Schneider}},
\bauthor{\binits{J.} \bsnm{Ehlers}},
\bauthor{\binits{E.E.} \bsnm{Falco}},
\bbtitle{Gravitational Lenses}
(\bpublisher{Springer}, \blocation{???}, \byear{1992}).
doi:\doiurl{10.1007/978-3-662-03758-4}
\end{bbook}
\endbibitem

\bibitem[\protect\citeauthoryear{Schneider et~al.}{2006}]{bib5:Schneider2006}
\begin{botherref}
\oauthor{\binits{P.} \bsnm{Schneider}},
\oauthor{\binits{C.S.} \bsnm{Kochanek}},
\oauthor{\binits{J.} \bsnm{Wambsganss}},
Gravitational Lensing: Strong, Weak, Micro,
ed. by G. Meylan, P. Jetzer, P. North
(Springer, 2006)
\end{botherref}
\endbibitem

\bibitem[\protect\citeauthoryear{Schneider}{1993}]{bib5:Schneider1993}
\begin{barticle}
\bauthor{\binits{P.} \bsnm{Schneider}},
\batitle{Upper bounds on the cosmological density of compact objects with
  sub-solar masses from the variability of qsos}.
\bjtitle{Astronomy \& Astrophysics}
\bvolume{279},
\bfpage{1}--\blpage{20}
(\byear{1993})
\end{barticle}
\endbibitem

\bibitem[\protect\citeauthoryear{Shakura and Sunyaev}{1973}]{bib5:Shakura1973}
\begin{barticle}
\bauthor{\binits{N.I.} \bsnm{Shakura}},
\bauthor{\binits{R.A.} \bsnm{Sunyaev}},
\batitle{Black holes in binary systems. observational appearance}.
\bjtitle{Astronomy and Astrophysics}
\bvolume{24},
\bfpage{337}
(\byear{1973})
\end{barticle}
\endbibitem

\bibitem[\protect\citeauthoryear{Shalyapin and
  Goicoechea}{2014}]{bib5:Shalyapin2014}
\begin{botherref}
\oauthor{\binits{V.N.} \bsnm{Shalyapin}},
\oauthor{\binits{L.J.} \bsnm{Goicoechea}},
Deep optical imaging and spectroscopy of the lens system sdss j1339+1310.
Astronomy and Astrophysics
\textbf{568}
(2014).
doi:\doiurl{10.1051/0004-6361/201323360}
\end{botherref}
\endbibitem

\bibitem[\protect\citeauthoryear{Shalyapin et~al.}{2021}]{bib5:Shalyapin2021}
\begin{botherref}
\oauthor{\binits{V.N.} \bsnm{Shalyapin}},
\oauthor{\binits{R.} \bsnm{Gil{-}Merino}},
\oauthor{\binits{L.J.} \bsnm{Goicoechea}},
Fast simulations of extragalactic microlensing.
Astronomy and Astrophysics
\textbf{653}
(2021).
doi:\doiurl{10.1051/0004-6361/202140527}
\end{botherref}
\endbibitem

\bibitem[\protect\citeauthoryear{{Shapiro} et~al.}{1976}]{bib5:Shapiro1976}
\begin{barticle}
\bauthor{\binits{S.L.} \bsnm{{Shapiro}}},
\bauthor{\binits{A.P.} \bsnm{{Lightman}}},
\bauthor{\binits{D.M.} \bsnm{{Eardley}}},
\batitle{A two-temperature accretion disk model for cygnus x-1: structure and
  spectrum.}
\bjtitle{\apj}
\bvolume{204},
\bfpage{187}--\blpage{199}
(\byear{1976}).
doi:\doiurl{10.1086/154162}
\end{barticle}
\endbibitem

\bibitem[\protect\citeauthoryear{Shen and Ho}{2014}]{bib5:Shen2014}
\begin{barticle}
\bauthor{\binits{Y.} \bsnm{Shen}},
\bauthor{\binits{L.C.} \bsnm{Ho}},
\batitle{The diversity of quasars unified by accretion and orientation}.
\bjtitle{Nature}
\bvolume{513},
\bfpage{210}--\blpage{213}
(\byear{2014}).
doi:\doiurl{10.1038/nature13712}
\end{barticle}
\endbibitem

\bibitem[\protect\citeauthoryear{Shen et~al.}{2008}]{bib5:Shen2008}
\begin{barticle}
\bauthor{\binits{Y.} \bsnm{Shen}},
\bauthor{\binits{J.E.} \bsnm{Greene}},
\bauthor{\binits{M.A.} \bsnm{Strauss}},
\bauthor{\binits{G.T.} \bsnm{Richards}},
\bauthor{\binits{D.P.} \bsnm{Schneider}},
\batitle{Biases in virial black hole masses: An sdss perspective}.
\bjtitle{The Astrophysical Journal}
\bvolume{680},
\bfpage{169}
(\byear{2008})
\end{barticle}
\endbibitem

\bibitem[\protect\citeauthoryear{Silpa et~al.}{2020}]{bib5:Silpa2020}
\begin{botherref}
\oauthor{\binits{S.} \bsnm{Silpa}},
\oauthor{\binits{P.} \bsnm{Kharb}},
\oauthor{\binits{L.C.} \bsnm{Ho}},
\oauthor{\binits{C.H.} \bsnm{Ishwara-Chandra}},
\oauthor{\binits{M.E.} \bsnm{Jarvis}},
\oauthor{\binits{C.} \bsnm{Harrison}},
Probing the origin of low-frequency radio emission in pg quasars with the ugmrt
  - i
\textbf{499},
5826--5839
(2020).
doi:\doiurl{10.1093/mnras/staa2970}
\end{botherref}
\endbibitem

\bibitem[\protect\citeauthoryear{{S{\k{a}}dowski}
  et~al.}{2011}]{bib5:Sadowski2011}
\begin{barticle}
\bauthor{\binits{A.} \bsnm{{S{\k{a}}dowski}}},
\bauthor{\binits{M.} \bsnm{{Abramowicz}}},
\bauthor{\binits{M.} \bsnm{{Bursa}}},
\bauthor{\binits{W.} \bsnm{{Klu{\'z}niak}}},
\bauthor{\binits{J.-P.} \bsnm{{Lasota}}},
\bauthor{\binits{A.} \bsnm{{R{\'o}{\.z}a{\'n}ska}}},
\batitle{Relativistic slim disks with vertical structure}.
\bjtitle{Astronomy \& Astrophysics}
\bvolume{527},
\bfpage{17}
(\byear{2011}).
doi:\doiurl{10.1051/0004-6361/201015256}
\end{barticle}
\endbibitem

\bibitem[\protect\citeauthoryear{Sluse and Tewes}{2014}]{bib5:Sluse2014}
\begin{barticle}
\bauthor{\binits{D.} \bsnm{Sluse}},
\bauthor{\binits{M.} \bsnm{Tewes}},
\batitle{Imprints of the quasar structure in time-delay light curves:
  Microlensing-aided reverberation mapping}.
\bjtitle{Astronomy and Astrophysics}
\bvolume{571},
\bfpage{1}--\blpage{10}
(\byear{2014}).
doi:\doiurl{10.1051/0004-6361/201424776}
\end{barticle}
\endbibitem

\bibitem[\protect\citeauthoryear{Sluse et~al.}{2007}]{bib5:Sluse2007}
\begin{barticle}
\bauthor{\binits{D.} \bsnm{Sluse}},
\bauthor{\binits{J.F.} \bsnm{Claeskens}},
\bauthor{\binits{D.} \bsnm{Hutsem{\'e}kers}},
\bauthor{\binits{J.} \bsnm{Surdej}},
\batitle{Multi-wavelength study of the gravitational lens system rxs j1131-1231
  iii. long slit spectroscopy: Micro-lensing probes the qso structure}.
\bjtitle{Astronomy and Astrophysics}
\bvolume{468},
\bfpage{885}--\blpage{901}
(\byear{2007}).
doi:\doiurl{10.1051/0004-6361:20066821}
\end{barticle}
\endbibitem

\bibitem[\protect\citeauthoryear{Sluse et~al.}{2011}]{bib5:Sluse2011}
\begin{barticle}
\bauthor{\binits{D.} \bsnm{Sluse}},
\bauthor{\binits{R.} \bsnm{Schmidt}},
\bauthor{\binits{F.} \bsnm{Courbin}},
\bauthor{\binits{D.} \bsnm{Hutsem{\'e}kers}},
\bauthor{\binits{G.} \bsnm{Meylan}},
\bauthor{\binits{A.} \bsnm{Eigenbrod}},
\bauthor{\binits{T.} \bsnm{Anguita}},
\bauthor{\binits{E.} \bsnm{Agol}},
\bauthor{\binits{J.} \bsnm{Wambsganss}},
\batitle{Zooming into the broad line region of the gravitationally lensed
  quasar qso 2237 $+$ 0305 - the einstein cross}.
\bjtitle{Astronomy \& Astrophysics}
\bvolume{528},
\bfpage{100}
(\byear{2011}).
doi:\doiurl{10.1051/0004-6361/201016110}.
\burl{http://www.aanda.org/10.1051/0004-6361/201016110}
\end{barticle}
\endbibitem

\bibitem[\protect\citeauthoryear{Sluse et~al.}{2012}]{bib5:Sluse2012}
\begin{barticle}
\bauthor{\binits{D.} \bsnm{Sluse}},
\bauthor{\binits{D.} \bsnm{Hutsem{\'e}kers}},
\bauthor{\binits{F.} \bsnm{Courbin}},
\bauthor{\binits{G.} \bsnm{Meylan}},
\bauthor{\binits{J.} \bsnm{Wambsganss}},
\batitle{Microlensing of the broad line region in 17 lensed quasars}.
\bjtitle{Astronomy \& Astrophysics}
\bvolume{544},
\bfpage{62}
(\byear{2012}).
doi:\doiurl{10.1051/0004-6361/201219125}.
\burl{http://www.aanda.org/10.1051/0004-6361/201219125}
\end{barticle}
\endbibitem

\bibitem[\protect\citeauthoryear{Sluse et~al.}{2013}]{bib5:Sluse2013}
\begin{botherref}
\oauthor{\binits{D.} \bsnm{Sluse}},
\oauthor{\binits{M.} \bsnm{Kishimoto}},
\oauthor{\binits{T.} \bsnm{Anguita}},
\oauthor{\binits{O.} \bsnm{Wucknitz}},
\oauthor{\binits{J.} \bsnm{Wambsganss}},
Mid-infrared microlensing of accretion disc and dusty torus in quasars: Effects
  on flux ratio anomalies.
Astronomy and Astrophysics
\textbf{553}
(2013).
doi:\doiurl{10.1051/0004-6361/201220843}
\end{botherref}
\endbibitem

\bibitem[\protect\citeauthoryear{Sluse et~al.}{2015}]{bib5:Sluse2015}
\begin{botherref}
\oauthor{\binits{D.} \bsnm{Sluse}},
\oauthor{\binits{D.} \bsnm{Hutsem{\'e}kers}},
\oauthor{\binits{T.} \bsnm{Anguita}},
\oauthor{\binits{L.} \bsnm{Braibant}},
\oauthor{\binits{P.} \bsnm{Riaud}},
Evidence for two spatially separated uv continuum emitting regions in the
  cloverleaf broad absorption line quasar.
Astronomy and Astrophysics
\textbf{582}
(2015).
doi:\doiurl{10.1051/0004-6361/201526832}
\end{botherref}
\endbibitem

\bibitem[\protect\citeauthoryear{Smith et~al.}{2005}]{bib5:Smith2005}
\begin{barticle}
\bauthor{\binits{J.E.} \bsnm{Smith}},
\bauthor{\binits{A.} \bsnm{Robinson}},
\bauthor{\binits{S.} \bsnm{Young}},
\bauthor{\binits{D.J.} \bsnm{Axon}},
\bauthor{\binits{E.A.} \bsnm{Corbett}},
\batitle{Equatorial scattering and the structure of the broad-line region in
  seyfert nuclei: Evidence for a rotating disc}.
\bjtitle{Monthly Notices of the Royal Astronomical Society}
\bvolume{359},
\bfpage{846}--\blpage{864}
(\byear{2005}).
doi:\doiurl{10.1111/j.1365-2966.2005.08895.x}
\end{barticle}
\endbibitem

\bibitem[\protect\citeauthoryear{Speranza et~al.}{2021}]{bib5:Speranza2021}
\begin{botherref}
\oauthor{\binits{G.} \bsnm{Speranza}},
\oauthor{\binits{B.} \bsnm{Balmaverde}},
\oauthor{\binits{A.} \bsnm{Capetti}},
\oauthor{\binits{F.} \bsnm{Massaro}},
\oauthor{\binits{G.} \bsnm{Tremblay}},
\oauthor{\binits{A.} \bsnm{Marconi}},
\oauthor{\binits{G.} \bsnm{Venturi}},
\oauthor{\binits{M.} \bsnm{Chiaberge}},
\oauthor{\binits{R.D.} \bsnm{Baldi}},
\oauthor{\binits{S.} \bsnm{Baum}},
\oauthor{\binits{P.} \bsnm{Grandi}},
\oauthor{\binits{E.T.} \bsnm{Meyer}},
\oauthor{\binits{C.} \bsnm{O'Dea}},
\oauthor{\binits{W.} \bsnm{Sparks}},
\oauthor{\binits{B.A.} \bsnm{Terrazas}},
\oauthor{\binits{E.} \bsnm{Torresi}},
The murales survey. iv. searching for nuclear outflows in 3c radio galaxies at
  z < 0.3 with muse observations.
Astronomy and Astrophysics
\textbf{653}
(2021).
doi:\doiurl{10.1051/0004-6361/202140686}
\end{botherref}
\endbibitem

\bibitem[\protect\citeauthoryear{Stalevski et~al.}{2012}]{bib5:Stalevski2012b}
\begin{barticle}
\bauthor{\binits{M.} \bsnm{Stalevski}},
\bauthor{\binits{P.} \bsnm{Jovanovi{\'c}}},
\bauthor{\binits{L.{\'C}.} \bsnm{Popovi{\'c}}},
\bauthor{\binits{M.} \bsnm{Baes}},
\batitle{Gravitational microlensing of active galactic nuclei dusty tori}.
\bjtitle{Monthly Notices of the Royal Astronomical Society}
\bvolume{425},
\bfpage{1576}--\blpage{1584}
(\byear{2012}).
doi:\doiurl{10.1111/j.1365-2966.2012.21611.x}
\end{barticle}
\endbibitem

\bibitem[\protect\citeauthoryear{Suganuma et~al.}{2006}]{bib5:Suganuma2006}
\begin{barticle}
\bauthor{\binits{M.} \bsnm{Suganuma}},
\bauthor{\binits{Y.} \bsnm{Yoshii}},
\bauthor{\binits{Y.} \bsnm{Kobayashi}},
\bauthor{\binits{T.} \bsnm{Minezaki}},
\bauthor{\binits{K.} \bsnm{Enya}},
\bauthor{\binits{H.} \bsnm{Tomita}},
\bauthor{\binits{T.} \bsnm{Aoki}},
\bauthor{\binits{S.} \bsnm{Koshida}},
\bauthor{\binits{B.A.} \bsnm{Peterson}},
\batitle{Reverberation measurements of the inner radius of the dust torus in
  nearby seyfert 1 galaxies}.
\bjtitle{The Astrophysical Journal}
\bvolume{639},
\bfpage{46}
(\byear{2006})
\end{barticle}
\endbibitem

\bibitem[\protect\citeauthoryear{{Sunyaev} and
  {Titarchuk}}{1980}]{bib5:Sunyaev1980}
\begin{barticle}
\bauthor{\binits{R.A.} \bsnm{{Sunyaev}}},
\bauthor{\binits{L.G.} \bsnm{{Titarchuk}}},
\batitle{Comptonization of x-rays in plasma clouds - typical radiation
  spectra}.
\bjtitle{\aap}
\bvolume{86},
\bfpage{121}
(\byear{1980})
\end{barticle}
\endbibitem

\bibitem[\protect\citeauthoryear{{Sunyaev} and
  {Truemper}}{1979}]{bib5:Sunyaev1979}
\begin{barticle}
\bauthor{\binits{R.A.} \bsnm{{Sunyaev}}},
\bauthor{\binits{J.} \bsnm{{Truemper}}},
\batitle{Hard x-ray spectrum of cyg x-1}.
\bjtitle{\nat}
\bvolume{279},
\bfpage{506}--\blpage{508}
(\byear{1979}).
doi:\doiurl{10.1038/279506a0}
\end{barticle}
\endbibitem

\bibitem[\protect\citeauthoryear{Tewes et~al.}{2013}]{bib5:Tewes2013}
\begin{barticle}
\bauthor{\binits{M.} \bsnm{Tewes}},
\bauthor{\binits{F.} \bsnm{Courbin}},
\bauthor{\binits{G.} \bsnm{Meylan}},
\batitle{Cosmograil: the cosmological monitoring of gravitational lenses. xi.
  techniques for time delay measurement in presence of microlensing}.
\bjtitle{Astronomy \& Astrophysics}
\bvolume{553},
\bfpage{120}
(\byear{2013}).
doi:\doiurl{10.1051/0004-6361/201220123}
\end{barticle}
\endbibitem

\bibitem[\protect\citeauthoryear{Thompson et~al.}{2014}]{bib5:Thompson2014}
\begin{botherref}
\oauthor{\binits{A.C.} \bsnm{Thompson}},
\oauthor{\binits{G.} \bsnm{Vernardos}},
\oauthor{\binits{C.J.} \bsnm{Fluke}},
\oauthor{\binits{B.R.} \bsnm{Barsdell}},
Gpu-d.
Astrophysics Source Code Library
(2014)
\end{botherref}
\endbibitem

\bibitem[\protect\citeauthoryear{Thompson et~al.}{2010}]{bib5:Thompson2010}
\begin{barticle}
\bauthor{\binits{A.C.} \bsnm{Thompson}},
\bauthor{\binits{C.J.} \bsnm{Fluke}},
\bauthor{\binits{D.G.} \bsnm{Barnes}},
\bauthor{\binits{B.R.} \bsnm{Barsdell}},
\batitle{Teraflop per second gravitational lensing ray-shooting using graphics
  processing units 1}.
\bjtitle{New Astronomy}
\bvolume{15},
\bfpage{16}
(\byear{2010})
\end{barticle}
\endbibitem

\bibitem[\protect\citeauthoryear{Thorne}{1974}]{bib5:Thorne1974}
\begin{barticle}
\bauthor{\binits{K.S.} \bsnm{Thorne}},
\batitle{Disk-accretion onto a black hole. ii. evolution of the hole}.
\bjtitle{The Astrophysical Journal}
\bvolume{191},
\bfpage{507}--\blpage{519}
(\byear{1974})
\end{barticle}
\endbibitem

\bibitem[\protect\citeauthoryear{Tie and Kochanek}{2018}]{bib5:Tie2018a}
\begin{barticle}
\bauthor{\binits{S.S.} \bsnm{Tie}},
\bauthor{\binits{C.S.} \bsnm{Kochanek}},
\batitle{Microlensing makes lensed quasar time delays significantly time
  variable}.
\bjtitle{Monthly Notices of the Royal Astronomical Society}
\bvolume{473},
\bfpage{80}--\blpage{90}
(\byear{2018}).
doi:\doiurl{10.1093/mnras/stx2348}.
\burl{http://arxiv.org/abs/1707.01908\%0Ahttp://dx.doi.org/10.1093/mnras/stx2348}
\end{barticle}
\endbibitem

\bibitem[\protect\citeauthoryear{{Treyer} and
  {Wambsganss}}{2004}]{bib5:Treyer2004}
\begin{barticle}
\bauthor{\binits{M.} \bsnm{{Treyer}}},
\bauthor{\binits{J.} \bsnm{{Wambsganss}}},
\batitle{Astrometric microlensing of quasars. dependence on surface mass
  density and external shear}.
\bjtitle{\aap}
\bvolume{416},
\bfpage{19}--\blpage{34}
(\byear{2004}).
doi:\doiurl{10.1051/0004-6361:20034284}
\end{barticle}
\endbibitem

\bibitem[\protect\citeauthoryear{{Turner} and {Pounds}}{1989}]{bib5:Turner1989}
\begin{barticle}
\bauthor{\binits{T.J.} \bsnm{{Turner}}},
\bauthor{\binits{K.A.} \bsnm{{Pounds}}},
\batitle{The exosat spectral survey of agn.}
\bjtitle{\mnras}
\bvolume{240},
\bfpage{833}--\blpage{880}
(\byear{1989}).
doi:\doiurl{10.1093/mnras/240.4.833}
\end{barticle}
\endbibitem

\bibitem[\protect\citeauthoryear{Udalski et~al.}{2006}]{bib5:Udalski2006}
\begin{barticle}
\bauthor{\binits{A.} \bsnm{Udalski}},
\bauthor{\binits{M.} \bsnm{Szymanski}},
\bauthor{\binits{M.} \bsnm{Kubiak}},
\bauthor{\binits{E.} \bsnm{Al.}},
\batitle{The optical gravitational lensing experiment. ogle-iii long term
  monitoring of the gravitational lens qso 2237$+$0305}.
\bjtitle{Acta Astronomica}
\bvolume{56},
\bfpage{293}--\blpage{305}
(\byear{2006})
\end{barticle}
\endbibitem

\bibitem[\protect\citeauthoryear{Urry and Padovani}{1995}]{bib5:Urry1995}
\begin{barticle}
\bauthor{\binits{C.M.} \bsnm{Urry}},
\bauthor{\binits{P.} \bsnm{Padovani}},
\batitle{Unified schemes for radio-loud active galactic nuclei}.
\bjtitle{Publications of the Astronomical Society of the Pacific}
\bvolume{107},
\bfpage{803}--\blpage{845}
(\byear{1995})
\end{barticle}
\endbibitem

\bibitem[\protect\citeauthoryear{{Van de Vyvere}
  et~al.}{2022}]{bib5:VanDeVyvere2022a}
\begin{botherref}
\oauthor{\binits{L.} \bsnm{{Van de Vyvere}}},
\oauthor{\binits{M.R.} \bsnm{{Gomer}}},
\oauthor{\binits{D.} \bsnm{{Sluse}}},
\oauthor{\binits{D.} \bsnm{{Xu}}},
\oauthor{\binits{S.} \bsnm{{Birrer}}},
\oauthor{\binits{A.} \bsnm{{Galan}}},
\oauthor{\binits{G.} \bsnm{{Vernardos}}},
Tdcosmo: Vii. boxyness/discyness in lensing galaxies: Detectability and impact
  on h 0.
Astronomy and Astrophysics
\textbf{659}
(2022).
doi:\doiurl{10.1051/0004-6361/202141551}
\end{botherref}
\endbibitem

\bibitem[\protect\citeauthoryear{Venumadhav et~al.}{2017}]{bib5:Venumadhav2017}
\begin{barticle}
\bauthor{\binits{T.} \bsnm{Venumadhav}},
\bauthor{\binits{L.} \bsnm{Dai}},
\bauthor{\binits{J.} \bsnm{Miralda-Escud{\'e}}},
\batitle{Microlensing of extremely magnified stars near caustics of galaxy
  clusters}.
\bjtitle{The Astrophysical Journal}
\bvolume{850},
\bfpage{49}
(\byear{2017}).
doi:\doiurl{10.3847/1538-4357/aa9575}
\end{barticle}
\endbibitem

\bibitem[\protect\citeauthoryear{Vernardos}{2022}]{bib5:Vernardos2022molet}
\begin{barticle}
\bauthor{\binits{G.} \bsnm{Vernardos}},
\batitle{Simualting time-vayring strong lenses}.
\bjtitle{Monthly Notices of the Royal Astronomical Society}
\bvolume{511},
\bfpage{4417}
(\byear{2022})
\end{barticle}
\endbibitem

\bibitem[\protect\citeauthoryear{Vernardos}{2018}]{bib5:Vernardos2018}
\begin{barticle}
\bauthor{\binits{G.} \bsnm{Vernardos}},
\batitle{A joint microlensing analysis of lensing mass and accretion disc
  models}.
\bjtitle{MNRAS}
\bvolume{480},
\bfpage{4675}
(\byear{2018})
\end{barticle}
\endbibitem

\bibitem[\protect\citeauthoryear{Vernardos}{2019}]{bib5:Vernardos2019a}
\begin{barticle}
\bauthor{\binits{G.} \bsnm{Vernardos}},
\batitle{Microlensing flux ratio predictions for euclid}.
\bjtitle{Monthly Notices of the Royal Astronomical Society}
\bvolume{483},
\bfpage{5583}
(\byear{2019})
\end{barticle}
\endbibitem

\bibitem[\protect\citeauthoryear{Vernardos and
  Fluke}{2013}]{bib5:Vernardos2013}
\begin{barticle}
\bauthor{\binits{G.} \bsnm{Vernardos}},
\bauthor{\binits{C.J.} \bsnm{Fluke}},
\batitle{A new parameter space study of cosmological microlensing}.
\bjtitle{Monthly Notices of the Royal Astronomical Society}
\bvolume{434},
\bfpage{832}
(\byear{2013})
\end{barticle}
\endbibitem

\bibitem[\protect\citeauthoryear{Vernardos and
  Fluke}{2014a}]{bib5:Vernardos2014b}
\begin{barticle}
\bauthor{\binits{G.} \bsnm{Vernardos}},
\bauthor{\binits{C.J.} \bsnm{Fluke}},
\batitle{Adventures in the microlensing cloud: large datasets, eresearch tools,
  and gpus}.
\bjtitle{Astronomy and Computing}
\bvolume{6},
\bfpage{1}
(\byear{2014}a)
\end{barticle}
\endbibitem

\bibitem[\protect\citeauthoryear{Vernardos and
  Fluke}{2014b}]{bib5:Vernardos2014c}
\begin{barticle}
\bauthor{\binits{G.} \bsnm{Vernardos}},
\bauthor{\binits{C.J.} \bsnm{Fluke}},
\batitle{The effect of macromodel uncertainties on microlensing modelling of
  lensed quasars}.
\bjtitle{Monthly Notices of the Royal Astronomical Society}
\bvolume{445},
\bfpage{1223}
(\byear{2014}b)
\end{barticle}
\endbibitem

\bibitem[\protect\citeauthoryear{Vernardos and
  Tsagkatakis}{2019}]{bib5:Vernardos2019b}
\begin{barticle}
\bauthor{\binits{G.} \bsnm{Vernardos}},
\bauthor{\binits{G.} \bsnm{Tsagkatakis}},
\batitle{Quasar microlensing light-curve analysis using deep machine learning}.
\bjtitle{Monthly Notices of the Royal Astronomical Society}
\bvolume{486},
\bfpage{1944}--\blpage{1952}
(\byear{2019}).
doi:\doiurl{10.1093/mnras/stz868}
\end{barticle}
\endbibitem

\bibitem[\protect\citeauthoryear{Vernardos et~al.}{2014}]{bib5:Vernardos2014a}
\begin{barticle}
\bauthor{\binits{G.} \bsnm{Vernardos}},
\bauthor{\binits{C.J.} \bsnm{Fluke}},
\bauthor{\binits{N.F.} \bsnm{Bate}},
\bauthor{\binits{D.J.} \bsnm{Croton}},
\batitle{Gerlumph data release 1: High-resolution cosmological microlensing
  magnification maps and eresearch tools}.
\bjtitle{The Astrophysical Journal Supplement Series}
\bvolume{211},
\bfpage{16}
(\byear{2014})
\end{barticle}
\endbibitem

\bibitem[\protect\citeauthoryear{Vernardos et~al.}{2015}]{bib5:Vernardos2015}
\begin{barticle}
\bauthor{\binits{G.} \bsnm{Vernardos}},
\bauthor{\binits{C.J.} \bsnm{Fluke}},
\bauthor{\binits{N.F.} \bsnm{Bate}},
\bauthor{\binits{D.J.} \bsnm{Croton}},
\bauthor{\binits{D.} \bsnm{Vohl}},
\batitle{Gerlumph data release 2: 2.5 bilion microlensing light curves}.
\bjtitle{The Astrophysical Journal Supplement Series}
\bvolume{217},
\bfpage{23}
(\byear{2015})
\end{barticle}
\endbibitem

\bibitem[\protect\citeauthoryear{{Villafa{\~n}a}
  et~al.}{2022}]{bib5:Villafana2022}
\begin{barticle}
\bauthor{\binits{L.} \bsnm{{Villafa{\~n}a}}},
\bauthor{\binits{P.R.} \bsnm{{Williams}}},
\bauthor{\binits{T.} \bsnm{{Treu}}},
\bauthor{\binits{B.J.} \bsnm{{Brewer}}},
\bauthor{\binits{A.J.} \bsnm{{Barth}}},
\bauthor{\binits{V.} \bsnm{{U}}},
\bauthor{\binits{V.N.} \bsnm{{Bennert}}},
\bauthor{\binits{H.A.} \bsnm{{Vogler}}},
\bauthor{\binits{H.} \bsnm{{Guo}}},
\bauthor{\binits{M.C.} \bsnm{{Bentz}}},
\bauthor{\binits{G.} \bsnm{{Canalizo}}},
\bauthor{\binits{A.V.} \bsnm{{Filippenko}}},
\bauthor{\binits{E.} \bsnm{{Gates}}},
\bauthor{\binits{F.} \bsnm{{Hamann}}},
\bauthor{\binits{M.D.} \bsnm{{Joner}}},
\bauthor{\binits{M.A.} \bsnm{{Malkan}}},
\bauthor{\binits{J.-H.} \bsnm{{Woo}}},
\bauthor{\binits{B.} \bsnm{{Abolfathi}}},
\bauthor{\binits{L.E.} \bsnm{{Abramson}}},
\bauthor{\binits{S.F.} \bsnm{{Armen}}},
\bauthor{\binits{H.-J.} \bsnm{{Bae}}},
\bauthor{\binits{T.} \bsnm{{Bohn}}},
\bauthor{\binits{B.D.} \bsnm{{Boizelle}}},
\bauthor{\binits{A.} \bsnm{{Bostroem}}},
\bauthor{\binits{A.} \bsnm{{Brandel}}},
\bauthor{\binits{T.G.} \bsnm{{Brink}}},
\bauthor{\binits{S.} \bsnm{{Channa}}},
\bauthor{\binits{M.C.} \bsnm{{Cooper}}},
\bauthor{\binits{M.} \bsnm{{Cosens}}},
\bauthor{\binits{E.} \bsnm{{Donohue}}},
\bauthor{\binits{S.P.} \bsnm{{Fillingham}}},
\bauthor{\binits{D.} \bsnm{{Gonzalez-Buitrago}}},
\bauthor{\binits{G.} \bsnm{{Halevi}}},
\bauthor{\binits{A.} \bsnm{{Halle}}},
\bauthor{\binits{C.E.} \bsnm{{Hood}}},
\bauthor{\binits{K.} \bsnm{{Horne}}},
\bauthor{\binits{J.C.} \bsnm{{Horst}}},
\bauthor{\binits{M.} \bsnm{{de Kouchkovsky}}},
\bauthor{\binits{B.} \bsnm{{Kuhn}}},
\bauthor{\binits{S.} \bsnm{{Kumar}}},
\bauthor{\binits{D.C.} \bsnm{{Leonard}}},
\bauthor{\binits{D.} \bsnm{{Loveland}}},
\bauthor{\binits{C.} \bsnm{{Manzano-King}}},
\bauthor{\binits{I.} \bsnm{{McHardy}}},
\bauthor{\binits{R.} \bsnm{{Michel}}},
\bauthor{\binits{M.K.B.} \bsnm{{Olaes}}},
\bauthor{\binits{D.} \bsnm{{Park}}},
\bauthor{\binits{S.} \bsnm{{Park}}},
\bauthor{\binits{L.} \bsnm{{Pei}}},
\bauthor{\binits{T.W.} \bsnm{{Ross}}},
\bauthor{\binits{J.N.} \bsnm{{Runco}}},
\bauthor{\binits{J.} \bsnm{{Samuel}}},
\bauthor{\binits{J.} \bsnm{{Sanchez}}},
\bauthor{\binits{B.} \bsnm{{Scott}}},
\bauthor{\binits{R.O.} \bsnm{{Sexton}}},
\bauthor{\binits{J.} \bsnm{{Shin}}},
\bauthor{\binits{I.} \bsnm{{Shivvers}}},
\bauthor{\binits{C.L.} \bsnm{{Spencer}}},
\bauthor{\binits{B.E.} \bsnm{{Stahl}}},
\bauthor{\binits{S.} \bsnm{{Stegman}}},
\bauthor{\binits{I.} \bsnm{{Stomberg}}},
\bauthor{\binits{S.} \bsnm{{Valenti}}},
\bauthor{\binits{J.L.} \bsnm{{Walsh}}},
\bauthor{\binits{H.} \bsnm{{Yuk}}},
\bauthor{\binits{W.} \bsnm{{Zheng}}},
\batitle{The lick agn monitoring project 2016: Dynamical modeling of
  velocity-resolved h$\beta$ lags in luminous seyfert galaxies}.
\bjtitle{The Astrophysical Journal}
\bvolume{930},
\bfpage{52}
(\byear{2022}).
doi:\doiurl{10.3847/1538-4357/ac6171}
\end{barticle}
\endbibitem

\bibitem[\protect\citeauthoryear{Vives-Arias
  et~al.}{2016}]{bib5:VivesArias2016}
\begin{barticle}
\bauthor{\binits{H.} \bsnm{Vives-Arias}},
\bauthor{\binits{J.A.} \bsnm{Mu{\~n}oz}},
\bauthor{\binits{C.S.} \bsnm{Kochanek}},
\bauthor{\binits{E.} \bsnm{Mediavilla}},
\bauthor{\binits{J.} \bsnm{Jim{\'e}nez-Vicente}},
\batitle{Observations of the lensed quasar q2237+0305 with canaricam at gtc}.
\bjtitle{The Astrophysical Journal}
\bvolume{831},
\bfpage{43}
(\byear{2016}).
doi:\doiurl{10.3847/0004-637x/831/1/43}
\end{barticle}
\endbibitem

\bibitem[\protect\citeauthoryear{Walsh et~al.}{1979}]{bib5:Walsh1979}
\begin{barticle}
\bauthor{\binits{D.} \bsnm{Walsh}},
\bauthor{\binits{R.F.} \bsnm{Carswell}},
\bauthor{\binits{R.J.} \bsnm{Weymann}},
\batitle{0957$+$ 561 a, b - twin quasistellar objects or gravitational lens}.
\bjtitle{Nature}
\bvolume{279},
\bfpage{381}
(\byear{1979})
\end{barticle}
\endbibitem

\bibitem[\protect\citeauthoryear{Wambsganss}{1992a}]{bib5:Wambsganss1992}
\begin{barticle}
\bauthor{\binits{J.} \bsnm{Wambsganss}},
\batitle{Probability distributions for the magnification of quasars due to
  microlensing}.
\bjtitle{The Astrophysical Journal}
\bvolume{386},
\bfpage{19}
(\byear{1992}a)
\end{barticle}
\endbibitem

\bibitem[\protect\citeauthoryear{Wambsganss}{1992b}]{bib5:Wambsganss1992a}
\begin{barticle}
\bauthor{\binits{J.} \bsnm{Wambsganss}},
\batitle{Probability distributions for the magnification of quasars due to
  microlensing}.
\bjtitle{The Astrophysical Journal}
\bvolume{386},
\bfpage{19}
(\byear{1992}b)
\end{barticle}
\endbibitem

\bibitem[\protect\citeauthoryear{Wambsganss}{1999}]{bib5:Wambsganss1999}
\begin{barticle}
\bauthor{\binits{J.} \bsnm{Wambsganss}},
\batitle{Gravitational lensing: numerical simulations with a hierarchical tree
  code}.
\bjtitle{J. Comput. Appl. Math.}
\bvolume{109},
\bfpage{353}
(\byear{1999})
\end{barticle}
\endbibitem

\bibitem[\protect\citeauthoryear{Wambsganss
  et~al.}{1990}]{bib5:Wambsganss1990b}
\begin{barticle}
\bauthor{\binits{J.} \bsnm{Wambsganss}},
\bauthor{\binits{B.} \bsnm{Paczynski}},
\bauthor{\binits{P.} \bsnm{Schneider}},
\batitle{Interpretation of the microlensing event in qso2237$+$0305}.
\bjtitle{The Astrophysical Journal}
\bvolume{358},
\bfpage{33}
(\byear{1990})
\end{barticle}
\endbibitem

\bibitem[\protect\citeauthoryear{{Wambsgan{\ss}}}{1990}]{bib5:Wambsganss1990a}
\begin{botherref}
\oauthor{\binits{J.} \bsnm{{Wambsgan{\ss}}}},
{Gravitational microlensing},
PhD thesis,
Ludwig-Maximilians University of Munich, Germany,
1990
\end{botherref}
\endbibitem

\bibitem[\protect\citeauthoryear{White et~al.}{2015}]{bib5:White2015}
\begin{barticle}
\bauthor{\binits{S.V.} \bsnm{White}},
\bauthor{\binits{M.J.} \bsnm{Jarvis}},
\bauthor{\binits{B.} \bsnm{Haussler}},
\bauthor{\binits{N.} \bsnm{Maddox}},
\batitle{Radio-quiet quasars in the video survey: Evidence for agn-powered
  radio emission at s1.4 ghz < 1 mjy}.
\bjtitle{Monthly Notices of the Royal Astronomical Society}
\bvolume{448},
\bfpage{2665}--\blpage{2686}
(\byear{2015}).
doi:\doiurl{10.1093/mnras/stv134}
\end{barticle}
\endbibitem

\bibitem[\protect\citeauthoryear{Wielgus et~al.}{2022}]{bib5:Wielgus2022}
\begin{barticle}
\bauthor{\binits{M.} \bsnm{Wielgus}},
\bauthor{\binits{D.} \bsnm{Lan{\'C}ov{\'a}}},
\bauthor{\binits{O.} \bsnm{Straub}},
\bauthor{\binits{W.} \bsnm{Kluźniak}},
\bauthor{\binits{R.} \bsnm{Narayan}},
\bauthor{\binits{D.} \bsnm{Abarca}},
\bauthor{\binits{A.} \bsnm{R{\'o}zanska}},
\bauthor{\binits{F.} \bsnm{Vincent}},
\bauthor{\binits{G.} \bsnm{T{\"o}r{\"o}k}},
\bauthor{\binits{M.} \bsnm{Abramowicz}},
\batitle{Observational properties of puffy discs: radiative grmhd spectra of
  mildly sub-eddington accretion}.
\bjtitle{Monthly Notices of the Royal Astronomical Society}
\bvolume{514},
\bfpage{780}--\blpage{789}
(\byear{2022}).
doi:\doiurl{10.1093/mnras/stac1317}
\end{barticle}
\endbibitem

\bibitem[\protect\citeauthoryear{{Wilkins} and
  {Gallo}}{2015}]{bib5:Wilkins2015}
\begin{barticle}
\bauthor{\binits{D.R.} \bsnm{{Wilkins}}},
\bauthor{\binits{L.C.} \bsnm{{Gallo}}},
\batitle{Driving extreme variability: the evolving corona and evidence for jet
  launching in markarian 335}.
\bjtitle{\mnras}
\bvolume{449}(\bissue{1}),
\bfpage{129}--\blpage{146}
(\byear{2015}).
doi:\doiurl{10.1093/mnras/stv162}
\end{barticle}
\endbibitem

\bibitem[\protect\citeauthoryear{Witt}{1991}]{bib5:Witt1991}
\begin{bbook}
\bauthor{\binits{H.J.} \bsnm{Witt}},
\bbtitle{Der Mikrogravitationslinseneffekt - Theorie und Anwendungen.}
\byear{1991}
\end{bbook}
\endbibitem

\bibitem[\protect\citeauthoryear{Witt and Mao}{1994}]{bib5:Witt1994}
\begin{barticle}
\bauthor{\binits{H.J.} \bsnm{Witt}},
\bauthor{\binits{S.} \bsnm{Mao}},
\batitle{Interpretation of microlensing events in q2237$+$0305}.
\bjtitle{The Astrophysical Journal}
\bvolume{429},
\bfpage{66}
(\byear{1994})
\end{barticle}
\endbibitem

\bibitem[\protect\citeauthoryear{Witt}{1990}]{bib5:Witt1990}
\begin{barticle}
\bauthor{\binits{H.J.} \bsnm{Witt}},
\batitle{Investigation of high amplification events in light curves of
  gravitationally lensed quasars}.
\bjtitle{Astronomy \& Astrophysics}
\bvolume{236},
\bfpage{311}
(\byear{1990})
\end{barticle}
\endbibitem

\bibitem[\protect\citeauthoryear{Witt}{1993}]{bib5:Witt1993}
\begin{barticle}
\bauthor{\binits{H.J.} \bsnm{Witt}},
\batitle{An efficient method to compute microlensed light curves for point
  sources}.
\bjtitle{The Astrophysical Journal}
\bvolume{403},
\bfpage{530}
(\byear{1993})
\end{barticle}
\endbibitem

\bibitem[\protect\citeauthoryear{Witt et~al.}{1995}]{bib5:Witt1995}
\begin{barticle}
\bauthor{\binits{H.J.} \bsnm{Witt}},
\bauthor{\binits{S.} \bsnm{Mao}},
\bauthor{\binits{P.L.} \bsnm{Schechter}},
\batitle{On the universality of microlensing in quadruple gravitational
  lenses}.
\bjtitle{The Astrophysical Journal}
\bvolume{443},
\bfpage{18}
(\byear{1995})
\end{barticle}
\endbibitem

\bibitem[\protect\citeauthoryear{{Wo{\'z}niak} et~al.}{2000}]{bib5:Wozniak2000}
\begin{barticle}
\bauthor{\binits{P.R.} \bsnm{{Wo{\'z}niak}}},
\bauthor{\binits{A.} \bsnm{{Udalski}}},
\bauthor{\binits{M.} \bsnm{{Szyma{\'n}ski}}},
\bauthor{\binits{M.} \bsnm{{Kubiak}}},
\bauthor{\binits{G.} \bsnm{{Pietrzy{\'n}ski}}},
\bauthor{\binits{I.} \bsnm{{Soszy{\'n}ski}}},
\bauthor{\binits{K.} \bsnm{{{\.Z}ebru{\'n}}}},
\batitle{The optical gravitational lensing experiment: A hunt for caustic
  crossings in qso 22370305 1}.
\bjtitle{The Astrophysical Journal}
\bvolume{540},
\bfpage{65}--\blpage{67}
(\byear{2000}).
\burl{http://www.astrouw.edu.pl/}
\end{barticle}
\endbibitem

\bibitem[\protect\citeauthoryear{Wyithe and Loeb}{2002}]{bib5:Wyithe2002a}
\begin{botherref}
\oauthor{\binits{J.S.B.} \bsnm{Wyithe}},
\oauthor{\binits{A.} \bsnm{Loeb}},
Measuring the size of quasar broad-line clouds through time-delay light-curve
  anomalies of gravitational lenses.
The Astrophysical Journal,
615--625
(2002)
\end{botherref}
\endbibitem

\bibitem[\protect\citeauthoryear{Wyithe and Turner}{2001}]{bib5:Wyithe2001}
\begin{barticle}
\bauthor{\binits{J.S.B.} \bsnm{Wyithe}},
\bauthor{\binits{E.L.} \bsnm{Turner}},
\batitle{Determining the microlens mass function from quasar microlensing
  statistics}.
\bjtitle{Monthly Notices of the Royal Astronomical Society}
\bvolume{320},
\bfpage{21}--\blpage{30}
(\byear{2001}).
doi:\doiurl{10.1046/j.1365-8711.2001.03917.x}.
\burl{http://doi.wiley.com/10.1046/j.1365-8711.2001.03917.x}
\end{barticle}
\endbibitem

\bibitem[\protect\citeauthoryear{Wyithe et~al.}{2000a}]{bib5:Wyithe2000a}
\begin{barticle}
\bauthor{\binits{J.S.B.} \bsnm{Wyithe}},
\bauthor{\binits{R.L.} \bsnm{Webster}},
\bauthor{\binits{E.L.} \bsnm{Turner}},
\batitle{The distribution of microlensed light-curve derivatives : the
  relationship between stellar proper motions and transverse velocity}.
\bjtitle{Monthly Notices of the Royal Astronomical Society}
\bvolume{312},
\bfpage{843}--\blpage{852}
(\byear{2000}a)
\end{barticle}
\endbibitem

\bibitem[\protect\citeauthoryear{Wyithe et~al.}{2000b}]{bib5:Wyithe2000b}
\begin{barticle}
\bauthor{\binits{J.S.B.} \bsnm{Wyithe}},
\bauthor{\binits{R.L.} \bsnm{Webster}},
\bauthor{\binits{E.L.} \bsnm{Turner}},
\batitle{Limits on the microlens mass function of q2237$+$0305}.
\bjtitle{Monthly Notices of the Royal Astronomical Society}
\bvolume{315},
\bfpage{51}--\blpage{61}
(\byear{2000}b)
\end{barticle}
\endbibitem

\bibitem[\protect\citeauthoryear{Wyithe et~al.}{2002}]{bib5:Wyithe2002b}
\begin{barticle}
\bauthor{\binits{J.S.B.} \bsnm{Wyithe}},
\bauthor{\binits{E.} \bsnm{Agol}},
\bauthor{\binits{E.L.} \bsnm{Turner}},
\bauthor{\binits{R.W.} \bsnm{Schmidt}},
\batitle{Constraints on the mass profile of the lens galaxy g2237$+$0305}.
\bjtitle{Monthly Notices of the Royal Astronomical Society}
\bvolume{330},
\bfpage{575}
(\byear{2002})
\end{barticle}
\endbibitem

\bibitem[\protect\citeauthoryear{Yang et~al.}{2020}]{bib5:Yang2020}
\begin{barticle}
\bauthor{\binits{Q.} \bsnm{Yang}},
\bauthor{\binits{Y.} \bsnm{Shen}},
\bauthor{\binits{X.} \bsnm{Liu}},
\bauthor{\binits{M.} \bsnm{Aguena}},
\bauthor{\binits{J.} \bsnm{Annis}},
\bauthor{\binits{S.} \bsnm{Avila}},
\bauthor{\binits{M.} \bsnm{Banerji}},
\bauthor{\binits{E.} \bsnm{Bertin}},
\bauthor{\binits{D.} \bsnm{Brooks}},
\bauthor{\binits{D.} \bsnm{Burke}},
\bauthor{\binits{A.C.} \bsnm{Rosell}},
\bauthor{\binits{M.C.} \bsnm{Kind}},
\bauthor{\binits{L.} \bparticle{da} \bsnm{Costa}},
\bauthor{\binits{J.D.} \bsnm{Vicente}},
\bauthor{\binits{S.} \bsnm{Desai}},
\bauthor{\binits{H.T.} \bsnm{Diehl}},
\bauthor{\binits{P.} \bsnm{Doel}},
\bauthor{\binits{B.} \bsnm{Flaugher}},
\bauthor{\binits{P.} \bsnm{Fosalba}},
\bauthor{\binits{J.} \bsnm{Frieman}},
\bauthor{\binits{J.} \bsnm{Garcia-Bellido}},
\bauthor{\binits{D.} \bsnm{Gerdes}},
\bauthor{\binits{D.} \bsnm{Gruen}},
\bauthor{\binits{R.} \bsnm{Gruendl}},
\bauthor{\binits{J.} \bsnm{Gschwend}},
\bauthor{\binits{G.} \bsnm{Gutierrez}},
\bauthor{\binits{S.} \bsnm{Hinton}},
\bauthor{\binits{D.L.} \bsnm{Hollowood}},
\bauthor{\binits{K.} \bsnm{Honscheid}},
\bauthor{\binits{N.} \bsnm{Kuropatkin}},
\bauthor{\binits{M.} \bsnm{Maia}},
\bauthor{\binits{M.} \bsnm{March}},
\bauthor{\binits{J.} \bsnm{Marshall}},
\bauthor{\binits{P.} \bsnm{Martini}},
\bauthor{\binits{P.} \bsnm{Melchior}},
\bauthor{\binits{F.} \bsnm{Menanteau}},
\bauthor{\binits{R.} \bsnm{Miquel}},
\bauthor{\binits{F.} \bsnm{Paz-Chinchon}},
\bauthor{\binits{A.P.} \bsnm{Malag{\'o}n}},
\bauthor{\binits{K.} \bsnm{Romer}},
\bauthor{\binits{E.} \bsnm{Sanchez}},
\bauthor{\binits{V.} \bsnm{Scarpine}},
\bauthor{\binits{M.} \bsnm{Schubnell}},
\bauthor{\binits{S.} \bsnm{Serrano}},
\bauthor{\binits{I.} \bsnm{Sevilla}},
\bauthor{\binits{M.} \bsnm{Smith}},
\bauthor{\binits{E.} \bsnm{Suchyta}},
\bauthor{\binits{G.} \bsnm{Tarle}},
\bauthor{\binits{T.N.} \bsnm{Varga}},
\bauthor{\binits{R.} \bsnm{Wilkinson}},
\batitle{Dust reverberation mapping in distant quasars from optical and
  mid-infrared imaging surveys}.
\bjtitle{The Astrophysical Journal}
\bvolume{900},
\bfpage{58}
(\byear{2020}).
doi:\doiurl{10.3847/1538-4357/aba59b}
\end{barticle}
\endbibitem

\bibitem[\protect\citeauthoryear{{Yonehara} et~al.}{2008}]{bib5:Yonehara2008}
\begin{barticle}
\bauthor{\binits{A.} \bsnm{{Yonehara}}},
\bauthor{\binits{H.} \bsnm{{Hirashita}}},
\bauthor{\binits{P.} \bsnm{{Richter}}},
\batitle{Origin of chromatic features in multiple quasars. variability, dust,
  or microlensing}.
\bjtitle{\aap}
\bvolume{478}(\bissue{1}),
\bfpage{95}--\blpage{109}
(\byear{2008}).
doi:\doiurl{10.1051/0004-6361:20067014}
\end{barticle}
\endbibitem

\bibitem[\protect\citeauthoryear{Young}{1981}]{bib5:Young1981}
\begin{barticle}
\bauthor{\binits{P.} \bsnm{Young}},
\batitle{Q0957$+$561 - effects of random stars on the gravitational lens}.
\bjtitle{The Astrophysical Journal}
\bvolume{244},
\bfpage{756}--\blpage{767}
(\byear{1981})
\end{barticle}
\endbibitem

\bibitem[\protect\citeauthoryear{Yu et~al.}{2022}]{bib5:Yu2022}
\begin{botherref}
\oauthor{\binits{Z.} \bsnm{Yu}},
\oauthor{\binits{P.} \bsnm{Martini}},
\oauthor{\binits{A.} \bsnm{Penton}},
\oauthor{\binits{T.M.} \bsnm{Davis}},
\oauthor{\binits{C.S.} \bsnm{Kochanek}},
\oauthor{\binits{G.F.} \bsnm{Lewis}},
\oauthor{\binits{C.} \bsnm{Lidman}},
\oauthor{\binits{U.} \bsnm{Malik}},
\oauthor{\binits{R.} \bsnm{Sharp}},
\oauthor{\binits{B.E.} \bsnm{Tucker}},
\oauthor{\binits{M.} \bsnm{Aguena}},
\oauthor{\binits{J.} \bsnm{Annis}},
\oauthor{\binits{E.} \bsnm{Bertin}},
\oauthor{\binits{S.} \bsnm{Bocquet}},
\oauthor{\binits{D.} \bsnm{Brooks}},
\oauthor{\binits{A.C.} \bsnm{Rosell}},
\oauthor{\binits{D.} \bsnm{Carollo}},
\oauthor{\binits{M.C.} \bsnm{Kind}},
\oauthor{\binits{J.} \bsnm{Carretero}},
\oauthor{\binits{M.} \bsnm{Costanzi}},
\oauthor{\binits{L.N.} \bparticle{da} \bsnm{Costa}},
\oauthor{\binits{M.E.S.} \bsnm{Pereira}},
\oauthor{\binits{J.D.} \bsnm{Vicente}},
\oauthor{\binits{H.T.} \bsnm{Diehl}},
\oauthor{\binits{P.} \bsnm{Doel}},
\oauthor{\binits{S.} \bsnm{Everett}},
\oauthor{\binits{I.} \bsnm{Ferrero}},
\oauthor{\binits{J.} \bsnm{Garc{\'i}a-Bellido}},
\oauthor{\binits{M.} \bsnm{Gatti}},
\oauthor{\binits{D.W.} \bsnm{Gerdes}},
\oauthor{\binits{D.} \bsnm{Gruen}},
\oauthor{\binits{R.A.} \bsnm{Gruendl}},
\oauthor{\binits{J.} \bsnm{Gschwend}},
\oauthor{\binits{G.} \bsnm{Gutierrez}},
\oauthor{\binits{S.R.} \bsnm{Hinton}},
\oauthor{\binits{D.L.} \bsnm{Hollowood}},
\oauthor{\binits{K.} \bsnm{Honscheid}},
\oauthor{\binits{D.J.} \bsnm{James}},
\oauthor{\binits{K.} \bsnm{Kuehn}},
\oauthor{\binits{J.} \bsnm{Mena-Fern{\'a}ndez}},
\oauthor{\binits{F.} \bsnm{Menanteau}},
\oauthor{\binits{R.} \bsnm{Miquel}},
\oauthor{\binits{B.} \bsnm{Nichol}},
\oauthor{\binits{F.} \bsnm{Paz-Chinch{\'o}n}},
\oauthor{\binits{A.} \bsnm{Pieres}},
\oauthor{\binits{A.A.P.} \bsnm{Malag{\'o}n}},
\oauthor{\binits{M.} \bsnm{Raveri}},
\oauthor{\binits{A.K.} \bsnm{Romer}},
\oauthor{\binits{E.} \bsnm{Sanchez}},
\oauthor{\binits{V.} \bsnm{Scarpine}},
\oauthor{\binits{I.} \bsnm{Sevilla-Noarbe}},
\oauthor{\binits{M.} \bsnm{Smith}},
\oauthor{\binits{E.} \bsnm{Suchyta}},
\oauthor{\binits{M.E.C.} \bsnm{Swanson}},
\oauthor{\binits{G.} \bsnm{Tarle}},
\oauthor{\binits{M.} \bsnm{Vincenzi}},
\oauthor{\binits{A.R.} \bsnm{Walker}},
\oauthor{\binits{N.} \bsnm{Weaverdyck}},
Ozdes reverberation mapping program: Mg ii lags and r-l relation
(2022).
\url{http://arxiv.org/abs/2208.05491}
\end{botherref}
\endbibitem

\bibitem[\protect\citeauthoryear{Zackrisson et~al.}{2003}]{bib5:Zackrisson2003}
\begin{barticle}
\bauthor{\binits{E.} \bsnm{Zackrisson}},
\bauthor{\binits{N.} \bsnm{Bergvall}},
\bauthor{\binits{T.} \bsnm{Marquart}},
\bauthor{\binits{P.} \bsnm{Helbig}},
\batitle{Can microlensing explain the long-term optical variability of
  quasars?}
\bjtitle{Astronomy and Astrophysics}
\bvolume{408},
\bfpage{17}--\blpage{25}
(\byear{2003}).
doi:\doiurl{10.1051/0004-6361:20030895}
\end{barticle}
\endbibitem

\bibitem[\protect\citeauthoryear{Zamaninasab
  et~al.}{2014}]{bib5:Zamaninasab2014}
\begin{barticle}
\bauthor{\binits{M.} \bsnm{Zamaninasab}},
\bauthor{\binits{E.} \bsnm{Clausen-Brown}},
\bauthor{\binits{T.} \bsnm{Savolainen}},
\bauthor{\binits{A.} \bsnm{Tchekhovskoy}},
\batitle{Dynamically important magnetic fields near accreting supermassive
  black holes}.
\bjtitle{Nature}
\bvolume{510},
\bfpage{126}--\blpage{128}
(\byear{2014}).
doi:\doiurl{10.1038/nature13399}
\end{barticle}
\endbibitem

\bibitem[\protect\citeauthoryear{Zdziarski et~al.}{2022}]{bib5:Zdziarski2022}
\begin{barticle}
\bauthor{\binits{A.A.} \bsnm{Zdziarski}},
\bauthor{\binits{B.} \bsnm{You}},
\bauthor{\binits{M.} \bsnm{Szanecki}},
\batitle{Corrections to estimated accretion disk size due to color correction,
  disk truncation, and disk wind}.
\bjtitle{The Astrophysical Journal Letters}
\bvolume{939},
\bfpage{2}
(\byear{2022}).
doi:\doiurl{10.3847/2041-8213/ac9474}
\end{barticle}
\endbibitem

\bibitem[\protect\citeauthoryear{Zheng et~al.}{2022}]{bib5:Zheng2022}
\begin{barticle}
\bauthor{\binits{W.} \bsnm{Zheng}},
\bauthor{\binits{X.} \bsnm{Chen}},
\bauthor{\binits{G.} \bsnm{Li}},
\bauthor{\binits{H.-Z.} \bsnm{Chen}},
\batitle{An improved gpu-based ray-shooting code for gravitational
  microlensing}.
\bjtitle{The Astrophysical Journal}
\bvolume{931},
\bfpage{114}
(\byear{2022}).
doi:\doiurl{10.3847/1538-4357/ac68ea}
\end{barticle}
\endbibitem

\end{thebibliography}

\end{document}